\documentclass[twocolumn,showpacs,aps,prd,superscriptaddress,floatfix]{revtex4}
\usepackage{graphicx}
\usepackage{dcolumn}
\usepackage{multirow}
\usepackage{amsmath}
\usepackage{epsfig}
\usepackage{color}
\usepackage{verbatim} 
\usepackage{rotating} 

\RequirePackage{xspace}

\usepackage{relsize}
\def\babar{\mbox{\slshape B\kern-0.1em{\smaller A}\kern-0.1em
    B\kern-0.1em{\smaller A\kern-0.2em R}}}

\def\en         {\ensuremath{e^-}\xspace}   
\def\ep         {\ensuremath{e^+}\xspace}
 
\def\epem       {\ensuremath{e^+e^-}\xspace}

\def\mup        {\ensuremath{\mu^+}\xspace}
\def\mumu       {\ensuremath{\mu^+\mu^-}\xspace}

\def\tautau     {\ensuremath{\tau^+\tau^-}\xspace}

\def\ellell     {\ensuremath{\ell^+ \ell^-}\xspace}

\def\q     {\ensuremath{q}\xspace}

\def\qqbar {\ensuremath{q\overline q}\xspace}

\def\ccbar {\ensuremath{c\overline c}\xspace}

\def\piz   {\ensuremath{\pi^0}\xspace}

\def\pip   {\ensuremath{\pi^+}\xspace}
\def\pim   {\ensuremath{\pi^-}\xspace}

\def\Kbar  {\kern 0.2em\overline{\kern -0.2em K}{}\xspace}

\def\Kz    {\ensuremath{K^0}\xspace}
\def\Kzb   {\ensuremath{\Kbar^0}\xspace}
\def\KzKzb {\ensuremath{\Kz \kern -0.16em \Kzb}\xspace}
\def\Kp    {\ensuremath{K^+}\xspace}
\def\Km    {\ensuremath{K^-}\xspace}

\def\KpKm  {\ensuremath{\Kp \kern -0.16em \Km}\xspace}

\def\Dbar    {\kern 0.2em\overline{\kern -0.2em D}{}\xspace}

\def\Dz      {\ensuremath{D^0}\xspace}
\def\Dzb     {\ensuremath{\Dbar^0}\xspace}
\def\DzDzb   {\ensuremath{\Dz {\kern -0.16em \Dzb}}\xspace}
\def\Dp      {\ensuremath{D^+}\xspace}
\def\Dm      {\ensuremath{D^-}\xspace}

\def\DpDm    {\ensuremath{\Dp {\kern -0.16em \Dm}}\xspace}

\def\B       {\ensuremath{B}\xspace}
\def\Bbar    {\kern 0.18em\overline{\kern -0.18em B}{}\xspace}

\def\BB      {\ensuremath{B\Bbar}\xspace} 
\def\Bz      {\ensuremath{B^0}\xspace}
\def\Bzb     {\ensuremath{\Bbar^0}\xspace}
\def\BzBzb   {\ensuremath{\Bz {\kern -0.16em \Bzb}}\xspace}
\def\Bu      {\ensuremath{B^+}\xspace}
\def\Bub     {\ensuremath{B^-}\xspace}
\def\Bp      {\ensuremath{\Bu}\xspace}

\def\BpBm    {\ensuremath{\Bu {\kern -0.16em \Bub}}\xspace}

\def\BorBbar    {\kern 0.18em\optbar{\kern -0.18em B}{}\xspace}
\def\DorDbar    {\kern 0.18em\optbar{\kern -0.18em D}{}\xspace}
\def\KorKbar    {\kern 0.18em\optbar{\kern -0.18em K}{}\xspace}

\def\jpsi     {\ensuremath{{J\mskip -3mu/\mskip -2mu\psi\mskip 2mu}}\xspace}

\mathchardef\Upsilon="7107
\def\Y#1S{\ensuremath{\Upsilon{(#1S)}}\xspace}

\def\FourS {\Y4S}

\mathchardef\Deltares="7101
\mathchardef\Xi="7104
\mathchardef\Lambda="7103
\mathchardef\Sigma="7106
\mathchardef\Omega="710A

\def\Deltabar{\kern 0.25em\overline{\kern -0.25em \Deltares}{}\xspace}
\def\Lbar{\kern 0.2em\overline{\kern -0.2em\Lambda\kern 0.05em}\kern-0.05em{}\xspace}
\def\Sigbar{\kern 0.2em\overline{\kern -0.2em \Sigma}{}\xspace}
\def\Xibar{\kern 0.2em\overline{\kern -0.2em \Xi}{}\xspace}
\def\Obar{\kern 0.2em\overline{\kern -0.2em \Omega}{}\xspace}
\def\Nbar{\kern 0.2em\overline{\kern -0.2em N}{}\xspace}
\def\Xb{\kern 0.2em\overline{\kern -0.2em X}{}\xspace}

\def\Btorholnu  {\ensuremath{B \to \rho \ell \nu}\xspace}
\def\Btopilnu   {\ensuremath{B \to \pi \ell \nu}\xspace}

\def\mes        {\mbox{$m_{\rm ES}$}\xspace}

\def\DeltaE     {\mbox{$\Delta E$}\xspace}

\newcommand{\tev}{\ensuremath{\mathrm{\,Te\kern -0.1em V}}\xspace}
\newcommand{\gev}{\ensuremath{\mathrm{\,Ge\kern -0.1em V}}\xspace}
\newcommand{\mev}{\ensuremath{\mathrm{\,Me\kern -0.1em V}}\xspace}
\newcommand{\kev}{\ensuremath{\mathrm{\,ke\kern -0.1em V}}\xspace}
\newcommand{\ev}{\ensuremath{\mathrm{\,e\kern -0.1em V}}\xspace}
\newcommand{\gevc}{\ensuremath{{\mathrm{\,Ge\kern -0.1em V\!/}c}}\xspace}
\newcommand{\mevc}{\ensuremath{{\mathrm{\,Me\kern -0.1em V\!/}c}}\xspace}
\newcommand{\gevcc}{\ensuremath{{\mathrm{\,Ge\kern -0.1em V\!/}c^2}}\xspace}
\newcommand{\mevcc}{\ensuremath{{\mathrm{\,Me\kern -0.1em V\!/}c^2}}\xspace}

\def\nb         {\ensuremath{{\rm \,nb}}\xspace}

\def\mus  {\ensuremath{\rm \,\mus}\xspace}

\def\ps   {\ensuremath{\rm \,ps}\xspace}

\def\mus        {\ensuremath{\,\mu{\rm s}}\xspace}    
\def\ps         {\ensuremath{{\rm \,ps}}\xspace}  

\def\rad{\ensuremath{\rm \,rad}\xspace}

\def\ra                 {\ensuremath{\rightarrow}\xspace}
\def\to                 {\ensuremath{\rightarrow}\xspace}

\def\pep2{PEP-II}

\def\gsim{{~\raise.15em\hbox{$>$}\kern-.85em
          \lower.35em\hbox{$\sim$}~}\xspace}
\def\lsim{{~\raise.15em\hbox{$<$}\kern-.85em
          \lower.35em\hbox{$\sim$}~}\xspace}

\def\CP                {\ensuremath{C\!P}\xspace}

\def\Vub  {\ensuremath{|V_{ub}|}\xspace}
\def\Vcb  {\ensuremath{|V_{cb}|}\xspace}

\xspace

\def\jetset74   {\mbox{\tt Jetset \hspace{-0.5em}7.\hspace{-0.2em}4}\xspace}

\newcommand{\BABARPubYear}    {09}
\newcommand{\BABARPubNumber} {037}
\newcommand{\SLACPubNumber}{14106}

\newcommand{\etal}{\textit{et al.\@}\xspace}

\def\rhoz{\ensuremath{\rho^{0}}\xspace}

\def\qq{\ensuremath{q\bar{q}\xspace}}

\def\Bzdslnu{\ensuremath{B^{0} \rightarrow D^{*-} \ell^+\nu}\xspace}
\def\Bpilnu{\ensuremath{B \rightarrow \pi \ell\nu}\xspace}
\def\Brholnu{\ensuremath{B \rightarrow \rho \ell\nu}\xspace}

\def\Bzpilnu{\ensuremath{B^{0} \rightarrow \pi^-\ell^+\nu}\xspace}
\def\Bppizlnu{\ensuremath{B^{+} \rightarrow \pi^0\ell^+\nu}\xspace}
\def\Bzrholnu{\ensuremath{B^{0} \rightarrow \rho^-\ell^+\nu}\xspace}
\def\Bprhozlnu{\ensuremath{B^{+} \rightarrow \rho^0\ell^+\nu}\xspace}
\def\Bpomegalnu{\ensuremath{B^{+} \rightarrow \omega \ell^+\nu}\xspace}
\def\Bpetalnu{\ensuremath{B^{+} \rightarrow \eta \ell^+\nu}\xspace}
\def\Bpetaplnu{\ensuremath{B^{+} \rightarrow \eta^\prime \ell^+\nu}\xspace}

\def\BXulnu{\ensuremath{B \rightarrow X_u\ell\nu}\xspace}
\def\BXclnu{\ensuremath{B \rightarrow X_c\ell\nu}\xspace}

\def\bulnu{\ensuremath{B \rightarrow X_u\ell\nu}\xspace}
\def\bclnu{\ensuremath{B \rightarrow X_c\ell\nu}\xspace}
\def\control{\ensuremath{B^{0} \ra D^{*-} \ell^{+} \nu, \bar{D^{0}} \ra K^{+}\pi^{-}}\xspace}

\def\BRXulnu{\ensuremath{{\cal B}(B \rightarrow X_u\ell\nu})\xspace}

\def\BRBzpilnu{\ensuremath{{\cal B}(B^{0} \rightarrow \pi^-\ell^+\nu})\xspace}
\def\BRBppizlnu{\ensuremath{{\cal B}(B^{+} \rightarrow \pi^0\ell^+\nu})\xspace}
\def\BRBzrholnu{\ensuremath{{\cal B}(B^{0} \rightarrow \rho^-\ell^+\nu})\xspace}
\def\BRBprhozlnu{\ensuremath{{\cal B}(B^{+} \rightarrow \rho^0\ell^+\nu})\xspace}
\def\BRBpomlnu{\ensuremath{{\cal B}(B^{+} \rightarrow \omega\ell^+\nu})\xspace}
\def\BRBpetalnu{\ensuremath{{\cal B}(B^{+} \rightarrow \eta\ell^+\nu})\xspace}
\def\BRBpetaplnu{\ensuremath{{\cal B}(B^{+} \rightarrow \eta^\prime\ell^+\nu})\xspace}
\def\BRDlnu{\ensuremath{{\cal B}(B \rightarrow D \ell\nu})\xspace}
\def\BRDslnu{\ensuremath{{\cal B}(B \rightarrow D^*\ell\nu})\xspace}
\def\BRDssnlnu{\ensuremath{{\cal B}(B\rightarrow D^{**}\ell\nu})_{\rm narrow}\xspace}
\def\BRDssblnu{\ensuremath{{\cal B}(B \rightarrow D^{**}\ell\nu})_{\rm broad} \xspace}

\def\mmiss{\ensuremath{m_{\rm miss}^2}\xspace}
\def\Emiss{\ensuremath{E_{\rm miss}}\xspace}
\def\pmiss{\ensuremath{p_{\rm miss}}\xspace}
\def\thetamiss{\ensuremath{\theta_{\rm miss}}\xspace}
\def\mES{\ensuremath{m_{\rm ES}}\xspace}
\def\DeltaE{\ensuremath{\Delta E}\xspace}
\def\cosBY{\ensuremath{\cos\theta_{\rm BY}}\xspace}
\def\cosDThrust{\ensuremath{\cos\Delta\theta_{\rm thrust}}\xspace}
\def\costhetaWL{\ensuremath{\cos\theta_{W\ell}}\xspace}

\def\ThetamissCut{\ensuremath{0.3 < \theta_{\rm miss} < 2.2} \rm rad\xspace}

\def\mmissEmissCut{\ensuremath{|m_{\rm miss}^2 / 2 E_{\rm miss} | < 2.5~\gev }\xspace}
\def\cosBYCut{\ensuremath{-1.2 < \cosBY < 1.1}\xspace}
\def\costhetaWLCut{\ensuremath{|\costhetaWL| < 0.8}\xspace}

\def\SignalRegion{\ensuremath{-0.15 <|\Delta E| < 0.25\gev \, ; \, 5.255 < \mES < 5.295 \gev}\xspace}

\usepackage{subfigure}

\def\twoFigOneCol#1#2{%
  \includegraphics[width=.5\columnwidth]{#1}%
  \includegraphics[width=.5\columnwidth]{#2}%
}
\def\threeFig#1#2#3{%
  \includegraphics[width=.25\textwidth]{#1}%
  \includegraphics[width=.25\textwidth]{#2}%
  \includegraphics[width=.25\textwidth]{#3}%
}
\def\fourFig#1#2#3#4{%
  \includegraphics[width=.25\textwidth]{#1}%
  \includegraphics[width=.25\textwidth]{#2}%
  \includegraphics[width=.25\textwidth]{#3}%
  \includegraphics[width=.25\textwidth]{#4}%
}
\def\twoFigTwoCol#1#2{%
  \includegraphics[width=0.5\textwidth]{#1}%
  \includegraphics[width=0.5\textwidth]{#2}%
}
\def\threeFigTwoCol#1#2#3{%
  \includegraphics[width=.25\textwidth]{#1}%
  \includegraphics[width=.25\textwidth]{#2}%
  \includegraphics[width=.25\textwidth]{#3}%
}

\begin{document}

  \preprint{\babar-PUB-\BABARPubYear/\BABARPubNumber} 
  \preprint{SLAC-PUB-\SLACPubNumber} 
  
  \begin{flushleft}
    \babar-PUB-\BABARPubYear/\BABARPubNumber\\
    SLAC-PUB-\SLACPubNumber\\
  \end{flushleft}
  
\title{\large\bf Study of \boldmath \Btopilnu and \Btorholnu Decays and Determination of \Vub}
%
\author{P.~del~Amo~Sanchez}
\author{J.~P.~Lees}
\author{V.~Poireau}
\author{E.~Prencipe}
\author{V.~Tisserand}
\affiliation{Laboratoire d'Annecy-le-Vieux de Physique des Particules (LAPP), Universit\'e de Savoie, CNRS/IN2P3,  F-74941 Annecy-Le-Vieux, France}
\author{J.~Garra~Tico}
\author{E.~Grauges}
\affiliation{Universitat de Barcelona, Facultat de Fisica, Departament ECM, E-08028 Barcelona, Spain }
\author{M.~Martinelli$^{ab}$}
\author{A.~Palano$^{ab}$ }
\author{M.~Pappagallo$^{ab}$ }
\affiliation{INFN Sezione di Bari$^{a}$; Dipartimento di Fisica, Universit\`a di Bari$^{b}$, I-70126 Bari, Italy }
% ...
\collaboration{The \babar\ Collaboration}
\noaffiliation
\date{\today}

\begin{abstract}
\noindent 
We present an analysis of exclusive charmless semileptonic $B$-meson decays
based on 377 million $B\overline B$ pairs recorded with the 
\babar\ detector at the $\Upsilon(4S)$ resonance.
We select four event samples corresponding to the decay modes 
\Bzpilnu, \Bppizlnu, \Bzrholnu, and \Bprhozlnu, 
and find the measured branching fractions to be consistent with isospin symmetry.
Assuming isospin symmetry, we combine the two \Bpilnu\ samples, and similarly the two 
\Brholnu\ samples, and measure the branching fractions 
${\cal B}(B^0 \to \pi^-\ell^+\nu) = (1.41 \pm 0.05 \pm 0.07)  \times 10^{-4}$ 
and 
${\cal B}(B^0 \to \rho^-\ell^+\nu) = (1.75 \pm 0.15 \pm 0.27) \times 10^{-4}$, 
where the errors are statistical and systematic.
We compare the measured distribution in $q^2$, the momentum transfer
squared, with predictions for the form factors from QCD calculations
and determine the CKM matrix element \Vub.
Based on the measured partial branching fraction for $B \to \pi\ell\nu$ in the range $q^2 <12 \gev^2$
and the most recent LCSR calculations we obtain $\Vub  = (3.78 \pm 0.13 ^{+0.55}_{-0.40}) \times 10^{-3}$, 
where the errors refer to the experimental and theoretical
uncertainties.  From a simultaneous fit to the data over the full $q^2$ range and
the FNAL/MILC lattice QCD results, we obtain
$\Vub  = (2.95 \pm 0.31) \times 10^{-3}$ from $B \rightarrow \pi\ell\nu$,
where the error is the combined experimental and theoretical uncertainty.

\end{abstract}


\maketitle  


\section{Introduction}

The elements of the Cabibbo-Kobayashi-Maskawa (CKM) quark-mixing matrix 
are fundamental parameters of the Standard Model (SM) of electroweak interactions. 
With the increasingly precise measurements of decay-time-dependent 
\CP\ asymmetries in $B$-meson decays, in particular sin(2$\beta$)~\cite{beta,phi1},
improved measurements of the magnitude of $V_{ub}$ and $V_{cb}$ will allow 
for more stringent experimental tests of the SM mechanism for \CP\ violation~\cite{sm}.  This is best illustrated 
in terms of the unitarity triangle, the graphical representation of one of 
the unitarity conditions for the CKM matrix, for which the length of
the side that is opposite to the angle $\beta$ is proportional to the ratio
$\Vub/\Vcb$.  The best method to determine $\Vub$ and $\Vcb$
is to measure semileptonic decay rates for \bclnu\ and \bulnu\ ($X_c$ and $X_u$ 
refer to hadronic states with or without charm), which are  proportional to $\Vcb^2$ and $\Vub^2$, respectively.

There are two methods to extract these two CKM elements from $B$ decays, one based on inclusive and the other on exclusive semileptonic decays. Exclusive decays offer better kinematic constraints and thus more effective background suppression than inclusive decays, but the lower branching fractions result in lower event yields. Since the experimental and theoretical techniques for these two approaches are different and largely independent, they can provide important cross checks of our understanding
of the theory and the measurements.   
An overview of the determination of \Vub\ using both approaches can be found in a recent review~\cite{kowalewski_mannel}. 

In this paper, we present a study of four exclusive charmless semileptonic decay modes,
\Bzpilnu, \Bppizlnu, \Bzrholnu, and \Bprhozlnu~\cite{charge_conjugate}, and a determination  
of $\Vub$. Here $\ell$ refers to a charged lepton, either 
\ep\ or \mup , and $\nu$ refers to a neutrino, either $\nu_e$ or $\nu_{\mu}$.
This analysis represents an update of an earlier measurement~\cite{pilnu_jochen} that was based on a significantly smaller data set.  
For the current analysis,
the signal yields and background suppression have been improved and the systematic uncertainties have been reduced through the 
use of improved reconstruction and signal extraction methods, combined with more detailed background studies. 

The principal experimental challenge is the separation of the \bulnu\ from the dominant \bclnu\ decays, for which the inclusive branching fraction is a factor of 50 larger. Furthermore,   the isolation of individual exclusive charmless decays from all other \bulnu\ decays is difficult, because the exclusive branching ratios are typically only 10\% of $\BRXulnu = (2.29 \pm 0.34) \times 10^{-3}$~\cite{PDG2008}, the inclusive branching fraction for charmless semileptonic $B$ decays.

The reconstruction of signal decays in $\epem \to \FourS \to \BB$ events requires the identification of three types of particles,  
the hadronic state $X_u$ producing one or two charged and/or neutral final state pions, 
the charged lepton, and the neutrino.
The presence of the neutrino is inferred from the missing momentum 
and energy in the whole event. 

The event yields for each of the four signal decay modes are extracted from
a binned maximum-likelihood fit to the three-dimensional distributions of the variables \mES, the energy-substituted $B$-meson mass, \DeltaE, the difference between the reconstructed and the expected $B$-meson energy, and $q^2$, the momentum transfer squared from the $B$ meson to the final-state hadron.  
The measured differential decay rates in combination with recent form-factor 
calculations are used to determine \Vub. By measuring both \Bpilnu\ and \Brholnu\ decays simultaneously, we reduce the sensitivity to the cross feed 
between these two decay modes and some of the background contributions.

The most promising decay mode for a precise determination of \Vub, 
both experimentally and theoretically, is the $\Bpilnu$ decay for which
a number of measurements exist.  The first measurement of this type was performed by the CLEO Collaboration~\cite{pilnu_cleo}.  In addition to the earlier \babar\ measurement mentioned above~\cite{pilnu_jochen}, there is a more recent \babar\ measurement~\cite{pilnu_cote} in which somewhat looser criteria on the neutrino selection were applied, resulting in a larger signal sample but also substantially higher backgrounds.  
These analyses also rely on the measurement of the missing energy and momentum of the whole event to reconstruct the neutrino, without explicitly reconstructing the second $B$-meson decay in the event, but are based on smaller data sets than
the one presented here. 
Recently a number of measurements of both \Bpilnu\ and \Brholnu\ decays were published, in which the \BB\ events were tagged by a fully reconstructed hadronic or semileptonic decay of the second $B$ meson in the 
event~\cite{pilnu_babar_tag,pilnu_belle_tag}. These analyses have led to a simpler and 
more precise reconstruction of the neutrino and very low backgrounds.  However, this is achieved at the expense of much smaller signal samples, which limit the statistical precision of the form-factor measurement.
\section{Form Factors}
\label{sec:theory}

\subsection{Overview}
 
The advantage of charmless semileptonic decays over charmless hadronic decays of the $B$ meson is that the leptonic and hadronic components of the matrix element factorize.  The hadronic matrix element is difficult to calculate, since it must take into account physical mesons, rather than free quarks. Therefore higher-order perturbative corrections and non-perturbative long-distance hadronization processes cannot be ignored. To overcome these difficulties, a set of Lorentz-invariant form factors has been introduced that give a global description of these QCD processes.

A variety of theoretical predictions for these form factors exist. They are based on QCD calculations, such as lattice QCD and sum rules, in addition to quark models.  We 
will make use of a variety of these calculations to assess their impact on the determination of \Vub\ from measurements of the decay rates.  

The $V-A$ structure of the hadronic current is invoked, along with the knowledge of the transformation properties of the final-state meson, to formulate these form factors. 
They are functions of $q^2 = m^2_W$, the mass squared of the virtual $W$,  
\begin{eqnarray}                
q^2 &=&(P_{\ell}+P_{\nu})^2  \nonumber \\
    &=&(P_B-P_X)^2 = M_B^2+m^2_X-2M_B E_X.
\label{eq:q2}   
\end{eqnarray}  
Here $P_{\ell}$ and $P_{\nu}$ refer to the four-momenta of the charged lepton and the neutrino, $M_B$ and $P_B$ to the mass and the four-momentum of the $B$ meson,  and $m_X$ and $E_X$ are the mass and energy (in the $B$-meson rest frame) of the final-state meson $X_u$. 

We distinguish two main categories of exclusive
semileptonic decays: decays to pseudoscalar mesons, \Bpilnu\ or \Bpetalnu,  and 
decays to vector mesons, \Brholnu\ or  \Bpomegalnu.

\begin{figure}[htb]
\centering 
\epsfig{file=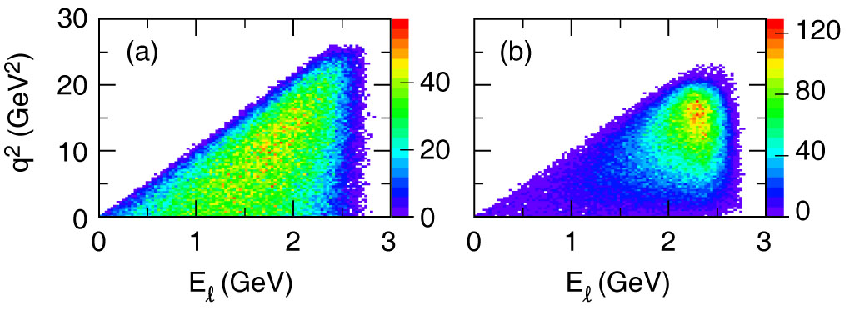,width=0.99\columnwidth}
\caption{
Simulated distributions of $q^2$ versus $E_{\ell}$ for a) $B \ra \pi\ell \nu$  and 
b) $B \ra \rho\ell \nu$ decays. $E_{\ell}$ is  the lepton energy 
in the $B$-meson rest frame.}
\label{fig:dalitz}
\end{figure}

Figure~\ref{fig:dalitz} shows the phase space for \Bpilnu\ and \Brholnu decays in terms of $q^2$ and $E_{\ell}$, the energy of the charged lepton in the $B$-meson rest frame.  The difference between the distributions is due to the different spin structure of the decays.

\subsection{Form Factors}
 
\subsubsection{{\bf $B$} Decays to Pseudoscalar Mesons: {\bf $B\ra \pi \ell \nu$}}

For decays to a final-state pseudoscalar meson, the hadronic matrix element is usually written in terms of two form factors, $f_+(q^2)$ and $f_0(q^2)$ 
~\cite{neubert94,richman-burchat},
\begin{eqnarray}
\lefteqn{\langle \pi(P_{\pi})| \overline{u} \gamma^{\mu} b |B(P_B)\rangle =}  \nonumber \\
& & f_+(q^2)\left[(P_B+P_{\pi})^\mu -
  \frac{M_B^2 - m_{\pi}^2}{q^2}\,q^\mu\right] + \nonumber \\ 
& & \mbox{} f_0(q^2)\,\frac{M_B^2 - m_{\pi}^2}{q^2}\,q^\mu ,
\label{eq:pplnuamp}
\end{eqnarray}
\noindent
where $P_{\pi}$ and $P_B$ are the four-momenta of the final-state pion and the parent $B$ meson, and $m_{\pi}$ and $M_B$ are their masses.
This expression can be simplified for leptons with small masses, such as 
electrons and muons, because in the limit of  $m_\ell \ll M_B$ 
the second term can be neglected.
We are left with a single form factor $f_+(q^2)$ and the differential decay rate becomes
\begin{eqnarray}
\frac{d\Gamma(B^0 \ra \pi^- \ell^+ \nu)}{dq^2 d \cos\theta_{W\ell}} =
|V_{ub}|^2  \frac{G^2_F \, p_{\pi}^3}
        {32 \pi^3} {\sin}^2\theta_{W\ell} |f_+(q^2)|^2,
\label{semilept:eq:pplnutdd}
\end{eqnarray}  
where $p_\pi$ is the momentum of the pion in the rest frame of the $B$ meson,
and $q^2$ varies from zero to $q^2_{max}= (M_B - m_{\pi})^2$.

The decay rate depends on the third power of the pion momentum, suppressing
the rate at high $q^2$.  The rate also depends on 
$\sin^2\theta_{W\ell}$, where $\theta_{W\ell}$ is the angle of the 
charged-lepton momentum in the $W$ rest frame with respect to direction of the $W$ 
boost from the $B$ rest frame. The combination of these two 
factors leads to a lepton-momentum spectrum that is peaked 
well below the kinematic limit 
(see Figure~\ref{fig:dalitz}).

\subsubsection{{\bf $B$} Decays to Vector Mesons: {\bf \Brholnu}}

For decays with a vector meson in the final state, the polarization vector $\epsilon$ of the vector meson plays an important role. The hadronic current is written in terms of four form factors, of which only three ($A_i$ with $i=0,1,2$) are independent ~\cite{neubert94,richman-burchat},

\begin{eqnarray}
\lefteqn{\langle \rho(P_{\rho},\epsilon)| V^\mu - A^\mu|B(P_B)\rangle  = 
  \frac{2iV(q^2)}{M_B+m_{\rho}}\epsilon^{\mu\nu\alpha\beta}
  \epsilon^*_\nu P_{\rho\alpha} P_{B\beta} } \nonumber \\
& &  -(M_B {+} m_{\rho}) A_1(q^2) \epsilon^{*\,\mu}
 + \frac{A_2(q^2)}{M_B {+} m_{\rho}} \epsilon^*\!\cdot P_B\,
  (P_B {+} P_{\rho})^\mu \nonumber \\
& &  + 2 m_{\rho} \frac{\epsilon^*\!\cdot P_B}{q^2} q^\mu 
     [A_3(q^2) -  A_0(q^2)] ,
 \label{eq:ffvadef}
\end{eqnarray}
where  $m_{\rho}$ and $P_{\rho}$ refer to the  vector-meson mass and four-momentum. 
\noindent
Again, a simplification can be made for low-mass charged leptons. The term with $q^{\mu}$ can be neglected, so there are effectively only three form factors for electrons and muons:  the axial-vector form factors, $A_1(q^2)$ and $A_2(q^2)$, and the vector form factor, $V(q^2)$. Instead of using these form factors, the full differential decay rate  is usually expressed in terms of the helicity amplitudes corresponding to the three helicity states of the $\rho$ meson,

\begin{eqnarray}
H_{\pm}(q^2) & = & (M_B+m_{\rho}) \bigg[ A_1(q^2)\mp \frac {2M_B\,p_{\rho}}{(M_B+m_{\rho})^2} V(q^2)\bigg]  , 
\nonumber \\
H_{0}(q^2) & = & \frac{M_B+m_{\rho}}{2m_{\rho} \sqrt{q^2}}  
\times \bigg[( M_B^2 - m_{\rho}^2 - q^2) A_1(q^2) \nonumber
\\
           &-&  \frac {4 M_B^2 \,p_{\rho}^2} {(M_B+m_{\rho})^2} A_2(q^2)\bigg] , 
\end{eqnarray}
\noindent
where $p_{\rho}$ is the momentum of the final-state $\rho$ meson in the $B$ rest frame.
While $A_1$ dominates the three helicity amplitudes, $A_2$ contributes only to $H_0$, and $V$ contributes only to $H_{\pm}$. 

Thus the differential decay rate can be written as
\begin{eqnarray}
\lefteqn{\frac{d\Gamma(B\rightarrow \rho \ell \nu)}{dq^2 
d\cos{\theta_{W\ell}}} = |V_{ub}|^2 \,  \frac{G^2_F \,p_{\rho} \,q^2}{128\pi^3 M_B^2} \times{} \bigg[ \sin {\theta^2_{W\ell}} |H_0|^2 
\nonumber }  \\
  &  + (1- \cos {\theta_{W\ell}})^2 \frac{|H_+|^2}{2} + (1+ \cos {\theta_{W\ell}})^2 \frac{|H_-|^2}{2} \bigg].
\label{equ:vdecrate}
\end{eqnarray}

\noindent
The $V{-}A$ nature of the charged weak current leads to a dominant contribution from  $H_-$ and a distribution of events characterized by a forward peak in $\cos\theta_{W\ell}$ and high lepton momenta (see Figure~\ref{fig:dalitz}).

\subsection{Form-Factor Calculations and Models}
\label{sect:FormFactorModels}

The $q^2$ dependence of the form factors can be extracted 
from the data. Since the differential decay rates are proportional 
to the product of $\Vub^2$ and the form-factor terms, we need at 
least one point in $q^2$ 
at which the form factor is predicted in order to extract \Vub\ from the 
measured branching fractions.

Currently predictions of form factors are based on
\begin{itemize} 
 \item quark-model calculations, (ISGW2)~\cite{isgw2},
 \item QCD light-cone sum rules (LCSR)~\cite{ball_pi,ball07,ball_rho,ball_eta,Siegen},
 \item lattice QCD calculations (LQCD)~\cite{uslattice,jplattice,fnal09,hpqcd04}. 
\end{itemize}
\noindent
These calculations will also be used to simulate the kinematics of the signal decay modes and thus might impact the detection efficiency and thereby the branching-fraction measurement. 
The two QCD calculations result in predictions for different regions of phase space. The lattice calculations are only available in the high-$q^2$ region, while LCSR provide information near $q^2=0$.  Interpolations between these two regions can be constrained by unitarity and analyticity requirements~\cite{BGL,BHill}.

\begin{figure}[!htbp]
\twoFigOneCol{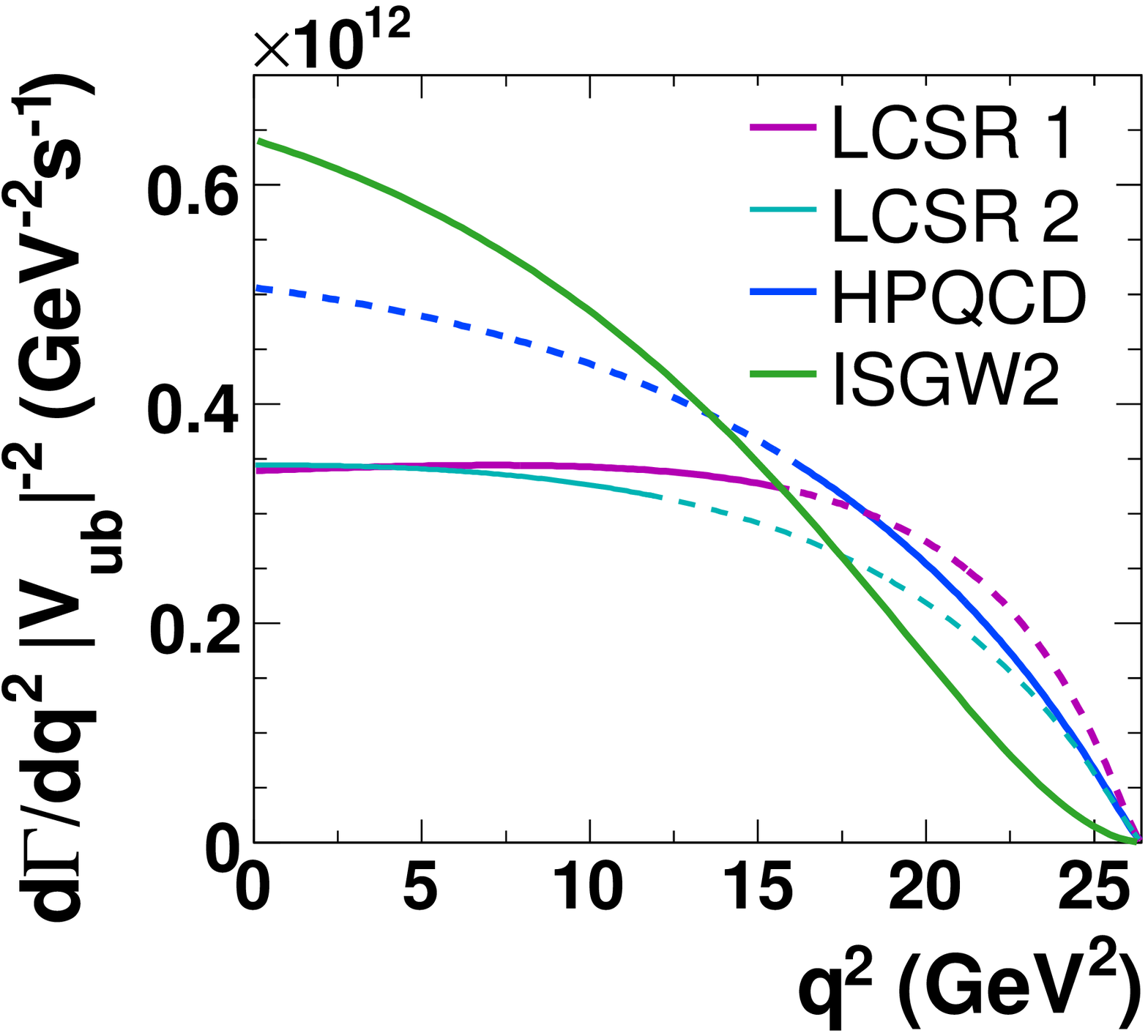} {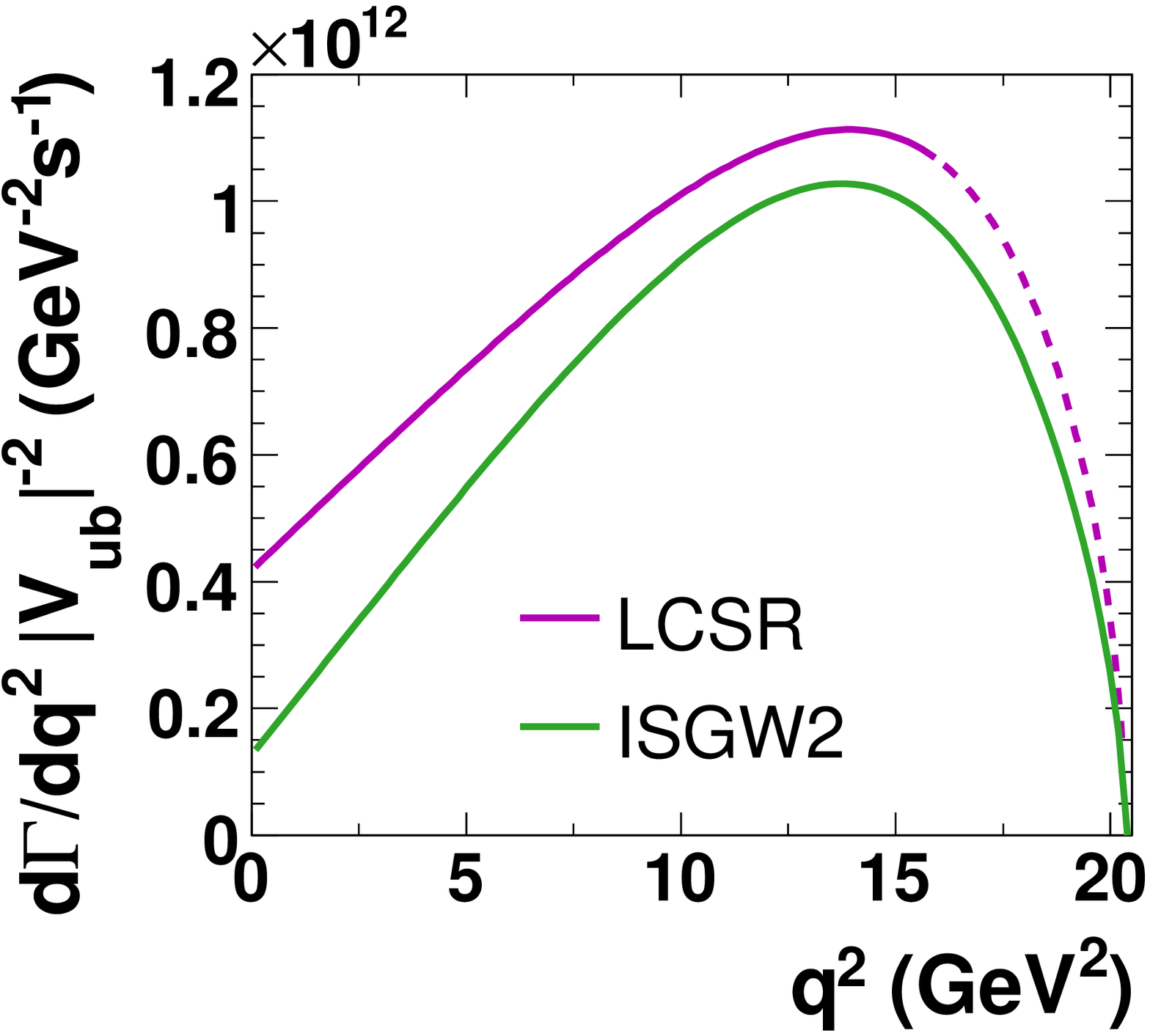} 
\caption{
$q^2$ distributions for \Bpilnu\ (left) and \Brholnu\ (right) 
decays, based on form-factor predictions from the ISGW2 model~\cite{isgw2},  LCSR calculations (LCSR~1~\cite{ball_pi} and LCSR~2~\cite{Siegen} for \Bpilnu\ and LCSR~\cite{ball_rho} for \Brholnu) 
and the HPQCD~\cite{hpqcd04} lattice calculation. 
The extrapolations of the QCD predictions to the full $q^2$ range are marked as dashed lines.}
\label{fig:q2mc}
\end{figure}

Figure~\ref{fig:q2mc} shows the $q^2$ distributions for \Bpilnu\ and  \Brholnu\ decays for various form-factor calculations.
The uncertainties in these predictions are not indicated. For \Bpilnu\ decays they are largest at low $q^2$ for LQCD predictions  and largest at high $q^2$ for LCSR calculations. 
Estimates of the uncertainties of the calculations are currently not available for \Brholnu\ decays.

The Isgur-Scora-Grinstein-Wise model (ISGW2)~\cite{isgw2} is a constituent quark model with relativistic corrections.   Predictions extend over the full $q^2$ range; they
are normalized at  $q^2 \approx q^2_{max}$.  The form factors are parameterized as
\begin{equation}
f_+(q^2) = f(q^2_{max}) \left( 1+ \frac{1}{6N} \xi^2 (q^2_{max} - q^2)\right)^{-N} \ ,
\end{equation}
where $\xi$ is the charge radius of the final-state meson, and $N=2$ ($N=3$) for decays to pseudoscalar (vector) mesons.
The uncertainties of the predictions by this model are difficult to quantify.

QCD light-cone sum-rule calculations are non-perturbative and combine the idea of QCD sum rules with twist expansions performed to ${\cal O}(\alpha_s)$. These calculations provide estimates of various form factors at low to intermediate $q^2$, for both pseudoscalar and vector decays.  
The overall normalization is predicted at low $q^2$ with typical uncertainties of 10-13\%~\cite{ball_pi,ball_rho}.

\begin{figure}
  \centering
  \epsfig{file=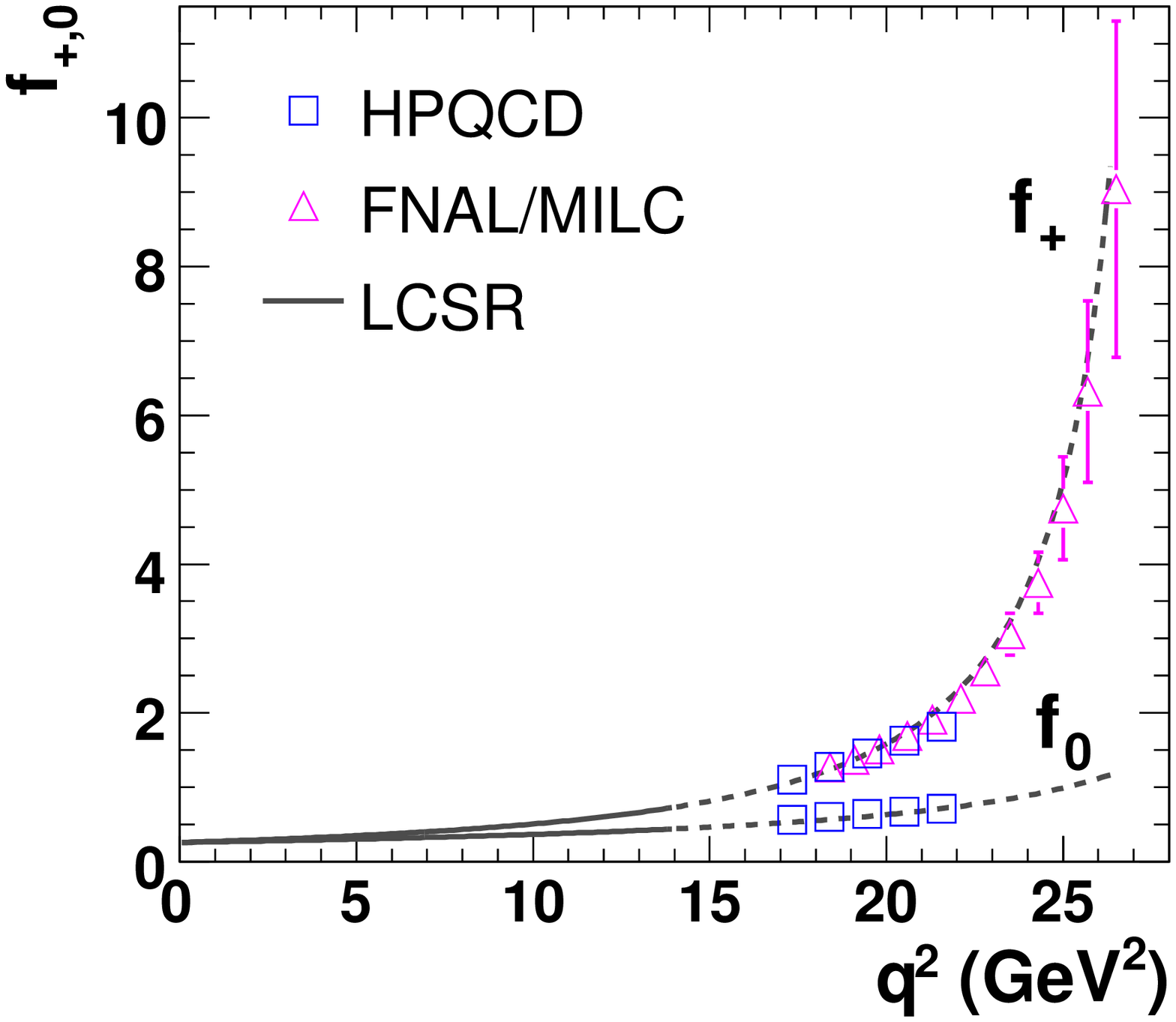,width=0.8\columnwidth}
  \caption{(color online)
Predictions of the form factors $f_+$ and $f_0$ for \Bpilnu\ decays based on unquenched LQCD calculations by the FNAL/MILC~\cite{fnal09} and HPQCD~\cite{hpqcd04} Collaborations (data points with combined statistical and systematic errors) and LCSR calculations~\cite{ball_pi} (solid black lines). The dashed lines indicate the extrapolations of the LCSR predictions to $q^2>16\gev^2$.}
\label{fig:unquenchedLQCD}
\end{figure}

Lattice QCD calculations can potentially provide heavy-to-light-quark form factors from first principles.  
Unquenched lattice calculations, in which quark-loop effects in the QCD vacuum are incorporated, are now available for the \Bpilnu\ form factors from the Fermilab/MILC ~\cite{fnal09} and the HPQCD~\cite{hpqcd04} Collaborations. Both calculations account for three dynamical quark flavors, the mass-degenerate
$u$ and $d$ quarks and a heavier $s$ quark, but they differ in the way the $b$ quark is simulated.
Predictions for $f_0(q^2)$ and  $f_+(q^2)$ are shown in 
Figure~\ref{fig:unquenchedLQCD}.  The two lattice calculations agree within the stated uncertainties, which are significantly smaller than those of earlier quenched approximations.

\subsection{Form-Factor Parameterizations}
\label{sec:FFparam}
Neither the lattice nor the LCSR QCD calculations predict the form factors over the full $q^2$ range. 
Lattice calculations are restricted to small hadron momenta, {\it i.e.}, to 
$q^2 \ge q^2_{max}/2$, while LCSR work best at small $q^2$.  If the 
$q^2$ spectrum is well measured, the shape of the form factors can be constrained, and the 
QCD calculations provide the normalization necessary to determine \Vub.

A number of parameterizations of the pseudoscalar form factor $f_+(q^2)$ are available in the literature.  
The following four parameterizations are commonly used. All of them include at least one pole 
term at $q^2=m^2_{B^*}$, with $m_{B^*} = 5.325 \gev < M_B +m_{\pi}$. 
\begin{enumerate}
\item{Becirevic-Kaidalov (BK)~\cite{BK} }:
\begin{eqnarray}
\label{eq:BK}
  f_+(q^2) &=& \frac{f_+(0)} {(1-q^2/m_{B^*}^2)(1-\alpha_{BK} q^2/m_{B^*}^2)} , \\
  f_0(q^2) &=& \frac{f_0(0)} {1-\beta_{BK}^{-1} q^2/m_{B^*}^2} \ ,
\end{eqnarray}
where $f_+(0)$ and $f_0(0)$ set the normalizations and $\alpha_{BK}$ and $\beta_{BK}$ define the shapes. 
The BK parameterization has been applied in fits to the HPQCD lattice predictions for form factors, with the constraint $f_+(0)=f_0(0)$.  

\item{Ball-Zwicky (BZ)~\cite{ball_pi,ball07}  }:
\begin{eqnarray}
  f_+(q^2) &=& f_+(0) \bigg[ \frac{1} {1-q^2/m_{B^*}^2} \nonumber \\
           &+& \frac{r_{BZ} q^2/m^2_{B^*} } { (1-q^2/m_{B^*}^2)(1-\alpha_{BZ} q^2/m_{B^*}^2)} \bigg]  , 
\end{eqnarray}
where $f_+(0)$ is the normalization, and $\alpha_{BZ}$ and $r_{BZ}$ determine the shape.   This is an extension of the BK ansatz, related by the simplification $\alpha_{BK}=\alpha_{BZ}=r_{BZ}$. This ansatz was used to extend the LCSR predictions
to higher $q^2$, as shown in Figure~\ref{fig:unquenchedLQCD}.

\item{Boyd, Grinstein, Lebed (BGL)~\cite{BGL,BHill} }:
\begin{equation}
\label{eq:BGL}
f_+(q^2) = \frac{1} {{\cal P}(q^2) \phi(q^2,q^2_0) } \sum_{k=0}^{k_{max}} a_k(q^2_0) [z(q^2,q^2_0)]^k \ ,
\end{equation}
\begin{equation}
\label{eq:z}
z(q^2,q^2_0) = \frac { \sqrt{m^2_+ - q^2} - \sqrt{m^2_+ - q^2_0} }
                   { \sqrt{m^2_+ - q^2} + \sqrt{m^2_+ - q^2_0} } \ ,
\end{equation}
\noindent
where $m_\pm = M_B \pm m_{\pi}$ and $q^2_0$ is a free parameter~\cite{q0sq}. 
The so-called Blaschke factor ${\cal P}(q^2)=z(q^2,m_{B^*}^2)$ accounts for 
the pole at $q^2=m_{B^*}^2$,  
and $\phi(q^2,q^2_0)$ is an
arbitrary analytic function~\cite{phi} 
whose choice only affects the particular values of the
series coefficients $a_k$. 
In this expansion in the variable~$z$, 
the shape is given by the values of $a_k$, with truncation at $k_{max}=2$ or $3$.
The expansion parameters are constrained  by unitarity, $\sum_k a_k^2 \le 1$.  
Becher and Hill~\cite{BHill} have
pointed out that due to the large $b$-quark mass, this bound is far from 
being saturated. Assuming that the ratio
$\Lambda/m_b$ is less than 0.1, the heavy-quark bound is approximately 30 times more constraining than the bound from unitarity alone, $\sum_k a_k^2 \sim (\Lambda/m_b)^3 \approx 0.001$.
For more details we refer to the literature~\cite{BGL,BHill}.

\item{Bourrely, Caprini, Lellouch (BCL)~\cite{BCL} }:
\begin{eqnarray}
\label{eq:BCL}
f_+(q^2) = \frac{1} {1-q^2/m_{B^*}^2} \sum_{k=0}^{k_{max}} b_k(q^2_0) 
\{ [z(q^2,q^2_0)]^k - \nonumber \\
(-1)^{k-k_{max}-1} \frac{k}{k_{max}+1} [z(q^2,q^2_0)]^{k_{max}+1}\} \ ,
\end{eqnarray}
\noindent
where the variable~$z$ is defined as in Eq.~\ref{eq:z} with free parameter $q^2_0$~\cite{q0sq}.
In this expansion the shape is given by the values of $b_k$, with truncation at $k_{max}=2$ or $3$.
The BCL parameterization exhibits the QCD scaling behavior $f_+(q^2) \propto 1/q^2$ at large $q^2$.
\end{enumerate}

The BK and BZ parameterizations are intuitive and have few free parameters. Fits to the previous \babar\ form-factor measurements using these parameterizations have shown that they describe the data quite well~\cite{pilnu_cote}.
The BGL and BCL parameterizations are based on fundamental theoretical concepts like analyticity and unitarity. The $z$-expansion avoids ad hoc assumptions about the number of poles and pole masses, and it can be adapted to the precision of the data. 


\section{Data Sample, Detector, and Simulation}
\label{sec:Detector}

\subsection{Data Sample}

The data used in this analysis were recorded with the 
\babar\ detector at the \pep2 energy-asymmetric $e^+e^-$ collider
operating at the \FourS\ resonance.
A sample of 377 million $\FourS \to \BB$ events,
corresponding to an integrated luminosity 
of 349~$\mathrm{fb}^{-1}$,  was collected.
An additional sample of 35.1~$\mathrm{fb}^{-1}$ was recorded 
at a center-of-mass (c.m.) energy approximately 40 \mev below the \FourS\ resonance, {\it i.e.,} just below the threshold for \BB\ production. 
This off-resonance data sample is used to subtract the non-\BB\ contributions
from the data collected at the \FourS\ resonance. The principal source
of these hadronic non-\BB\ events is \epem\ annihilation in the continuum to \qqbar\ pairs,
where $q=u,d,s,c$ refers to quarks.
The relative normalization of the  off-resonance and on-resonance data samples
is derived from luminosity 
measurements, which are based on the number of detected $\mu^+\mu^-$
pairs and the QED cross section for 
$e^+e^-\to \mu^+\mu^-$ 
production, adjusted for the small difference in c.m. energy. 
The systematic error on the relative normalization is estimated
to be $0.25\%$.  

\subsection{BABAR Detector}

The \babar\ detector and event reconstruction are described in
detail elsewhere~\cite{babar_NIM,babar_reconst}.  The momenta and angles of charged
particles are measured in a tracking system consisting of a five-layer
silicon vertex tracker (SVT) and a 40-layer drift chamber (DCH) filled with a helium-isobutane gas mixture.
Charged particles of different masses are distinguished by their
ionization energy loss in the tracking devices and by a ring-imaging Cerenkov
detector (DIRC). Electromagnetic showers from electrons and photons
are measured in a finely segmented CsI(Tl) calorimeter (EMC).  
These detector components are embedded in the $1.5$-$\mathrm{T}$ magnetic
field of the solenoid. The magnet flux return steel is segmented and 
instrumented (IFR) with planar resistive plate chambers and limited streamer tubes,
which detect particles penetrating the magnet coil and steel. 

The efficiency for the reconstruction of charged particles inside the fiducial volume of the tracking system exceeds 96\% and is well reproduced by MC simulation. An effort has been made to minimize fake charged tracks, caused by multiple counting of a single low-energy track curling in the DCH, split tracks, or background-generated tracks.  The average uncertainty in the track-reconstruction efficiency is estimated to range from 0.25\% to 0.5\% per track. 

To remove beam generated background and noise in the EMC, photon candidates 
are required to have an energy of more than 50 \mev\ and a shower shape that 
is consistent with an electromagnatic shower.  The photon efficiency and its
uncertainty are evaluated by comparing $\tau^{\pm} \to \pi^{\pm} \nu$ to  
$\tau^{\pm} \to \rho^{\pm} \nu$ samples and by studying $\epem \to \mumu (\gamma)$ events.   

Electron candidates are selected on the basis of the ratio of
the energy detected in the EMC and the track momentum, the
EMC shower shape, the energy loss in the SVT and DCH, and
the angle of the Cerenkov photons reconstructed in the DIRC.  
The energy of electrons is corrected for bremsstrahlung detected as photons emitted close to the electron direction. 
Muons are identified by using a neural network that combines the information from the IFR with the measured track momentum 
and the energy deposition in the EMC.

The electron and muon identification efficiencies and the
probabilities to misidentify a pion, kaon, or proton as an electron or
muon are measured as a function of the laboratory momentum and
angles using high-purity samples of particles selected from data.
These measurements are performed separately for positive and negative leptons. 
For the determination of misidentification probabilities, knowledge of the inclusive momentum spectra of positive and negative hadrons, and the measured fractions of pions, kaons and protons and their misidentification rates is used.

Within the acceptance of the SVT, DCH and EMC defined by the polar angle
in the laboratory frame, $-0.72 < \cos \theta_{\mathrm{lab}} < 0.92$, the average electron efficiency for laboratory momenta above 
0.5 \gev is $93\%$, largely
independent of momentum.  The average hadron
misidentification rate is less than 0.2\%. 
Within the same polar-angle acceptance,
the average muon efficiency rises with laboratory momentum to reach a plateau of about 70\% above 1.4 \gev.  The muon efficiency varies between 50\% and 80\% as a function of the polar angle. The average hadron misidentification rate is 2.5\%,
varying by about 1\% as a function of momentum and polar angle.

Neutral pions are reconstructed from pairs of photon candidates that are
detected in the EMC and  assumed to originate from
the primary vertex. 
Photon pairs with an invariant mass within 17.5~\mev of the nominal \piz\ mass are considered \piz\ candidates.  The overall detection efficiency, including solid angle restrictions, varies between 55\% and 65\% for \piz\ energies in the range of 0.2 to 2.5 \gev.

\subsection{Monte Carlo Simulation}

We assume that the \FourS\ resonance decays exclusively to \BB\ pairs~\cite{fours} and that the non-resonant cross section for $\epem \to \qqbar$ is 3.4\nb , compared to the \FourS\ peak cross section of 1.05\nb . 
We use Monte Carlo (MC) techniques~\cite{evtgen} to simulate the production
and decay of \BB\ and \qqbar\ pairs and the detector response~\cite{geant4},  
to estimate signal and background efficiencies,
and to extract the expected signal and background distributions. 
The size of the simulated sample of generic \BB\ events exceeds the \BB\ data sample by
about a factor of three, while the MC samples for inclusive and exclusive $B \to X_u \ell \nu$ decays 
exceed the data samples by factors of 15 or larger.  The MC sample for \qqbar\ events is comparable in size to the \qqbar\ data sample recorded 
at the \FourS\ resonance.

Information extracted from studies of selected data control samples on 
efficiencies and resolution is used to improve the accuracy of the simulation.
Specifically, comparisons of data with the MC simulations reveal small differences in the tracking efficiencies and calorimeter resolution. We apply corrections to account for these differences.  
The MC simulations include radiative effects such as bremsstrahlung
in the detector material and initial-state and final-state
radiation~\cite{photos}. Adjustments are made to take into account the small variations of the beam energies over time.

For this analysis, no attempt is made to reconstruct $K^0_L$ interacting in the EMC or IFR.  Since a $K^0_L$ deposits only a small
fraction of its energy in the EMC, $K^0_L$ production can have a significant impact on
the energy and momentum balance of the whole event and thereby the neutrino reconstruction.  It is therefore important to verify that the production rate of neutral kaons and their interactions in the detector are well reproduced.

From detailed studies of large data and MC samples of $D^0 \to K^0_L \pi^+\pi^-$ and $D^0 \to K^0_S \pi^+\pi^-$ decays, corrections to the simulation of the $K^0_L$ detection efficiency and energy deposition in the EMC are determined. The MC simulation reproduces the efficiencies well for $K^0_L$ laboratory momenta above 0.7\gev. At lower momenta, the difference between MC and data increases significantly; in this range the MC efficiencies are reduced by randomly eliminating a fraction of the associated EMC showers. The energy deposited by $K^0_L$ in the EMC is significantly underestimated by the simulation for momenta up to 1.5\gev. At higher momenta the differences decrease.  Thus the simulated energies are scaled by factors varying between 1.20  and 1.05 as a function of momentum.  Furthermore, assuming equal inclusive production rates for $K^0_L$ and $K^0_S$ we verify the production rate as a function of momentum, by comparing data and MC simulated $K^0_S$ momentum spectra.  We observe differences at small momenta; below 0.4\gev the data rate is lower by as much as $22\pm 7\%$ compared to the MC simulation. 
To account for this difference, we reduce  the rate of low momentum $K^0_L$ in the simulation by randomly transforming the excess $K^0_L$ into a fake photon, {\it i.e.}, we replace the energy deposited in the EMC by the total $K^0_L$ energy and set the mass to zero. Thus we correct the overall energy imbalance created by the excess in $K^0_L$ production.  

For reference, the values of the branching fractions, lifetimes, and parameters 
most relevant to the MC simulation are presented in Tables~\ref{tab:BF} 
and \ref{tab:phys_param}.
\begin{table}[hbt]
\renewcommand{\arraystretch}{1.2}
\centering
\caption{Branching fractions and their errors for the semileptonic $B$ decays used in this analysis.}
\begin{tabular}{lcccc} 
\hline\hline 
Decay                   &  Unit     &       $B^0$       &      $B^\pm$    &  Ref.   \\ \hline 
$B \to \eta \ell \nu$    & $ 10^{-4} $ &                  & $ 0.40\pm 0.09 $  & \cite{HFAG2009}\\  
$B \to \eta^\prime \ell \nu$  & $ 10^{-4} $ &                & $ 0.21 \pm 0.21 $  &\cite{HFAG2009}\\
$B \to \omega \ell \nu$  & $ 10^{-4} $ &                  & $ 1.15 \pm 0.16 $  & \cite{HFAG2009}\\ 
$B \to X_u \ell \nu $    & $ 10^{-3} $& $ 2.25 \pm 0.22 $ & $ 2.41 \pm 0.22 $  & \cite{PDG2008}\\ \hline
$B \to D  \ell \nu$      & $ 10^{-2} $& $ 2.17 \pm 0.08 $ & $ 2.32 \pm 0.09 $  & \cite{HFAG2009}\\ 
$B \to D^* \ell \nu$      & $ 10^{-2} $& $ 5.11 \pm 0.19 $ & $ 5.48 \pm 0.27 $  & \cite{HFAG2009}\\ 
$B \to D^*_1\ell \nu$      & $ 10^{-2} $& $ 0.69 \pm 0.14 $ & $ 0.77 \pm 0.15 $  & \cite{HFAG2009,DLVL}\\ 
$B \to D^*_2\ell \nu$      & $ 10^{-2} $& $ 0.56 \pm 0.11 $ & $ 0.59 \pm 0.12 $  & \cite{HFAG2009,DLVL}\\ 
$B \to D^*_0\ell \nu$      & $ 10^{-2} $& $ 0.81 \pm 0.24 $ & $ 0.88 \pm 0.26 $  & \cite{HFAG2009,DLVL}\\   
$B \to D_1^\prime \ell \nu$ & $ 10^{-2} $& $ 0.76 \pm 0.22 $ & $ 0.82 \pm 0.25 $  & \cite{HFAG2009,DLVL}\\ 
\hline\hline
\end{tabular}
\label{tab:BF}
\end{table}

\begin{table}[hbt]
\renewcommand{\arraystretch}{1.2}
\centering
\caption{Form factors used in the simulation of $B \to D \ell \nu$ and 
$B \to D^* \ell \nu$ decays, based on the parameterization of Caprini, Lellouch and Neubert~\cite{CLN}. The \Bz\ lifetime, the \Bz\ to \Bp\ lifetime ratio, and relative branching fraction at the \FourS\ resonance, used in the simulation.}
\begin{tabular}{lccc} 
\hline\hline 
Parameter                                 &     Value                           &   Ref. \\ \hline 
$B \to D   \ell \nu\, FF : \rho^2_{D}   $ & $ 1.18 \pm 0.04 \pm 0.04 $  &\cite{HFAG2009}\\ 
$B \to D^* \ell \nu\, FF : \rho^2_{D^*} $ & $ 1.191 \pm 0.048 \pm 0.028  $  &\cite{babarff}\\
$B \to D^* \ell \nu\, FF : R_1          $ & $ 1.429 \pm 0.061 \pm 0.044  $  &\cite{babarff}\\
$B \to D^* \ell \nu\, FF : R_2          $ & $ 0.827 \pm 0.038 \pm 0.022  $  &\cite{babarff}\\ 
\hline
$B^0$\, life time $\tau_0  (\ps)      $ & $ 1.530 \pm 0.009            $ &\cite{PDG2008} \\
$B$\,life time ratio $\tau_+/\tau_0    $ & $ 1.071 \pm 0.009            $  &\cite{PDG2008}\\ 
\FourS\ ratio $f_{+-}/f_{00}          $ & $ 1.065 \pm 0.026            $  &\cite{HFAG2009}
\\ \hline\hline
\end{tabular}
\label{tab:phys_param}
\end{table}
  
The simulation of inclusive charmless semileptonic decays \bulnu\ is
based on predictions of a heavy-quark expansion (HQE) (valid to 
${\cal O}(\alpha_s)$~\cite{dFN})  for the differential decay rates.  
This calculation produces a smooth hadronic mass spectrum. The
hadronization of $X_u$ with masses above $2 m_{\pi}$ 
is performed by JETSET~\cite{jetset}.  To describe the dynamics 
of the $b$ quark inside the $B$ meson we use HQE parameters extracted from global fits to moments of inclusive lepton-energy and hadron-mass distributions 
in $B \to X_c \ell \nu$ decays and moments of inclusive photon-energy distributions in $B \to X_s \gamma$ decays~\cite{moments}.  The specific values of the HQE parameters in the shape-function scheme are $m_b=4.631 \pm 0.034$ \gev
and $\mu^2_{\pi}= 0.184 \pm 0.36\gev^2$; they have a correlation of $\rho=-0.27$.
Samples of exclusive semileptonic decays involving low-mass charmless mesons
($\pi, \rho, \omega,\eta, \eta^\prime$) are simulated separately  
and then combined with samples of decays to non-resonant and higher-mass resonant states, so that the cumulative distributions of the hadron mass, the momentum transfer squared, and the lepton momentum reproduce the HQE predictions. 
The generated distributions are reweighted to accommodate variations due to specific choices of the parameters for the inclusive and exclusive decays.
The overall normalization is adjusted to reproduce the measured inclusive \bulnu\ branching fraction.

For the generation of decays involving charmless pseudo-scalar mesons we choose two approaches.  
For the signal decay \Btopilnu\ we use the ansatz  by Becirevic and Kaidalov~\cite{BK} for the $q^2$ dependence, with the single parameter $\alpha_{BK}$ set to the value determined in a previous \babar\ analysis~\cite{pilnu_cote} of \Bpilnu decays, $\alpha_{BK}=0.52 \pm 0.06$.
For decays to $\eta$ and $\eta^\prime$ we use the form factor parameterization of Ball and Zwicky with specific values reported in~\cite{ball_eta}.

Decays involving charmless vector mesons ($\rho,\omega$) 
are generated based on form factors determined from LCSR by Ball and Zwicky~\cite{ball_rho}.  We use the parameterization 
proposed by the authors to describe the $q^2$ dependence of the form factors in 
terms of a modified pole ansatz using up to three
independent parameters $r_1$, $r_2$ and $m_{\rm fit}$. 
Table~\ref{tab:srparamnew} shows the suggested values for these parameters. 
$m_{\rm fit}$ refers to an effective pole mass that accounts for contributions from higher-mass $B$ mesons with $J^P=1^-$, 
and $r_1$, and $r_2$  give the relative scale of the two pole terms.

\begin{table}[hbt]
\renewcommand{\arraystretch}{1.2}
\caption
{Parameterization of the LCSR form-factor calculations~\cite{ball_pi,ball_rho} for
decays to pseudo-scalar mesons $\eta$ and $\eta^\prime$ ($f_+$) and vector mesons $\rho$ and $\omega$ ($A_1, A_2, V$).
}
\begin{center}
\begin{tabular}{lccccccc} \hline\hline
Form factor              &  $f_+$     &  $A_1^{\rho} $ &  $ A_2^{\rho} $ & $V^{\rho} $
                                      & $ A_1^{\omega}$& $ A_2^{\omega}$ & $V^{\omega}$ \\ \hline
$F(0)$                     &   0.273    &  0.242  & 0.221   & 0.323   & 0.219   & 0.198   & 0.293  \\
$r_1$                    &   0.122    &   --    & 0.009   & 1.045   &   --    & 0.006   & 1.006  \\
$r_2$                    &   0.155    &  0.240  & 0.212   &-0.721   & 0.217  & 0.192   &  -0.713 \\
$m_{\rm fit}^2 (\gev^2)$ &   31.46    &  37.51  & 40.82   & 38.34   &  37.01  &  41.24   & 37.45  \\
\hline \hline
\end{tabular}
\end{center}
\label{tab:srparamnew}
\end{table}

For the simulation of the dominant \bclnu\ decays, we have chosen
a variety of models. 
For $B\to D \ell \nu$ and $B \ra D^* \ell \nu$ decays we use 
parameterizations~\cite{hqet,CLN} of the form factors based on heavy quark effective theory (HQET).  In the limit of negligible lepton masses, decays to pseudoscalar mesons are described by a single form factor for which the $q^2$ dependence is given by a slope parameter.  We use the world average~\cite{HFAG2009}, updated for recent precise measurements by the \babar\ Collaboration~\cite{kenji,david}. 
Decays to vector mesons are described by three form factors, of which the axial vector form factor dominates.  In the limit of heavy quark symmetry,
their $q^2$ dependence can be described by three parameters: $\rho_{D^*}^2$, $R_1$, and $R_2$.  
We use the most precise \babar\ measurement~\cite{babarff} of these parameters.

For the generation of the semileptonic decays to $D^{**}$ resonances (four $L=1$ states), we use the ISGW2~\cite{isgw2} model.  At present, the sum of the branching fractions for these four decays modes is measured to be 1.7\%, but so far only the decays $D^{**} \to D \pi$ and $D^{**} \to D^* \pi$ have been reconstructed, while the total individual branching fractions for these four states remain unknown. Since the measured inclusive  branching fraction for $B \to X_c \ell \nu$  exceeds the sum of the measured branching fractions of all exclusive semileptonic 
decays by about 1.0\%, and since non-resonant $B \ra D^{(*)} \pi \ell \nu$ decays have not been observed~\cite{DLVL}, we assume that the missing decays are due to $B \to D^{**} \ell \nu$, involving hadronic decays of the $D^{**}$ mesons that have not yet been measured.  To account for the observed deficit, we increase the $B \to D^{**} \ell \nu$ branching fractions by 60\% and inflate the errors by a factor of three.
\section{Event Reconstruction and Candidate Selection}
\label{sec:Selection}

In the following, we describe the selection and kinematic reconstruction of signal candidates, the definition of the various background classes, and the application of neural networks to suppress these backgrounds.

\subsection{Signal-Candidate Selection}

Signal candidates are selected from events having at least four charged tracks. 
The reconstruction of the four signal decay modes,
\Bzpilnu, \Bppizlnu, \Bzrholnu\ and \Bprhozlnu, requires the 
identification of a charged lepton,  the reconstruction
of the hadronic state consisting of one or more charged or neutral
pions, and the reconstruction of the neutrino from the missing energy
and missing momentum of the whole event.

\subsubsection{Lepton and Hadron Selection}
\label{sec:Ysystem}

Candidates for leptons, both $e^{\pm}$ and $\mu^{\pm}$, are required
to have high c.m.\ momenta, $p^*_{\ell} \ge 1.0 \gev$ for the \Bpilnu,
and $p^*_{\ell} \ge 1.8\gev$ for 
the \Brholnu\ sample. 
This requirement significantly reduces the background from hadrons
that are misidentified as leptons, and also removes a large fraction
of true leptons from secondary decays or photon conversions, and from
\bclnu\ decays. 

To suppress Bhabha scattering and two-photon processes in which an electron or a photon 
from initial-state or final state radiation interacts 
in the material of the detector and generates additional charged tracks and photons at small angles to the beam axis, we require
$\xi_z < 0.65$ for events with a candidate electron.
Here $\xi_z=\sum_i{p^z_i}/\sum_i {E_i}$, where the sum runs over all charged particles 
in the event and $p^z_i$ and $E_i$ are their longitudinal momentum components and 
energies measured in the laboratory frame. 

For the reconstruction of the signal hadron, we consider all charged
tracks that are not consistent with a signal lepton and not identified
as a kaon.
Neutral pions are reconstructed from pairs of photon candidates and
the \piz\ c.m.\ momentum is required to exceed 0.2\gev. Candidate $\rho^\pm
\rightarrow \pi^\pm \pi^0$ or $\rho^0 \rightarrow \pi^+ \pi^-$ decays
are required to have a two-pion mass within one full width of the
nominal $\rho$ mass,  $0.650 < M_{\pi\pi} < 0.850\gev$. To reduce the
combinatorial background, we also require that the c.m.\ momentum of one of
the pions exceed 0.4 \gev, and that the c.m.\ momentum of the other pion be
larger than 0.2 \gev.

Each charged lepton candidate is combined with a hadron candidate to
form a  so-called $Y$ candidate of charge zero or one. 
At this stage in the analysis we allow for more than one candidate per event.
Two or three charged tracks associated with the $Y$ candidate are fitted to a
common vertex. This vertex fit must yield a $\chi^2$~probability of at
least $0.1\%$.  To remove background from $\jpsi \to \ell^+\ell^-$
decays, we reject a $Y$ candidate if the invariant mass of the lepton
and any oppositely charged track in the event is consistent with this decay. 

To further reduce backgrounds without significant signal losses, we impose additional restrictions on the c.m.\ momenta of the lepton and hadron candidates by requiring at least one of the following conditions to be satisfied, \linebreak
for \Bpilnu\: 
\begin{eqnarray*}
p^*_{\mathrm hadron} \ge 1.3\gev  \quad {\rm or} \\
p^*_{\ell}   \ge 2.2\gev  \quad {\rm or} \\
p^*_{\mathrm hadron}+ p^*_{\ell}\ge 2.8\gev ,
\end{eqnarray*}
\noindent
and for \Brholnu\: 
\begin{eqnarray*} 
p^*_{\mathrm hadron} \ge 1.3\gev  \quad {\rm or} \\
p^*_{\ell}   \ge 2.0\gev \quad {\rm or} \\
p^*_{\mathrm hadron}+ p^*_{\ell}\ge 2.65\gev.
\end{eqnarray*}
These additional requirements on the lepton and hadron c.m.\ momenta primarily reject background candidates that are inconsistent with the phase space of the signal decay modes.
 
If a $Y$ candidate originates from a signal decay mode,
the cosine of the angle between the momentum vectors of the $B$ meson and the $Y$ candidate, $\cos\theta_{BY}$, can be calculated as follows,
\begin{equation}
\cosBY={\frac{2 E^*_B E^*_{Y} - M_B^2 -M_{Y}^2}{2 p^*_B\ p^*_{Y}}}, \label{eq:cosBY}
\end{equation}
\noindent 
and the condition $|\cosBY|\leq 1.0$ should be fulfilled.
The energy $E^*_B$ and momentum $p^*_B$  of the $B$ meson are not measured event-by-event.  Specifically, $E^*_B=\sqrt{s}/2$ is given by the average c.m.\ energy of the colliding beams, and the $B$ momentum is derived as $p^*_B=\sqrt{E^{*2}_B - M^2_B}$. To allow for the finite resolution in this variable, we impose the requirement $-1.2 < \cosBY < 1.1$.

\subsubsection{Neutrino Reconstruction}
\label{sec:Neutrino}

The neutrino four-momentum, $P_{\nu}\simeq (E_{\mathrm{miss}}, \vec{p}_{\mathrm{miss}})$, is inferred from the difference between the net four-momentum of the colliding-beam particles, $P_{\epem}=(E_{\epem}, \vec{p}_{\epem})$, and the sum of the measured four-vectors of all detected particles in the event, 
\begin{equation}
\label{eq:pmiss}
(E_{\mathrm{miss}},\vec{p}_{\mathrm{miss}}) = ( E_{\epem},\vec{p}_{\epem}) - (\sum_i E_i,\sum_i \vec{p}_i),
\end{equation}
\noindent
where  $ E_i$ and $\vec{p}_i$ are the energy and three-momentum 
of the $i^{\rm th}$ track or EMC shower, measured in the laboratory frame.
The energy calculation depends on the correct mass assignments for charged tracks.
For this reason we choose to calculate the missing momentum and energy in the laboratory frame rather than in the rest frame of the $\FourS$. By doing so, we keep this uncertainty confined to the missing energy.

If all particles in the event, except the neutrino,
are well measured, $P_{\nu} \simeq (E_{\mathrm{miss}},\vec{p}_{\mathrm{miss}})$ is a good approximation. However, 
particles that are undetected because of inefficiency or acceptance losses, in 
particular $K_L$ mesons and additional neutrinos, or spurious tracks or photons that do not originate from the $\BB$ event, impact the accuracy of this approximation. 
To reduce the effect of losses due to the limited detector acceptance, we require that the polar angle of the missing momentum in the laboratory frame be in the range 
$0.3 < \theta_{\mathrm miss} < 2.2 \rad$. We also require the missing momentum in the laboratory frame to exceed $0.5\gev$.
 
For the rejection of background events and signal decays that are poorly reconstructed as well as events with more than one missing particle, 
we make use of the missing mass squared of the whole event,
\begin{equation}
\label{eq:MM2}
P_{\nu}^2 \simeq \mmiss = \Emiss^2 - |\vec{p}_{\mathrm{miss}}|^2 .
\end{equation}
\noindent
For a correctly reconstructed event with a single semileptonic $B$ decay, \mmiss\ should be consistent with zero within measurement errors. Failure to detect one or more particles in the event creates a substantial tail at large positive values.
Since the resolution in \mmiss\ increases linearly with $\Emiss$, we use the variable
$\mmiss/2\Emiss \simeq \Emiss-p_{\mathrm{miss}}$ as a discriminator and require 
$ \mmiss/2\Emiss < 2.5 \gev$.

\subsubsection{Variables Used for Signal Extraction}
\label{sec:signalvariables}

The kinematic consistency of the candidate decay with a signal $B$ decay is ascertained using two variables, the beam-energy substituted \B~mass \mes, and the difference between the reconstructed and expected energy of the $B$ candidate \DeltaE.
In the laboratory frame, they are defined as
\begin{equation}
\Delta E = \frac {P_B \cdot P_{\epem} - s/2}{\sqrt{s}}
\end{equation}
and
\begin{equation}
\mES = \sqrt{ \frac{(s/2 + \vec{p}_B \cdot \vec{p}_{\epem})\,^2}{E_{\epem}^2} - p_B^2} \ ,
\end{equation}
\noindent
where 
$P_B=(E_B,\vec{p}_B)$ and $P_{\epem}$ denote the four-momenta of the $B$ meson and the colliding beam particles, respectively. 
The $B$-meson momentum vector $\vec{p}_B$ is determined from the measured three-momenta of the decay products, and $P_{\epem}$ is derived from the calibration and monitoring of the energies and angles of the stored beams. 
We extract the signal yields by a fit to the two-dimensional $\DeltaE - \mes$ distributions in bins of the momentum transfer squared $q^2$.
We define a region in the  $\DeltaE - \mes$ plane that contains almost all of the signal events and leaves sufficient phase space to constrain the different background contributions. This $\it {fit \, region}$ is defined as
\begin{equation}
|\DeltaE| < 0.95\gev  \ , \  5.095 < \mES < 5.295 \gev.
\label{eq:fitregion}
\end{equation}
Only candidates that fall inside the fit region are considered in the analysis. 
We also define a smaller region where the signal contribution 
is much enhanced relative to the background. This $\it {signal \, region}$ is defined as
\begin{equation}
-0.15 < \DeltaE < 0.25\gev \ , \ 5.255 < \mES < 5.295 \gev.
\label{eq:signalregion}
\end{equation}
The signal region is chosen to be slightly asymmetric in \DeltaE to avoid sizable \bclnu\ background, which peaks near $-0.2\gev$.
In the following, we refer to the phase space outside the
signal region, but inside the 
fit region, as the {\it side bands}.

As a measure of the momentum transfer squared $q^2$ we adopt the mass squared of the virtual $W$, {\it i.e.}, the invariant mass squared of the
four-vector sum of the reconstructed lepton and neutrino,
\begin{equation}
q^2_{\rm raw} = [(E_\ell,\vec{p}_{\ell}) + (E_{\mathrm{miss}}, \vec{p}_{\mathrm{miss}})]^2 .
\end{equation}
The resolution in $q^2_{\rm raw}$ is dominated by the measurement of the missing energy
which tends to have a poorer resolution than the measured missing momentum, because the missing momentum is a vector sum and contributions from particle losses (or additional tracks and EMC showers) do not add linearly as is the case for \Emiss.  
Thus for the definition of $q^2_{\rm raw}$ it is advantageous to replace $E_{\mathrm{miss}}$ by $p_{\mathrm{miss}}$, the absolute value of the measured missing momentum,

\begin{equation}
q^2_{\rm raw} = [(E_\ell,\vec{p}_{\ell}) + (p_{\mathrm{miss}}, \vec{p}_{\mathrm{miss}})]^2 .
\label{eq:q2raw}
\end{equation}

The resolution of $q^2_{\rm raw}$ can be further improved by scaling $p_{\mathrm{miss}}$ by a factor of $\alpha$, such that \DeltaE\ of the \B~candidate 
is forced to zero,
\begin{equation}
  \vec{p}_\nu = \alpha \vec{p}_{\mathrm{miss}} \quad \mbox{with} \ \alpha = 1 - \frac{\DeltaE}{\Emiss} \ ,
\label{eq:pmiss_force}
\end{equation}
and substituting $\vec{p}_{\nu}$  for $\vec{p}_{\mathrm{miss}}$ to obtain 
$q^2_{\mathrm{corr}}$.  Any candidates for which this $q^2$ correction yields unphysical values, $q^2_{\mathrm{corr}} < 0\gev^2$,
are rejected. This is the case for about $1\%$ of the background not associated with semileptonic decays.
The quantity $q^2_{\mathrm{corr}}$ is used as the measured $q^2$ throughout this analysis. 

\begin{figure*}[htb]
  \twoFigTwoCol{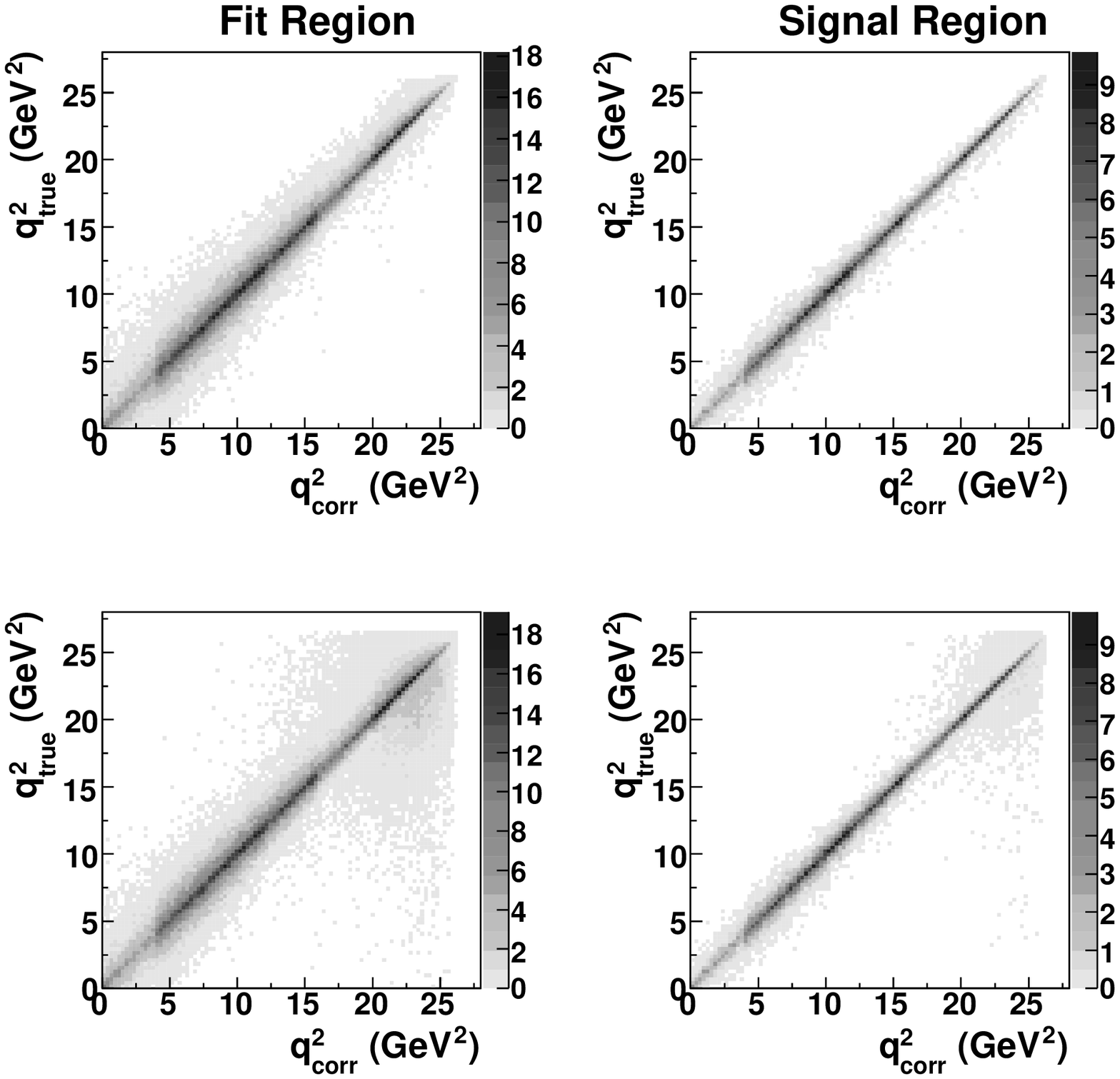} {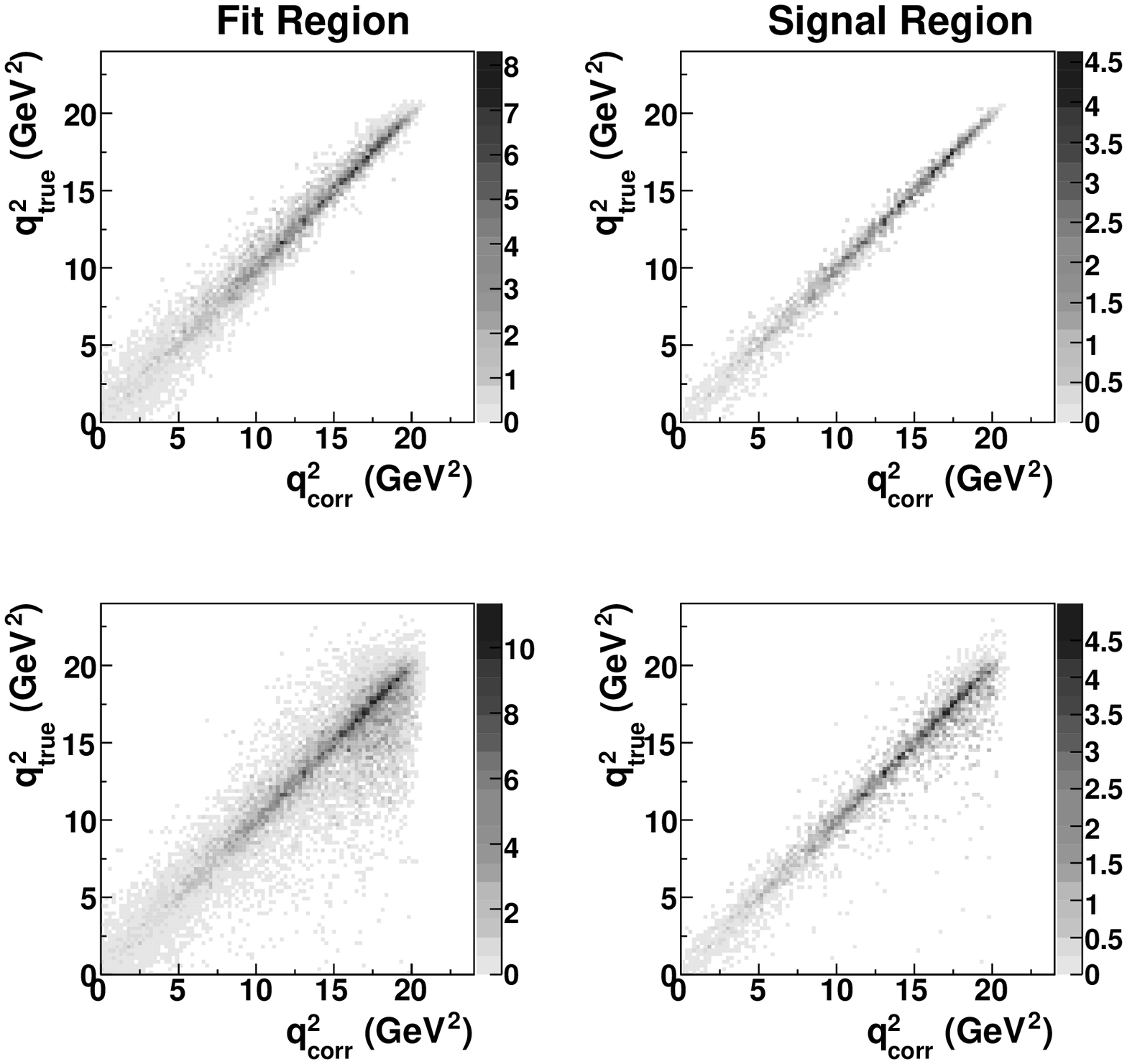}
  \twoFigTwoCol{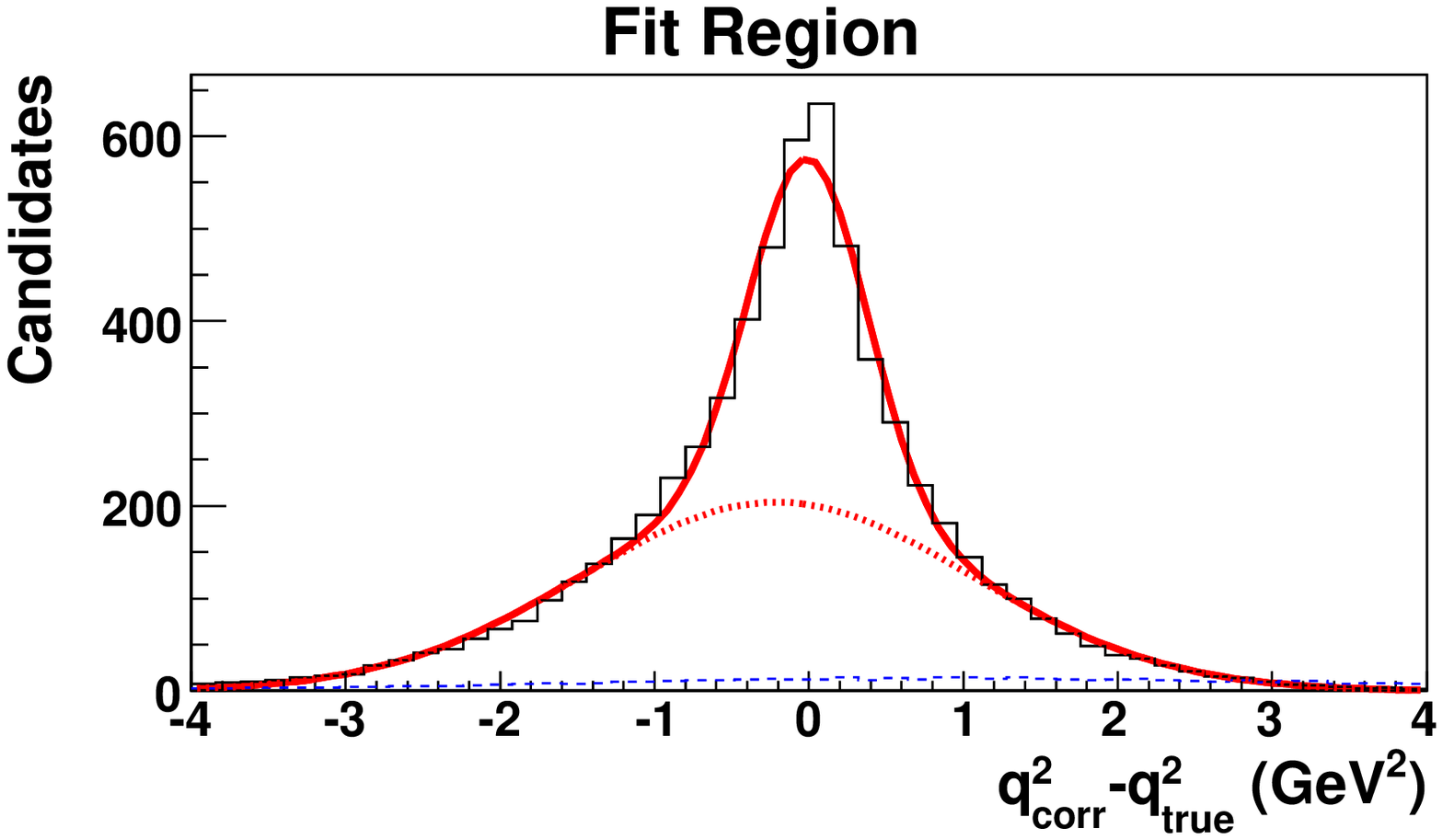} {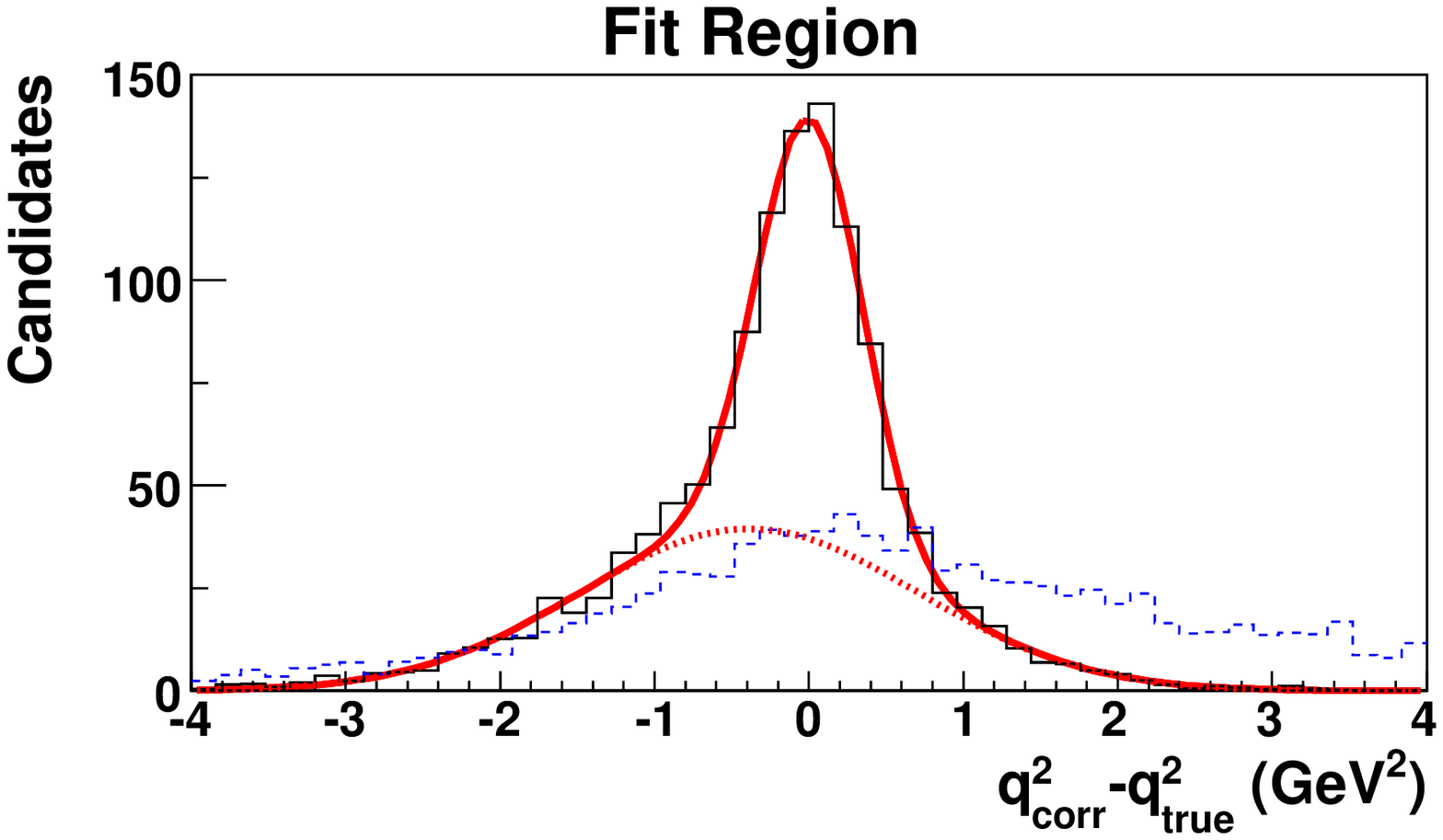}

  \caption{$q^2$ resolution for \Bzpilnu\ (left) and \Bzrholnu\ (right) samples after the full event selection:
   two-dimensional distribution of $q^2_{\mathrm{true}}$ versus $q^2_{\mathrm{corr}}$ in the 
   fit region and in the signal region.
   Top row: true signal decays, middle row: all signal decays (true and combinatorial), 
   bottom row: distribution of $q^2_{\mathrm{corr}}-q^2_{\mathrm{true}}$ for true signal (black, solid histogram) and combinatorial signal
   (blue, dashed histogram) in the fit region. The fit of the sum of two Gaussian functions to the
   true signal distribution is shown as a solid red line, the contribution of the broader of the two
   functions is shown as a dotted red line.}
  \label{fig:q2resolutions}
\end{figure*}

The $q^2$ resolution is critical for the measurements of the form factors. 
Figure~\ref{fig:q2resolutions}
shows the correlation between the true $q^2$ and the reconstructed $q^2_{\rm corr}$ for 
simulated samples of \Bpilnu\ and \Brholnu\ candidates passing the entire event selection,
which is described below. Correctly reconstructed signal events and combinatorial signal events, 
for which the hadron has been incorrectly selected, are shown.
For correctly reconstructed signal decays the resolution improves with higher $q^2$ and can be well described by the sum of two Gaussian resolution functions,
see Table~\ref{tab:q2resolutions}.  In the signal region, the widths of the core resolution are in the range $0.18-0.34 \gev^2$, and the tails can be approximated by a second Gaussian function with widths in the range $0.6-0.8 \gev^2$. As expected, the resolution is significantly worse in the larger fit region.  
Combinatorial signal events contribute primarily at high $q^2$.  We rely on the Monte Carlo simulation to reproduce the resolution in the reconstructed $q^2_{\rm corr}$ variable.

\begin{table*}
\centering
\caption{ 
Description of the $q^2$ resolution in terms of a sum of two Gaussian
resolution functions for true signal decays in the $\DeltaE-\mES$ fit  
region and in the signal region, integrated over $q^2$;
$\sigma_1$, $\mu_1$ and $\sigma_2$, $\mu_2$ denote the means and the widths of the two Gaussian functions, and the last column lists the fraction of the events characterized by the narrower resolution function.
}   

\begin{tabular}{llrrrcc} \hline \hline
              &              & \multicolumn{2}{c}{Gaussian Fct. 1} & \multicolumn{2}{c}{Gaussian Fct. 2}            &                           \\ 
              &  Signal mode & $\mu_1$($\gev^{2}$) & $\sigma_1$($\gev^{2}$) & 
$\mu_2$($\gev^{2}$) & $\sigma_2$($\gev^{2}$) & Fraction         \\ \hline
Fit region    &  \Bzpilnu    &-0.005                & 0.380                    &-0.021                     & 1.270              &   0.35  \\ 
              &  \Bppizlnu   & 0.076                & 0.468                    &-0.039                     & 1.343              &   0.43   \\ 
              &  \Bzrholnu   & 0.005                & 0.343                    &-0.386                     & 1.094              &   0.45  \\ 
              &  \Bprhozlnu  &-0.032                & 0.311                    &-0.498                     & 1.086              &   0.46   \\ \hline
Signal region &  \Bzpilnu    & 0.006                & 0.242                    &-0.020                     & 0.720              &   0.45   \\ 
              &  \Bppizlnu   & 0.058                & 0.338                    & 0.172                     & 0.807              &   0.58   \\ 
              &  \Bzrholnu   & 0.042                & 0.246                    & 0.036                     & 0.647              &   0.50 \\ 
              &  \Bprhozlnu  & 0.010                & 0.177                    &-0.078                     & 0.586              &   0.46   \\ \hline \hline 
\end{tabular}
\label{tab:q2resolutions}
\end{table*}

\subsection{Background Suppression}
\label{sec:background}

\subsubsection{Signal and Background Sources}

A variety of processes contribute to the four samples of selected candidates for the charmless semileptonic decay modes 
$\Bzpilnu$, $\Bppizlnu$, $\Bzrholnu$, and  $\Bprhozlnu$.
We divide the signal and background for each of the four candidate samples into a set of sources based on 
the origin of the charged lepton candidate.
\begin{itemize} 
\item{Signal:}
\quad We differentiate four classes of signal events; for all of them
the lepton originates from a signal decay under study:  
\begin{itemize}
\item{\it True signal: }\,\,
the hadron originates from the signal decay under study;  
\item{\it Combinatorial Signal: }\,\,
the hadron is incorrectly selected, in many cases from decay products of the second $B$ meson in the event;
\item{\it Isospin-conjugate signal:} \,\,
the lepton originates from the isospin conjugate of the signal decay;
\item{\it Cross-feed signal:} \,\,
the lepton originates 
from another signal decay mode, for instance $\Brholnu$ in a $\Bpilnu$ sample.
\end{itemize}

\item{Continuum background:}
\quad We differentiate two classes of continuum backgrounds:
\begin{itemize}
\item{\it True leptons:}\,\,
the lepton candidate originates from a leptonic or semileptonic decay of 
a hadron produced in $\epem \to \qqbar$ (mostly $\ccbar$)
or $\epem \to \ellell (\gamma)$  processes, where $\ellell$ stands for $\epem, \mumu$ or $\tautau$, or $\epem \to \gamma\gamma $;
\item{\it Fake leptons:}\,\,
the lepton candidate is a misidentified hadron; this is a sizable contribution to the muon sample.
\end{itemize}

\item{\bulnu\ background:}
\quad We differentiate two different sources of \bulnu\ background:
\begin{itemize}
\item{\it Exclusive \bulnu\ decays involving a single hadron with mass below 1 \gev:}\,\,
decays that are not analyzed as signal (\Bpomegalnu, \Bpetalnu, and \Bpetaplnu) ;
\item{\it Inclusive \bulnu\ decays:} \,\,
decays involving more than one hadron or a single hadron with mass above 1 \gev.
\end{itemize}

\item{\BB\ background:}
\quad We differentiate three classes of \BB\ background, excluding \bulnu\ decays:
\begin{itemize}
\item{\it Primary leptons, {\it i.e.}, \bclnu\ decays:} \,\,
the lepton originates from a charm semileptonic $B$ decay, either $B \to D \ell \nu$,
$B\to D^* \ell \nu$, or $B \to D^{(*)}(n\pi)\ell\nu$ with $n\ge1$ additional pions;
this class is dominated by $B \to D^* \ell \nu$ decays; the largest contributions 
involve hadrons that do not originate from the semileptonic decay;
\item{\it Secondary leptons:} \,\,
the lepton originates from the decay of a particle other than a $B$ meson,
for instance charm mesons, $\tau$ leptons, $\jpsi$, or from photon conversions;
\item{\it Fake leptons:} \,\,
the lepton candidate is not a lepton, but a misidentified charged hadron;
this background is dominated by fake muons.
\end{itemize}
Given that the secondary-lepton and fake-lepton \BB\ background contributions 
are relatively small in this analysis,
we combine them into one class (other~\BB).
\end{itemize}
For intermediate values of $q^2$ (in the range $4<q^2<20\gev^2$), \bclnu\ decays are by far the dominant background, whereas continuum background contributes mostly at low and high~$q^2$. The \bulnu\ decays have much smaller branching fractions, but their properties are very similar to the signal decays and thus
they are difficult to discriminate against.  They contribute mostly at high $q^2$, where they are the dominant background.

\subsubsection{Neural Networks}
\label{sec:Neuralnets}

To separate signal events from the background sources, continuum events, non-signal \bulnu\ decays and the remaining \BB\ events, 
we employ a neural-network technique based on a multi-layer perceptron (MLP)~\cite{MLP}. 
We have set up a network structure with seven input neurons and one hidden layer 
with three neurons and have adopted the method introduced by Broyden, Fletcher, Goldfarb, and Shanno~\cite{BFGS} to train the network. 
Some of the input variables are used as part of the event preselection that is designed to reduce the \BB\ and continuum backgrounds by cutting out regions where the
signal contribution is small or where there are spikes in distributions, which the neural network may not deal with effectively. 
The following variables are input to the neural networks:
\begin{itemize}
  \item 
$R2$, the second normalized Fox-Wolfram moment~\cite{Fox-wolfram} determined from all charged and neutral particles in the event; we require $R2 < 0.5$;  
  \item 
$L2 = \sum_i p^*_i {\cos}^2\theta^*_i$, 
where  the sum runs over all tracks in the event excluding the $Y$ candidate, 
and $p^*_i$ and $\theta^*_i$ refer to the c.m.\ momenta and the angles 
measured with respect to the thrust axis of the $Y$ candidate;
we set a loose restriction, $L2 < 3.0$ \gev .   
  \item $\cosDThrust$, where $\Delta\theta_{\rm thrust}$ is the angle between the	thrust axis of the $Y$ candidate and thrust axis of all other detected particles in the event; there is no preselection requirement for this variable;
  \item $\mmiss/(2\Emiss)= (\Emiss^2-\pmiss^2)/(2\Emiss)$; we require                  \mmissEmissCut;
  \item $\cosBY$; we require \cosBYCut;
  \item $\costhetaWL$, the helicity angle of the lepton;  
	we require \costhetaWLCut 
                for the \Bzpilnu\ and \Bppizlnu\ modes;
 \item $\thetamiss$, the polar angle of the missing momentum in the
	laboratory frame; we require \ThetamissCut.
\end{itemize}
\noindent
The first three input variables are sensitive to the topological difference between the jet-like continuum events and the more spherical \BB\ events.  Restrictions on these variables do not bias the $q^2$ distribution significantly. 

The restrictions placed on $\cosBY$, $\mmiss/(2\Emiss)$, and $\thetamiss$ do not significantly bias the $q^2$ distribution either.
However, the variable $\costhetaWL$ 
is correlated with the lepton momentum and thereby $q^2$. To ensure that the selection does not adversely affect the measurement of the $q^2$ spectrum, we have chosen rather moderate restrictions on
$\costhetaWL$. 

\begin{figure}[htb]
  \twoFigOneCol
      {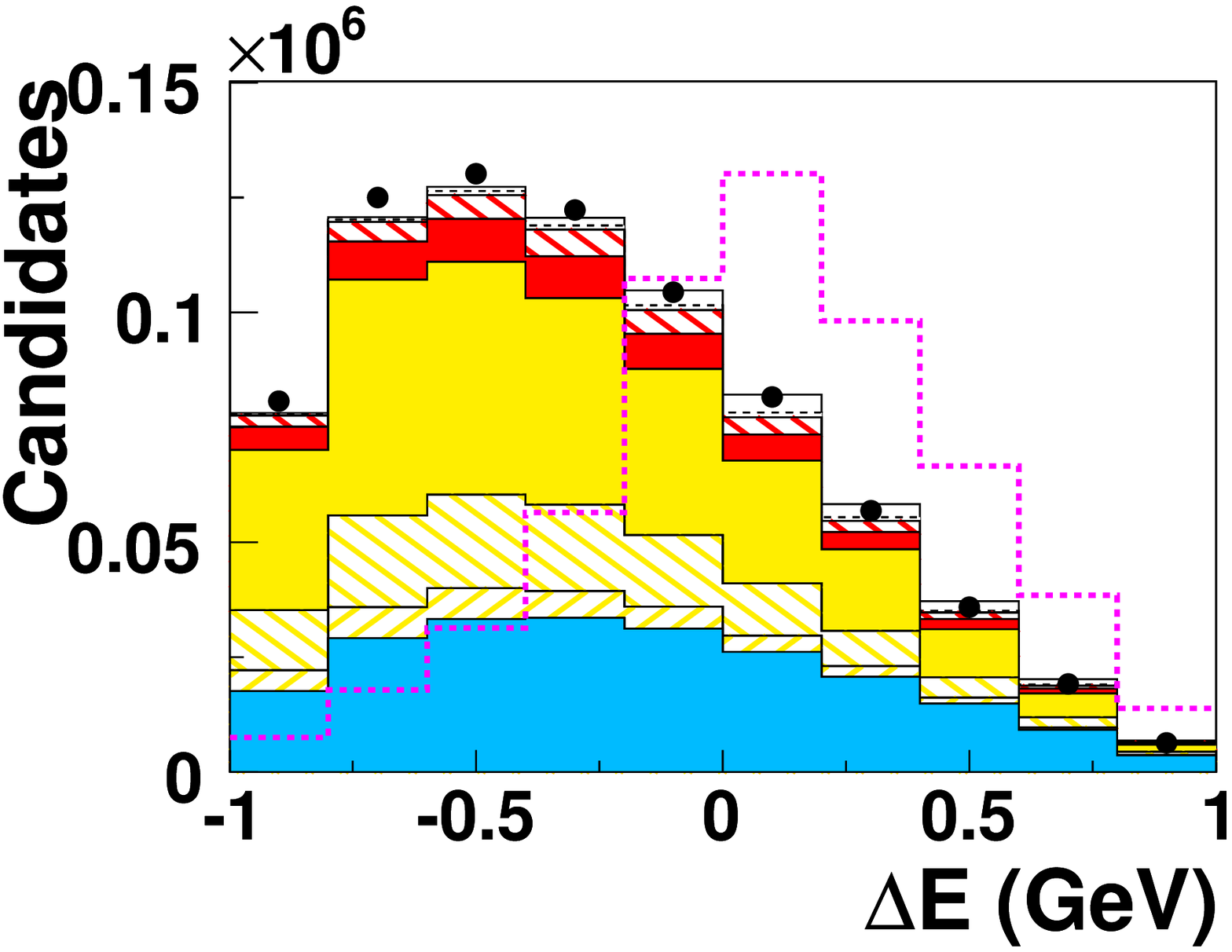}
      {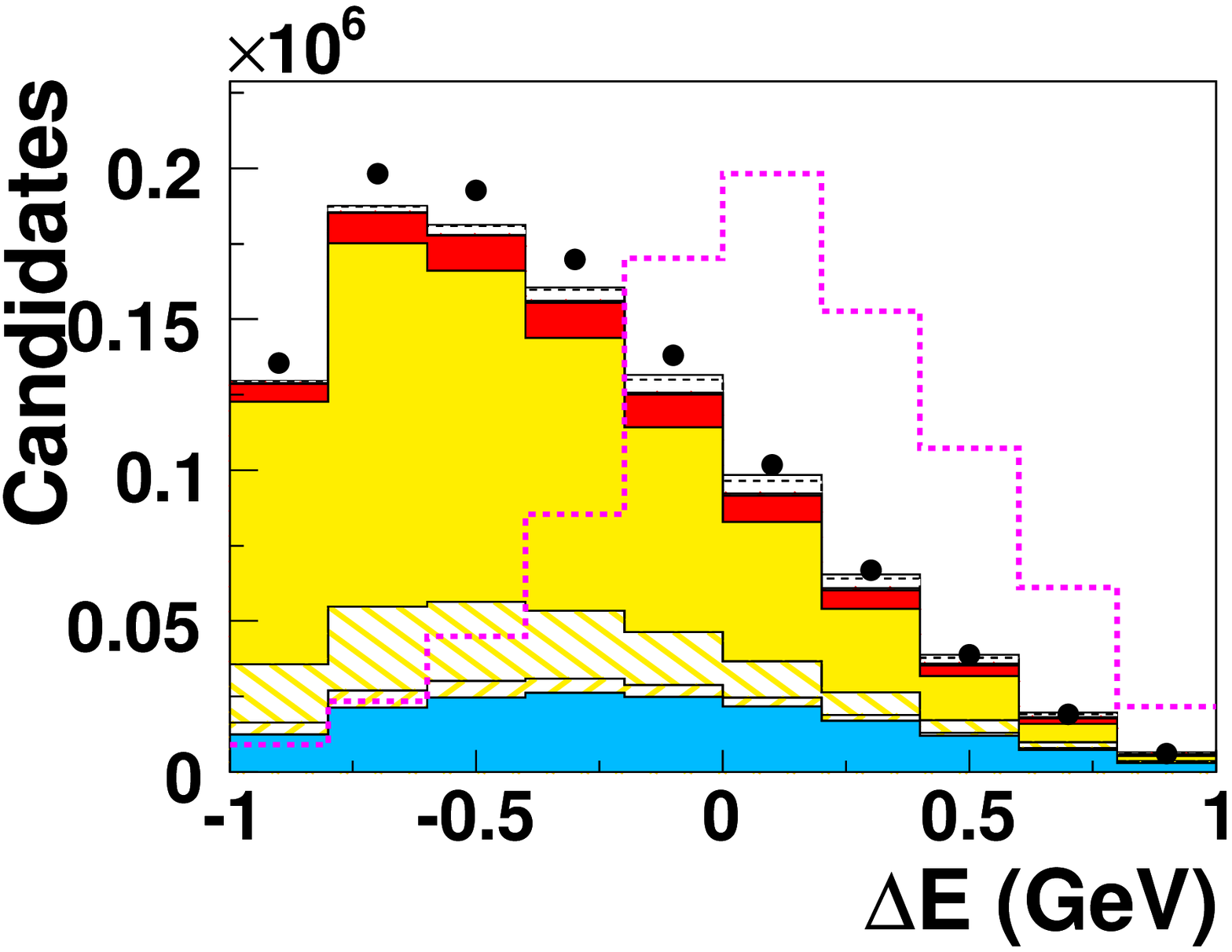}\\
  \twoFigOneCol
      {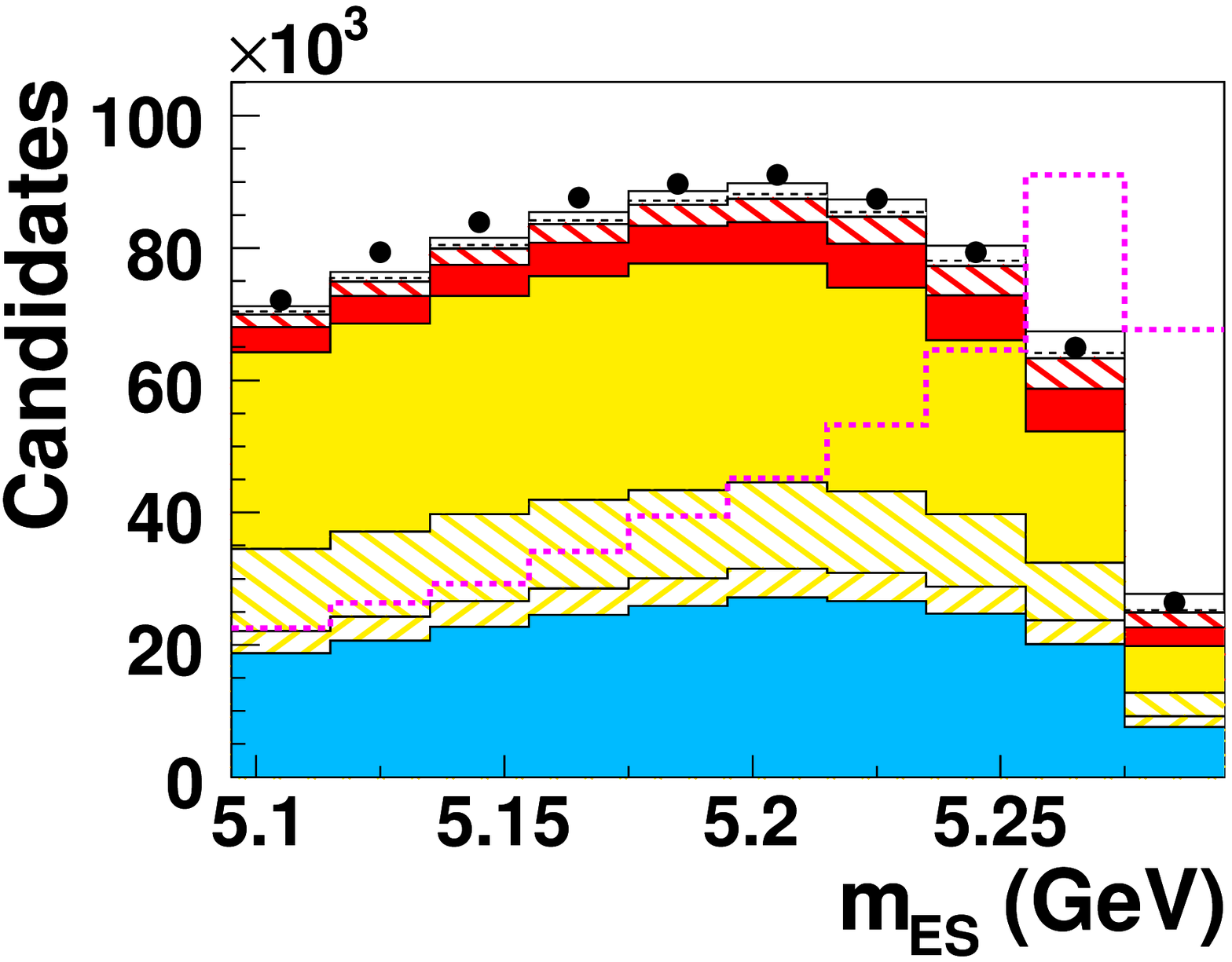}
      {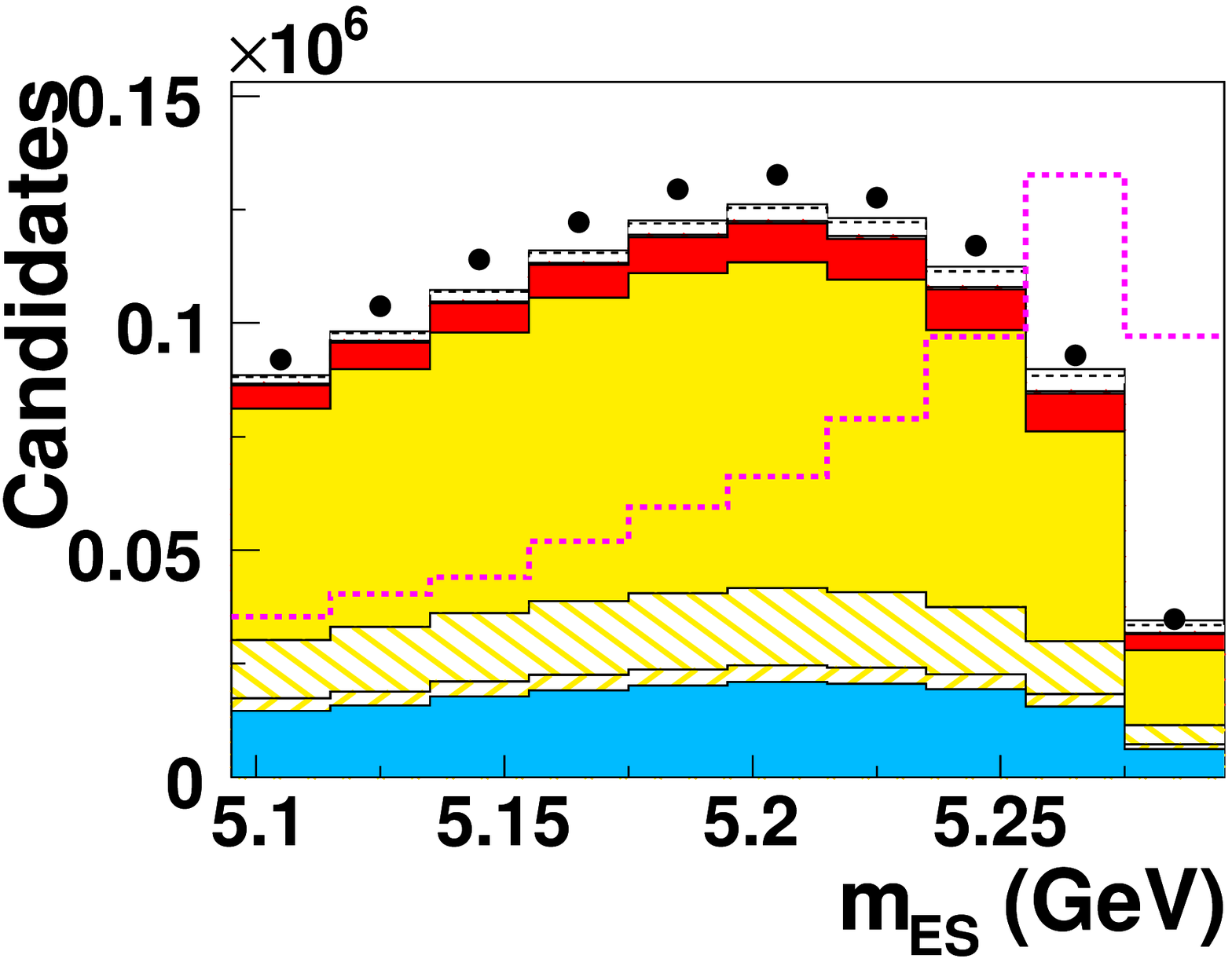}\\
  \vspace{0.2cm}    
  \includegraphics[width=0.4\columnwidth]{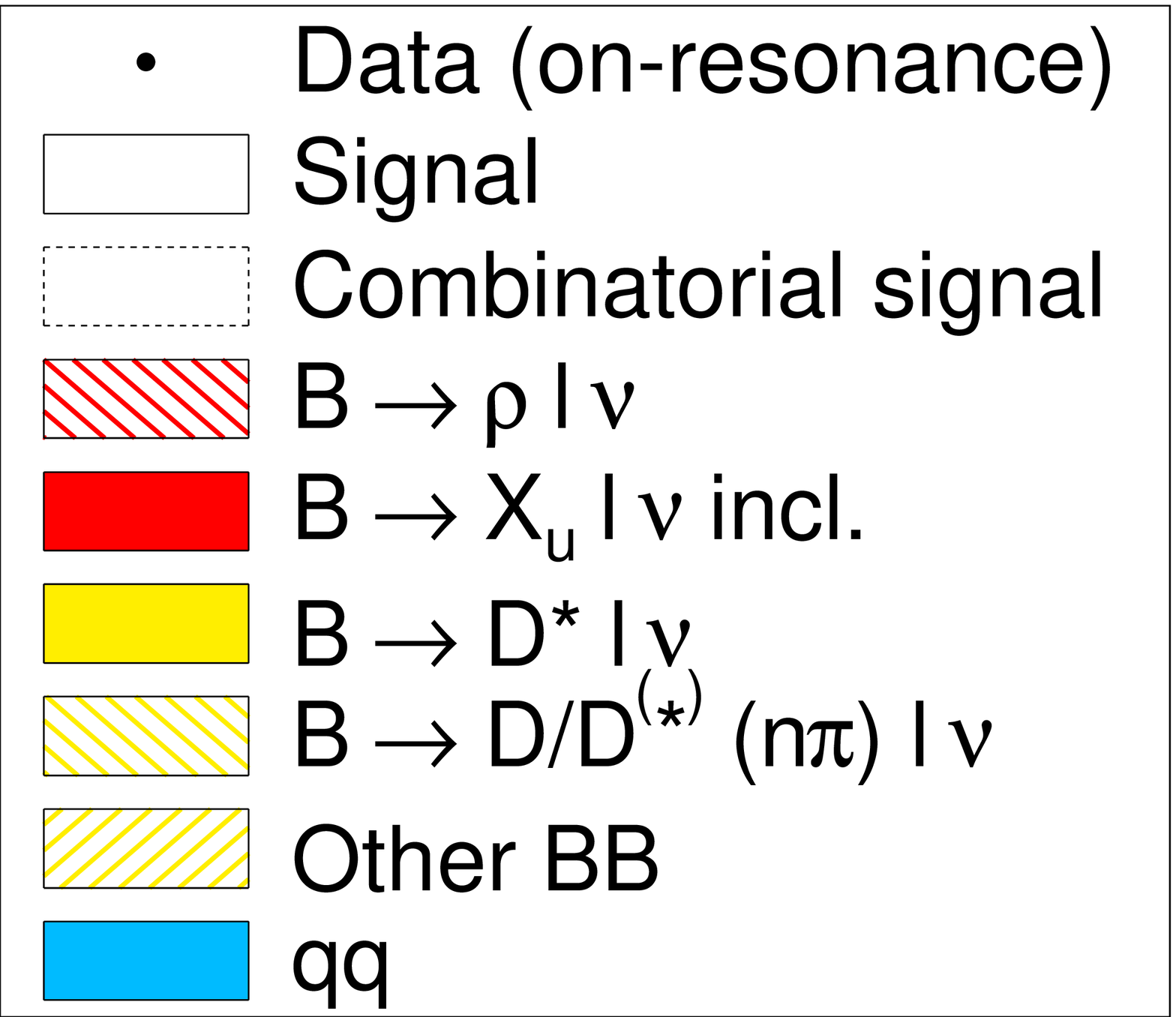}
  \hspace{0.05\columnwidth}
  \includegraphics[width=0.4\columnwidth]{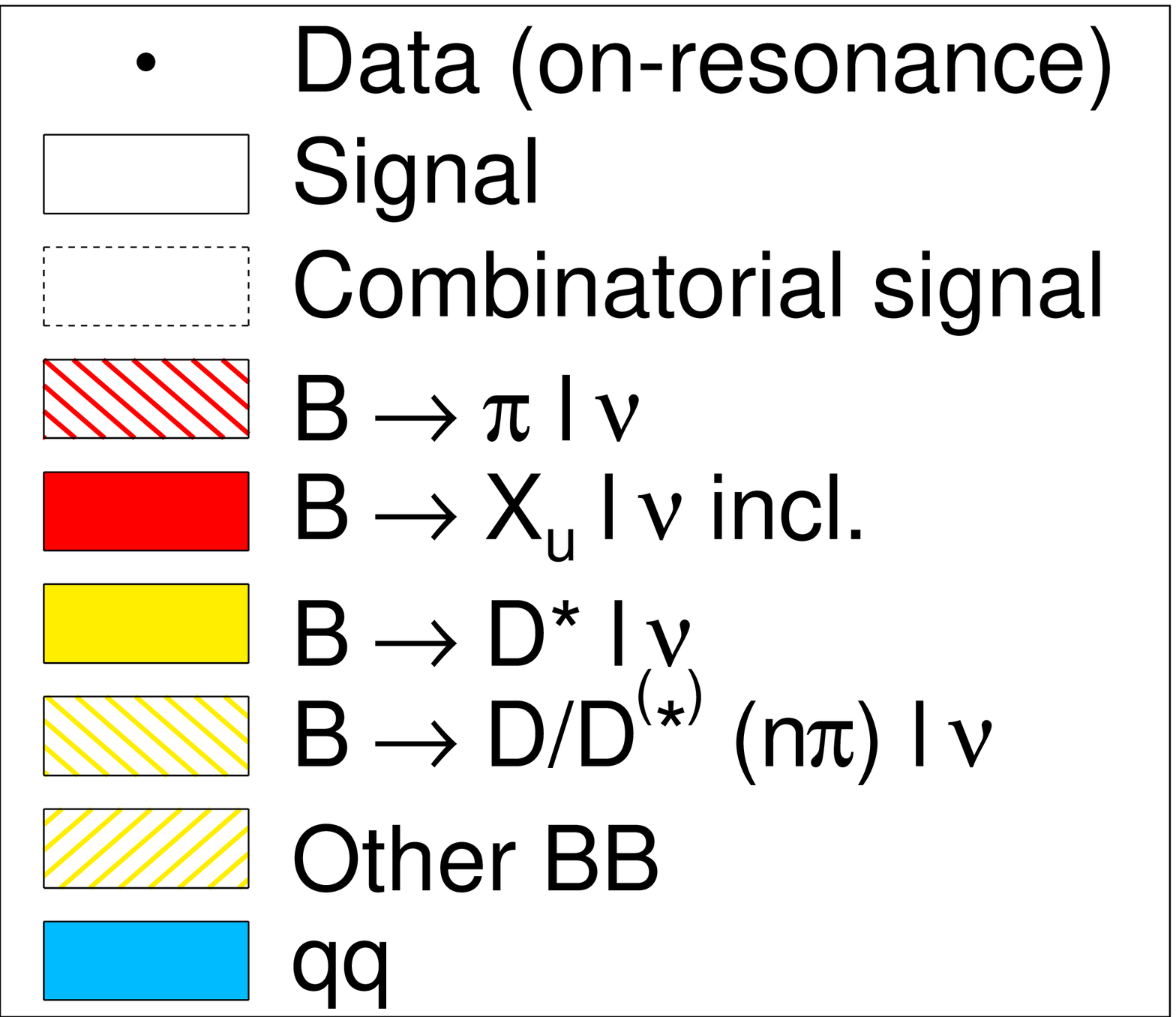}

  \caption{(color online)
  Distributions of \DeltaE and \mES for \Bzpilnu\ (left) and 
  \Bzrholnu\ (right) candidates after the preselection, {\it i.e.}, prior 
             to the neural-network application. 
             The stacked histograms show the predicted signal and background 
             contributions prior to the fit.
   The expected signal distribution (with arbitrary normalization) 
   is indicated as a magenta dashed histogram.
}
  \label{fig:legend}
\end{figure}

Figure~\ref{fig:legend} shows the \DeltaE and \mES distributions for samples of
\Bzpilnu\ and \Bzrholnu\ candidates (integrated over $q^2$) that have been preselected by the criteria described above. The stacked histograms show the signal and background contributions compared to the data, prior to the fit. 
The three dominant backgrounds are \bclnu\ decays (including $B \to D\ell\nu$, $B \to D^*\ell\nu$ and $B \to D^{(*)}(n\pi)\ell\nu$),
\qq\ continuum and \bulnu\ decays.
The signal contributions are very small by comparison and difficult to observe.

The neural networks are trained separately for the three background categories
and for different $q^2$ intervals. We introduce six bins in $q^2$ for \Bpilnu\ and
three bins for \Brholnu. The bin sizes are $4\gev^2$ for \Bpilnu\ and $8\gev^2$ for \Brholnu, 
except for the last bin, which extends to the kinematic limit of $26.4\gev^2$ and $20.3\gev^2$, respectively. 
Thus in total we train $3\times (2\times 6 + 2\times 3) = 54$ neural networks.
Since we aim for a good signal-to-background ratio in the region where
most of the signal is located,
we do not train the neural network with events in the whole fit region, 
but in an extended signal region, 
$-0.25 < \DeltaE < 0.35\gev,  5.240 < \mES < 5.295 \gev$.

For the training of the neural networks we use MC simulated events containing correctly reconstructed signal decays and the following simulated background samples:
\begin{enumerate}
\item a sample of continuum events, $\ep\en \to \qqbar$ with $q=u,d,s,c$ (\qq\ neural network); 
\item a combined sample of $\bclnu$ decays (\bclnu\ neural network); and
\item a sample of inclusive $\bulnu$ decays (\bulnu\ neural network).
\end{enumerate}
\noindent
The training of the neural networks and the subsequent background reduction is performed sequentially for the three background samples.
We use subsamples of typically less than half the total MC samples for training and validation of the neural networks.
Of these subsamples, one half of the events is used as training sample, and the other half for validation.

Studies of the neural-network performance for the $\bulnu$ background indicate that the separation of this background from the signal is very difficult because of the similarity in the shape of the distributions, especially for the
\Bppizlnu\ and the \Brholnu\ samples.   
Given these difficulties, 
we use the \bulnu\ neural network only for the 
$\Bzpilnu$ sample, and only for 
$q^2 > 12 \gev^2$, where the $\bulnu$ background becomes significant. 

\begin{figure*}    
  \centering
  \begin{tabular}{ccc}
    \begin{minipage}{0.33\linewidth}
      \epsfig{file=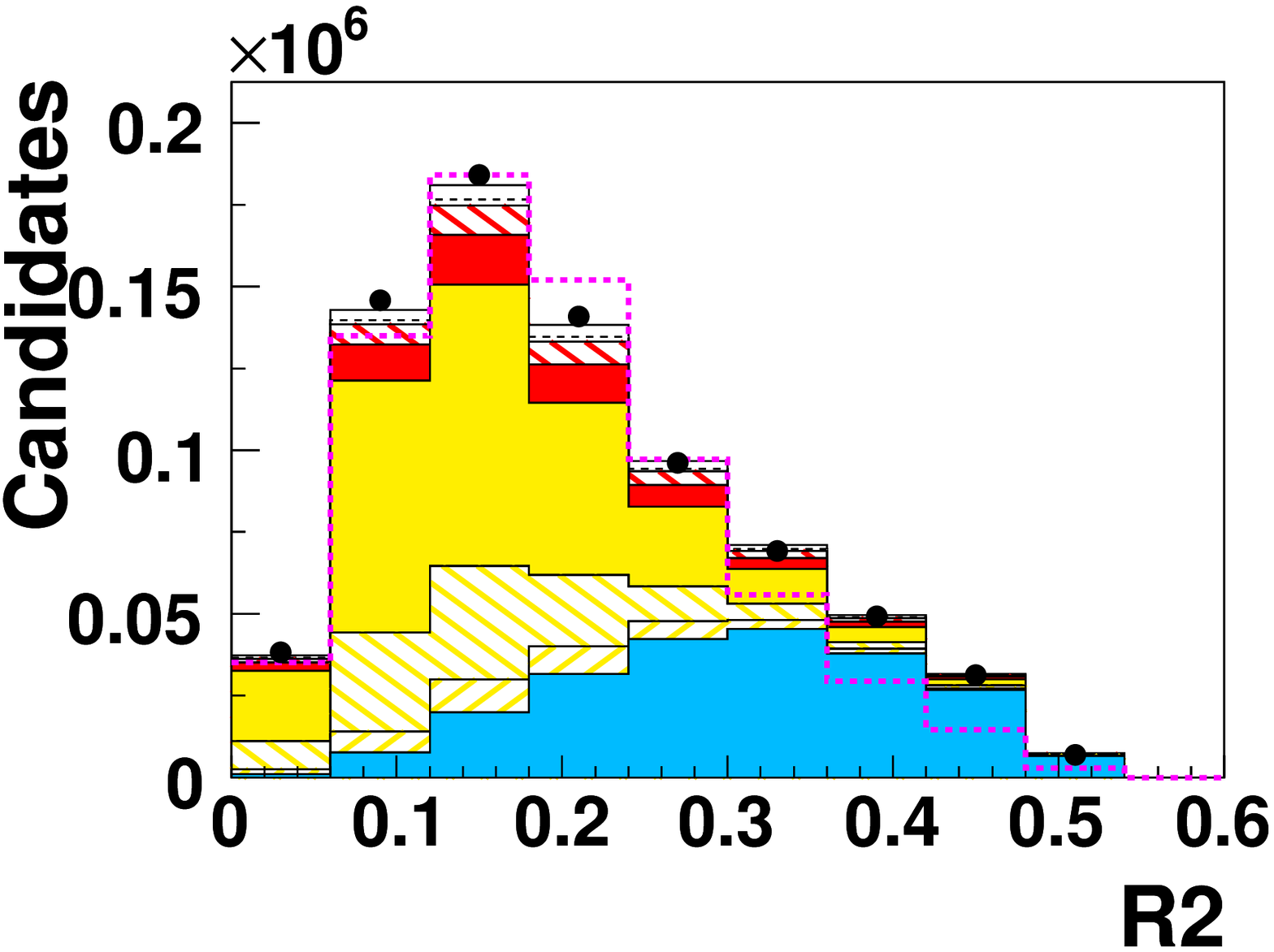, width = 4.5cm}
      \epsfig{file=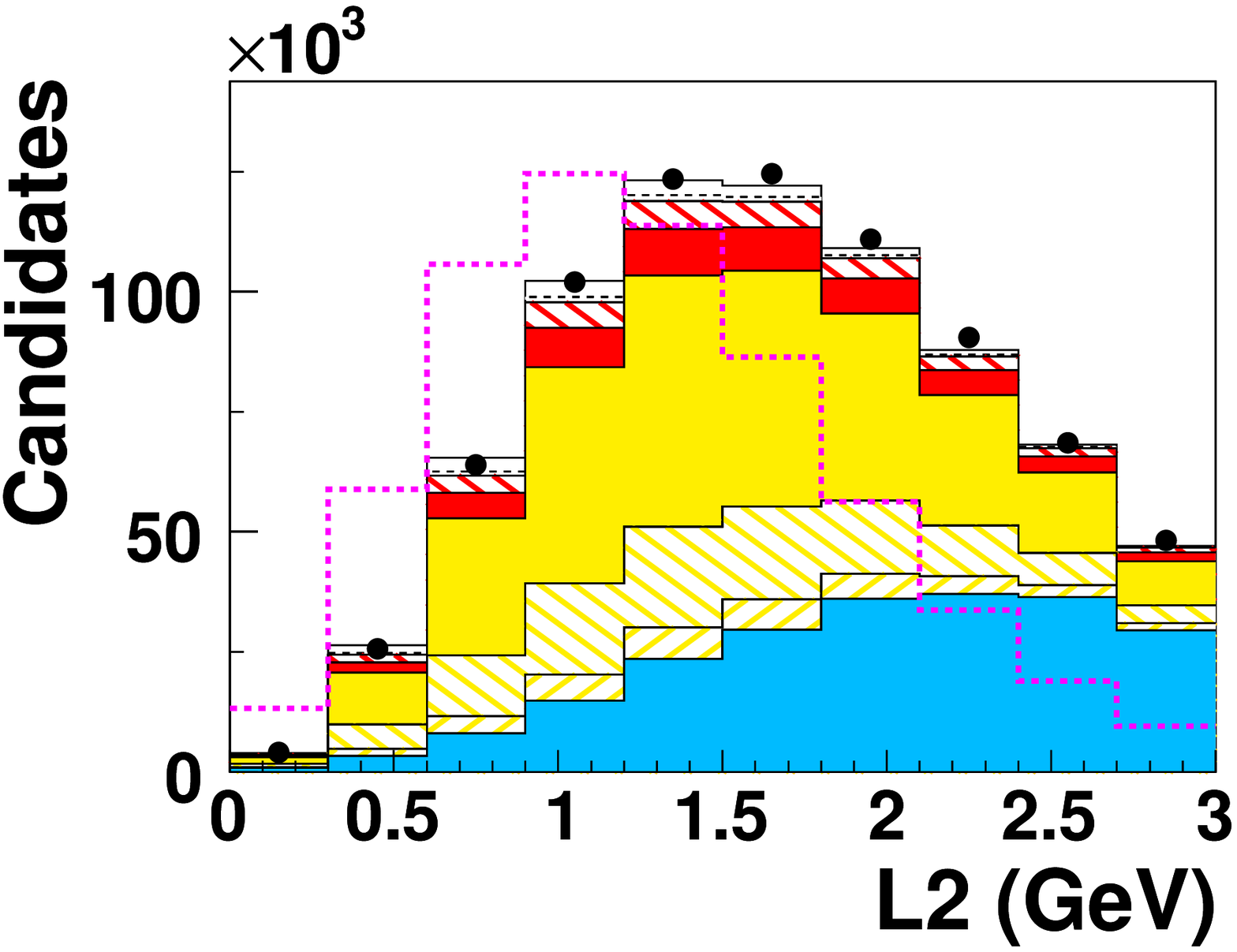, width = 4.5cm}
      \epsfig{file=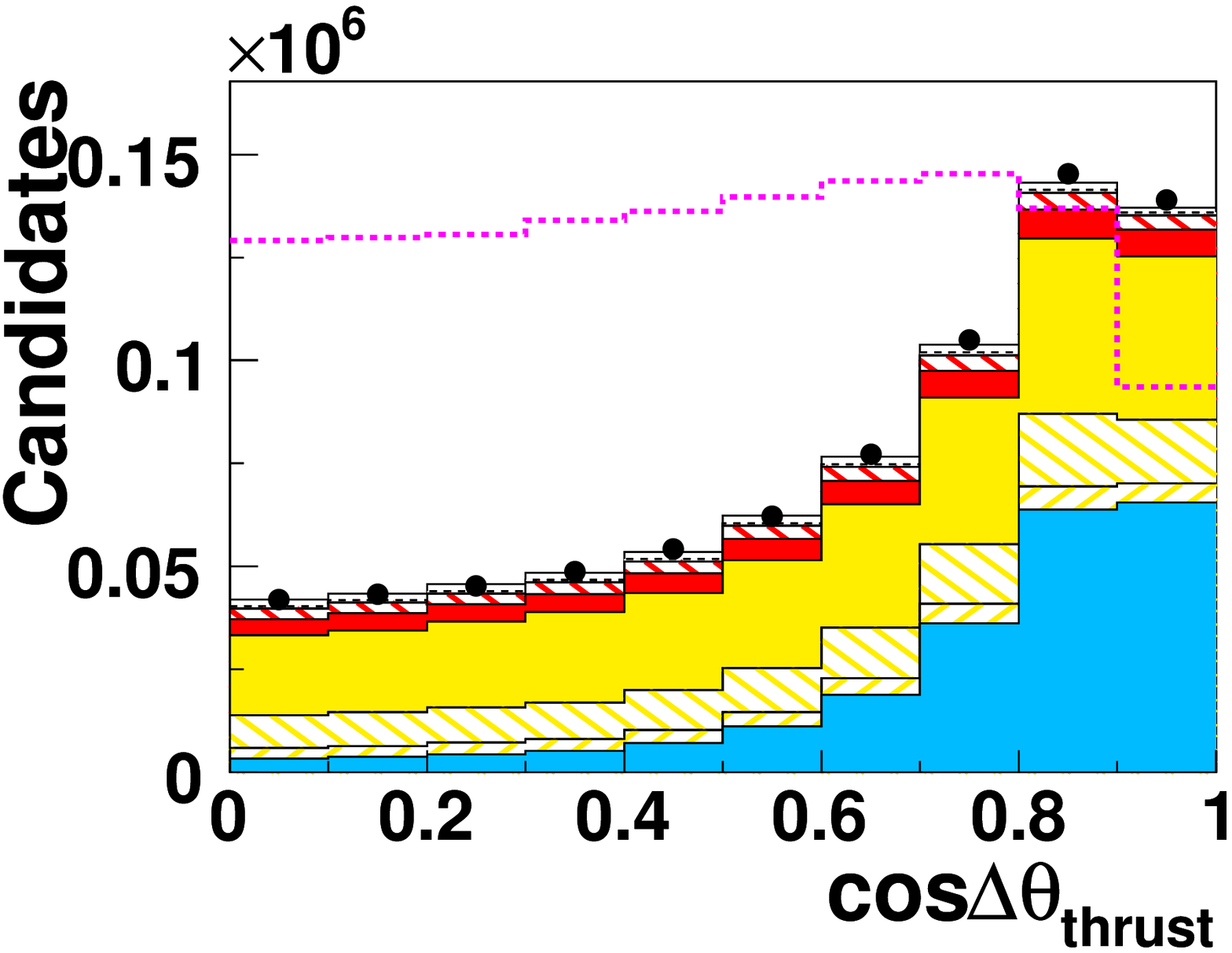, width = 4.5cm}
      \epsfig{file=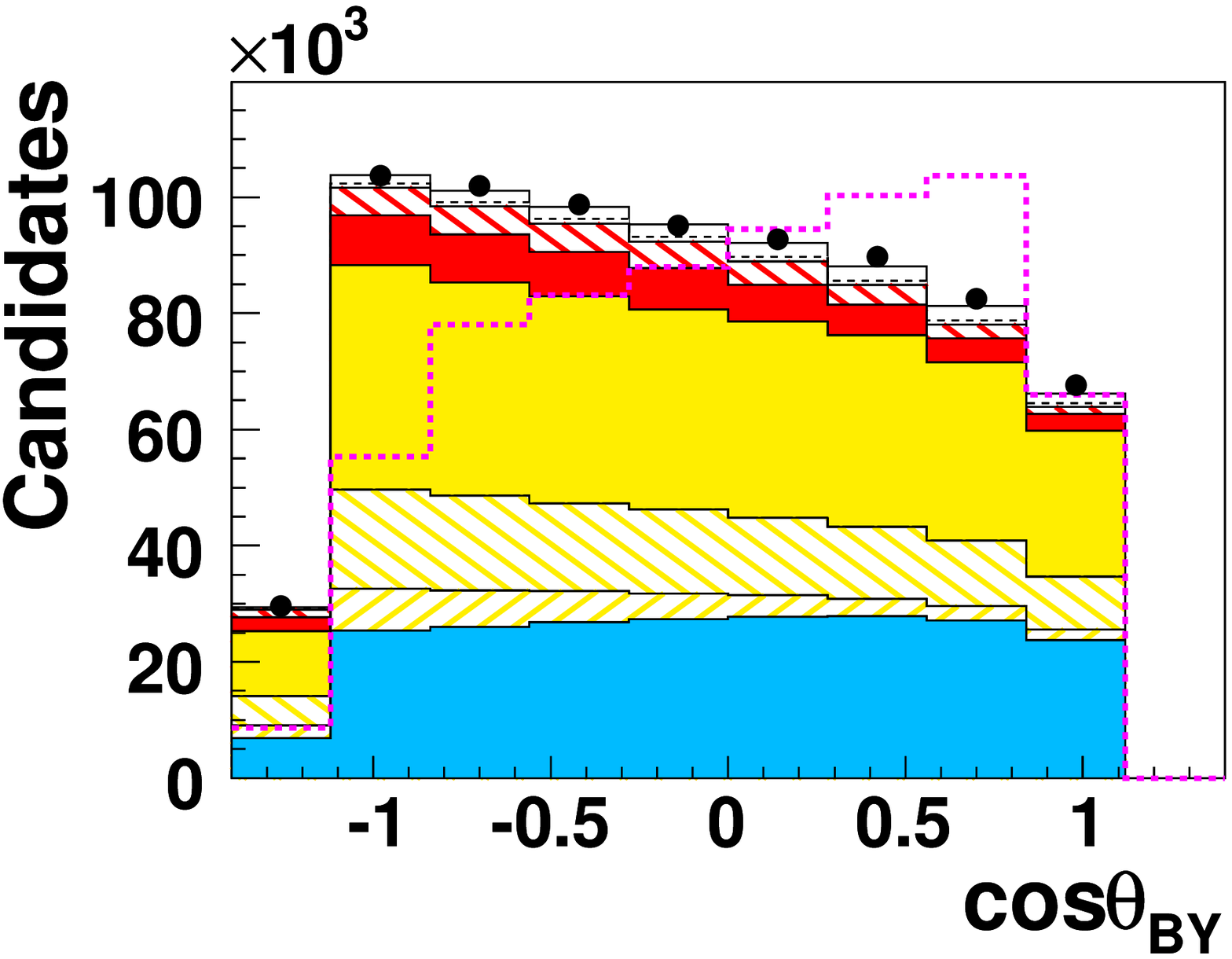, width = 4.5cm}
      \epsfig{file=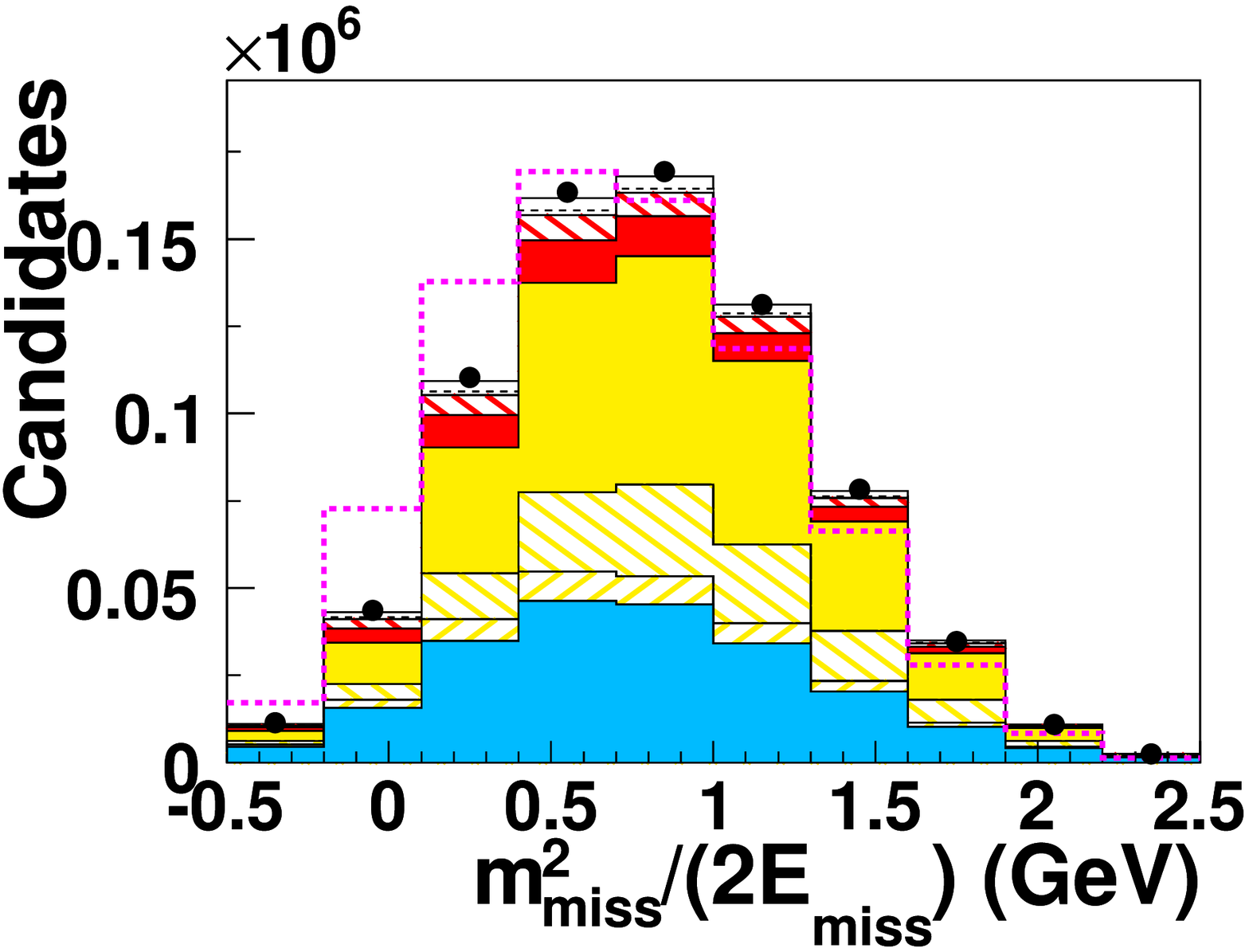, width = 4.5cm}
      \epsfig{file=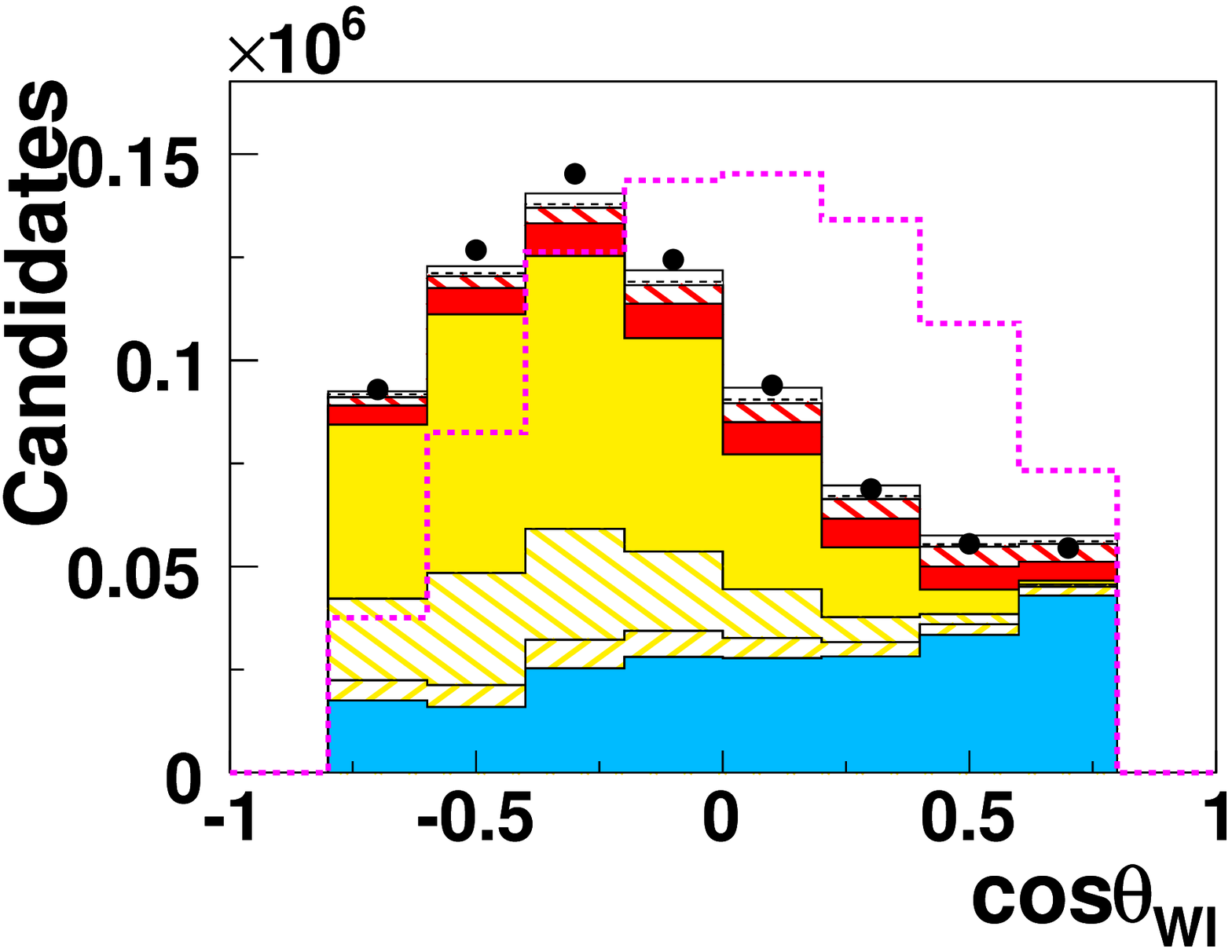, width = 4.5cm}
      \epsfig{file=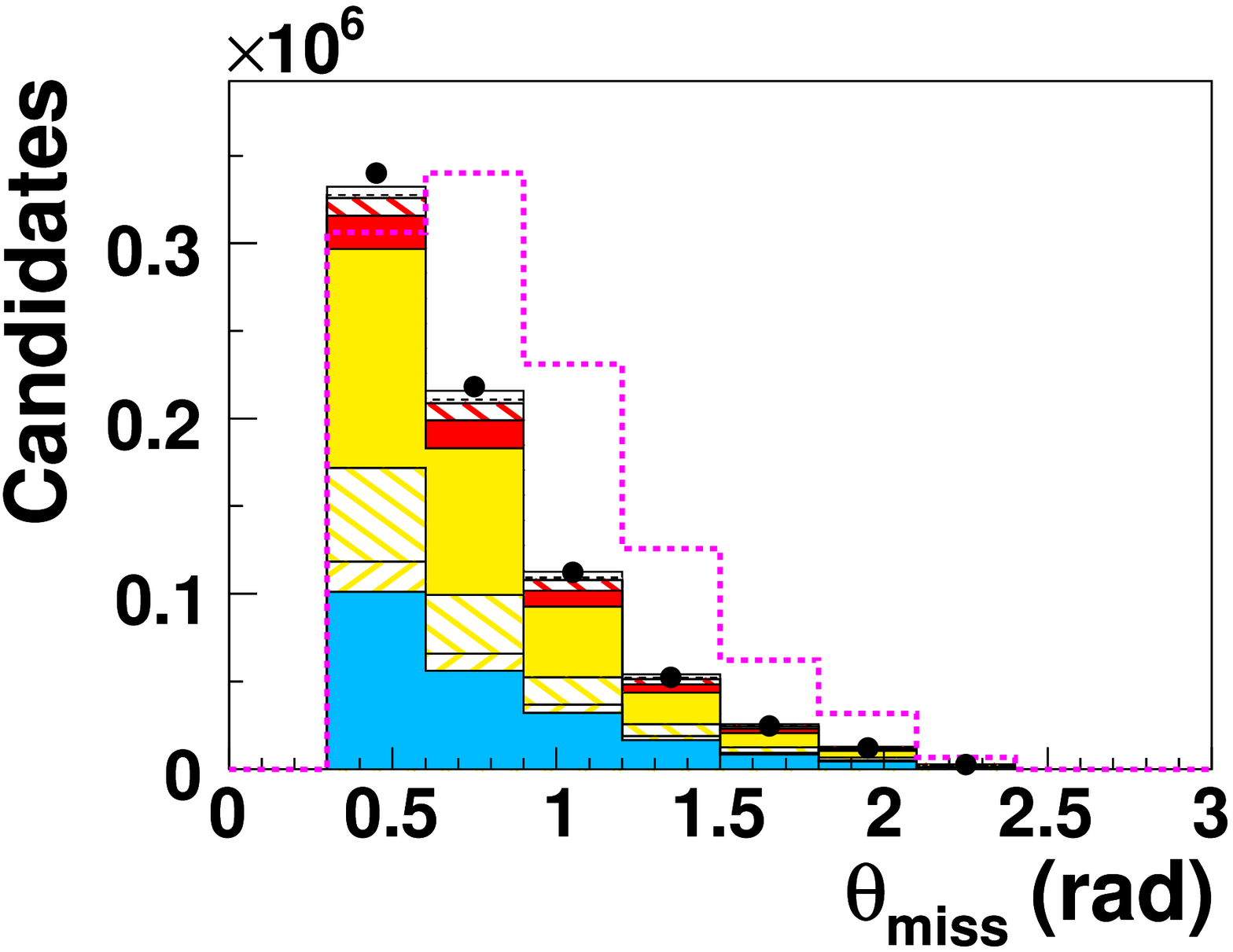, width = 4.5cm}
    \end{minipage}
    \begin{minipage}{0.33\linewidth}
      \epsfig{file=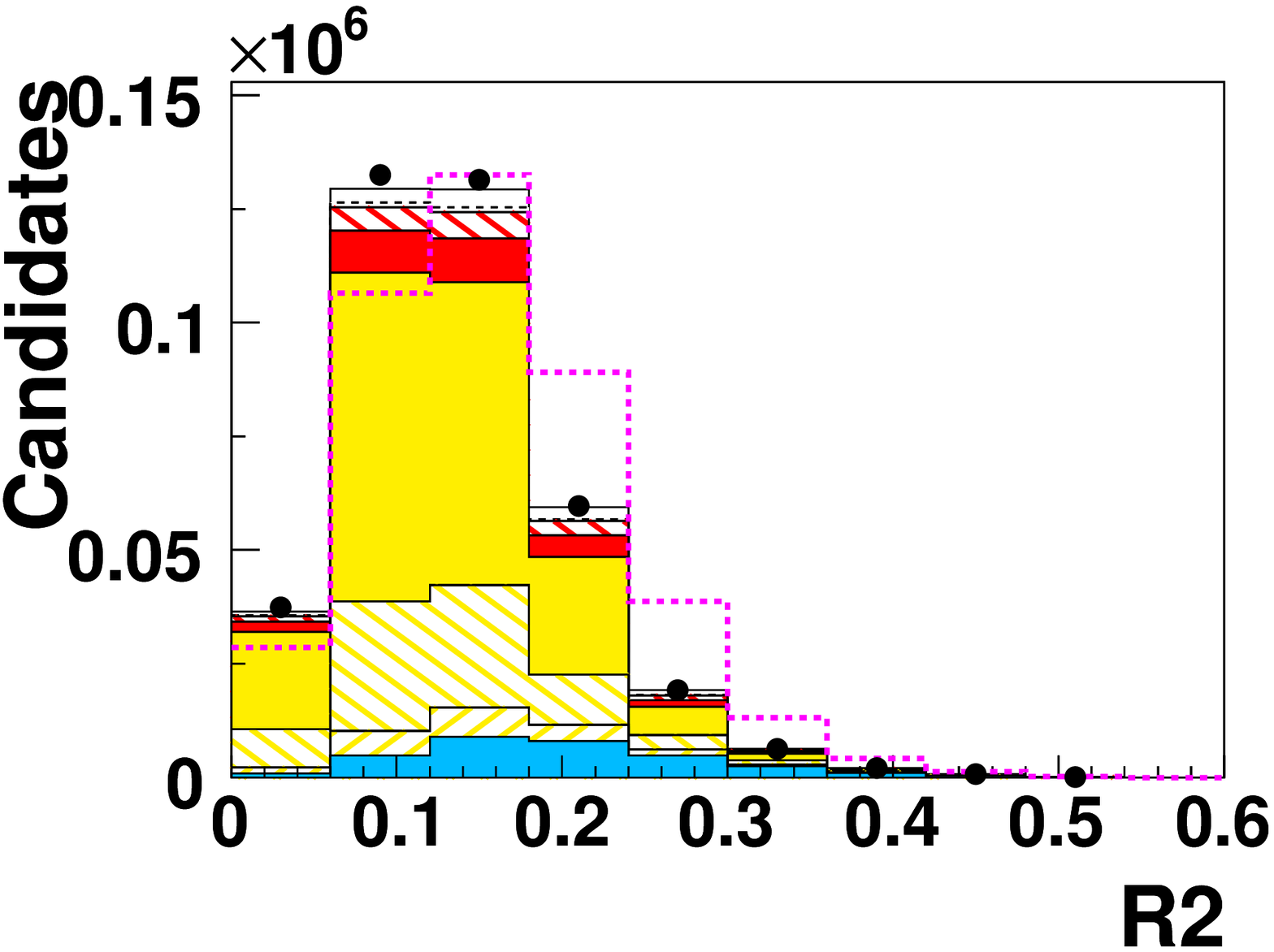, width = 4.5cm}
      \epsfig{file=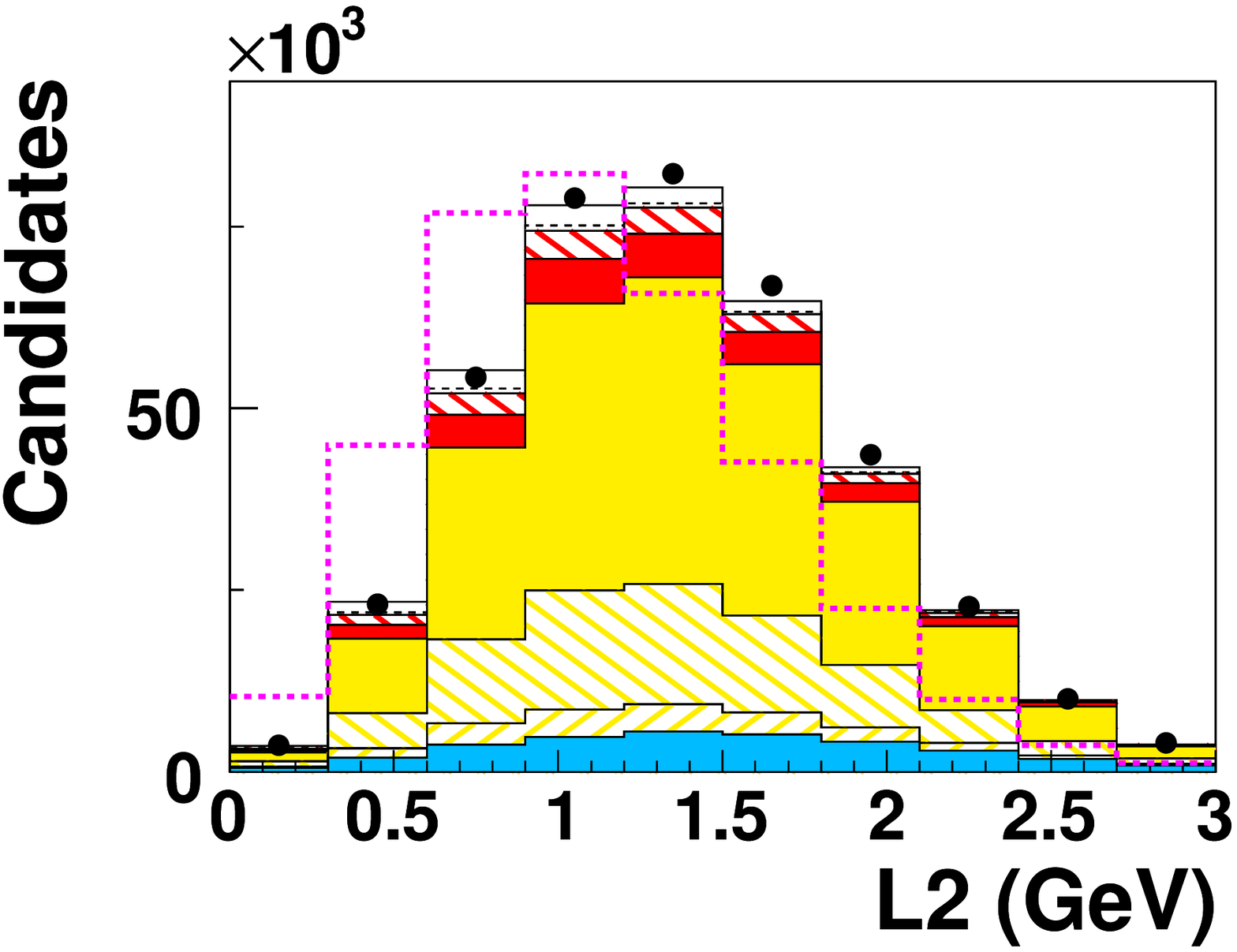, width = 4.5cm}
      \epsfig{file=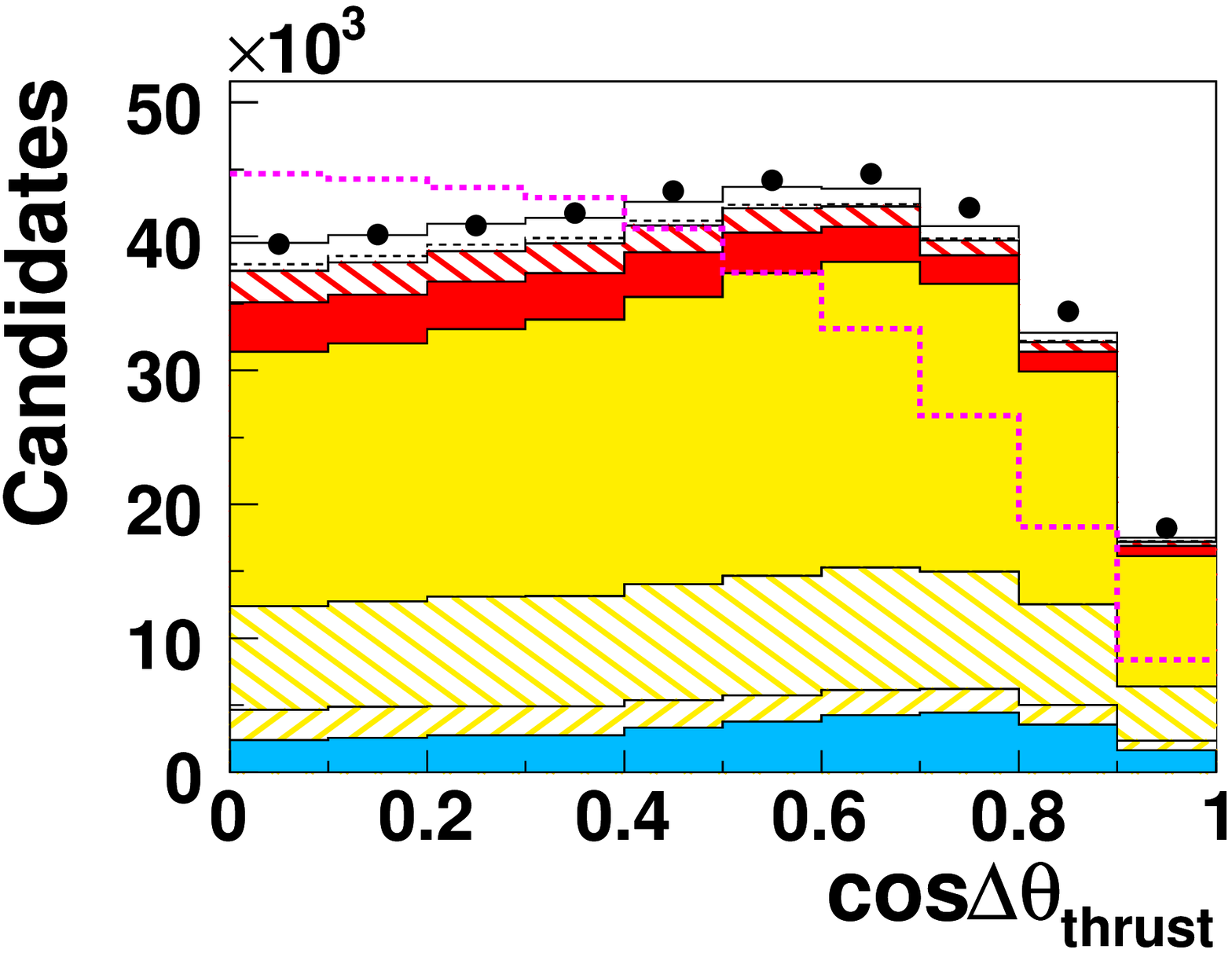, width = 4.5cm}
      \epsfig{file=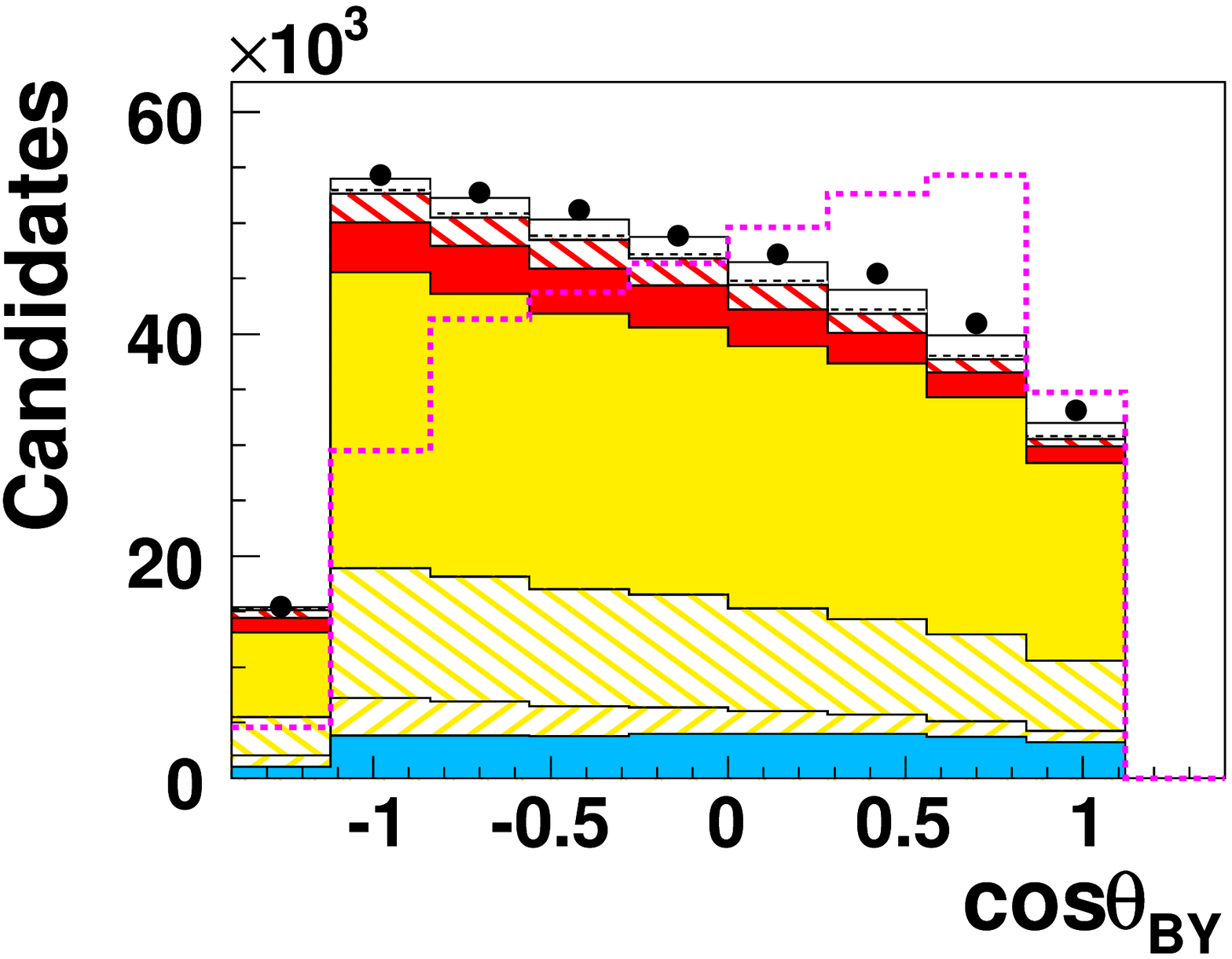, width = 4.5cm}
      \epsfig{file=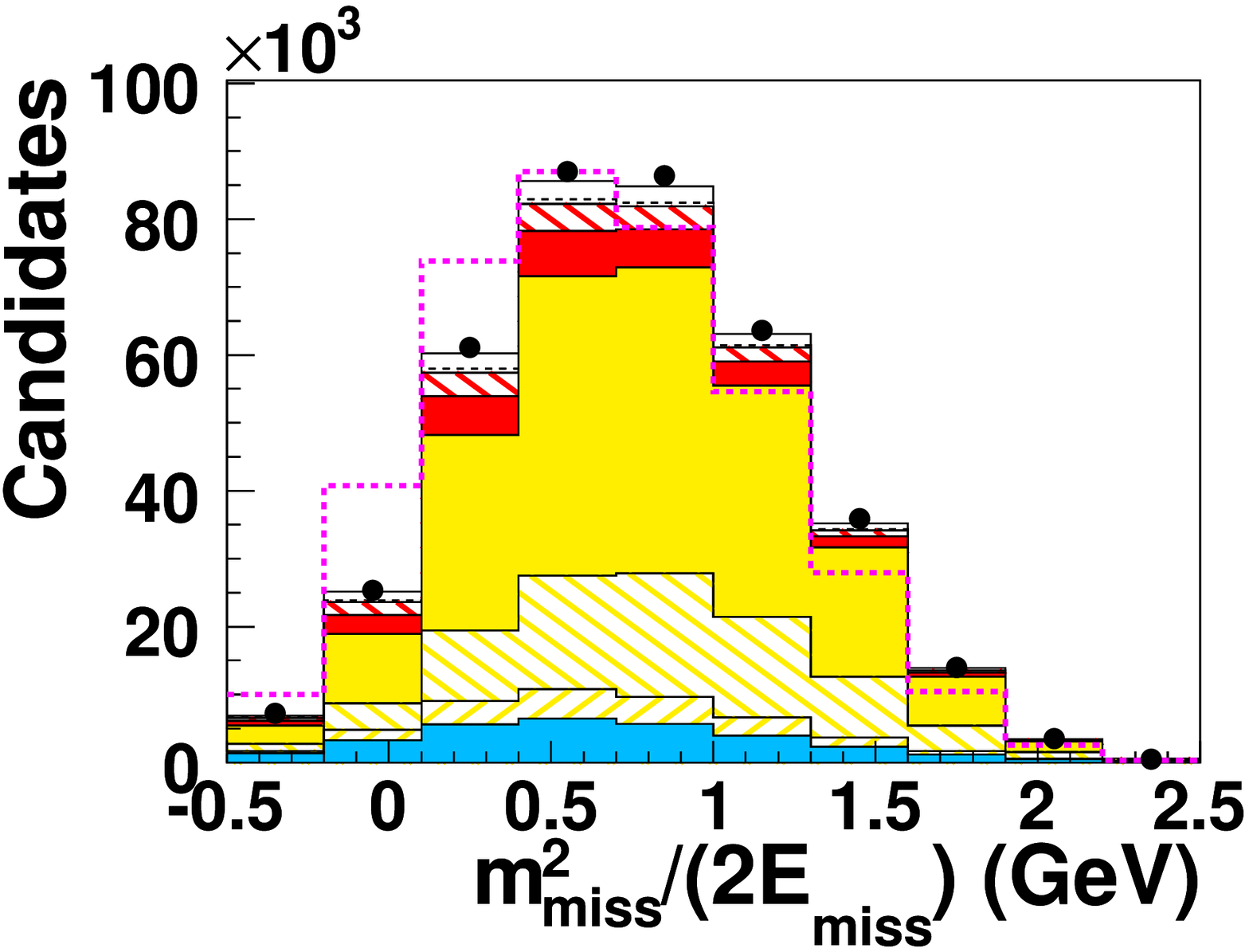, width = 4.5cm}
      \epsfig{file=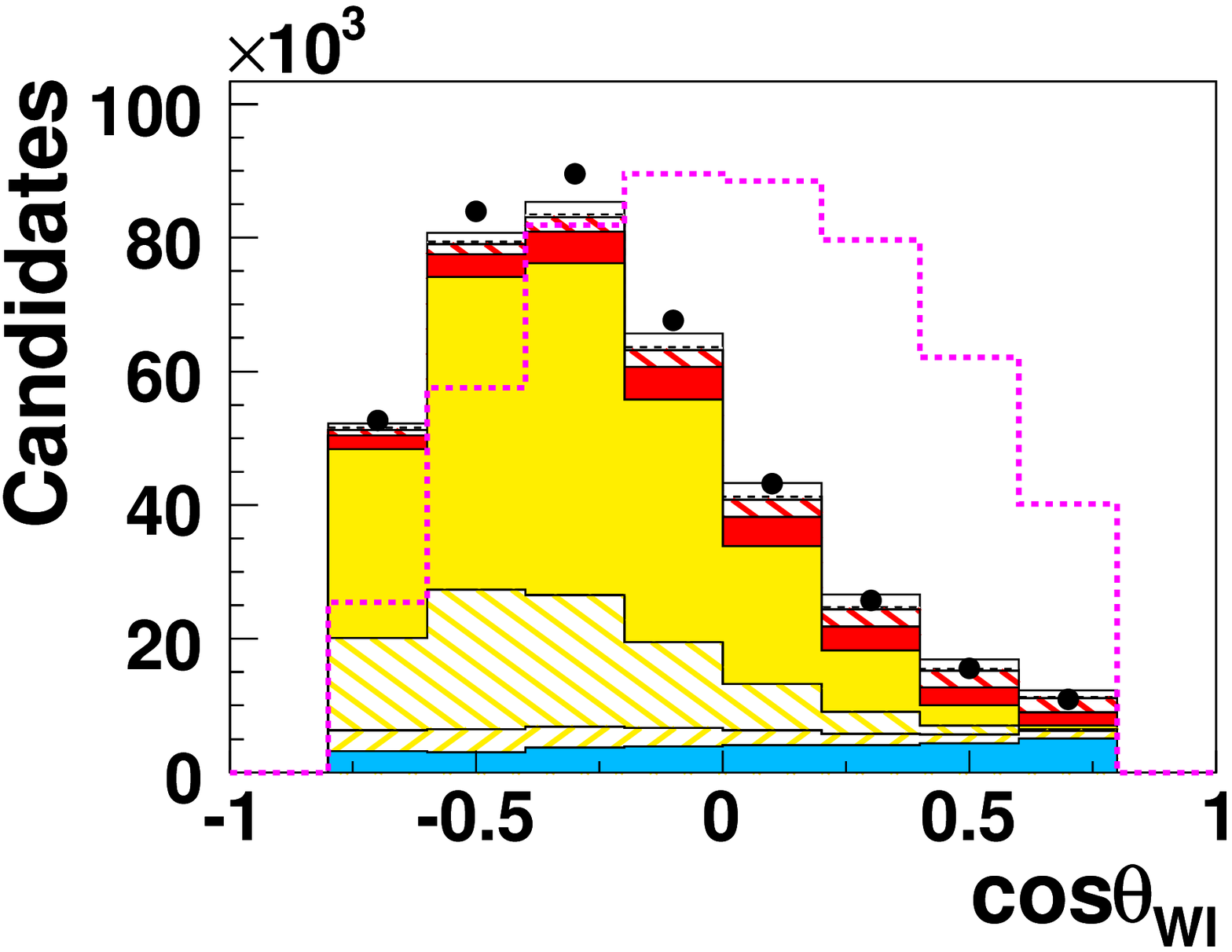, width = 4.5cm}
      \epsfig{file=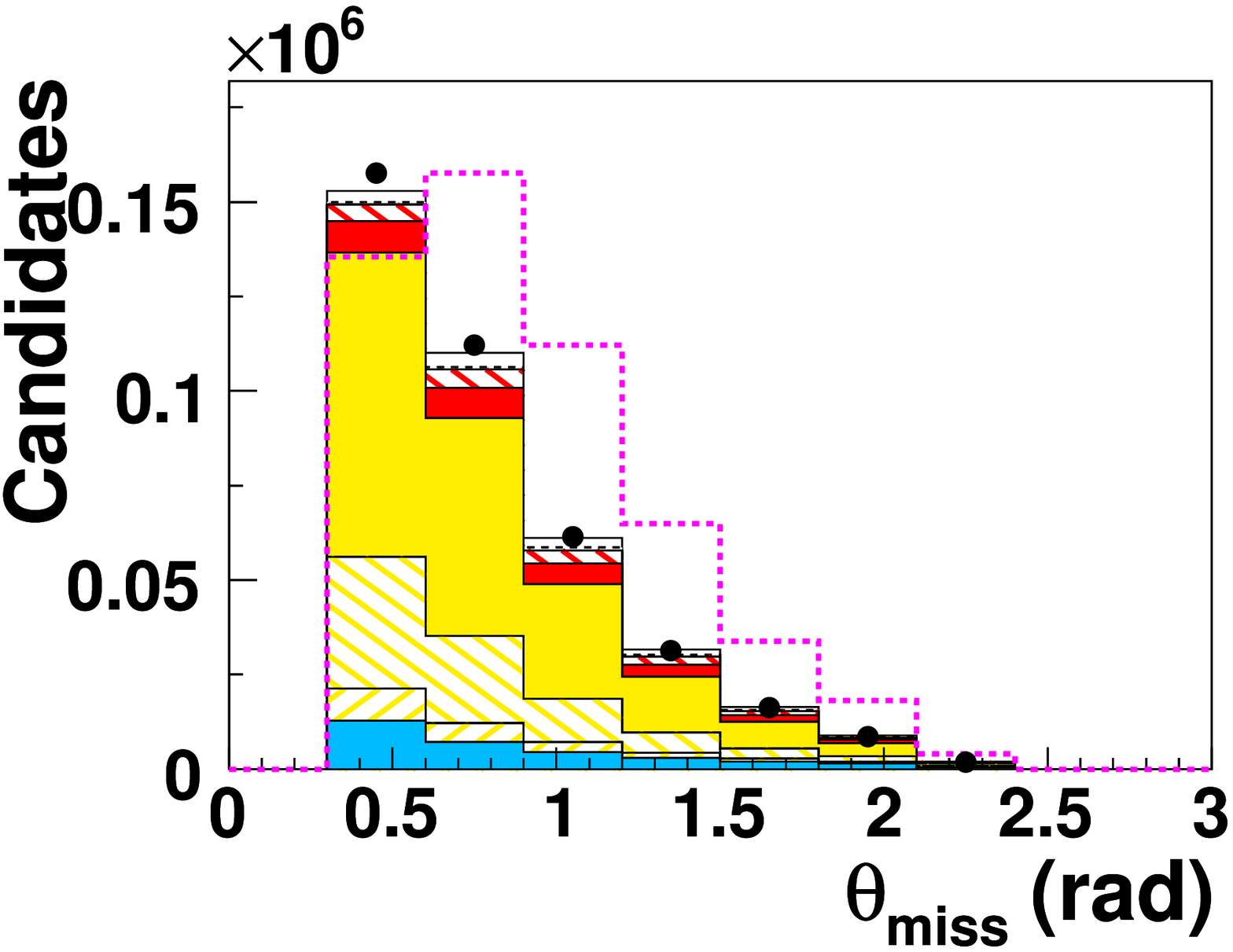, width = 4.5cm}
    \end{minipage}
    \begin{minipage}{0.33\linewidth}
      \epsfig{file=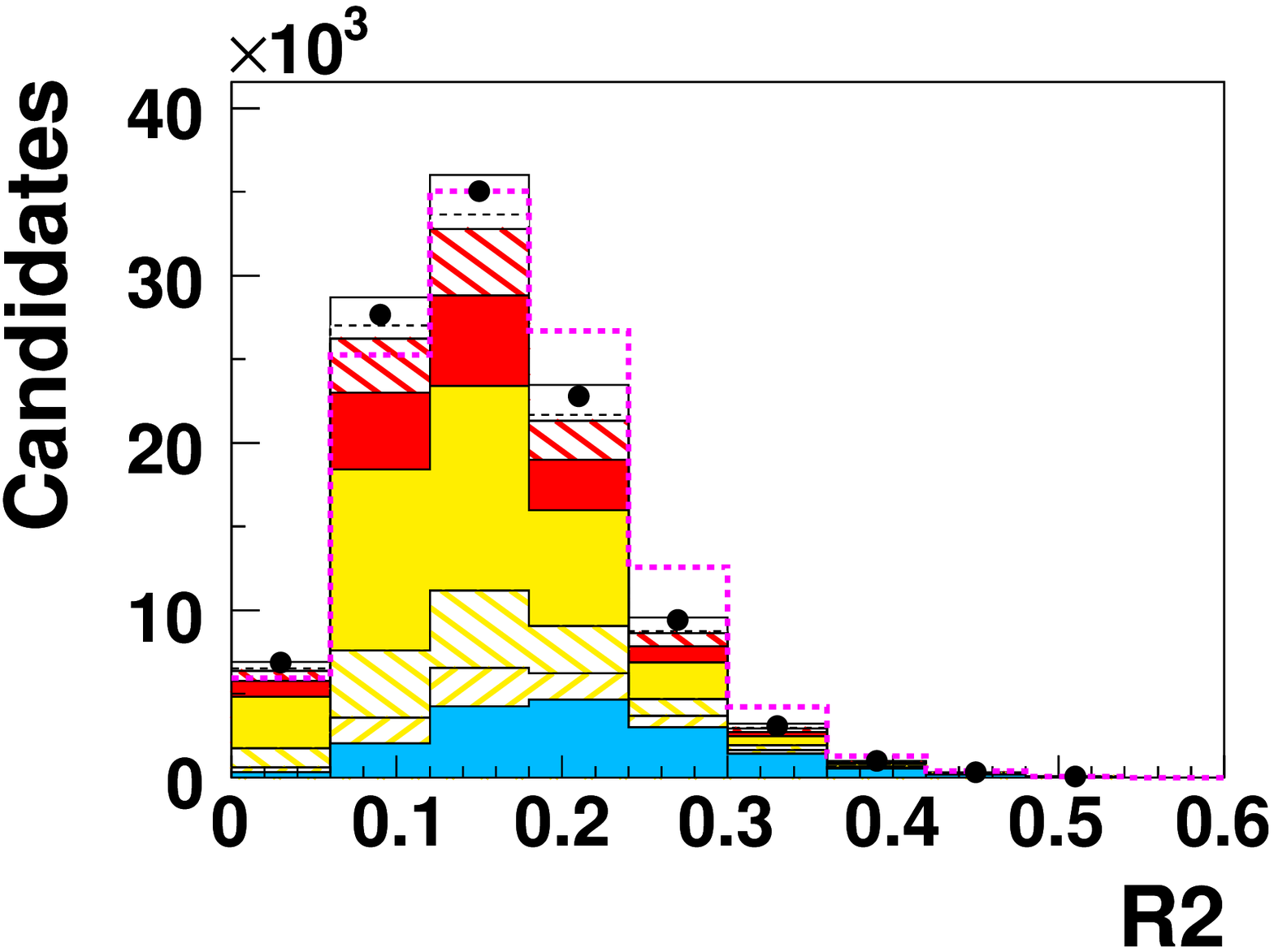, width = 4.5cm}
      \epsfig{file=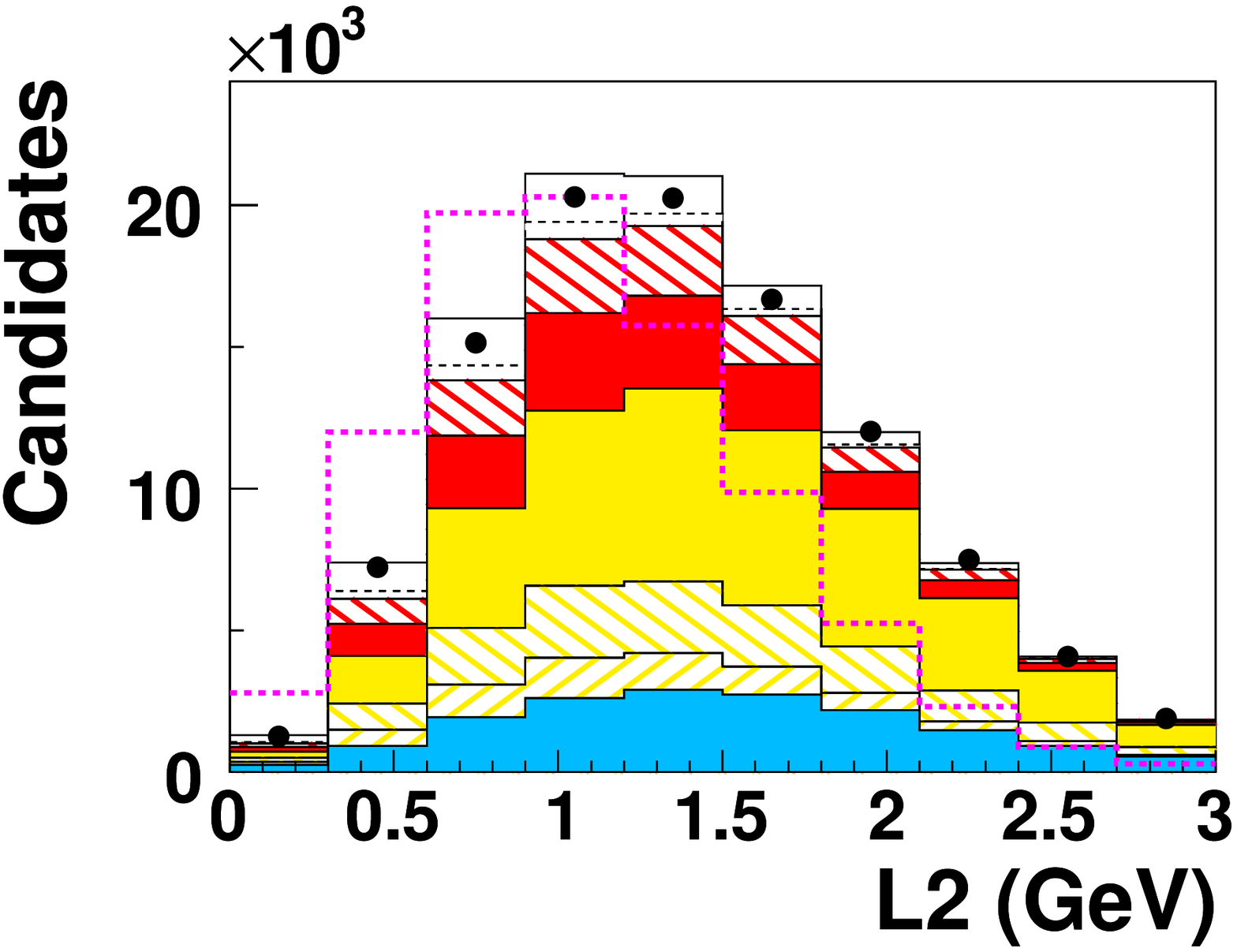, width = 4.5cm}
      \epsfig{file=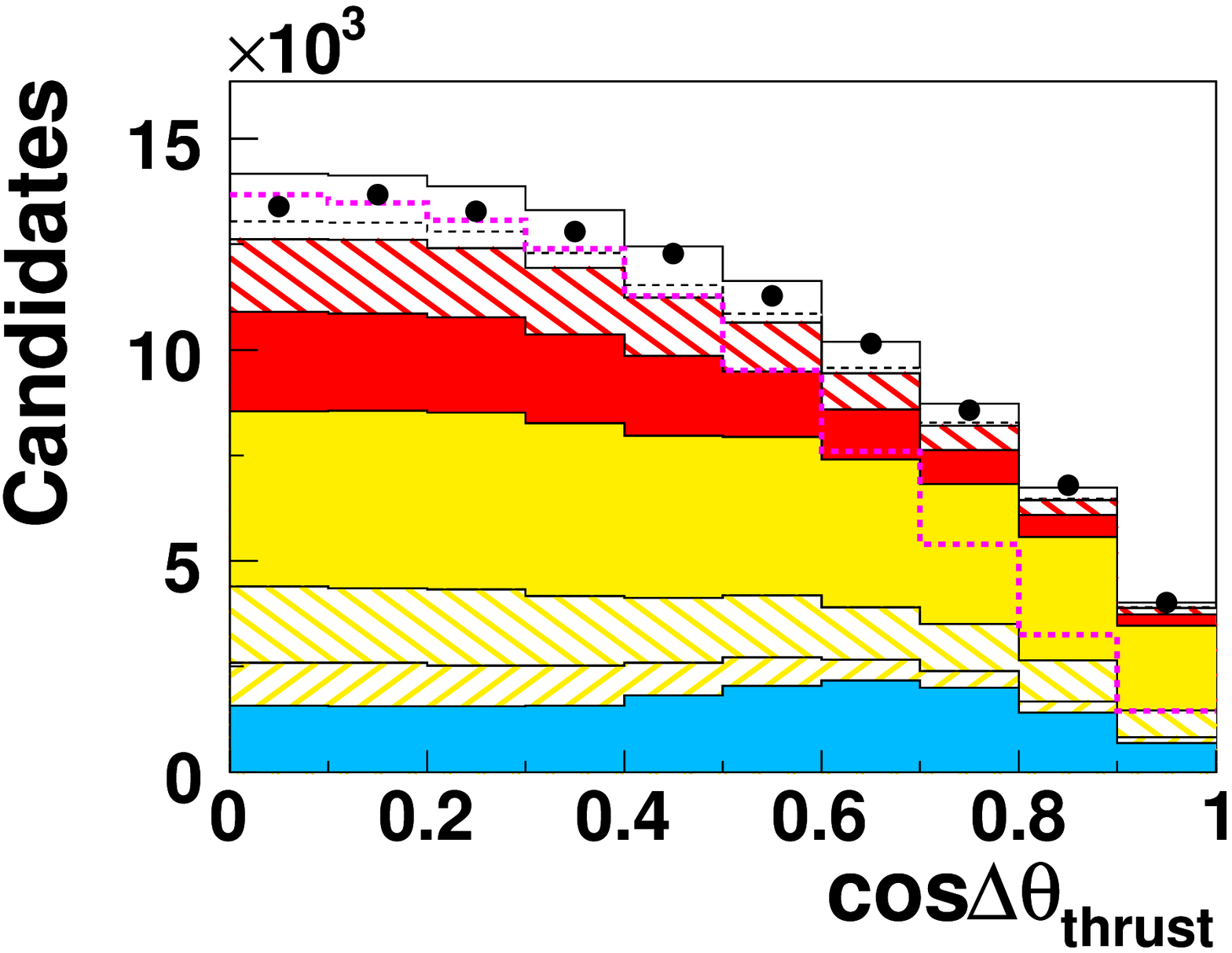, width = 4.5cm}
      \epsfig{file=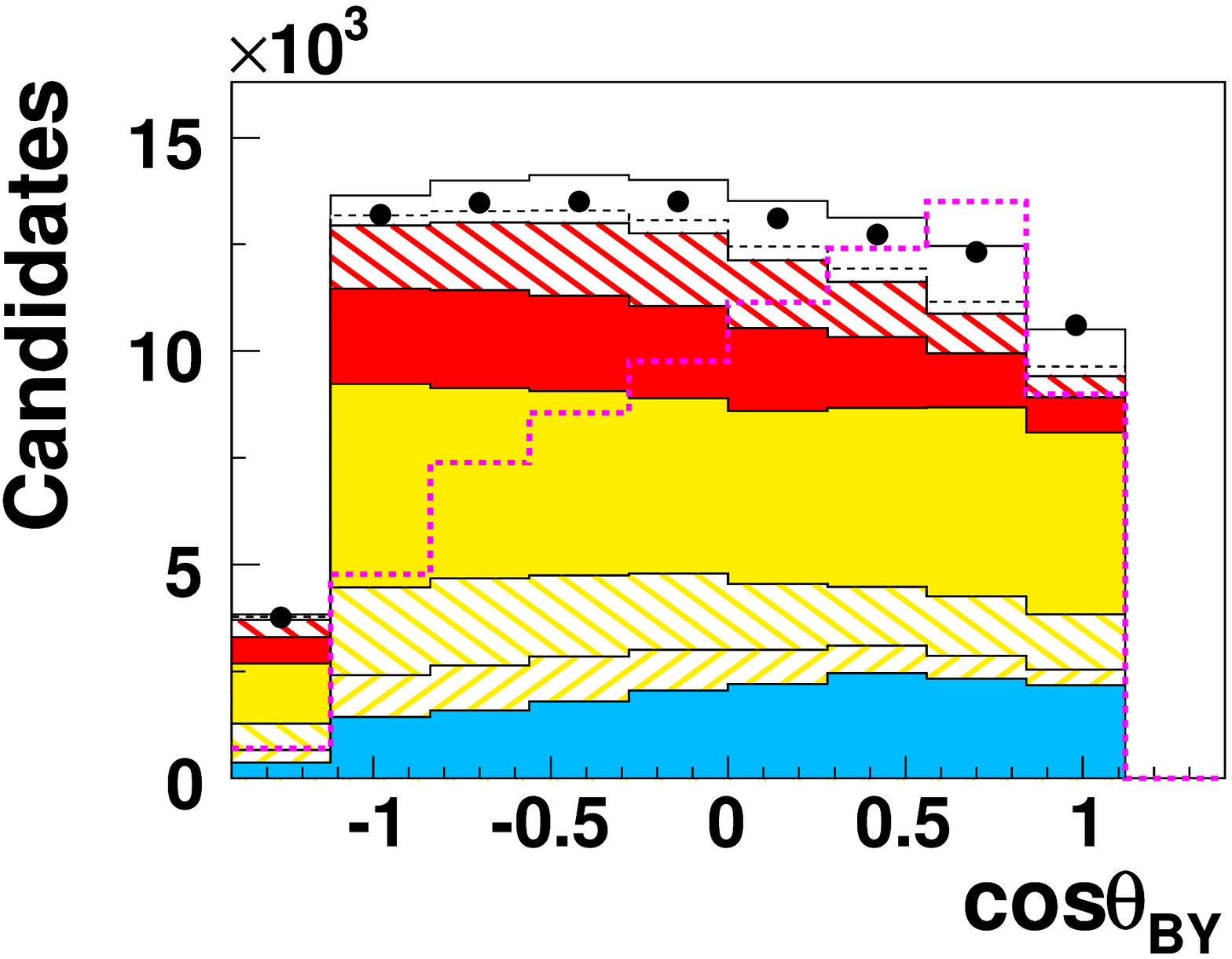, width = 4.5cm}
      \epsfig{file=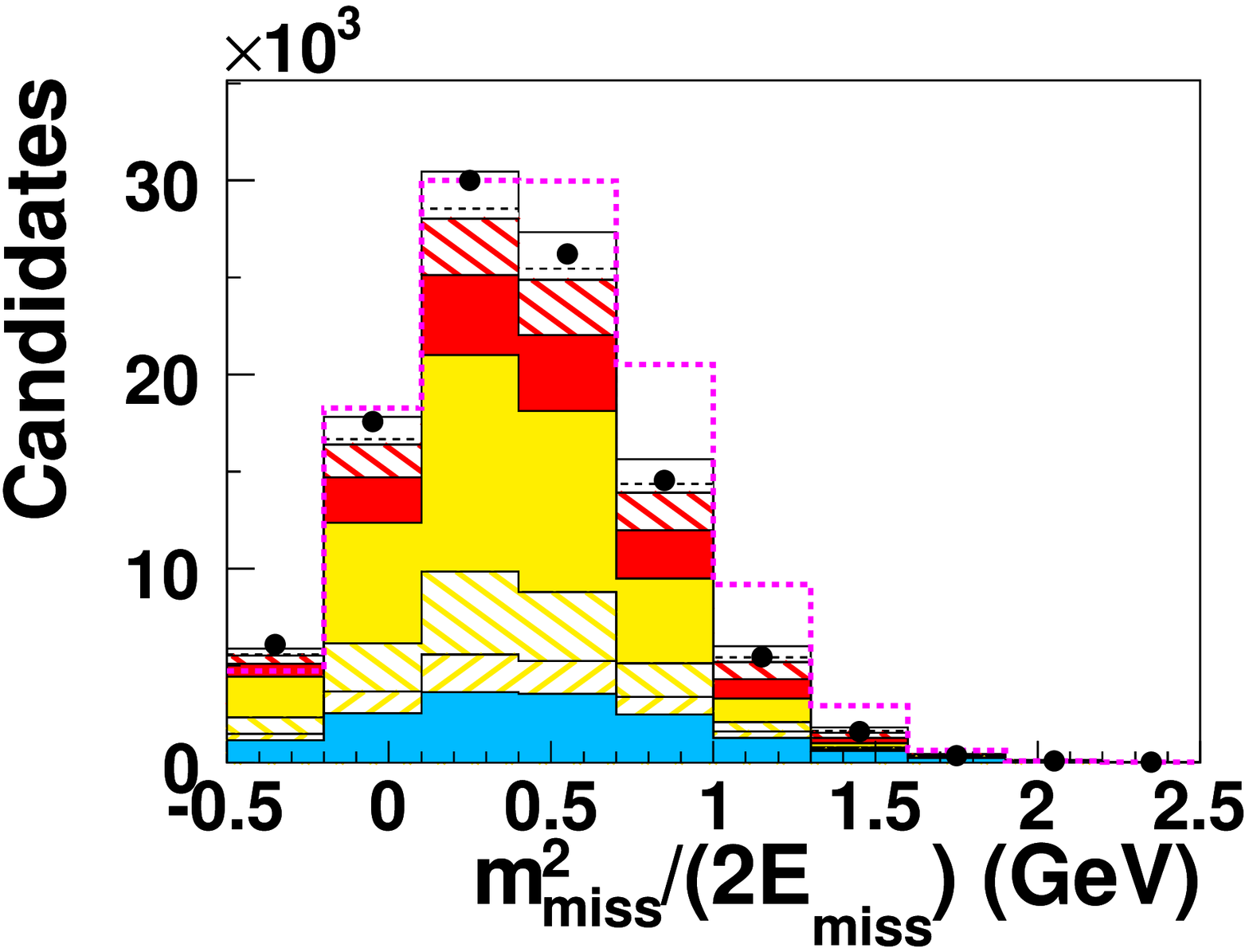, width = 4.5cm}
      \epsfig{file=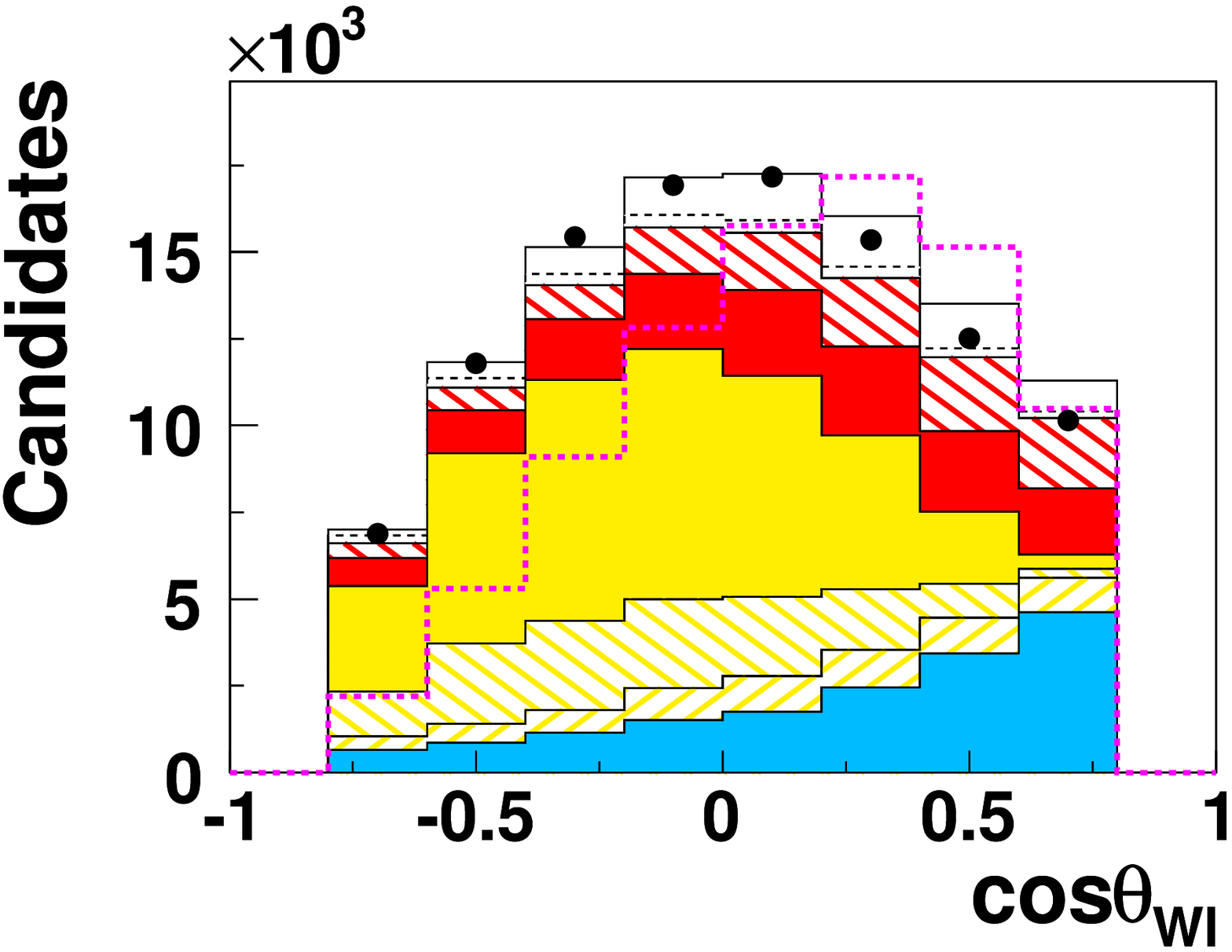, width = 4.5cm}
      \epsfig{file=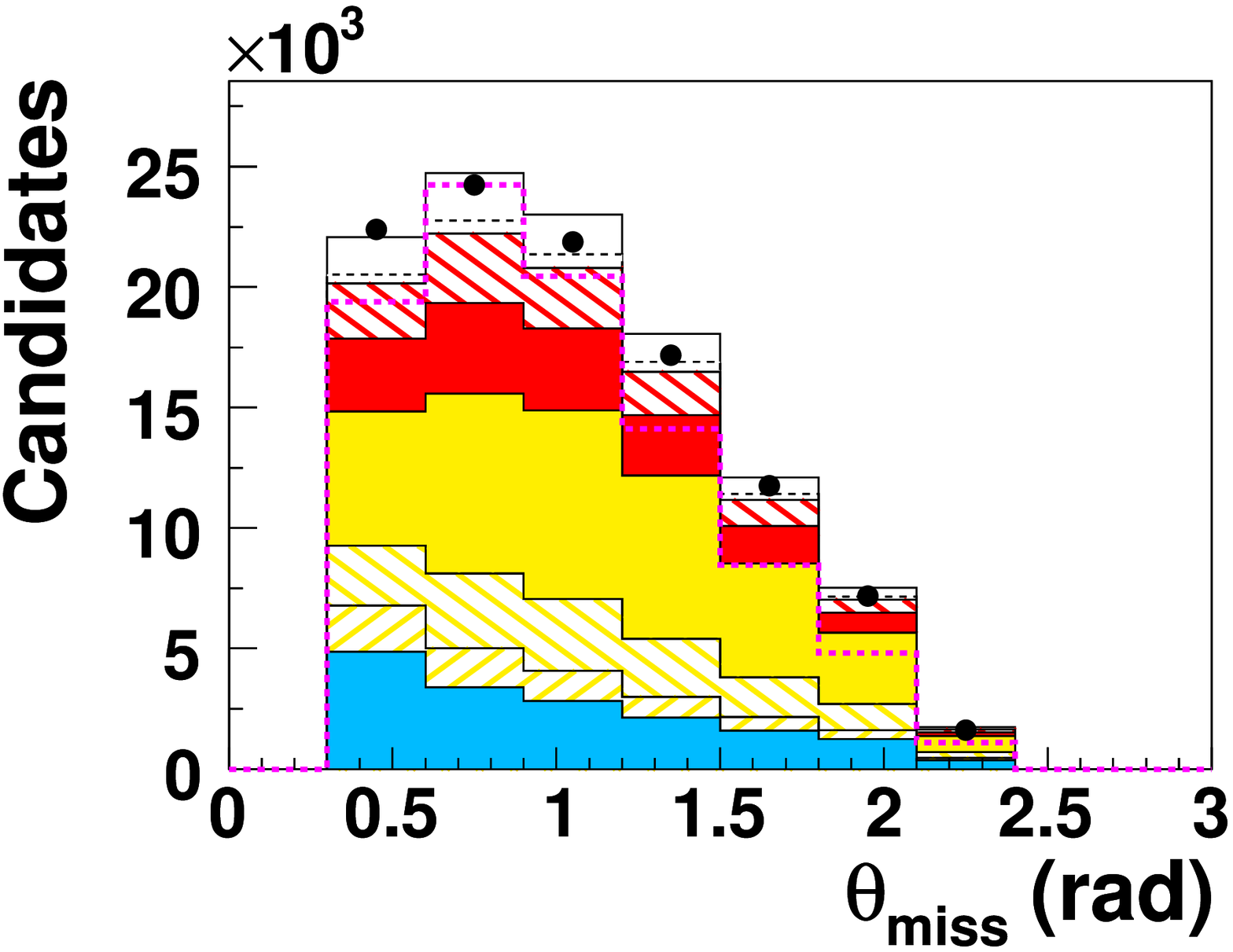, width = 4.5cm}
    \end{minipage}
  \end{tabular}
  \caption{(color online)
Background suppression for \Bzpilnu\ candidates.  Distributions of the seven input variables to the neural network:  after the preselection (left column),  after the \qq\ neural network (center column), and after the \bclnu\ neural network (right column). The data are compared to the sum of the MC-simulated background contributions; for a legend see Figure~\ref{fig:legend}.  The expected signal distribution is overlaid as a magenta, dashed histogram with arbitrary normalization.
}
  \label{fig:NNinputpilnu}
\end{figure*}

\begin{figure*}[htb]
  \fourFig
      {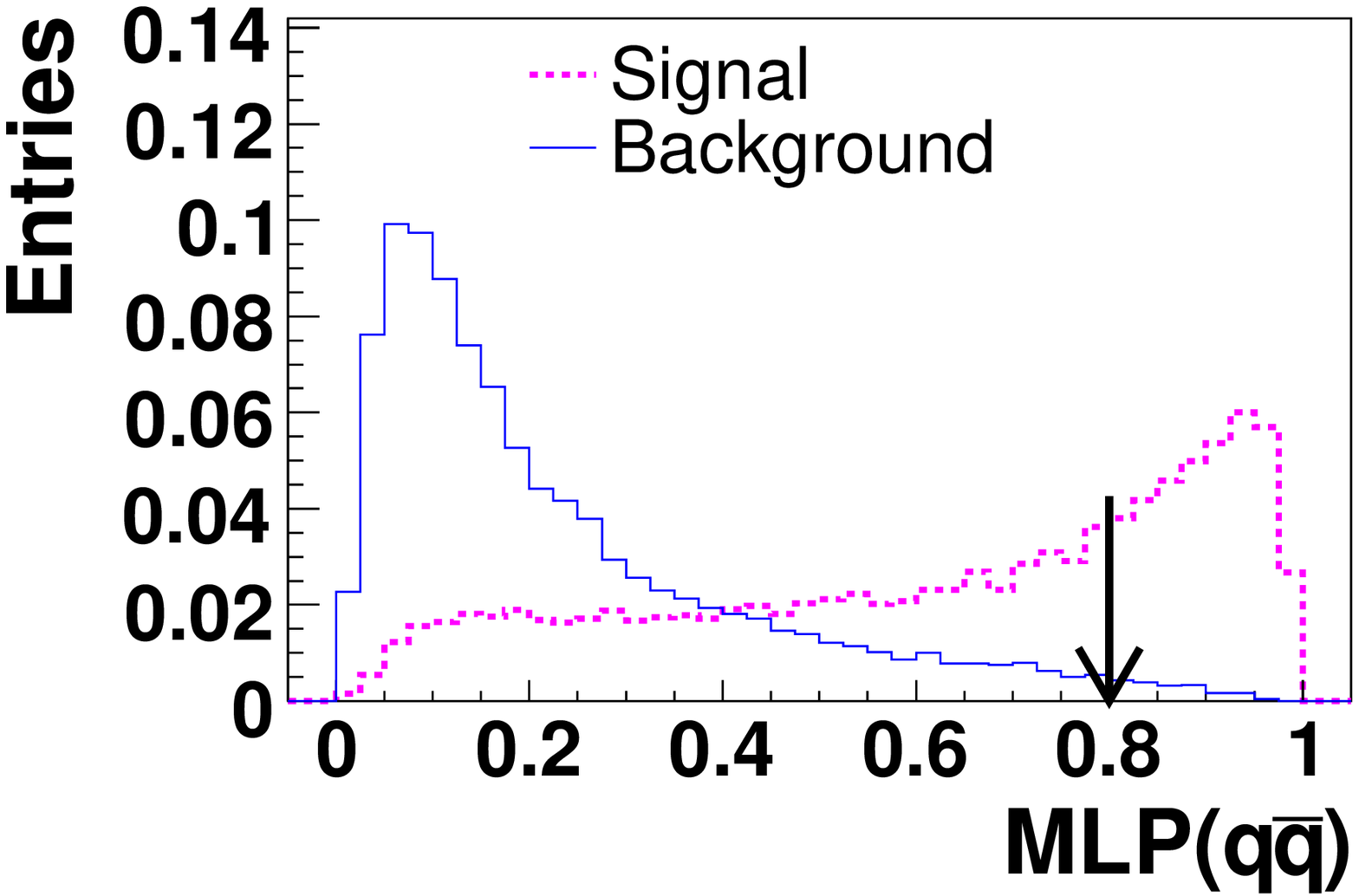}
      {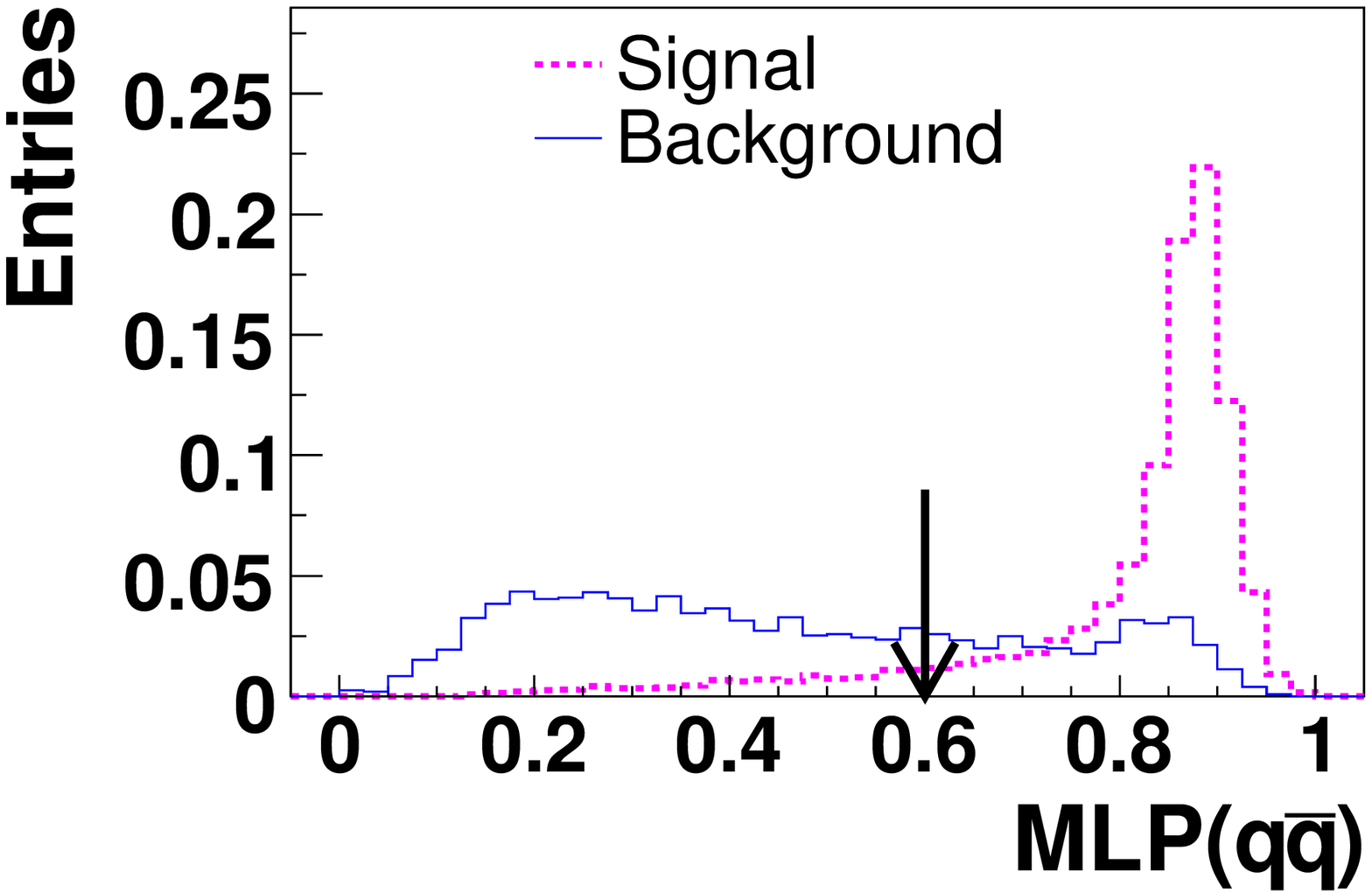}
      {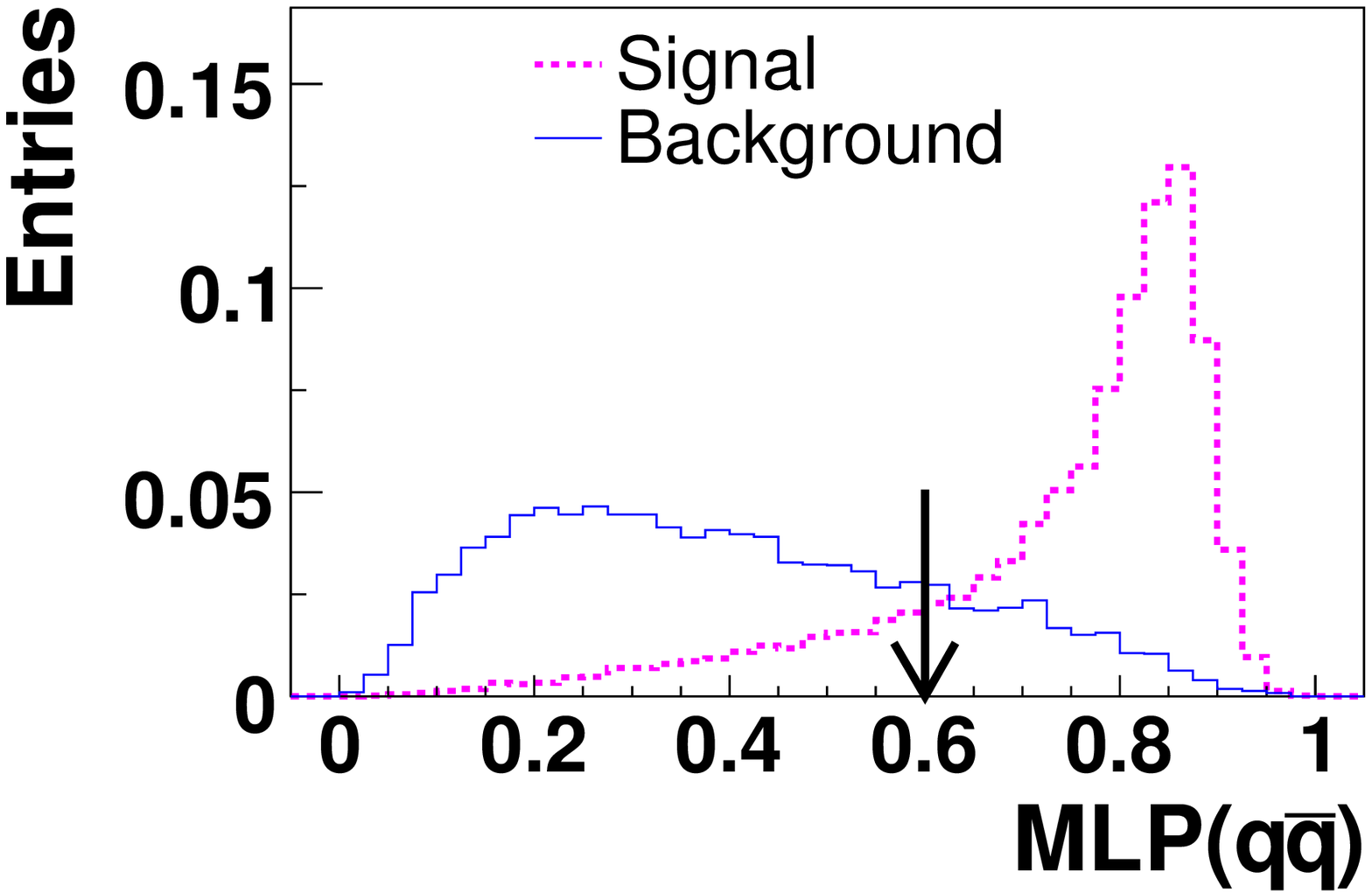}
      {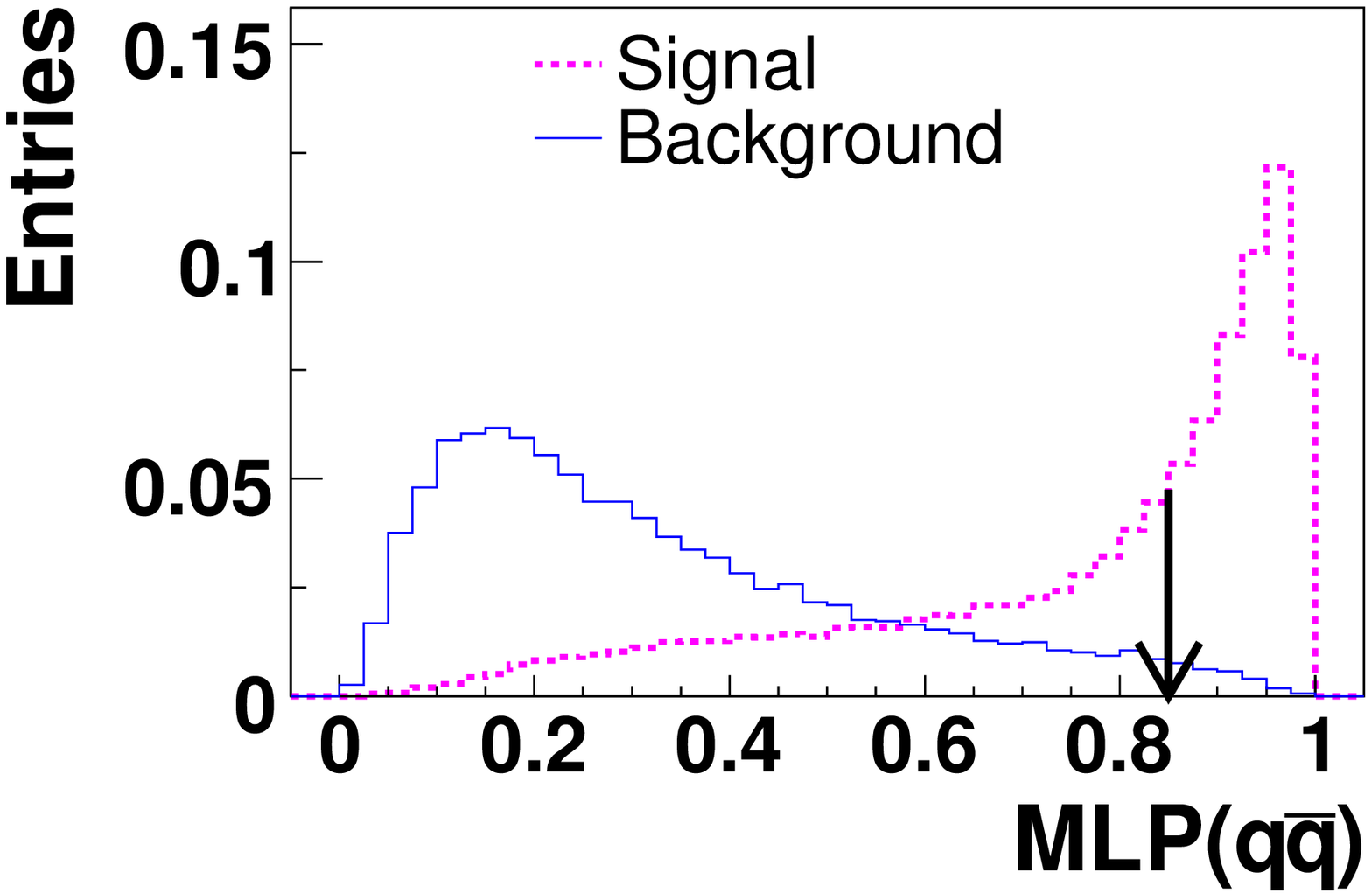}
  \fourFig
      {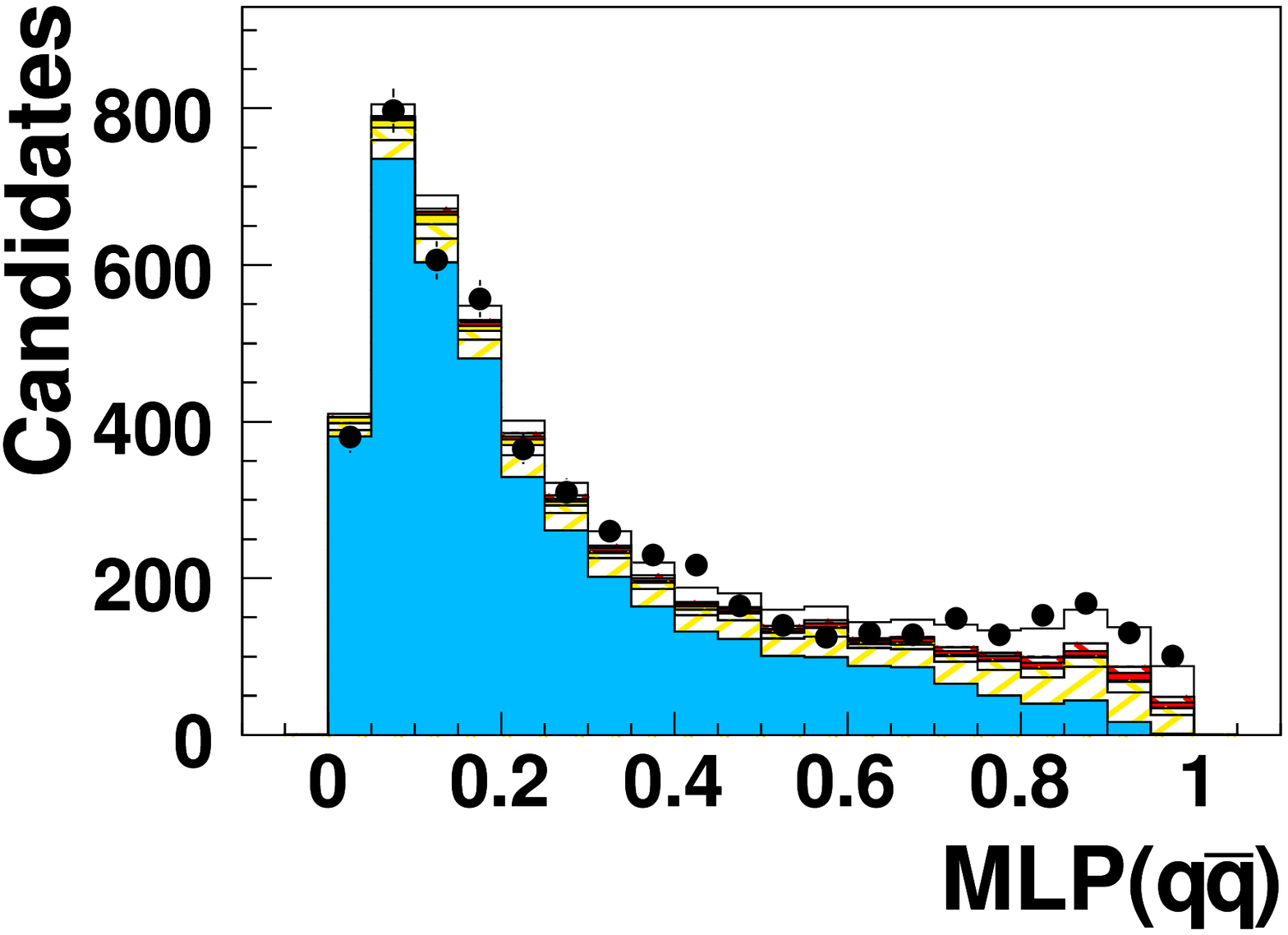}
      {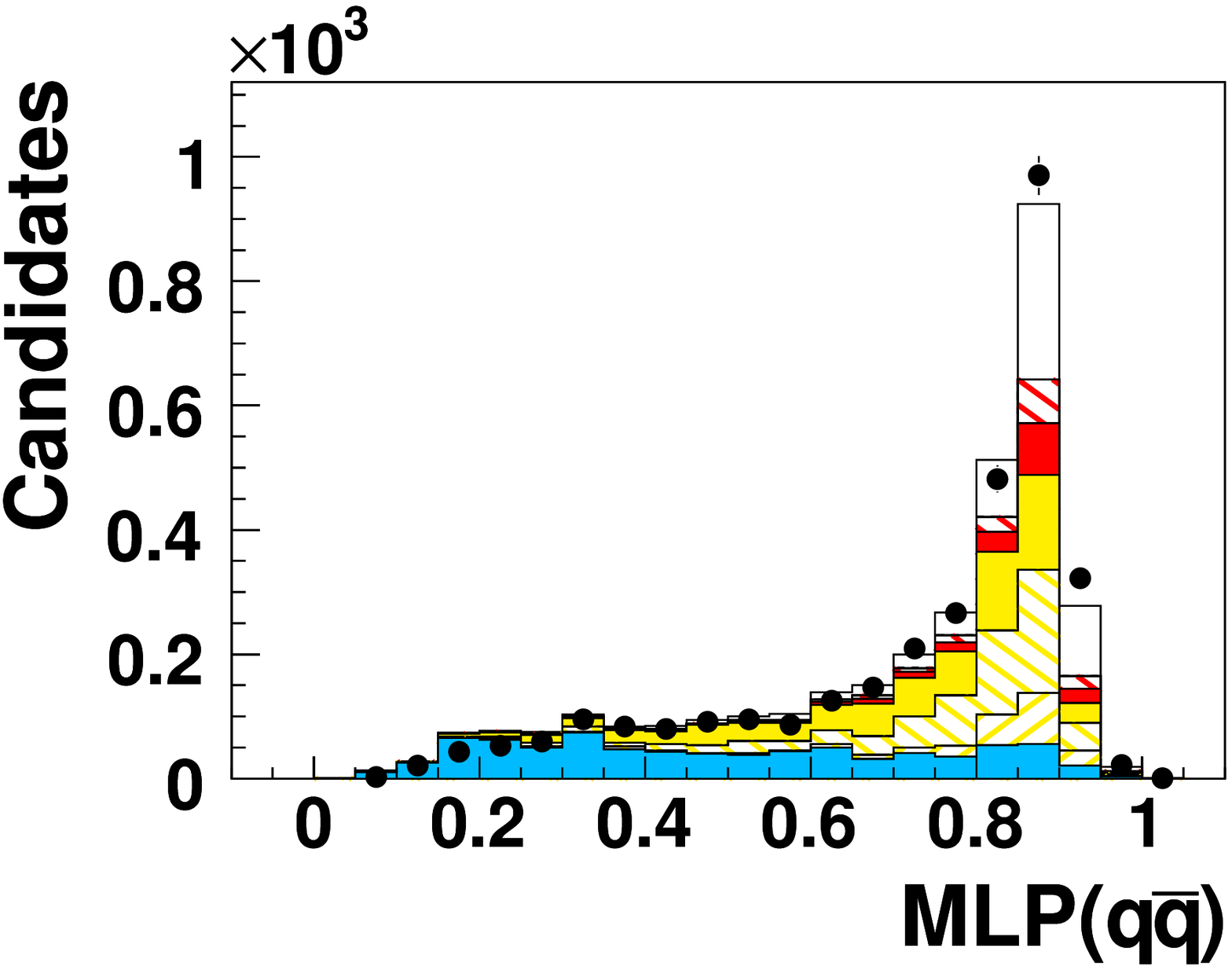}
      {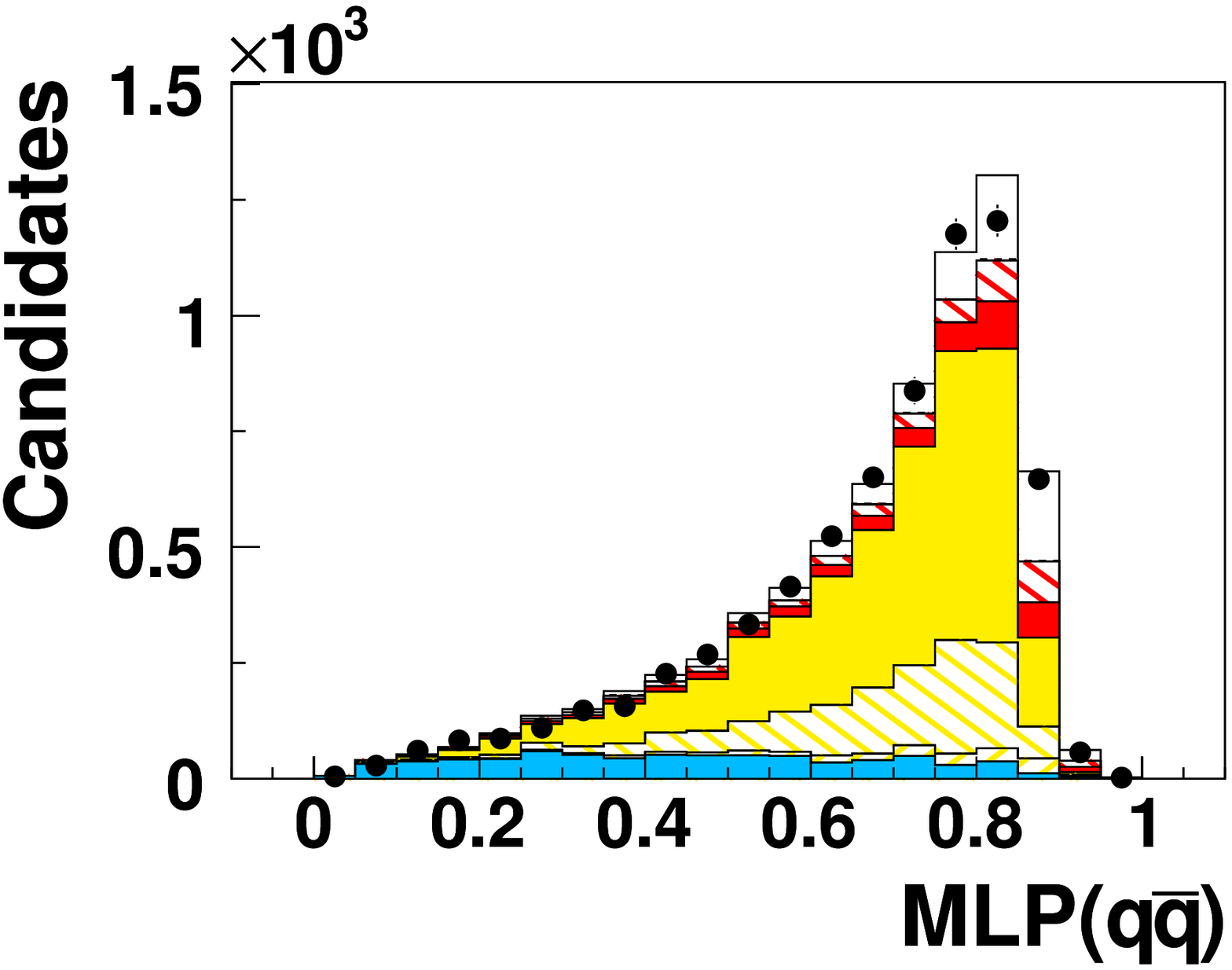}
      {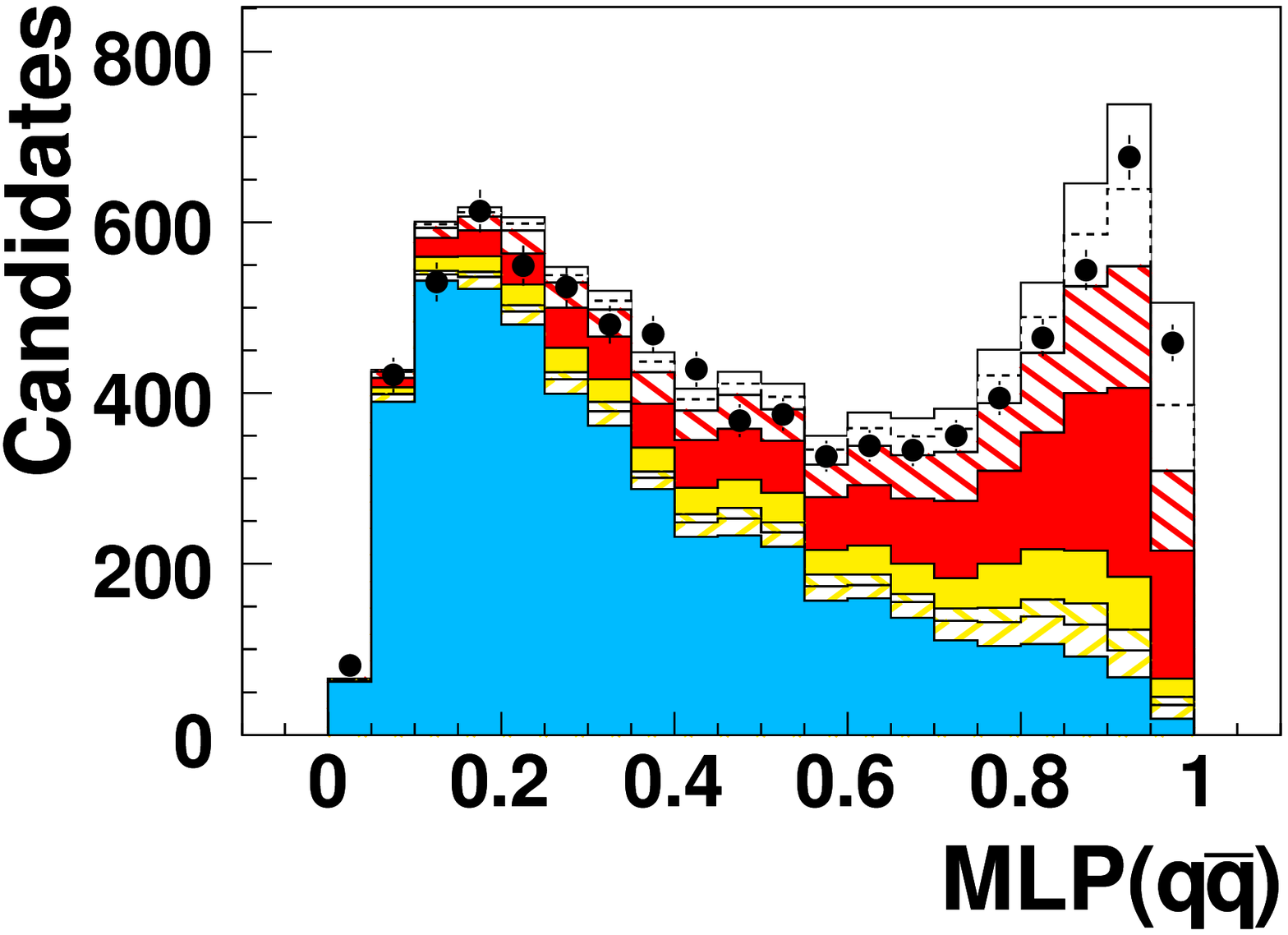}
   \caption{(color online)
The \qq\ neural-network discriminators for \Bzpilnu\ candidates in the signal region, \SignalRegion.
The distributions are shown for four different $q^2$ bins, columns from left to right:
   $0  <\q^2 < 4 \gev^2$,
   $4  <\q^2 < 8 \gev^2$,
   $12 <\q^2 < 16 \gev^2$,
   $    \q^2 > 20 \gev^2$.
Top row: Discriminator distributions
for signal (magneta, dashed) and $\qq$ background (blue, solid), normalized to the same area. 
The arrows indicate the chosen cuts.
Bottom row: Discriminator distributions for data compared with MC-simulated signal and background contributions.
For a legend see Figure~\ref{fig:legend}.
}
      \label{fig:NNoutputpilnu1}
\end{figure*}

\begin{figure*}[htb]
  \fourFig
      {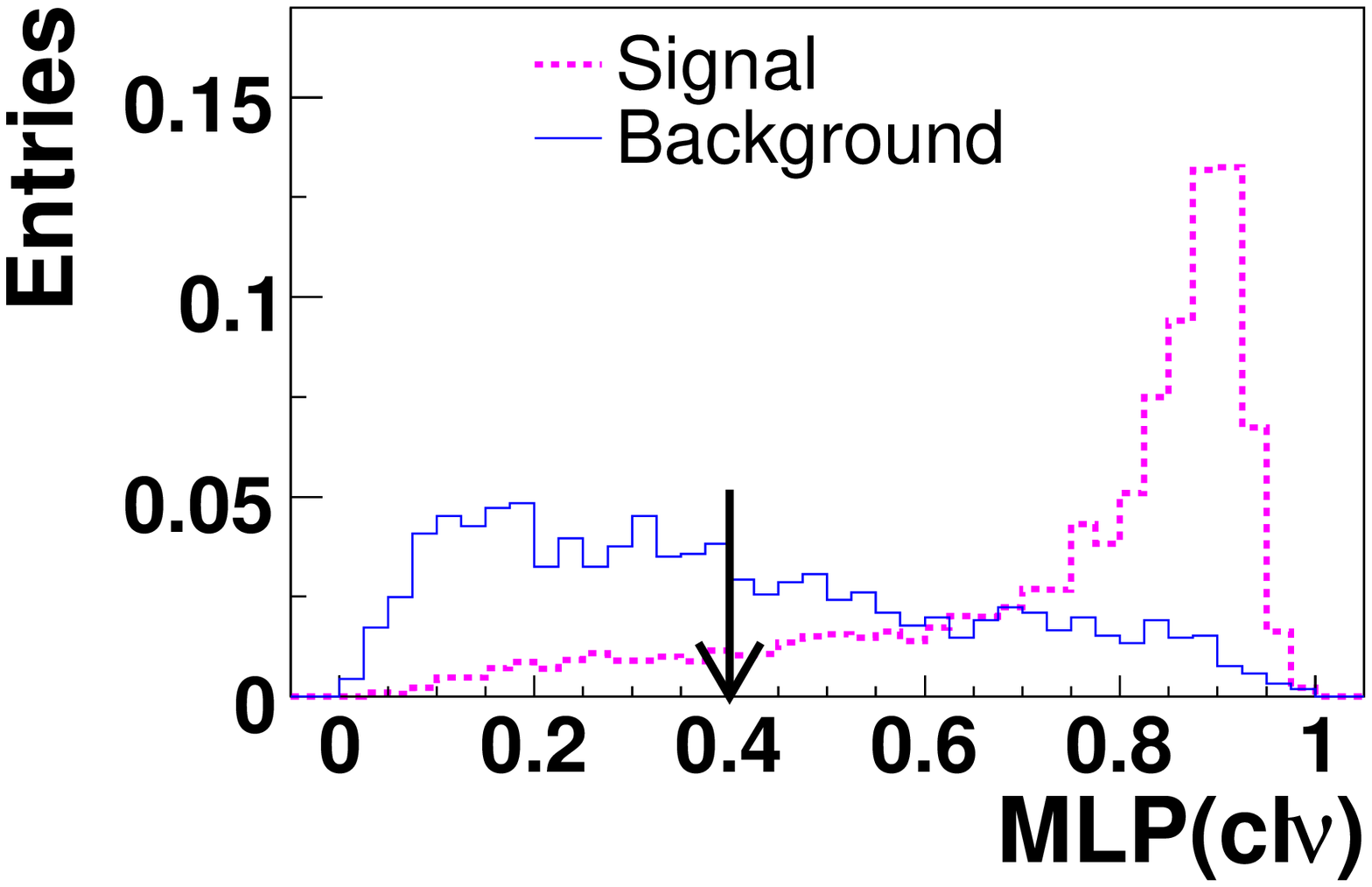}
      {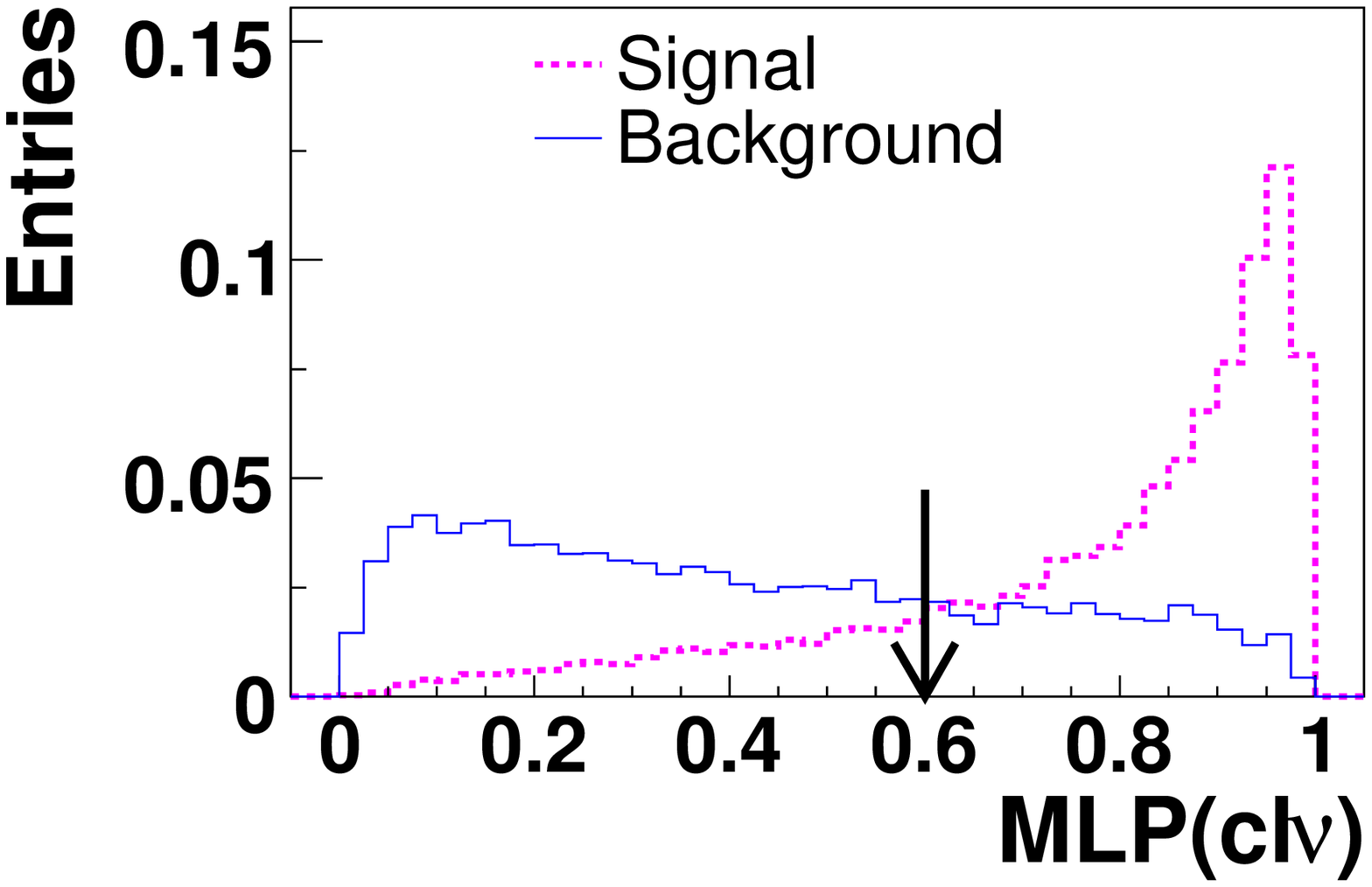}
      {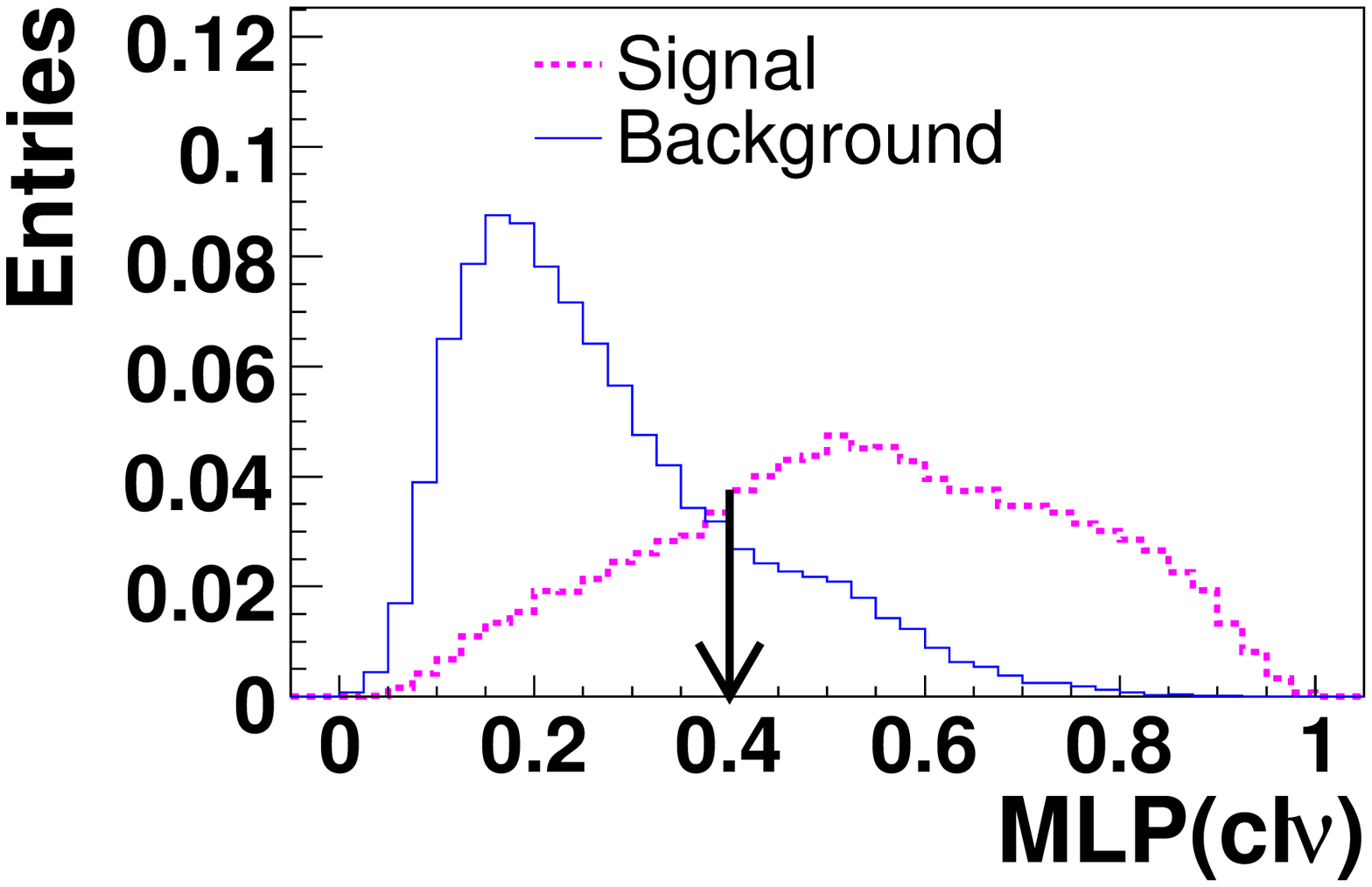}
      {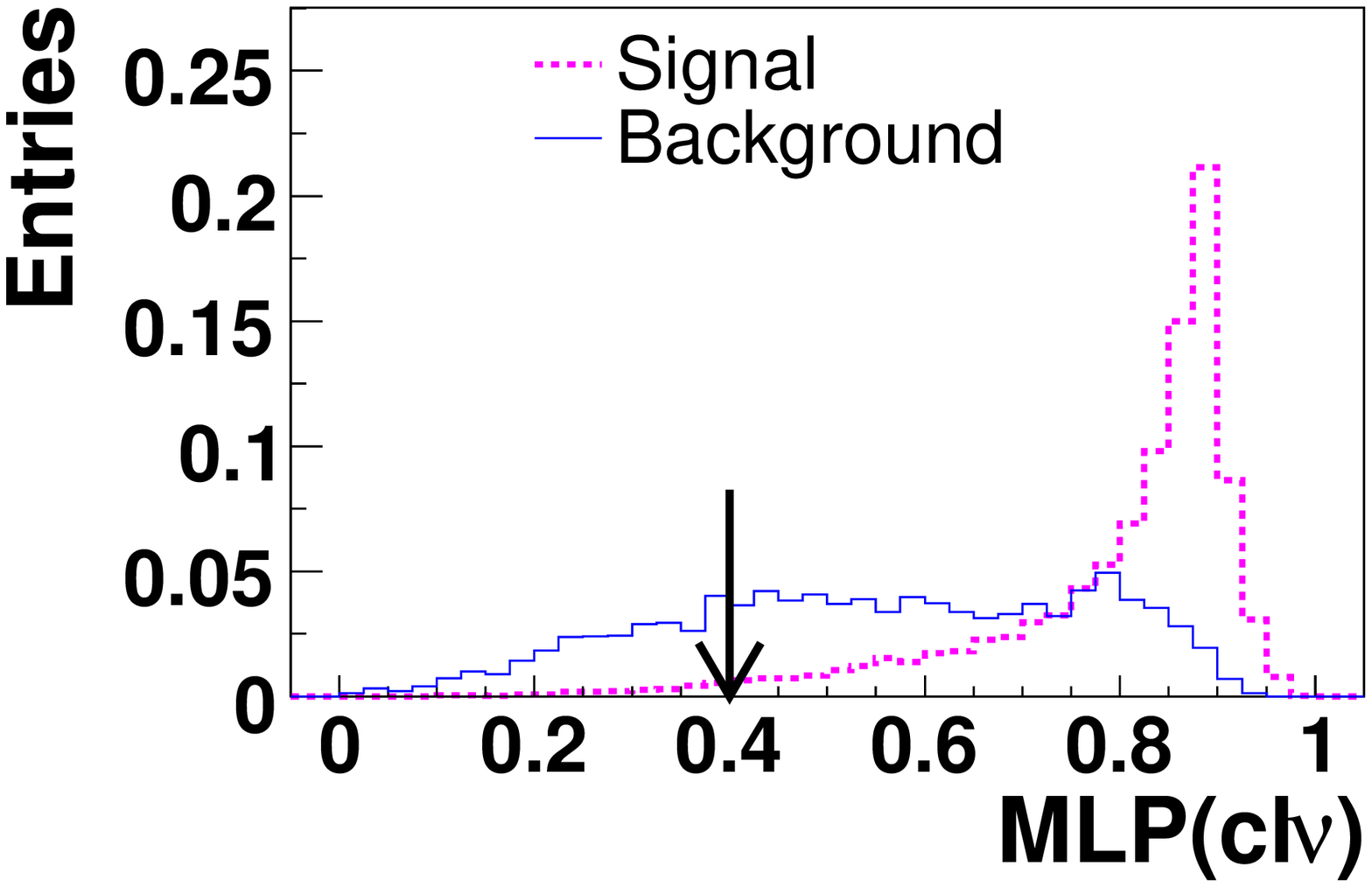}
  \fourFig
      {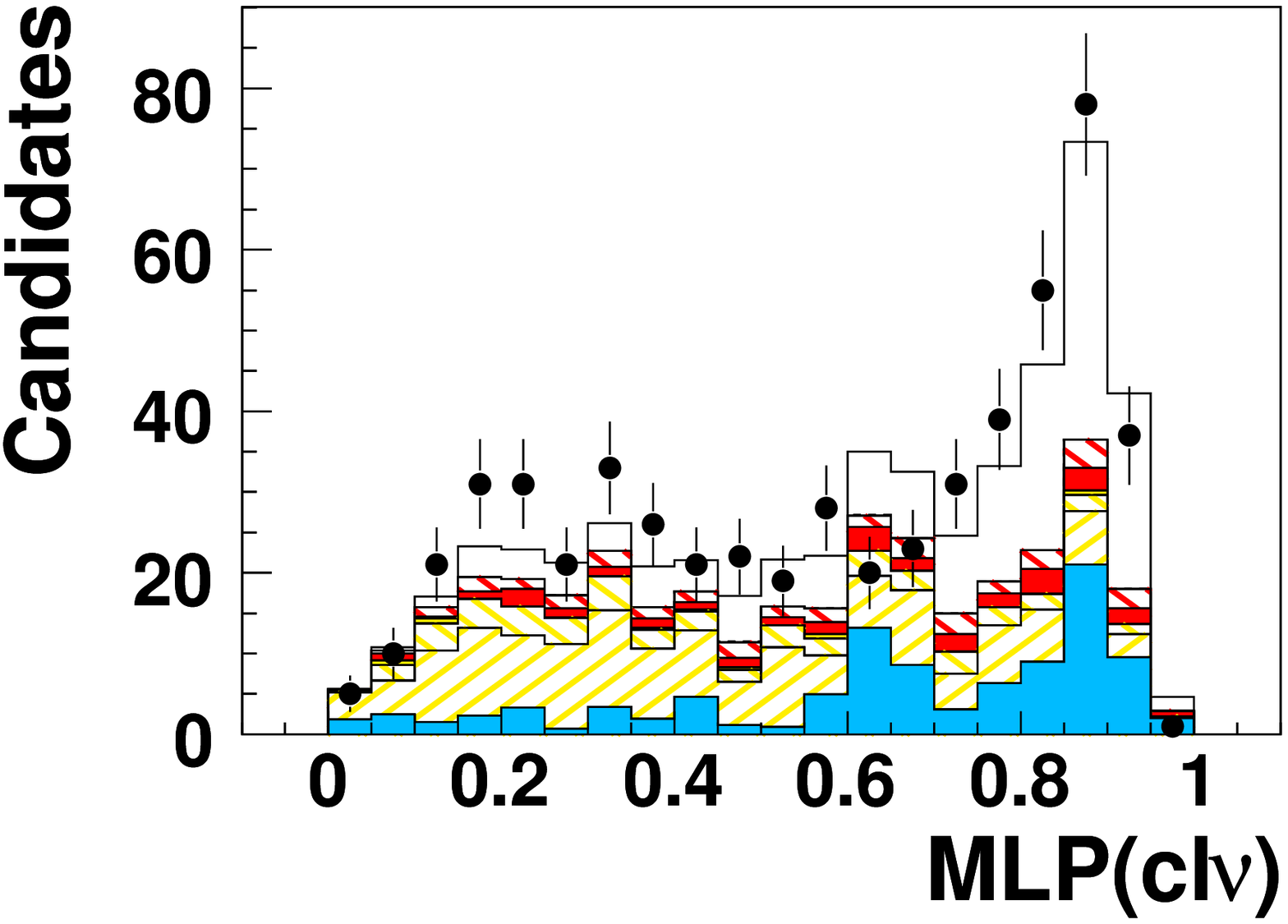}
      {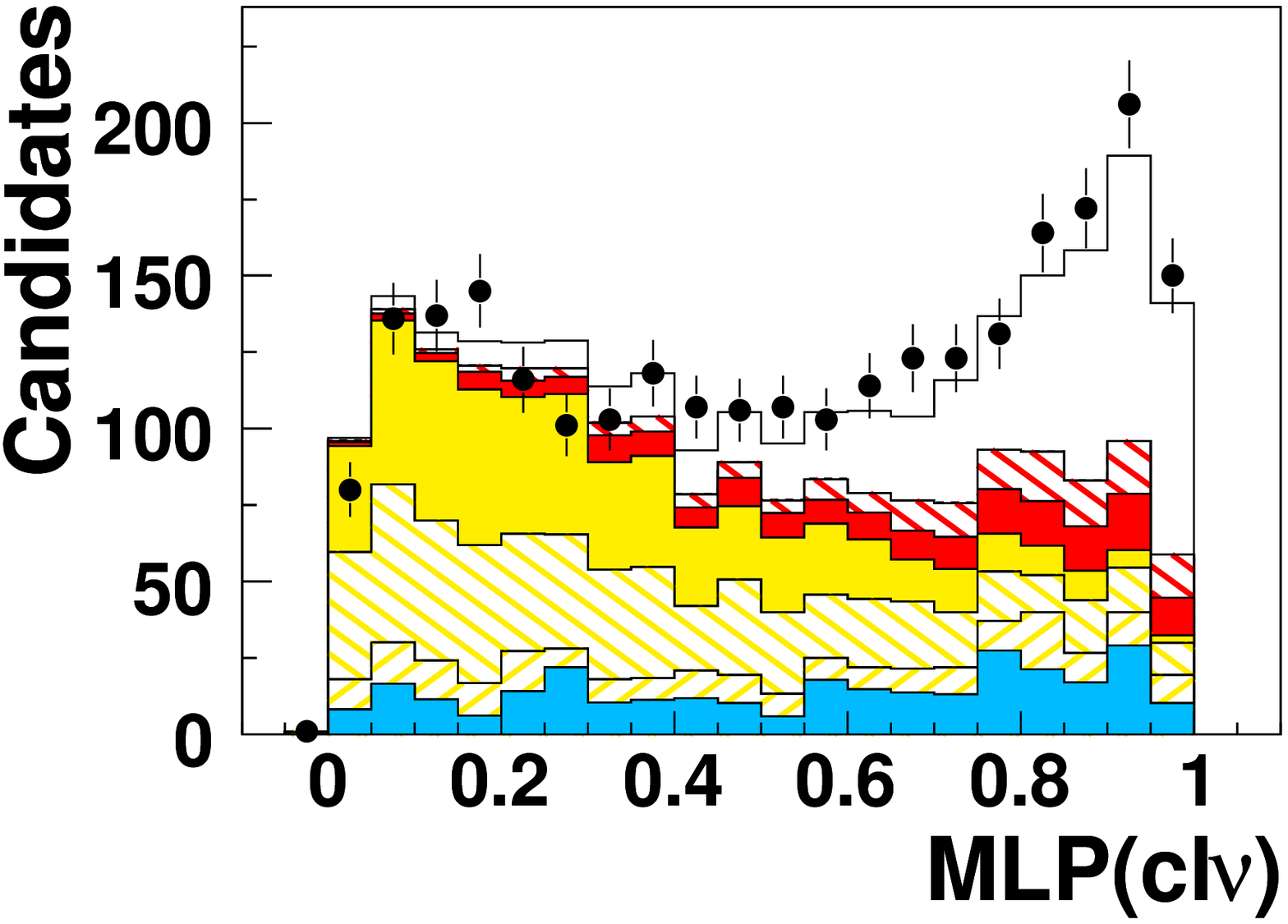}
      {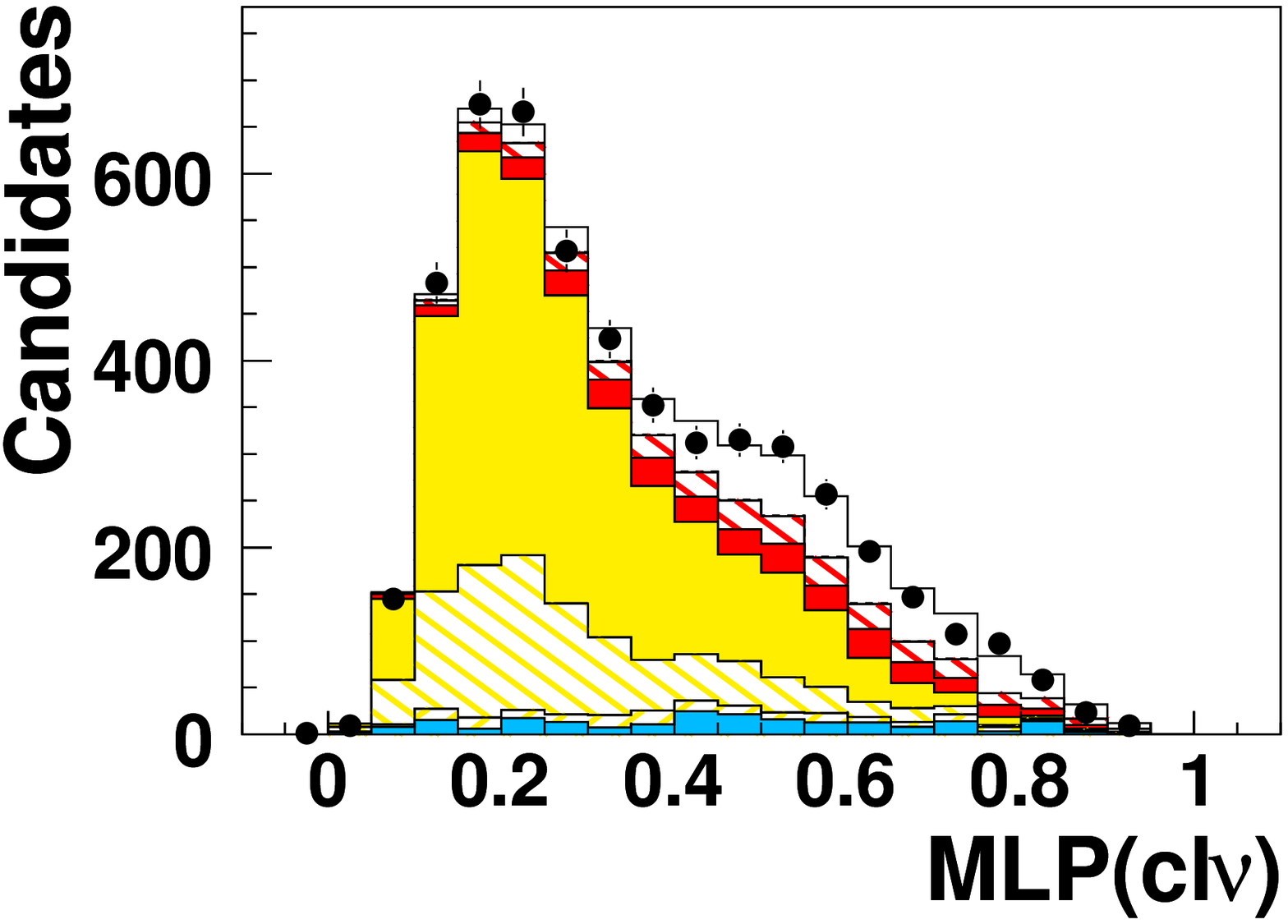}
      {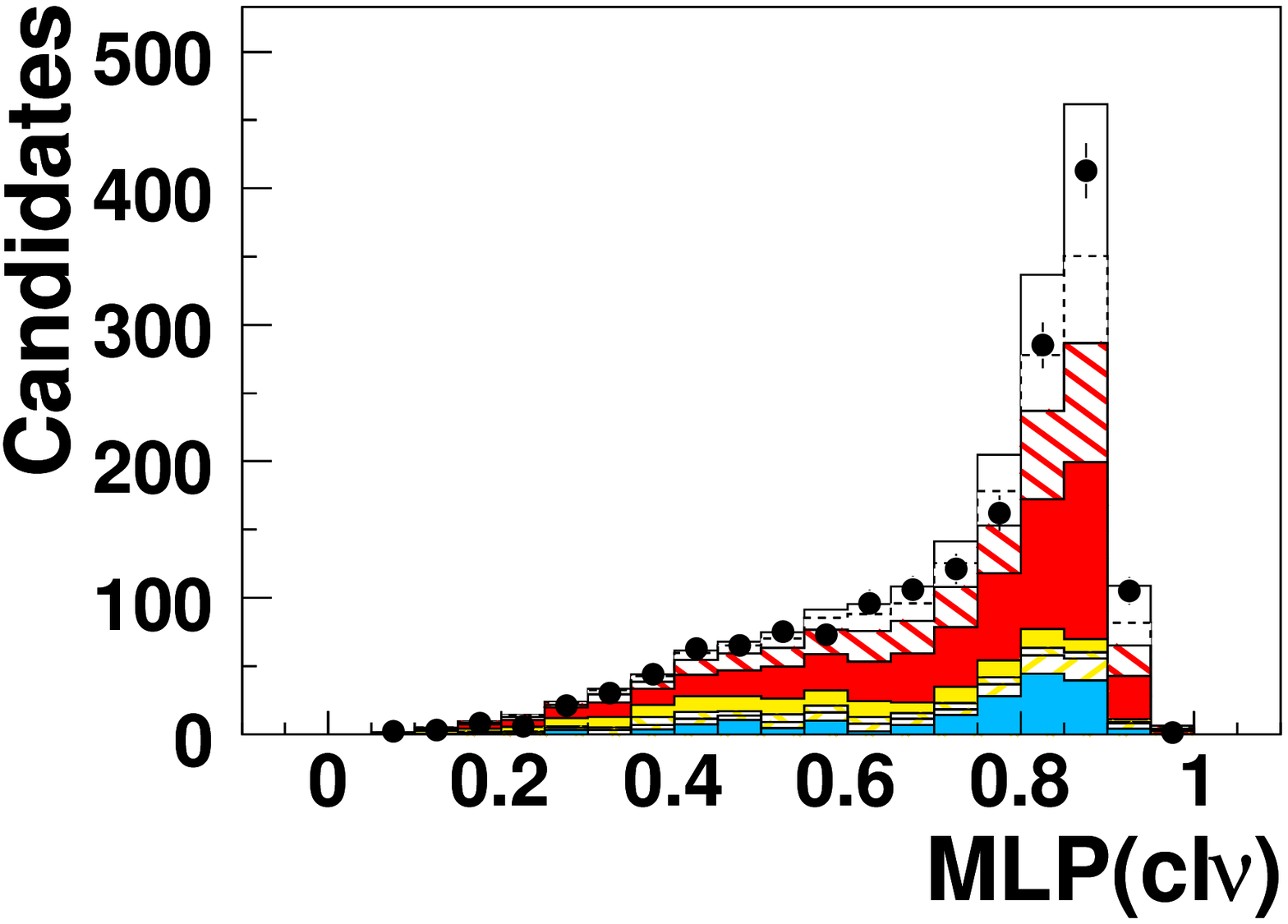}
  \caption{(color online)
The \bclnu\ neural-network discriminators for \Bzpilnu\ candidates in the signal region, \SignalRegion.
The distributions are shown for four different $q^2$ bins, columns from left to right:
   $0  <\q^2 < 4 \gev^2$,
   $4  <\q^2 < 8 \gev^2$,
   $12 <\q^2 < 16 \gev^2$,
   $    \q^2 > 20 \gev^2$.
Top row: Discriminator distributions
for signal (magneta, dashed) and \bclnu\ background (blue, solid), normalized to the same area. 
The arrows indicate the chosen cuts.
Bottom row: Discriminator distributions for data compared with MC-simulated signal and background contributions.
For a legend see Figure~\ref{fig:legend}.
}
      \label{fig:NNoutputpilnu2}
\end{figure*}

\begin{figure*}[htb]
  \threeFig
      {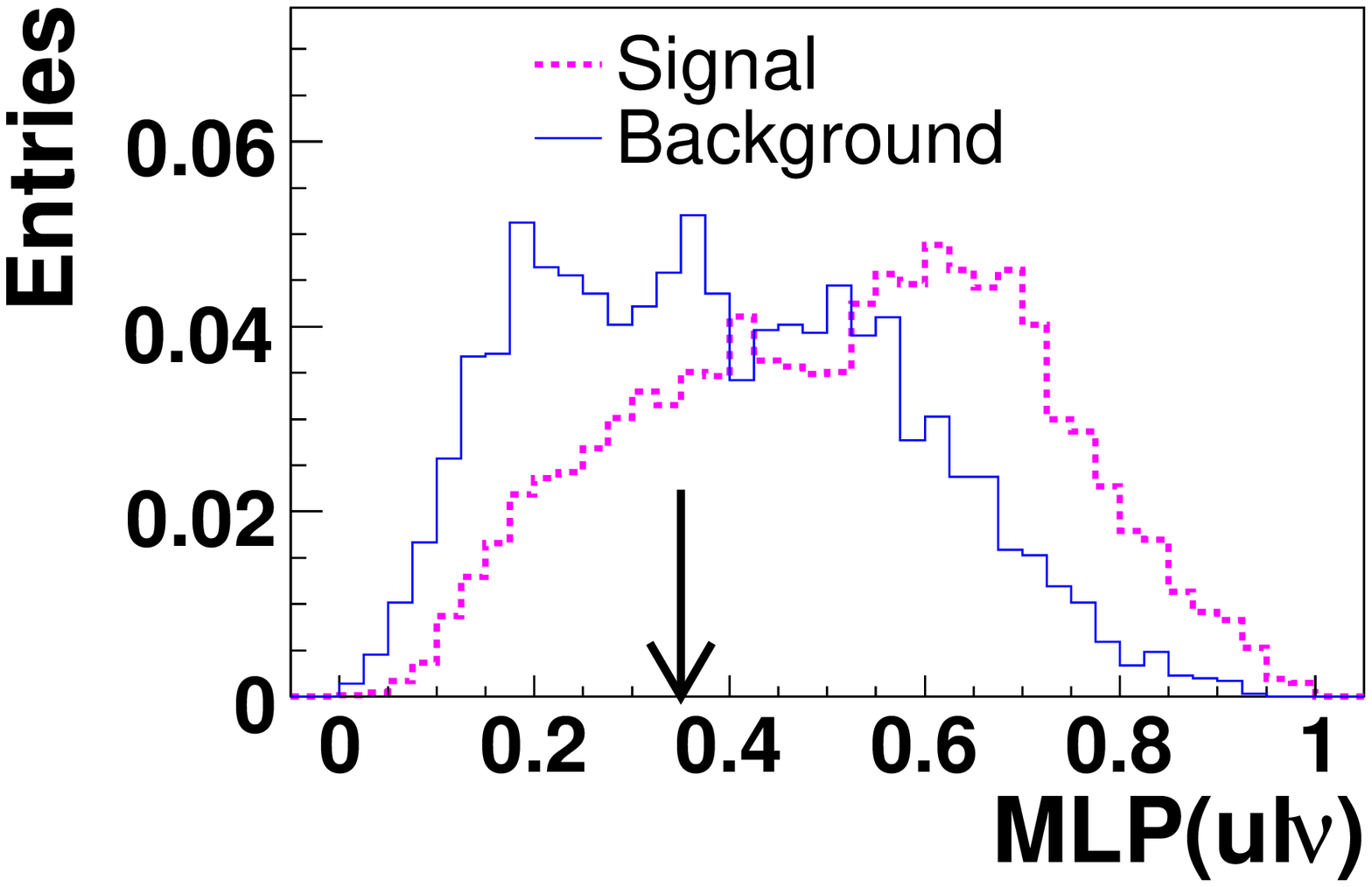}
      {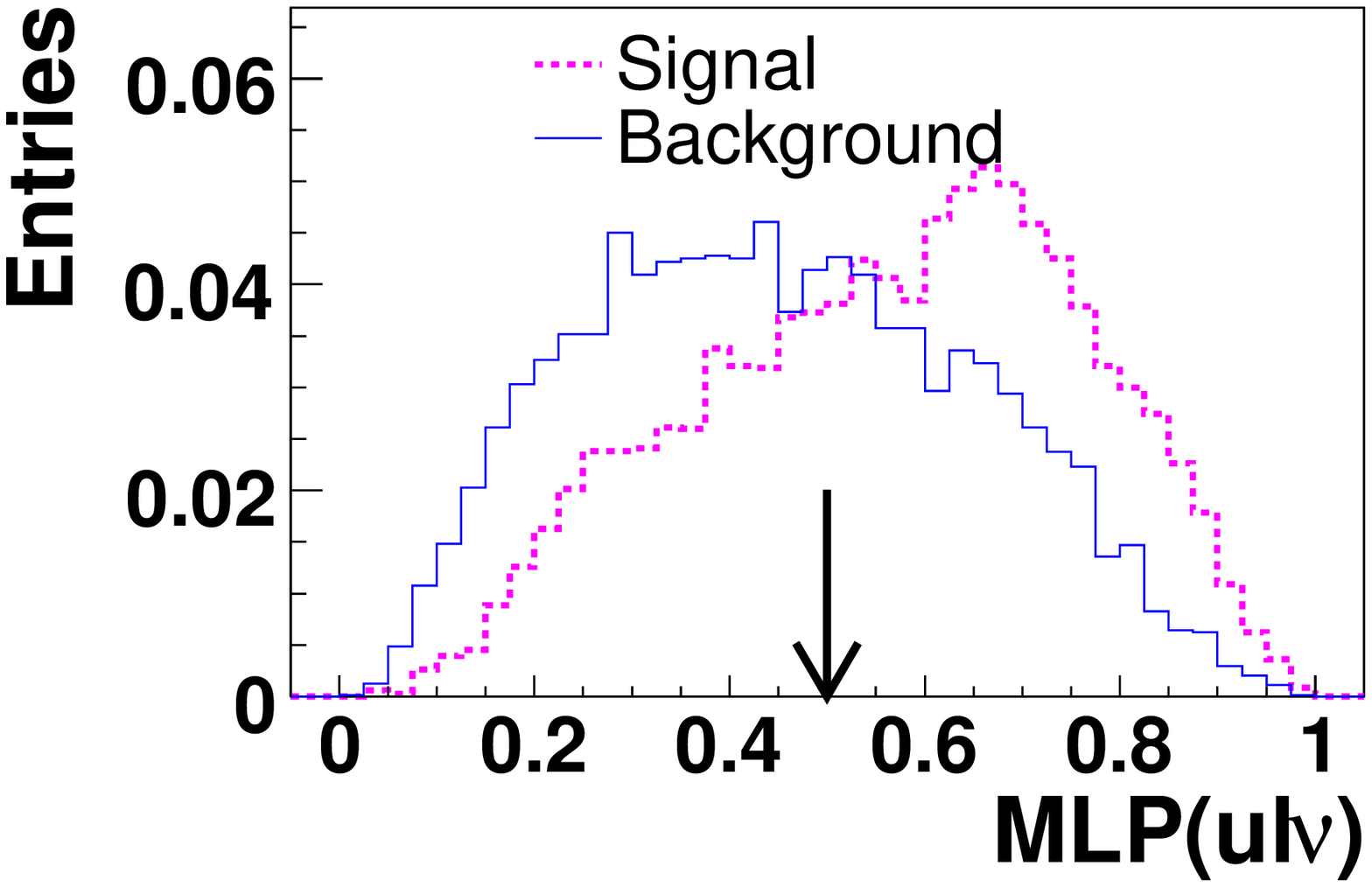}
      {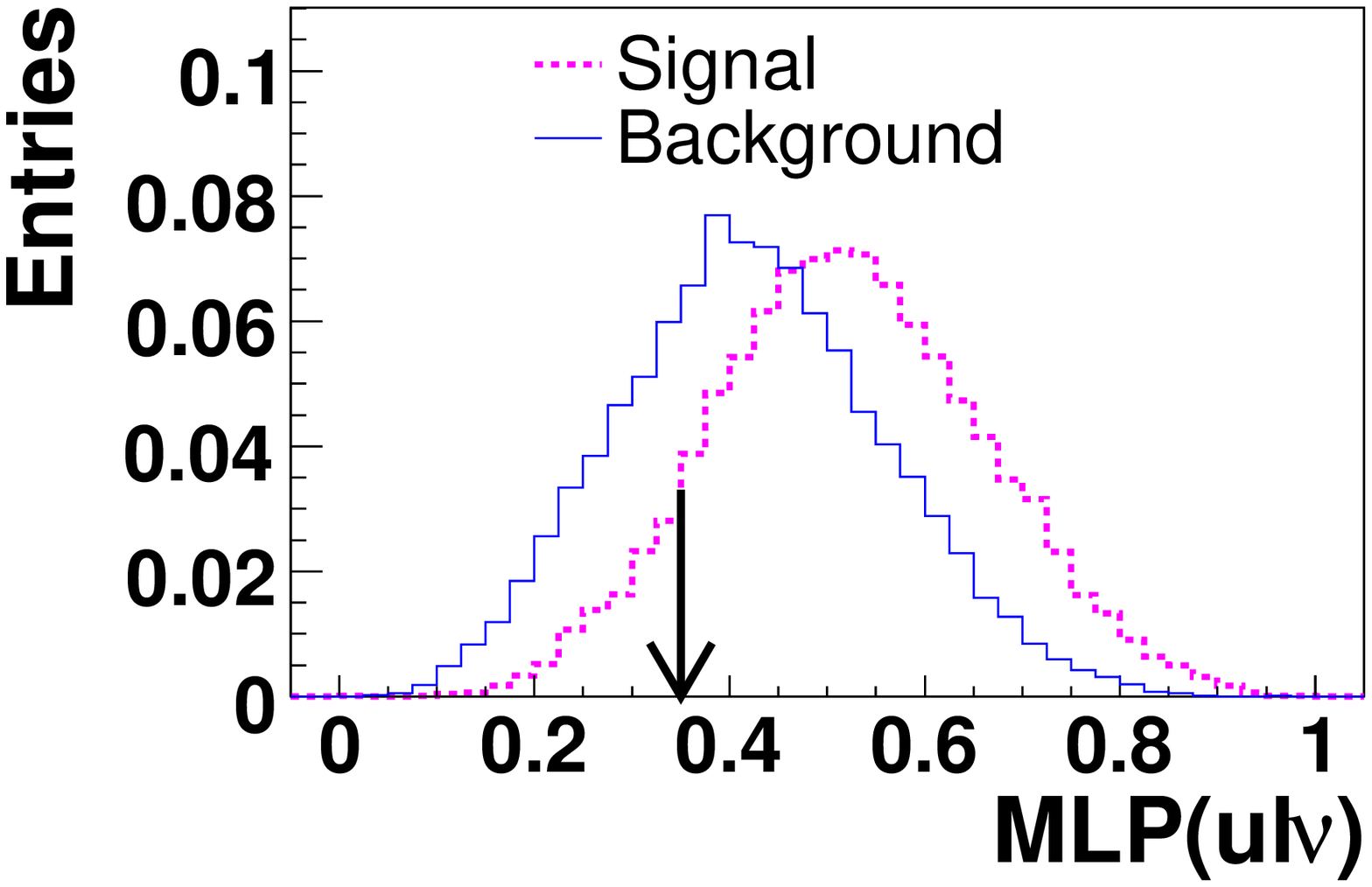}
  \threeFig
      {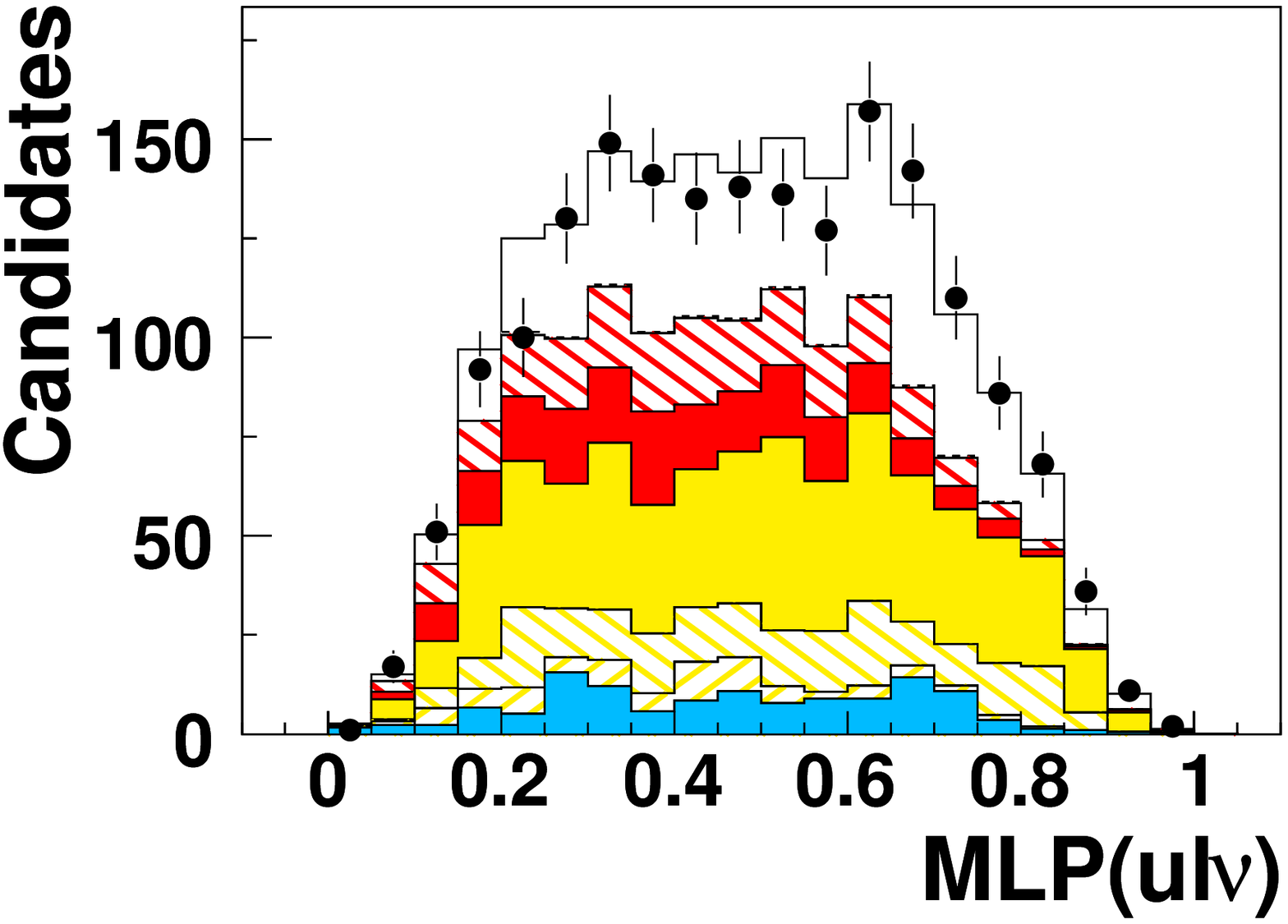}
      {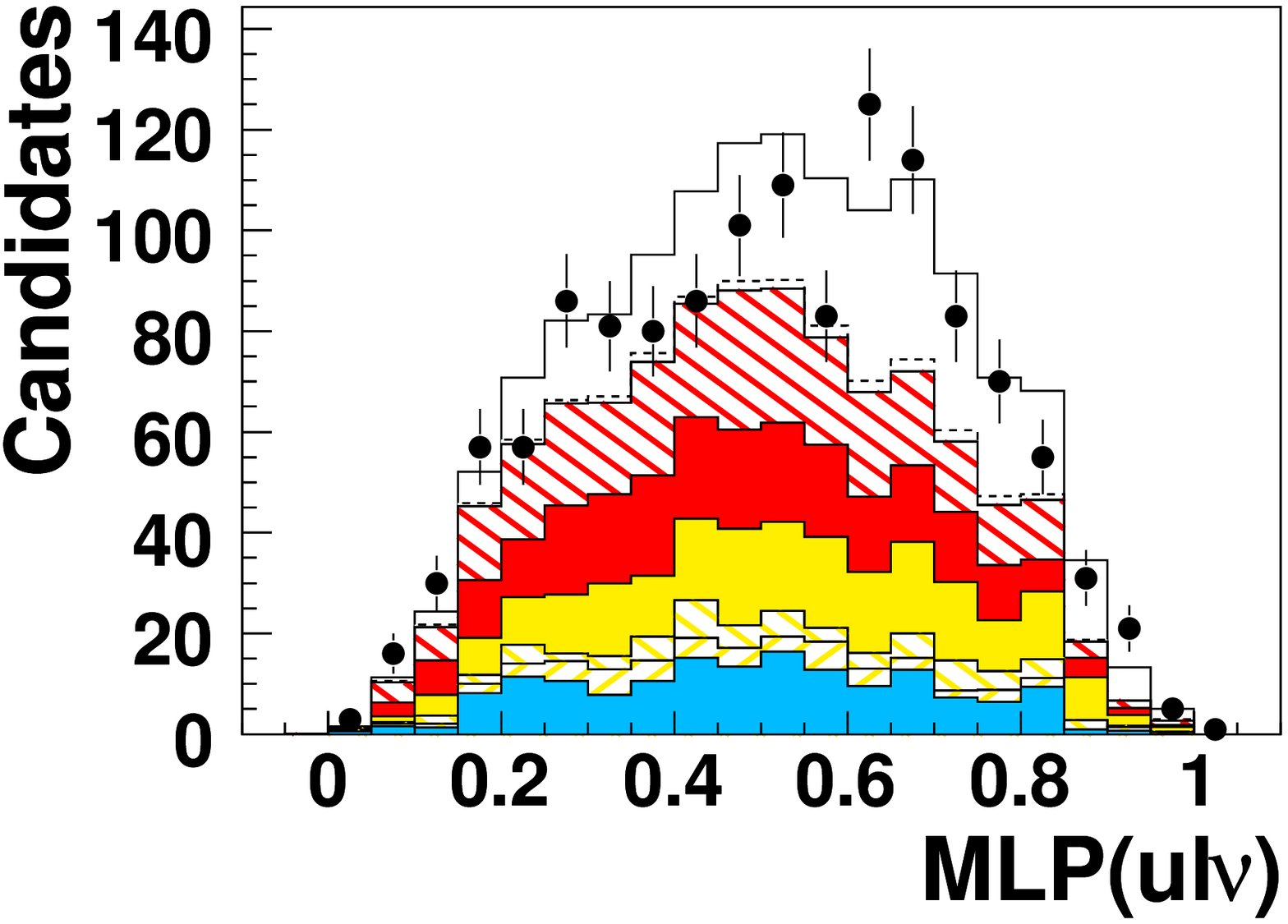}
      {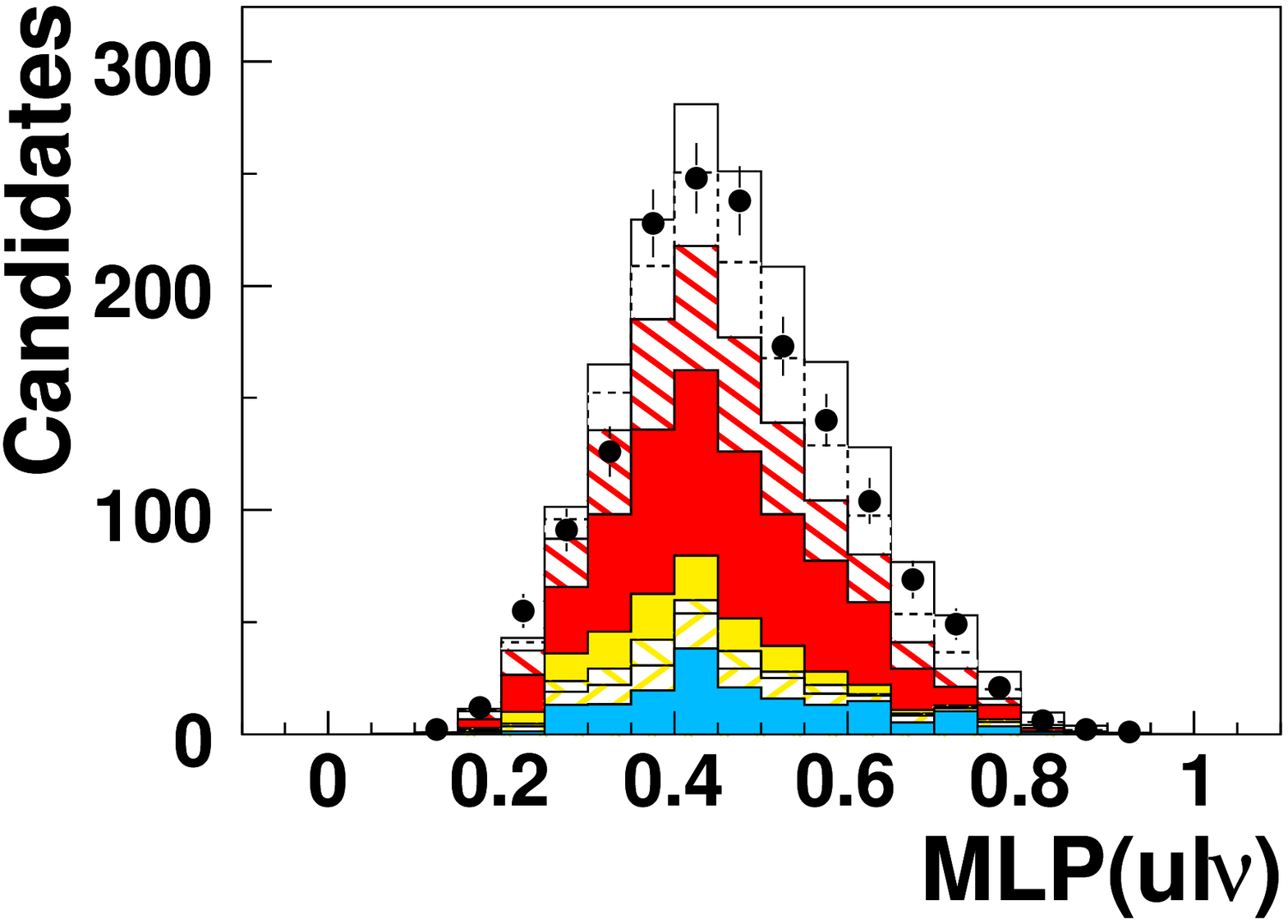}
  \caption{(color online)
The \bulnu\ neural-network discriminators for \Bzpilnu\ candidates in the signal region, \SignalRegion.
The distributions are shown for the three highest $q^2$ bins, columns from left to right:
   $12 <\q^2 < 16 \gev^2$,
   $16 <\q^2 < 20 \gev^2$,
   $    \q^2 > 20 \gev^2$.
Top row: Discriminator distributions
for signal (magneta, dashed) and \bulnu background (blue, solid), normalized to the same area. 
The arrows indicate the chosen cuts.
Bottom row: Discriminator distributions for data compared with MC-simulated signal and background contributions.
For a legend see Figure~\ref{fig:legend}.
}
      \label{fig:NNoutputpilnu3}
\end{figure*}

\begin{figure*}[htb]
  \threeFigTwoCol
      {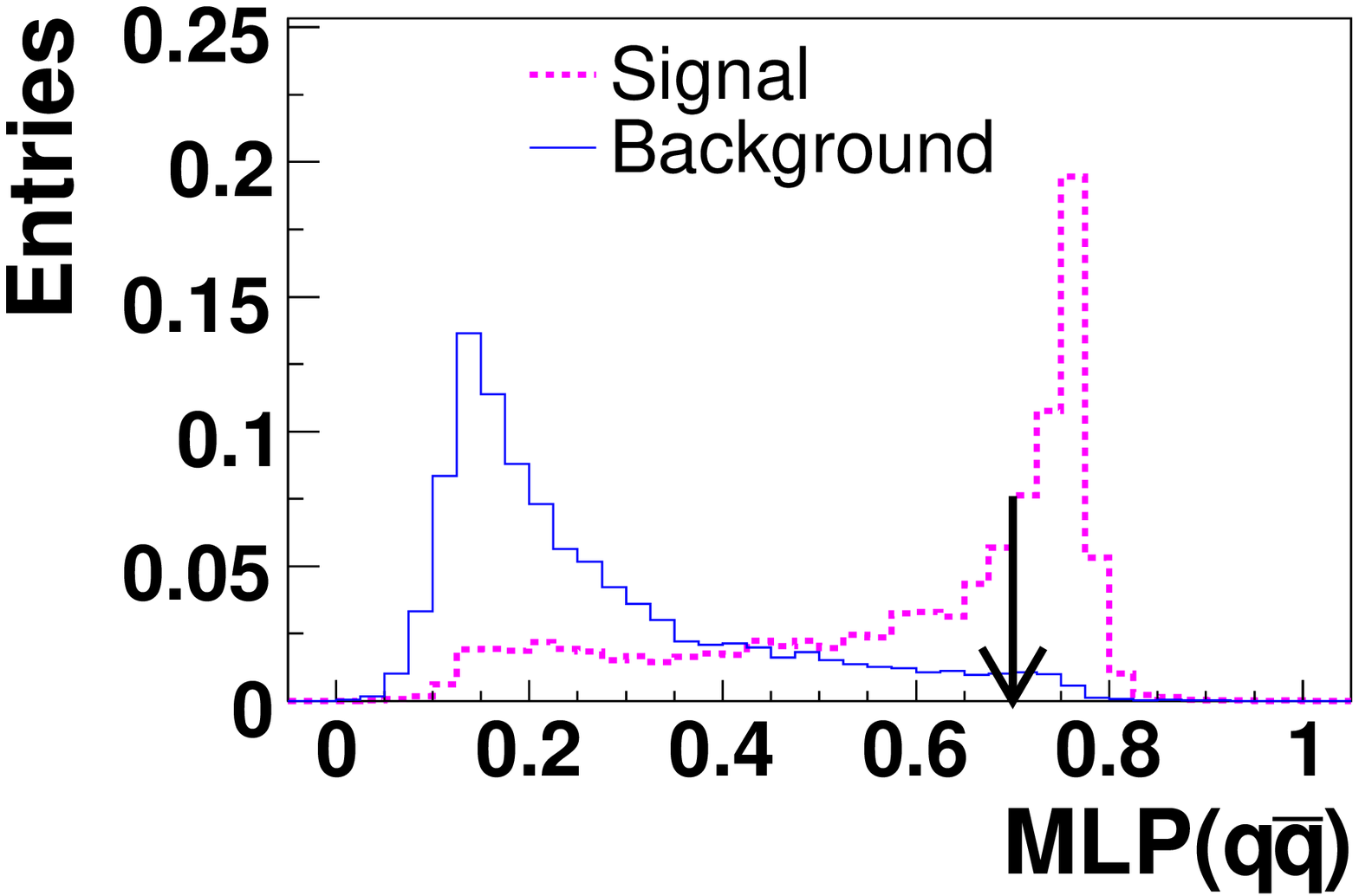}
      {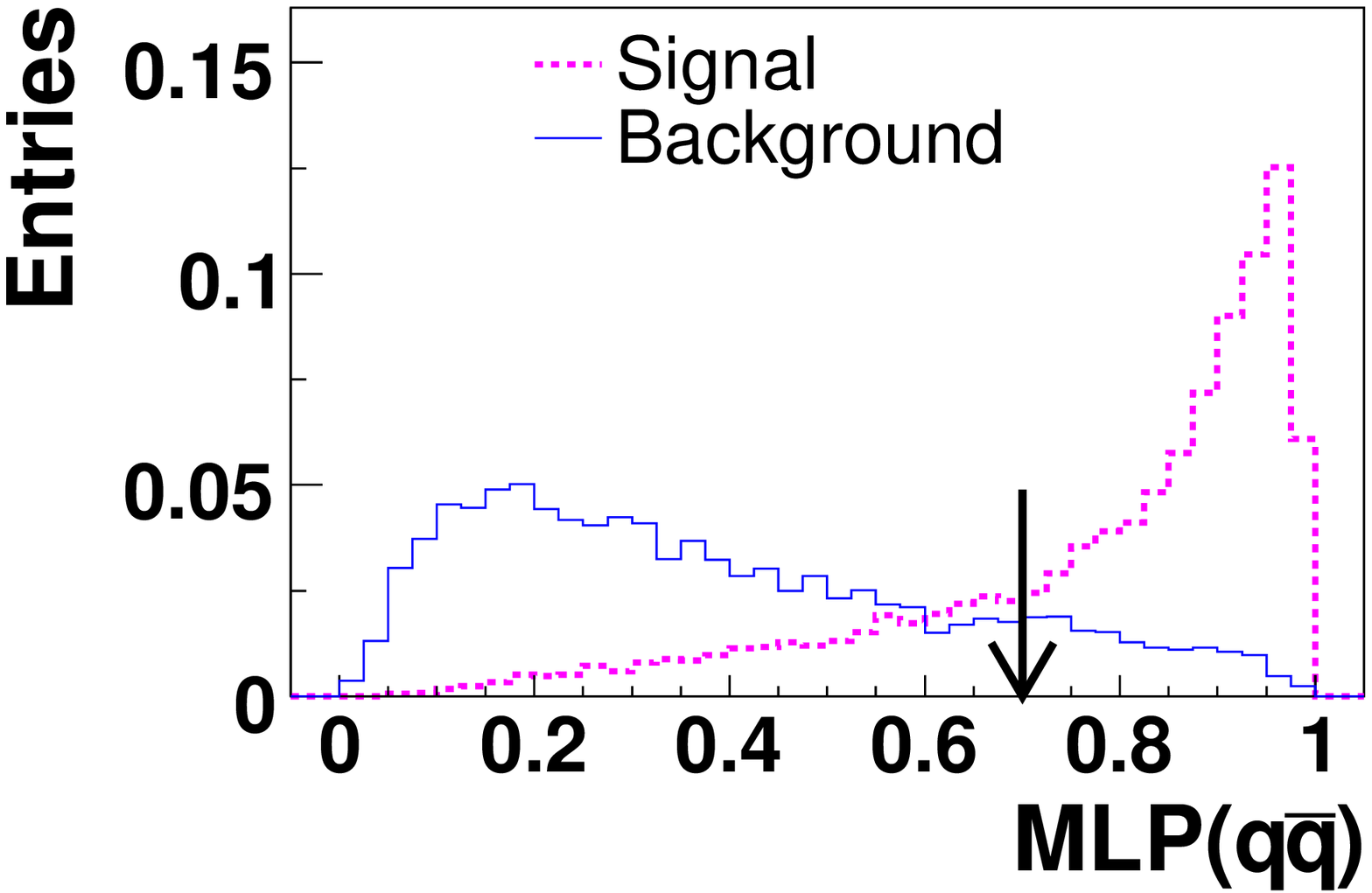}
      {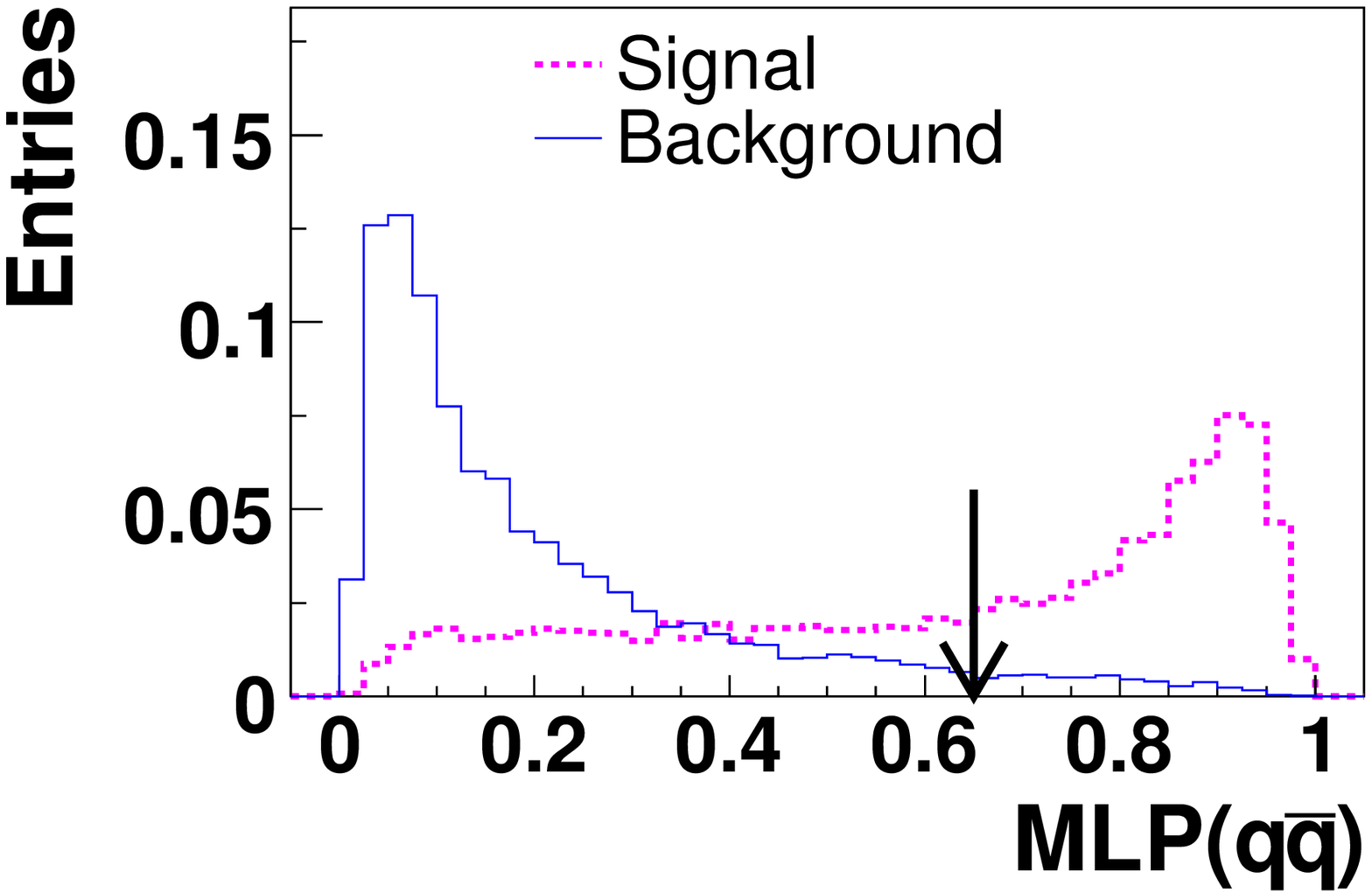}
  \threeFigTwoCol
	  {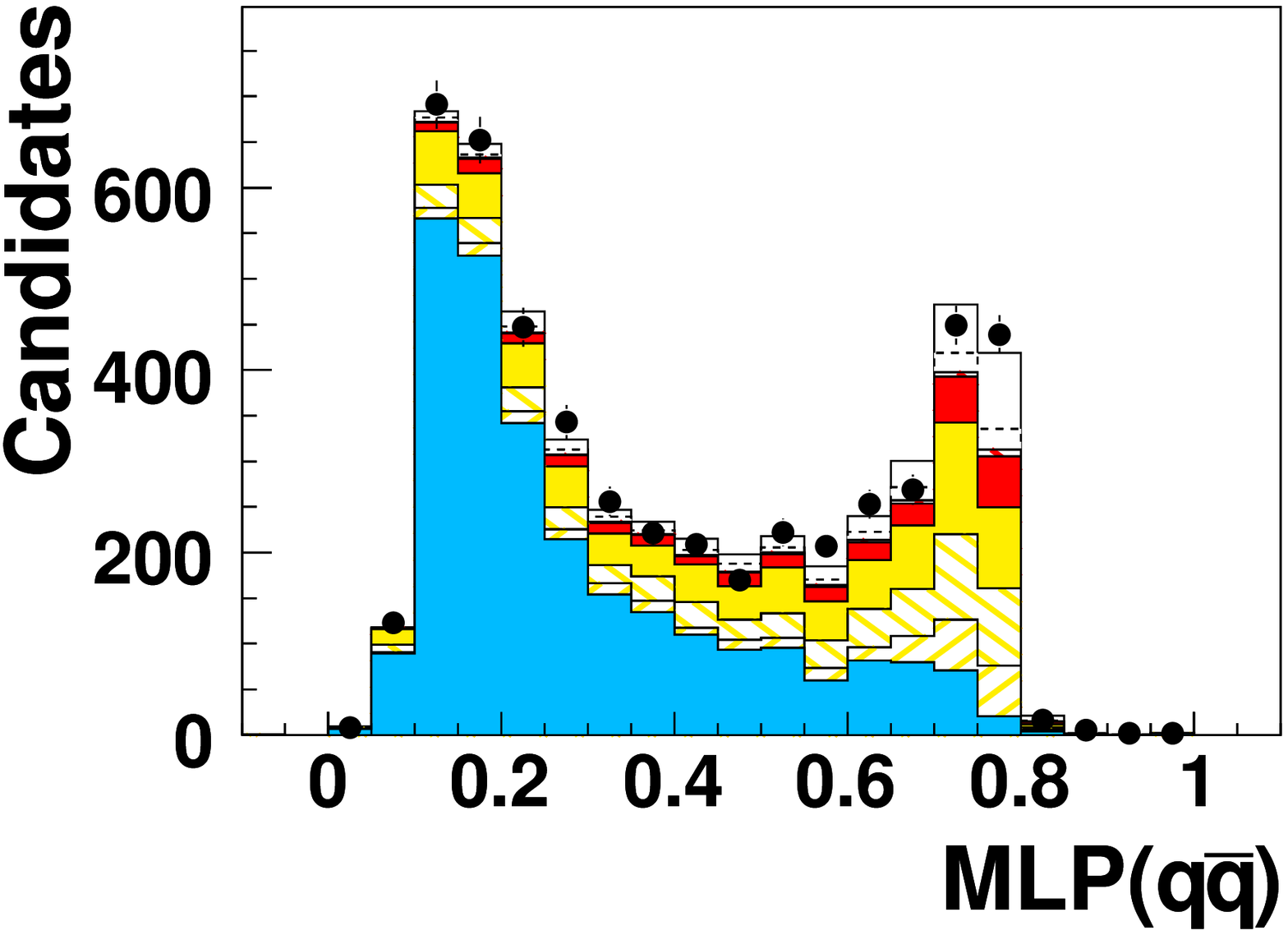}
	  {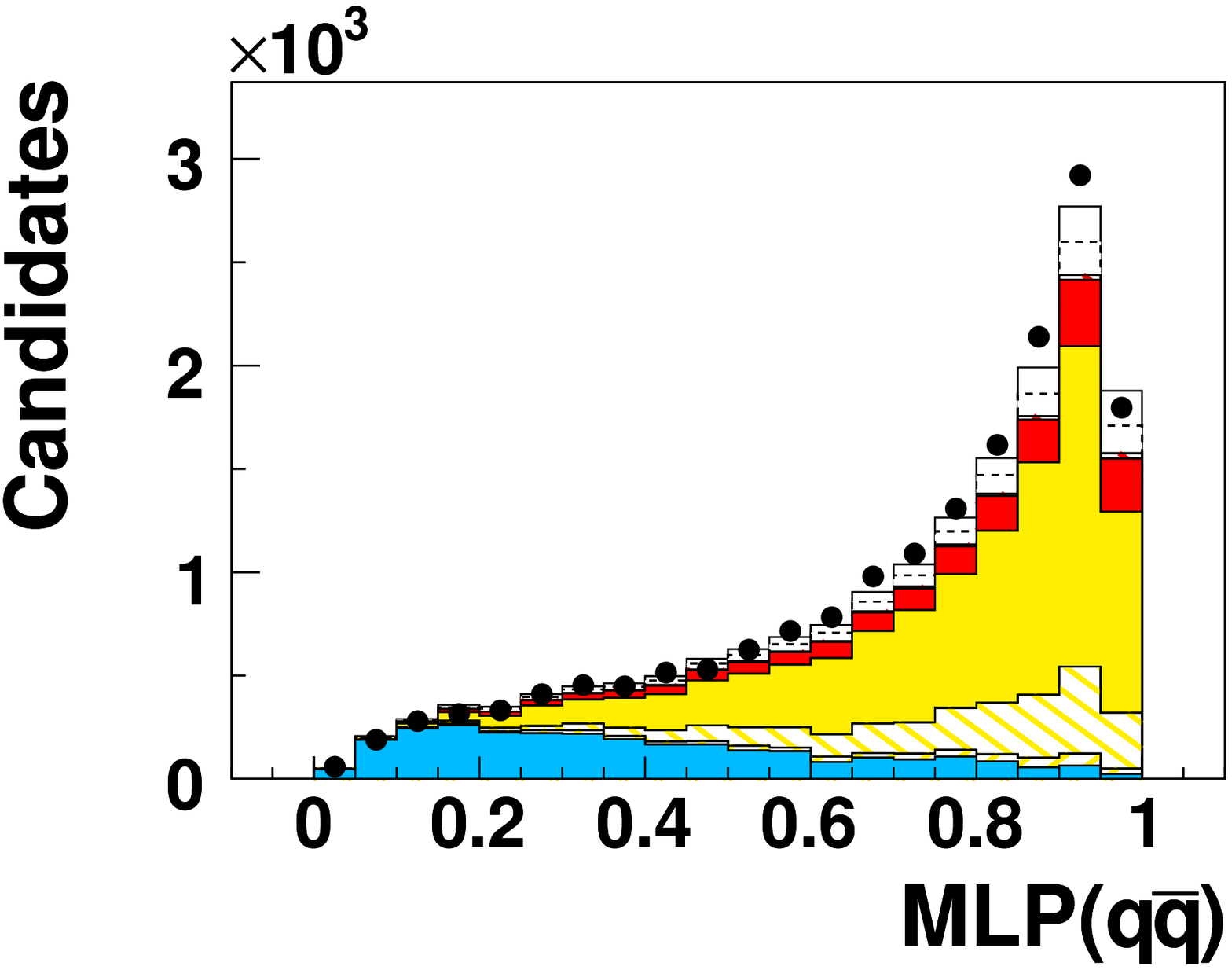}
	  {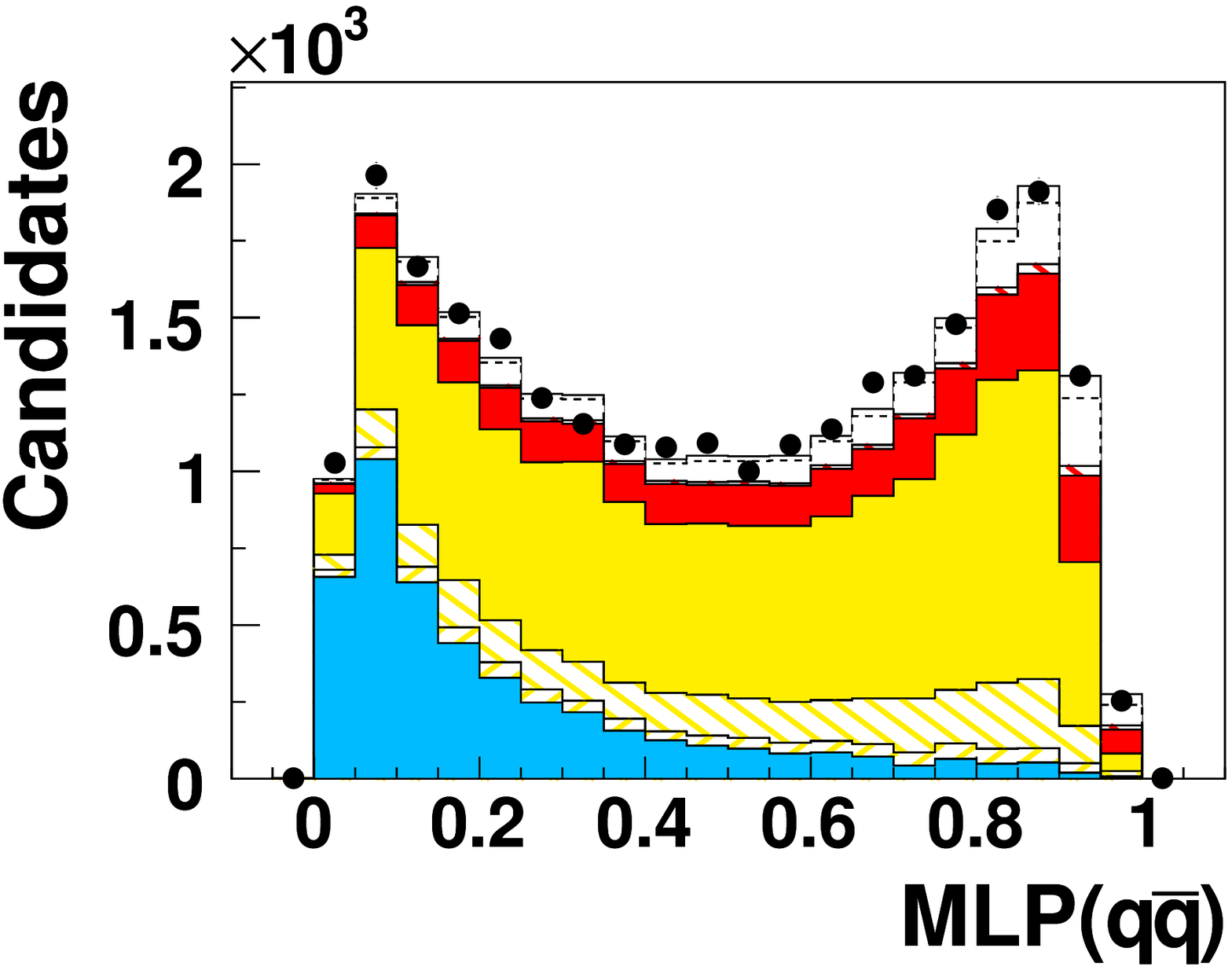}
  \caption{(color online)
The \qq\ neural-network discriminators for \Bzrholnu\ candidates in the signal region, \SignalRegion.
The distributions are shown for three different $q^2$ bins, columns from left to right:
   $0  <\q^2 < 8 \gev^2$,
   $8 <\q^2 < 16 \gev^2$,
   $    \q^2 > 16 \gev^2$.
Top row: Discriminator distributions
for signal (magneta, dashed) and $\qq$ background (blue, solid), normalized to the same area. 
The arrows indicate the chosen cuts.
Bottom row: Discriminator distributions for data compared with MC-simulated signal and background contributions.
For a legend see Figure~\ref{fig:legend}.
}
  \label{fig:NNoutputrholnu1}
\end{figure*}

\begin{figure*}[htb]
  \threeFigTwoCol
      {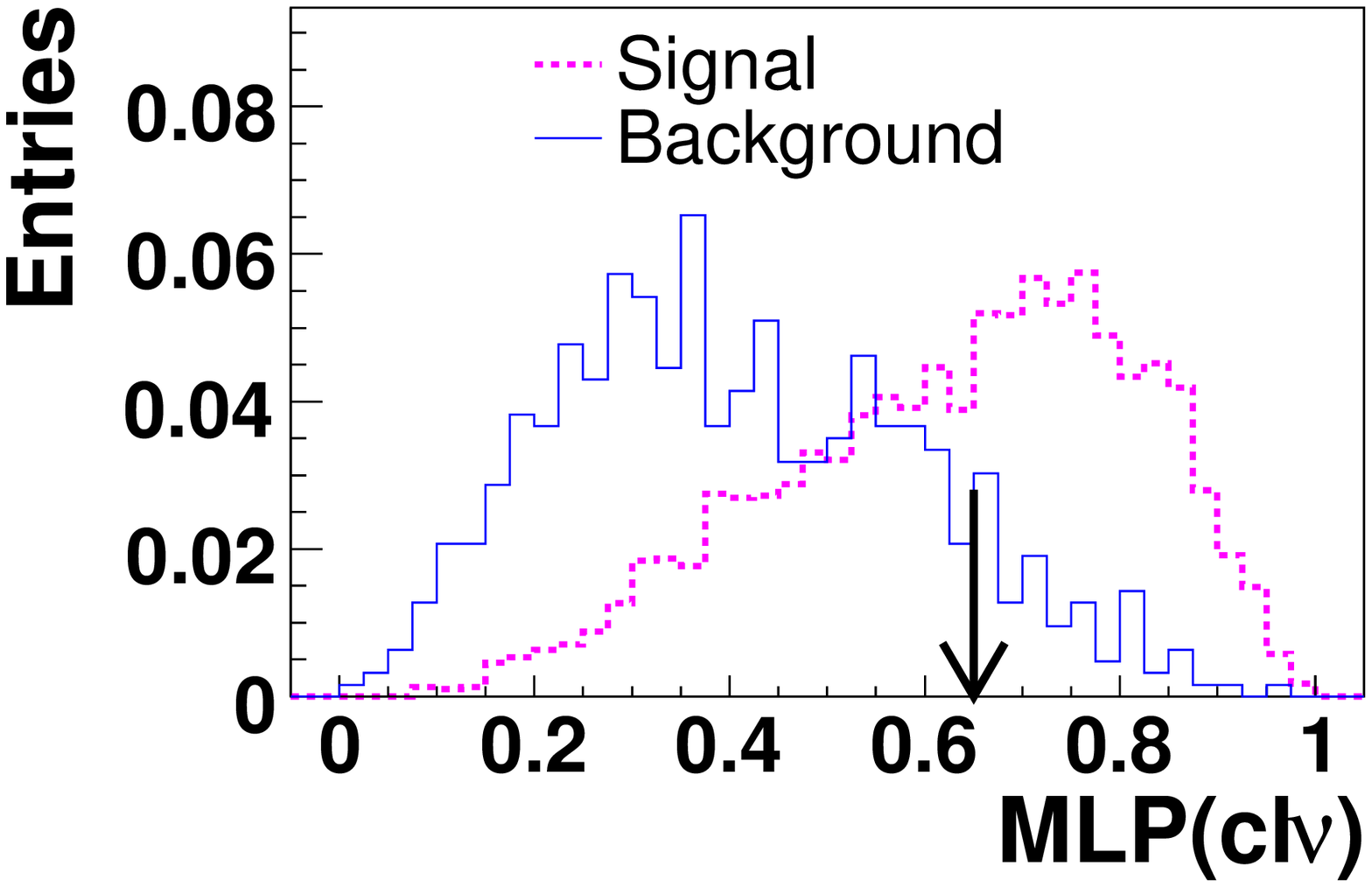}
      {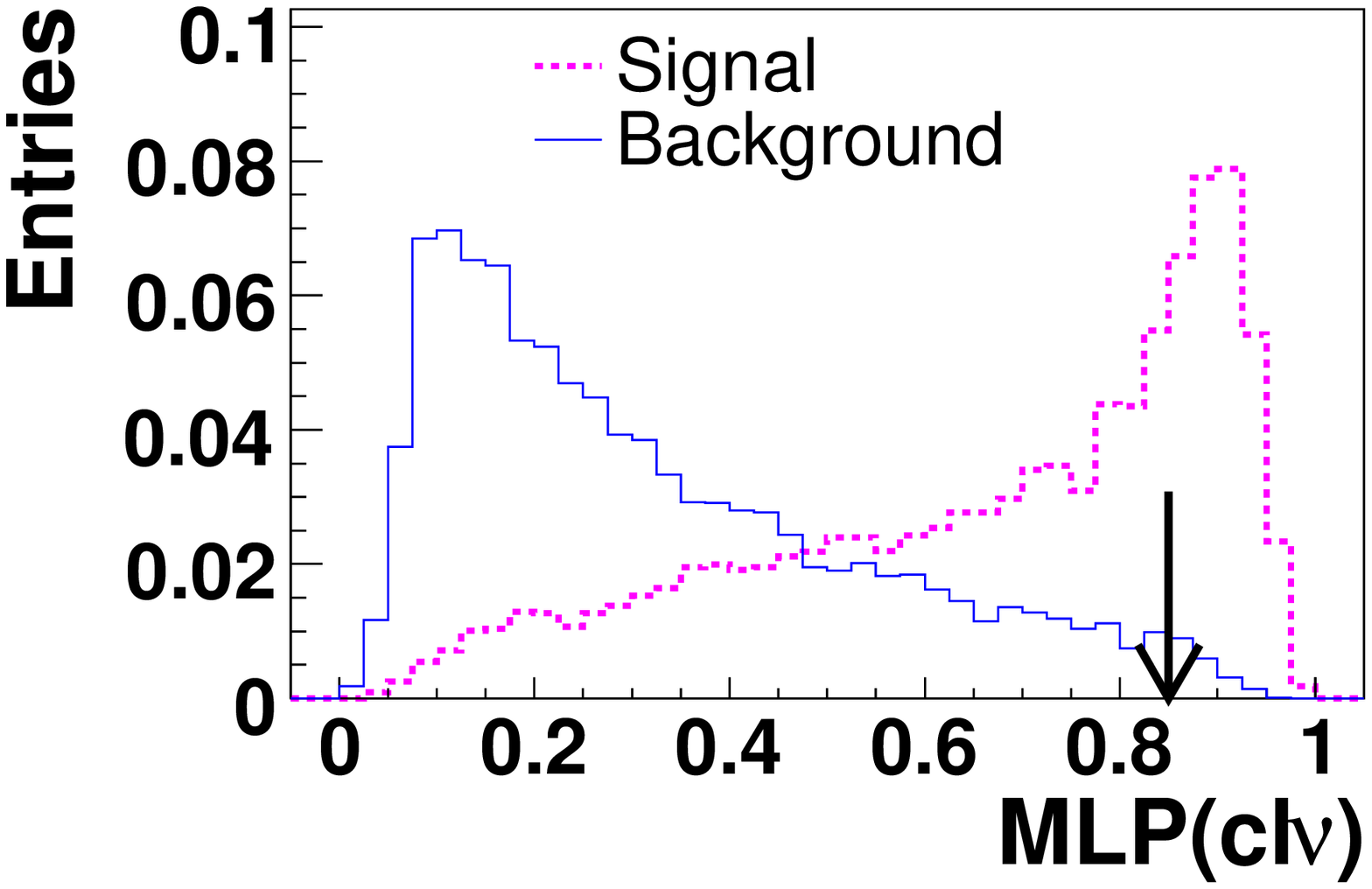}
      {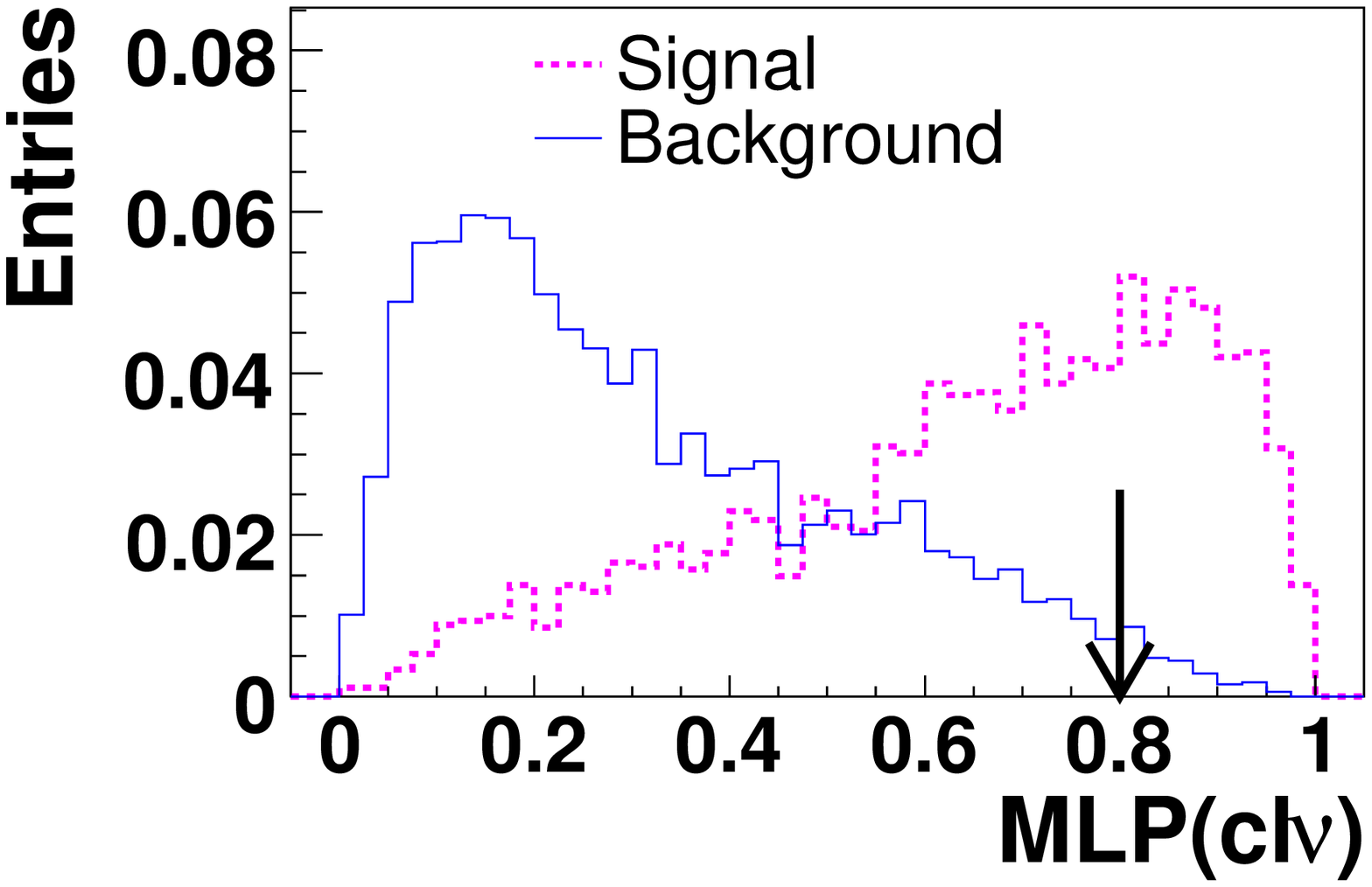}
      \threeFigTwoCol
	  {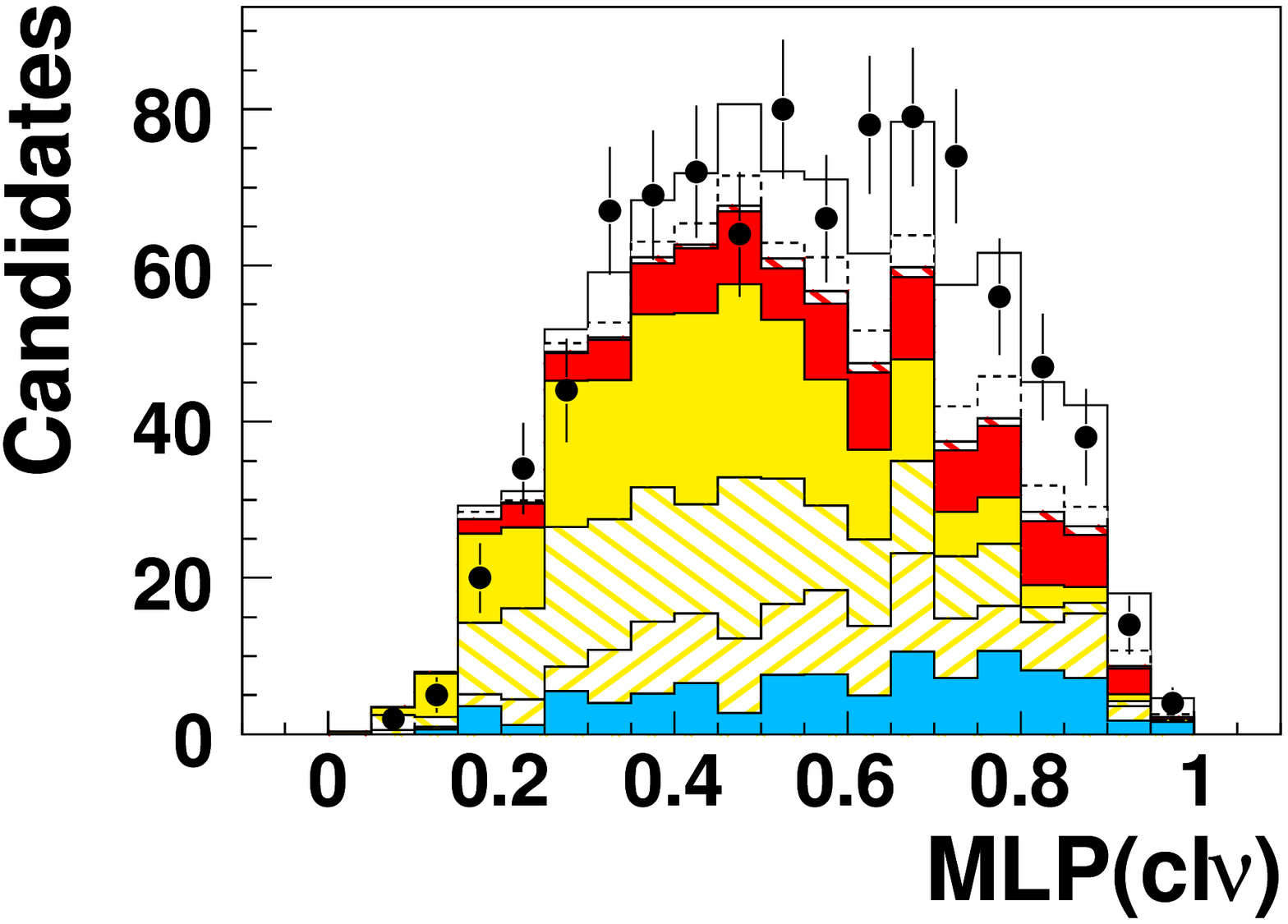}
	  {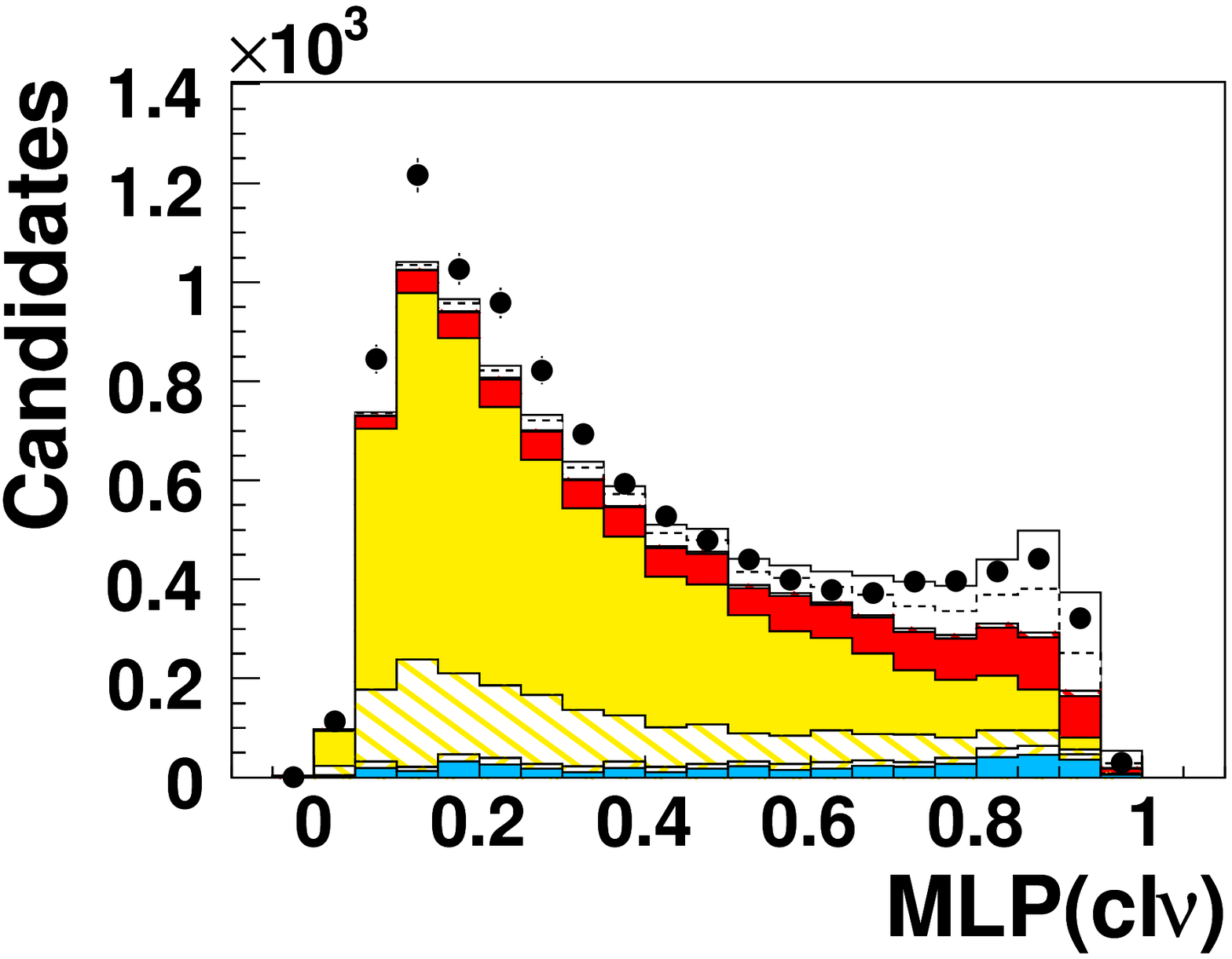}
	  {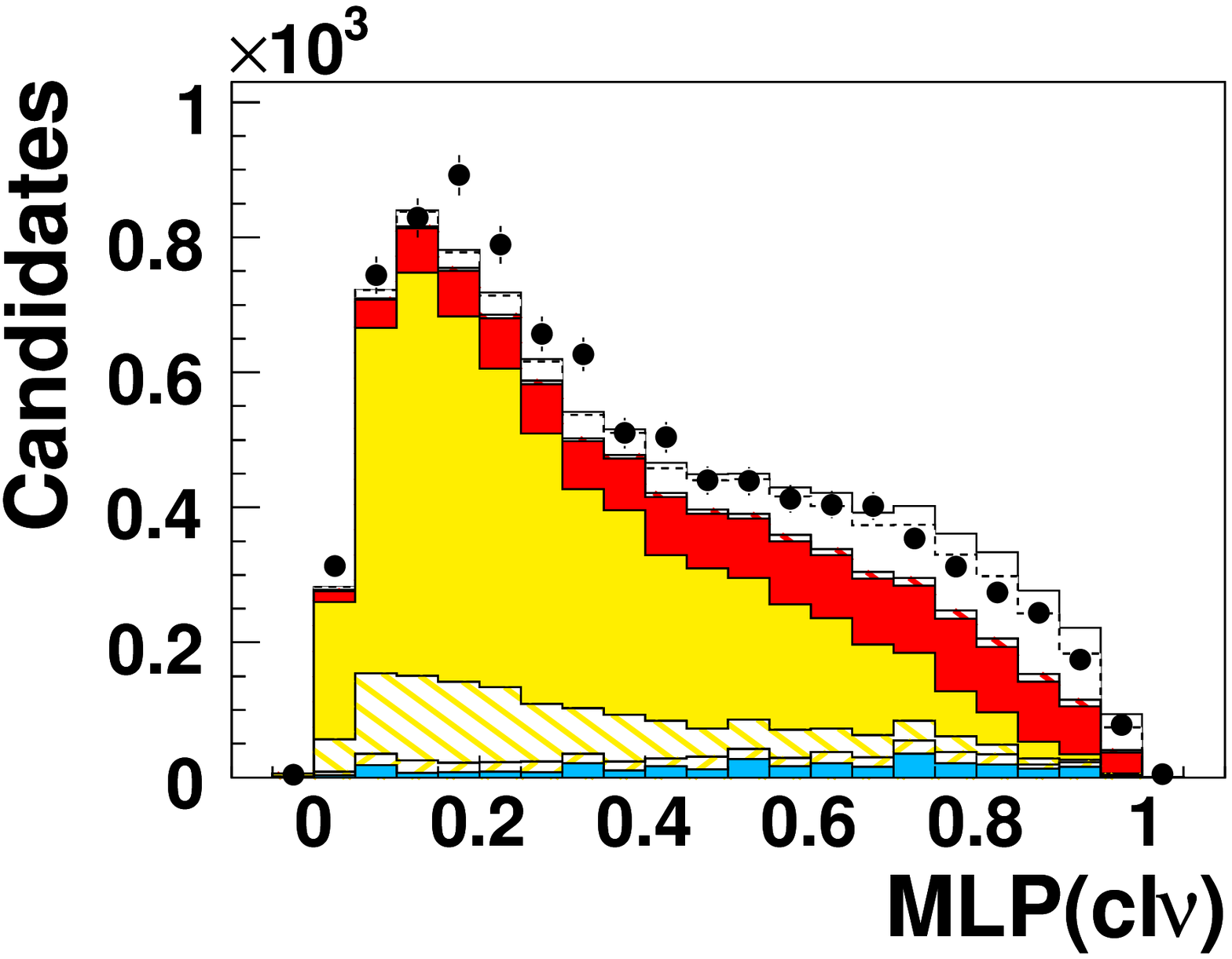}
\caption{(color online)
The \bclnu\ neural-network discriminators for \Bzrholnu\ candidates in the signal region, \SignalRegion.
The distributions are shown for three different $q^2$ bins, columns from left to right:
   $0  <\q^2 < 8 \gev^2$,
   $8 <\q^2 < 16 \gev^2$,
   $    \q^2 > 16 \gev^2$.
Top row: Discriminator distributions
for signal (magneta, dashed) and \bclnu\ background (blue, solid), normalized to the same area. 
The arrows indicate the chosen cuts.
Bottom row: Discriminator distributions for data compared with MC-simulated signal and background contributions.
For a legend see Figure~\ref{fig:legend}.
}
\label{fig:NNoutputrholnu2}
\end{figure*}

Figure~\ref{fig:NNinputpilnu} shows, for the sample of \Bzpilnu\ candidates,
the distributions of the seven input variables to the neural networks. 
The distributions are shown sequentially after application of the preselection, 
the \qq\ neural network and the \bclnu\ neural network to illustrate the change
in the sample composition. 
Figures~\ref{fig:NNoutputpilnu1} to~\ref{fig:NNoutputpilnu3} 
show the distributions of the three neural-network discriminators 
for the \Bzpilnu\ sample in four of the six $q^2$ bins. 
Figures~\ref{fig:NNoutputrholnu1} and~\ref{fig:NNoutputrholnu2} 
show the distributions of the two neural-network discriminators 
for the \Bzrholnu\ sample in all three $q^2$ bins.
The discriminator cuts are chosen to minimize the total error on the signal yield for each channel, using the sum in quadrature of the error obtained from the maximum-likelihood fit described in Section~\ref{sec:fit} 
and the estimated total systematic error of the partial signal branching fraction in each $q^2$ bin (see Section~\ref{sec:systematics}).  
The data-MC agreement is reasonably good for the input distributions
and the neural-network discriminators. One should keep in mind that at this stage
the distributions are taken directly from the simulation, without any adjustments or fit.

\begin{table}[hbt]
\centering
\caption{Overview of the selection efficiencies for the four signal decays 
(true and combinatorial signal combined) and 
their primary background sources, $\bulnu$, $\bclnu$, and non-\BB\ background.
}

\begin{tabular}{llrrrr} 
\hline \hline
&Selection       &  Signal   &$X_u\ell\nu$ & $X_c\ell\nu$ &  \qq\  \\
&Units           & $10^{-2}$ & $10^{-3}$ & $10^{-4}$   & $10^{-5}$ \\
\hline
 $\Bzpilnu$    & &           &           &             &             \\
& Preselection   &   18.1    &  25.6     &   26.4      &  19.4 \\
&  NN \qq\       &   11.6    &  13.4     &   18.3      &  2.6    \\  
&  NN \BXclnu\   &    7.8    &  8.1      &   3.4       &  1.3    \\
&  NN \BXulnu\   &    6.8    &  5.3      &   2.5       &  1.0    \\ 
&  Signal region &    1.8    &  0.5      &   0.1       &  $<$0.1 \\
\hline

\Bppizlnu\     & &           &           &             &         \\
&  Preselection  &   12.8    &  20.0     &   17.4      &  15.1   \\   
&  NN \qq\       &    8.4    &  11.3     &   13.3      &  2.7    \\
&  NN \BXclnu\   &    5.9    &   6.6     &    1.9      &  1.6    \\
&  Signal region &    1.6    &   0.5     &    $<$0.1   & $<$0.1  \\
\hline

\Bzrholnu\    &  &           &           &             &         \\
&  Preselection  &    8.9    &  23.9     &   35.8      &  13.2   \\
&  NN \qq\       &    4.8    &  11.9     &   18.5      &   1.0   \\
&  NN \BXclnu\   &    1.1    &   1.9     &    0.3      &   0.2   \\
&  Signal region &    0.3    &   0.3     &    $<$0.1   & $<$0.1  \\
\hline

\Bprhozlnu    &  &           &           &             &         \\
&  Preselection  &   11.1    &  22.1     &   30.0      &  12.6   \\
&  NN \qq\       &    6.8    &  12.4     &   17.7      &   1.5   \\
&  NN \BXclnu\   &    2.5    &   3.2     &    0.7      &   0.5   \\
&  Signal region &    0.8    &   0.6     &    $<$0.1   &  $<$0.1 \\ 
\hline\hline

\end{tabular}

\label{tab:efficiencies}
\end{table}

Table~\ref{tab:efficiencies} shows the selection efficiencies for the four signal samples compared to the efficiencies for the dominant background sources for these samples.  The total signal efficiencies are typically $6-7$\% for \Bpilnu\ decays and roughly $1-2.5$\% for \Brholnu\ decays in the fit region. The dominant \BB\ and \qq\ backgrounds are suppressed by factors of order $10^4$ and $10^5$, respectively.

\subsubsection{Candidate Multiplicity}
\label{sec:nCand}

After the neural-network selection there are on average 1.14 candidates per event in the \Bzpilnu\ sample, 1.46 in the \Bppizlnu\ sample, 1.30 in the \Bzrholnu\ sample, and 1.17 in the \Bprhozlnu\ sample. We observe fewer candidates for decay modes without neutral pions in the final state. For all four samples, the observed candidate multiplicity is well reproduced  by MC simulation. 

In case of multiple candidates for a given decay mode, 
we select the one with the highest probability of the vertex fit for the 
$Y$ candidate. Since this is not an option for \Bppizlnu\ decays, we select 
the photon pair with an invariant mass closest to the \piz\ mass.   
Simulations of signal events indicate that this procedure selects the correct
signal decay in $60 - 65$\% of the cases. By this selection we do not allow a single event to contribute more than one candidate to a given decay-mode sample, though we do allow an event to contribute candidates to more than one decay-mode sample.

\section{Data-Monte Carlo Comparisons}

The determination of the number of signal events relies heavily on the MC 
simulation to correctly describe the distributions for signal and background sources. Therefore a significant effort has been devoted to 
detailed comparisons of data and MC distributions 
for samples that have been selected to enhance a given source of
background.

\subsection{Comparison of Off-Resonance Data with {\boldmath \qqbar} MC}
\label{sec:offres}

Though we record data below \BB\ threshold (off-resonance data)
the total luminosity of this sample is only about 10\% 
of the \FourS\ data sample (on-resonance data), and
thus we need to rely on MC simulation to predict 
the shapes of these background distributions. 

To study the simulation of \qqbar\ events, we scale the MC
sample to match the integrated luminosity of the
off-resonance data. The study is performed separately for samples with electrons and muons. 
This background contains events with true leptons from
leptonic or semileptonic decays of hadrons, as well as hadrons 
misidentified as leptons.  The muon sample is dominated by misidentified hadrons,
whereas the electron sample contains small contributions from Dalitz pairs and photon conversions, as
well as some residual background from 
non-\qqbar\ processes.  We observe a clear difference in the normalization,
not only in the relatively small event sample passing the neural-network selection, but also for the much larger sample available before the neural-network suppression.  
To correct for this difference, we apply additional scale factors 
to the simulated \qqbar\ samples; they are different for electrons and muons. 

\begin{figure}[htb]
\twoFigOneCol{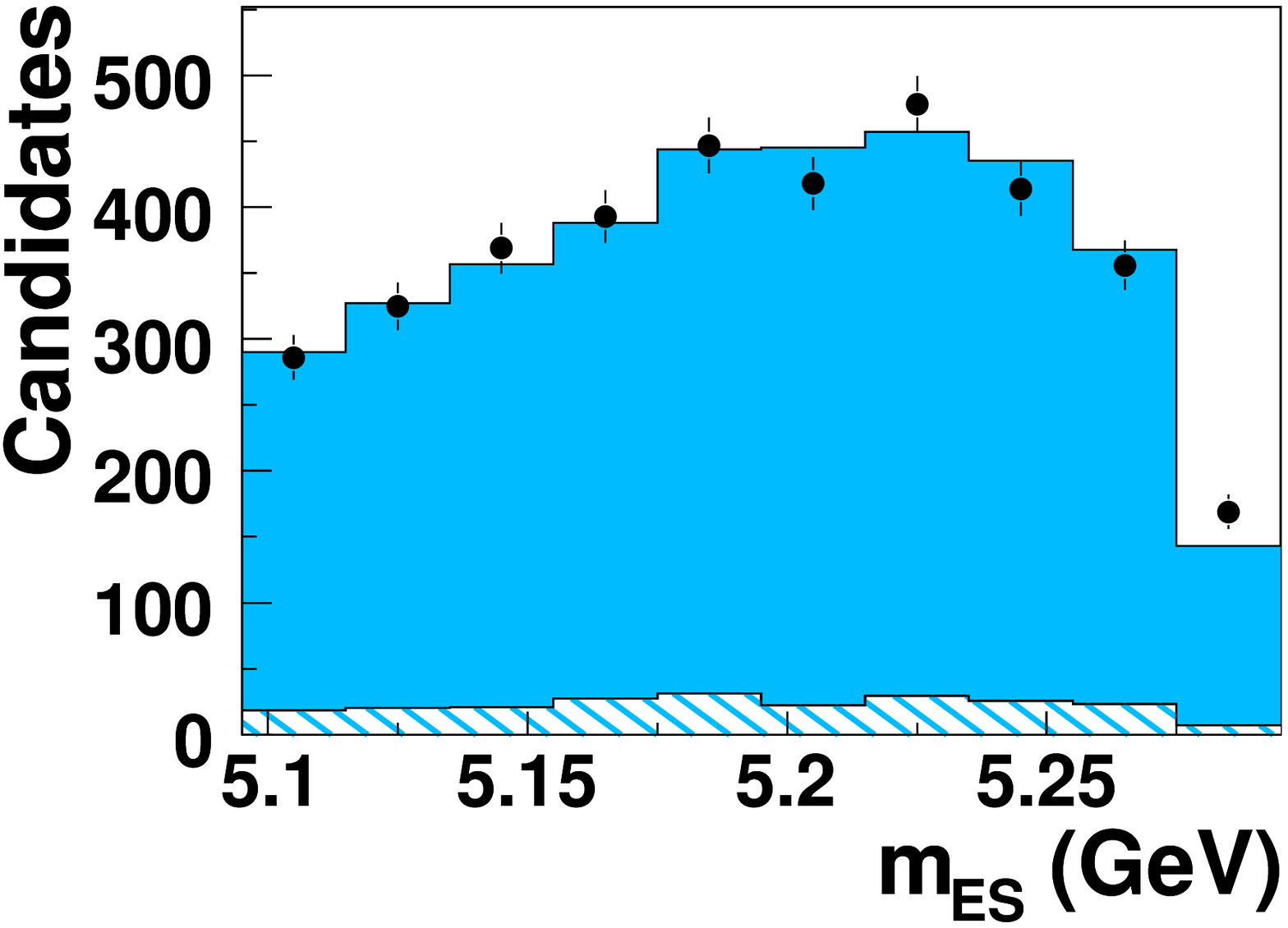}   {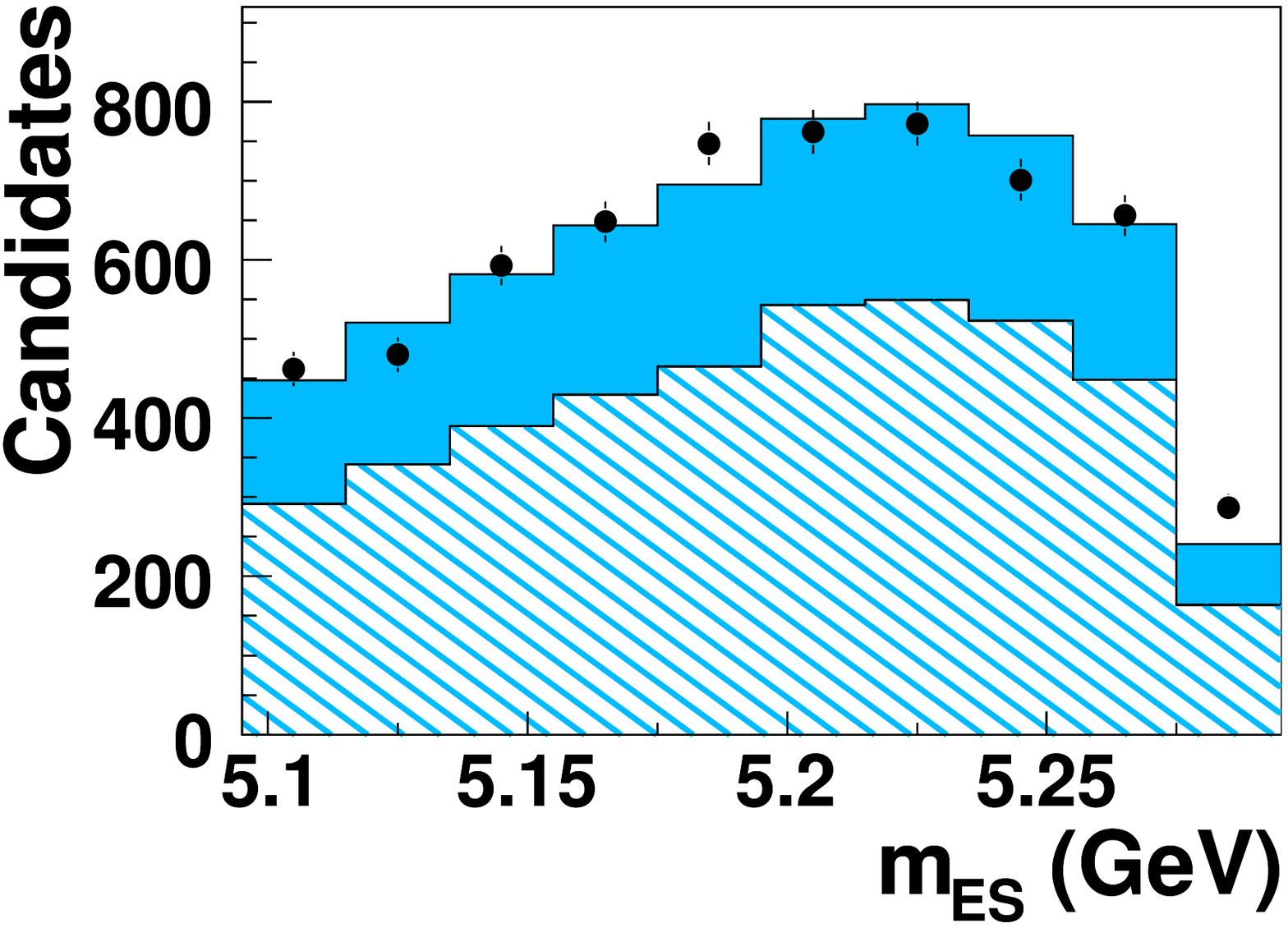}
\twoFigOneCol{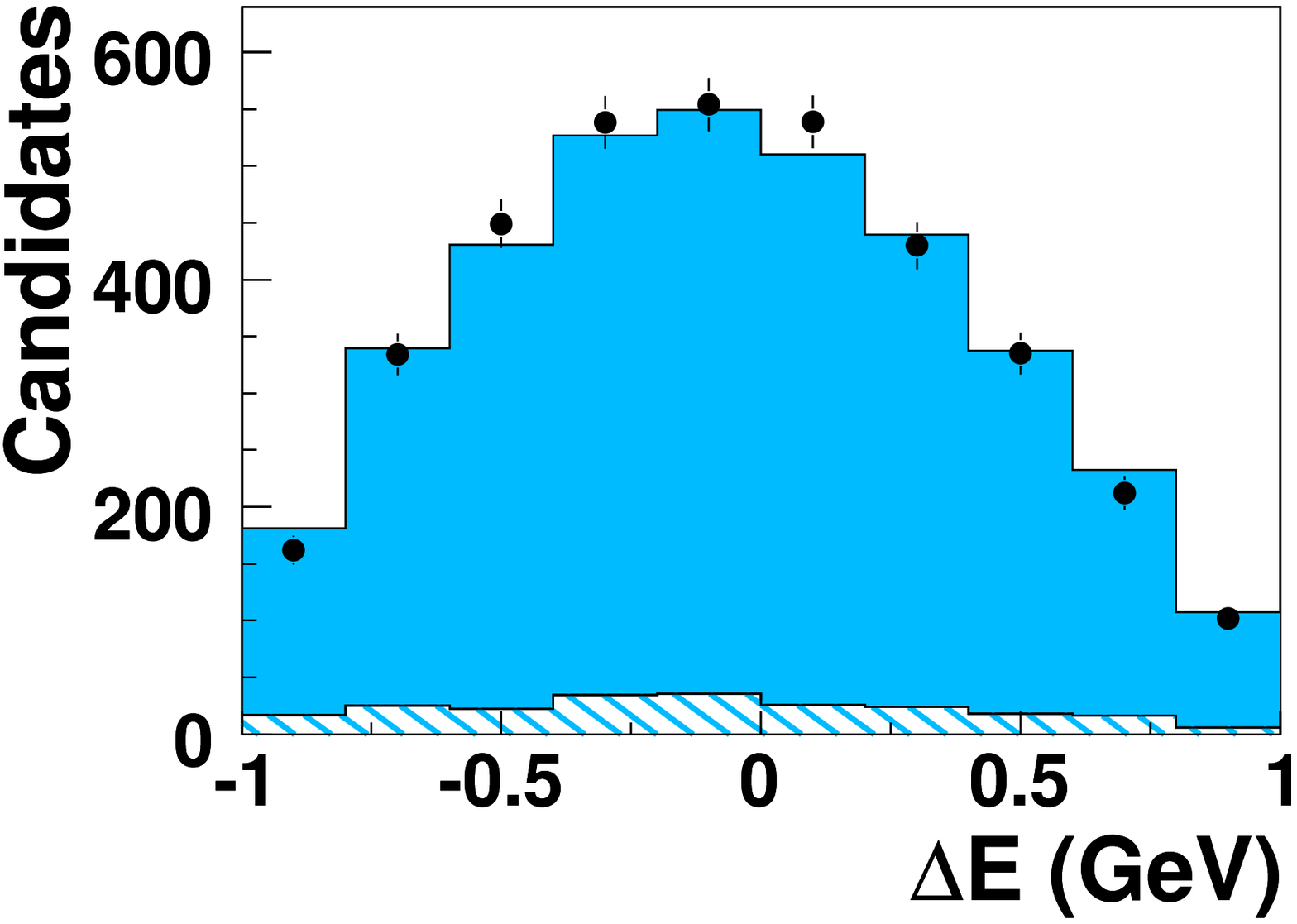}{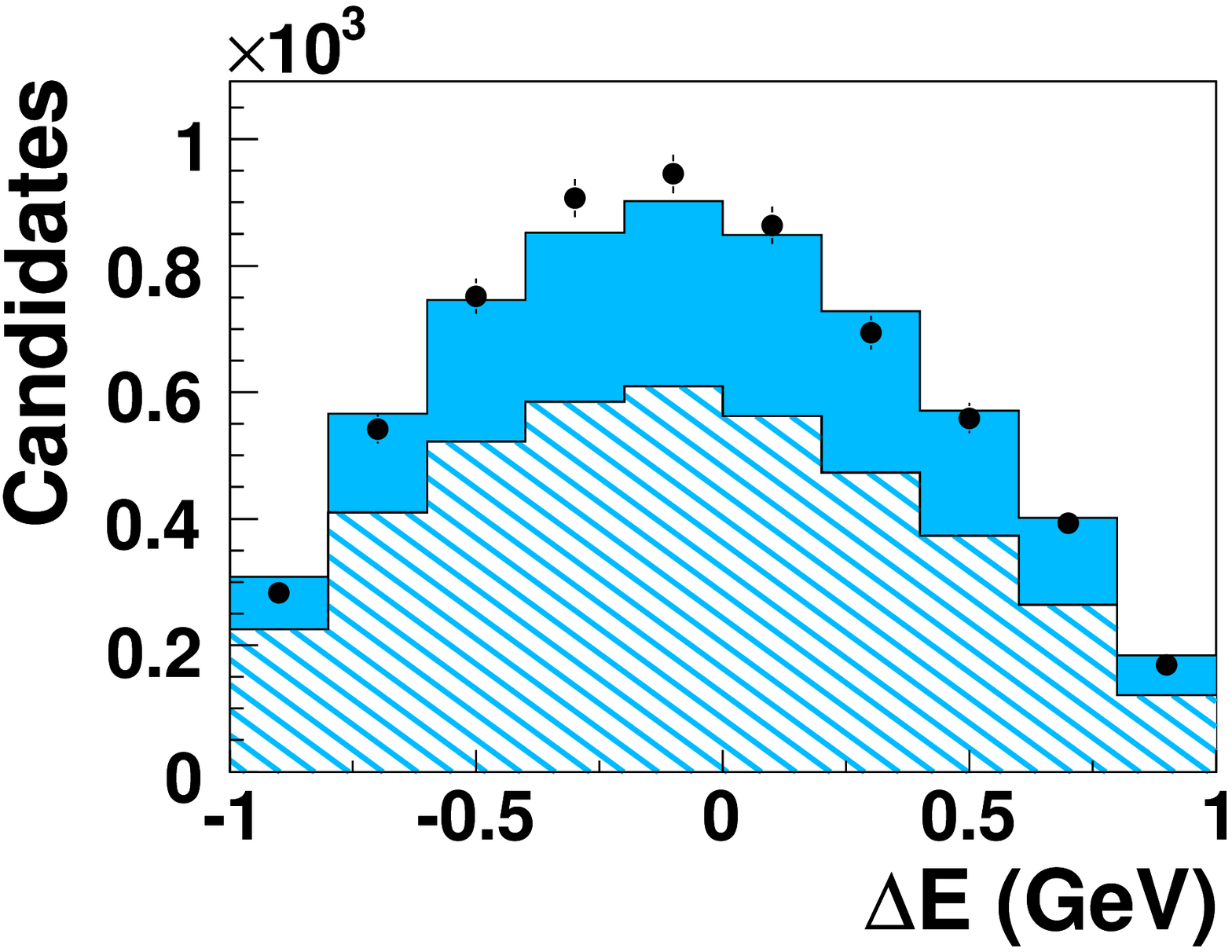}
\twoFigOneCol{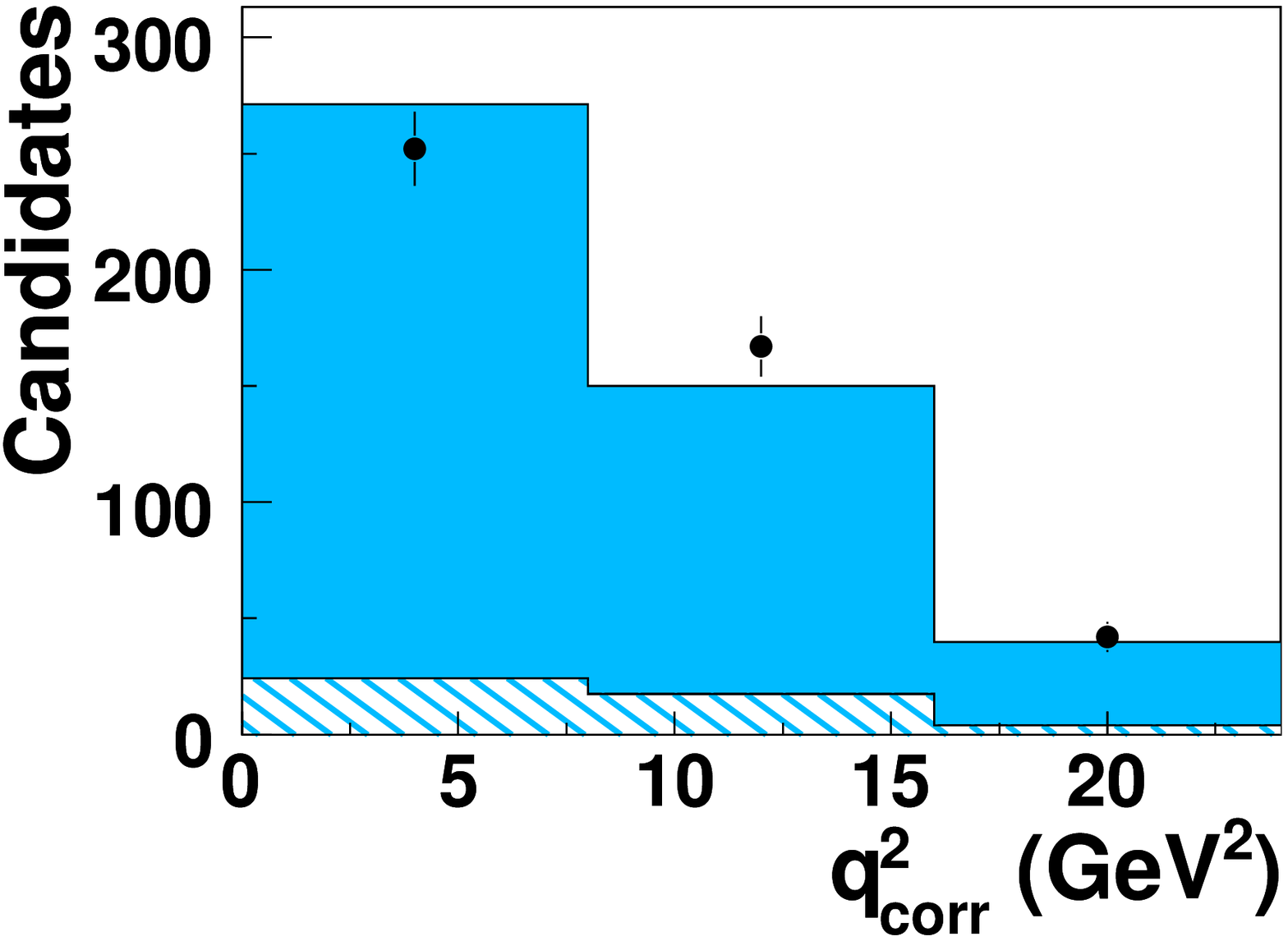}{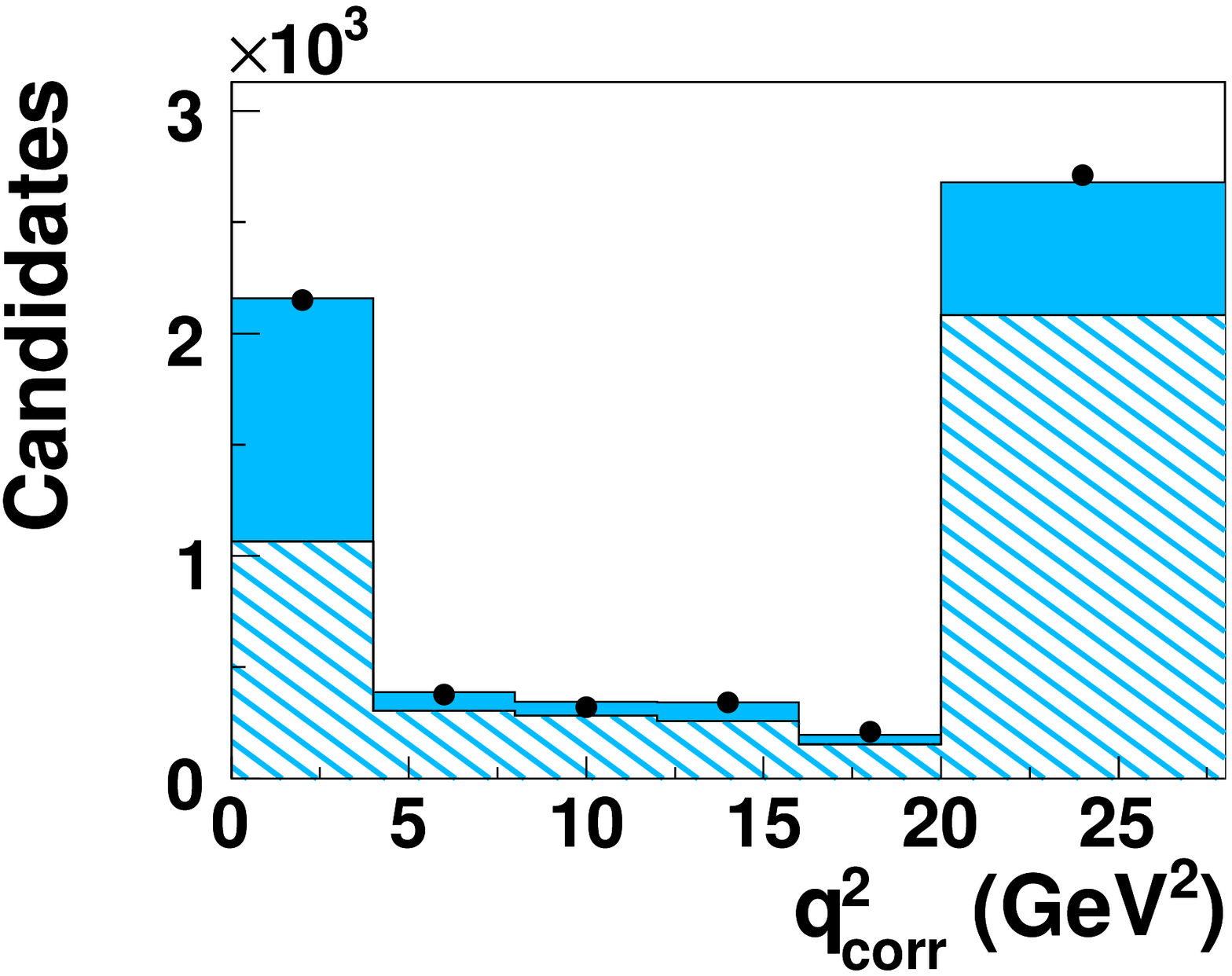}
\caption{(color online)
Comparison of off-resonance background to \Bzpilnu\ samples for data and MC distributions. 
  Top row: \mES, center row: \DeltaE, and bottom row: $q^2$, separately for the electron (left column) and muon (right column) samples. The shaded histograms indicate the true leptons, the hatched histograms indicate the fake leptons. 
The distributions are obtained from the full event selection, except for the \qq\ neural-network discrimination.
Linear corrections have been applied to the simulation. 
}
\label{fig:contcorrelpilnu}
\end{figure}

\begin{figure}[htb]
\twoFigOneCol{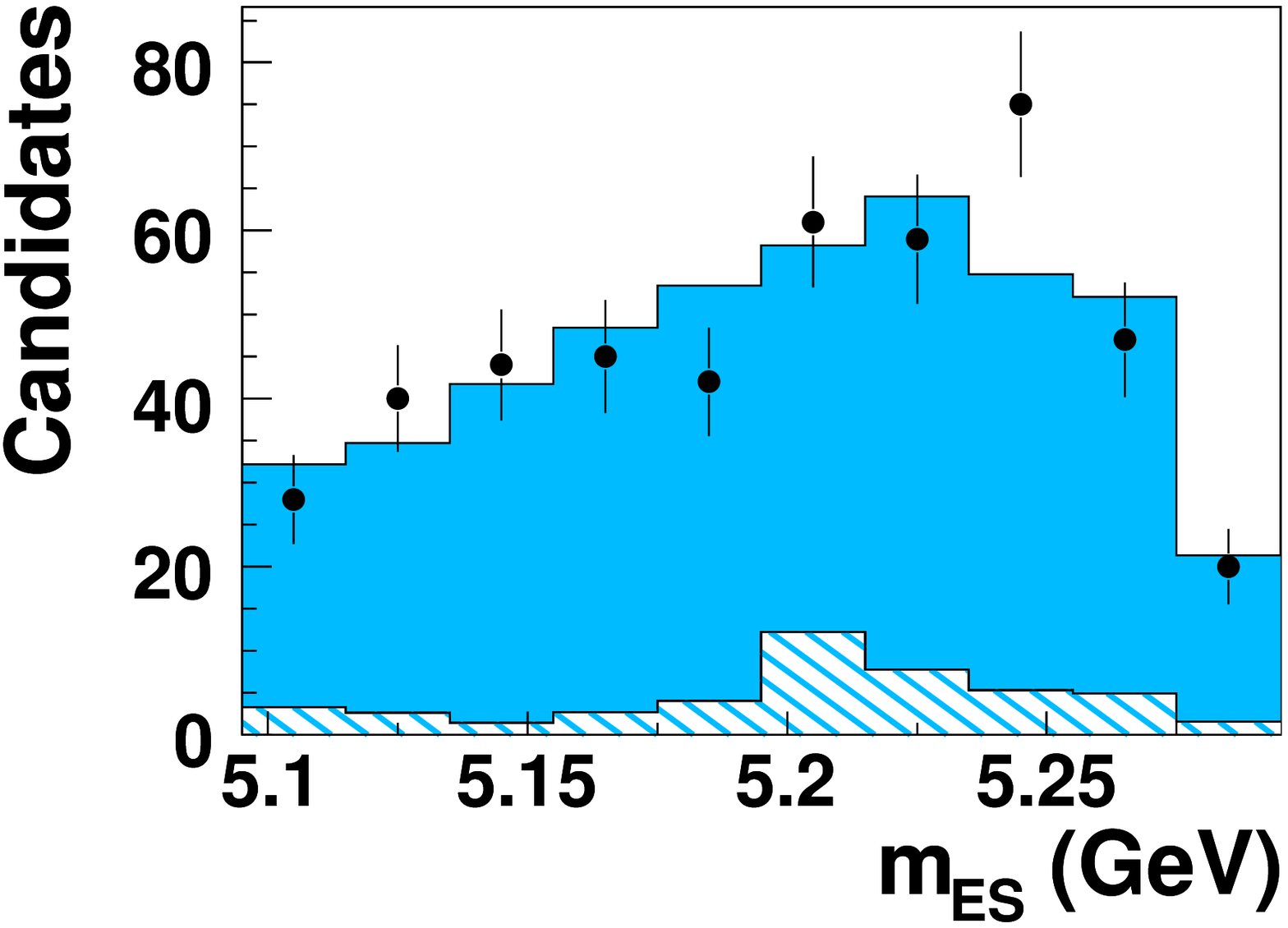}   {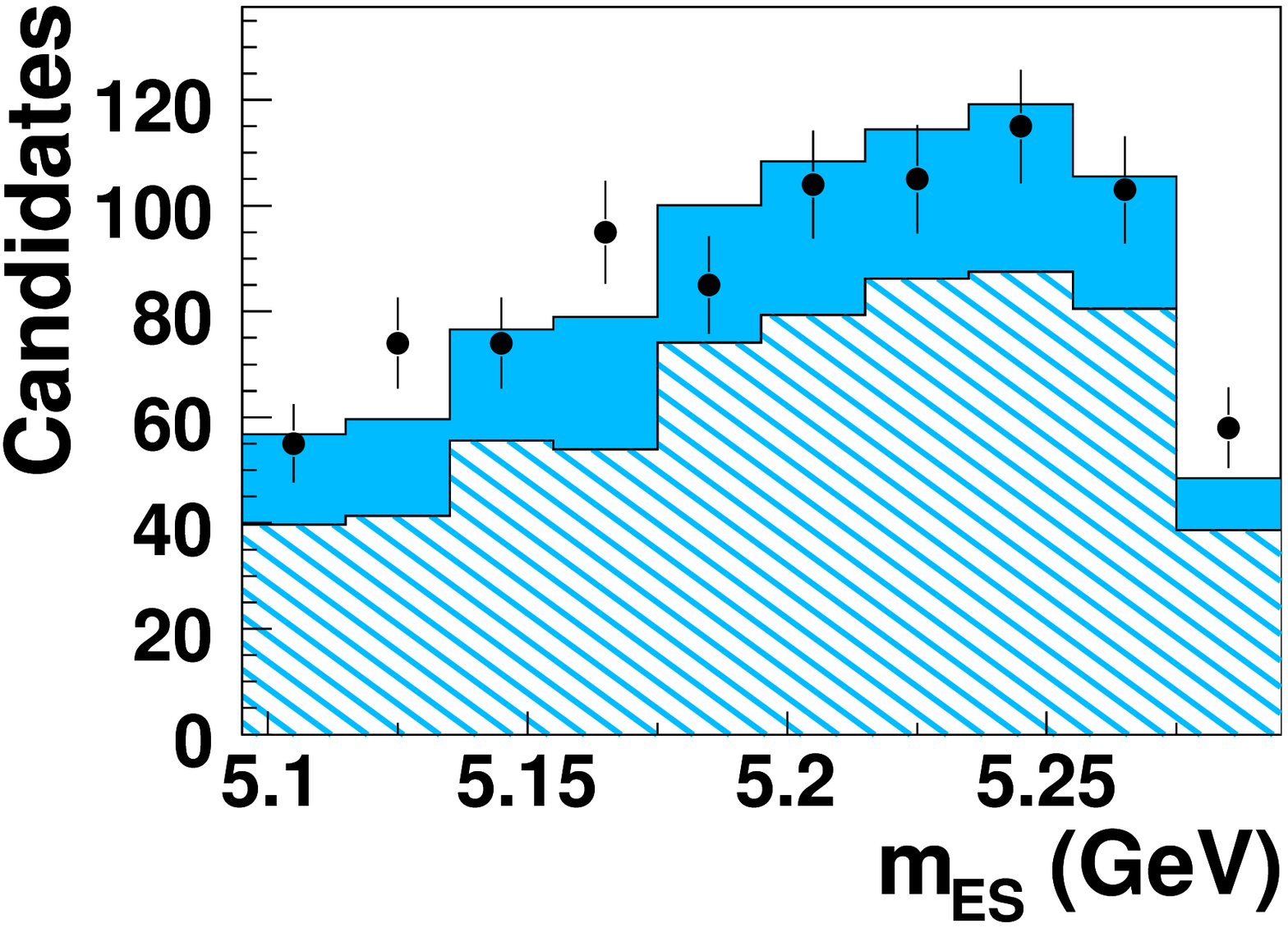}
\twoFigOneCol{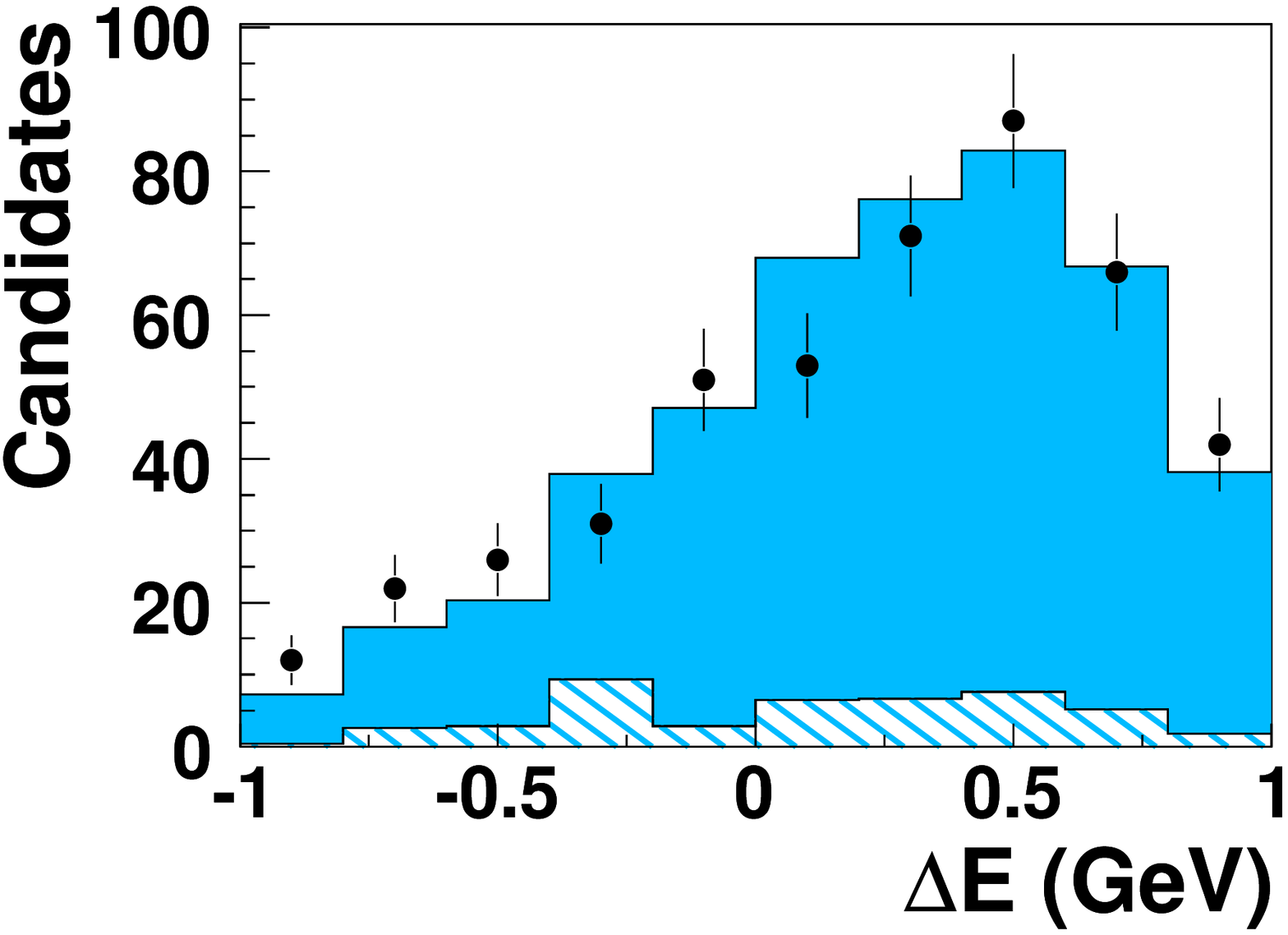}{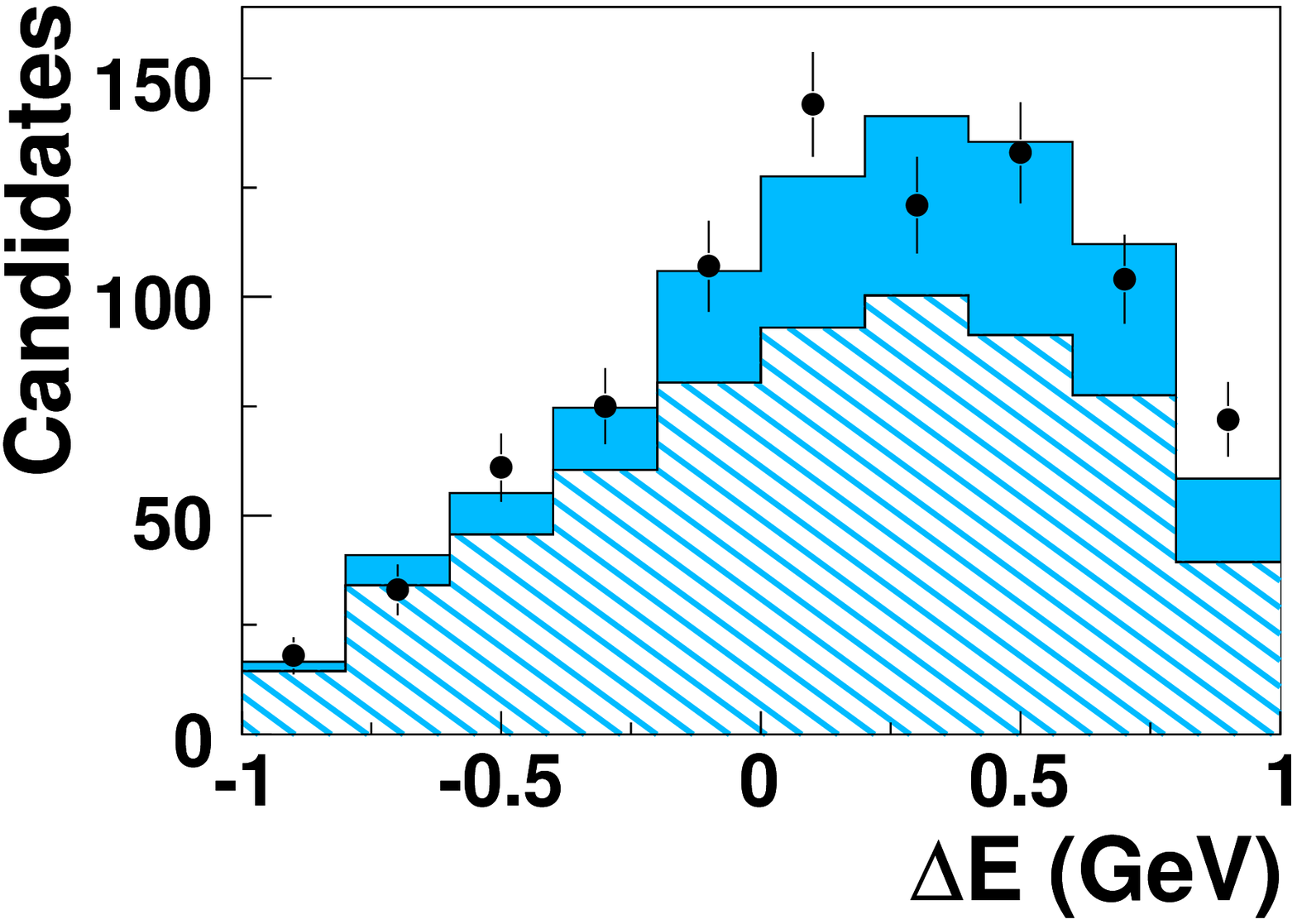}
\twoFigOneCol{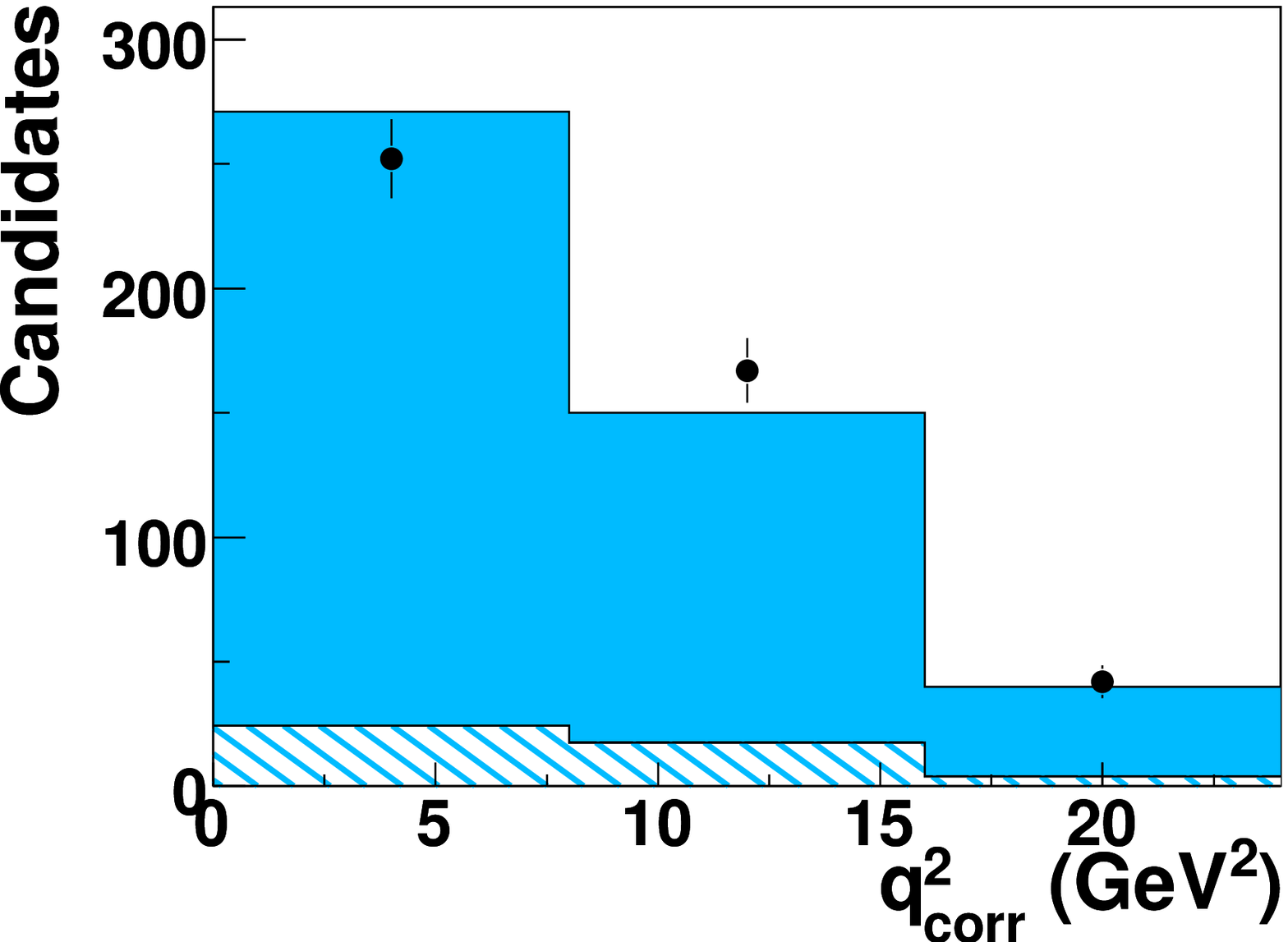}{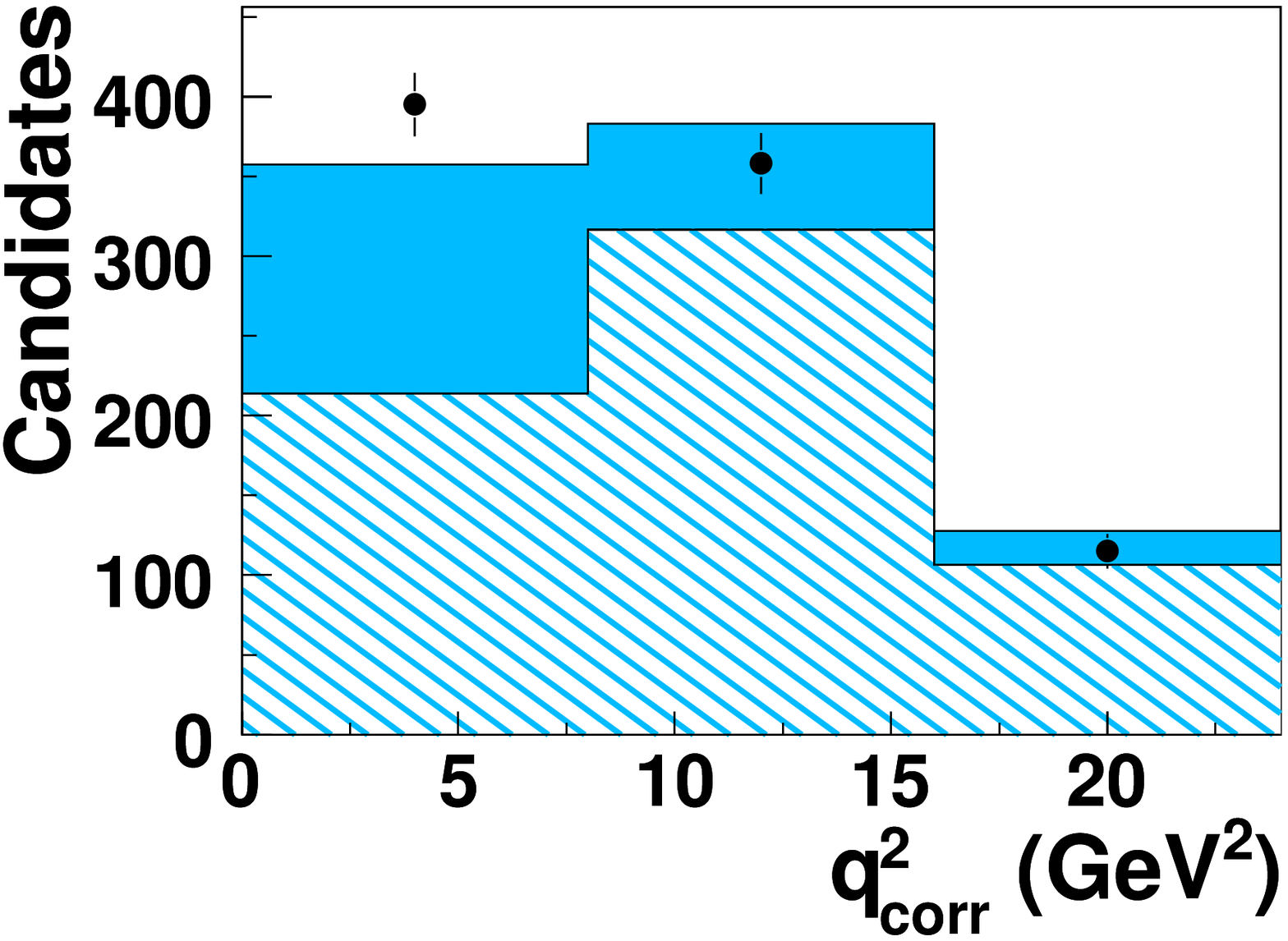}
\caption{(color online)
Comparison of off-resonance background to \Bzrholnu\ samples for data and MC distributions. 
  Top row: \mES, center row: \DeltaE, and bottom row: $q^2$, separately for the electron (left column) and muon (right column) samples. The shaded histograms indicate the true leptons, the hatched histograms indicate the fake leptons. 
The distributions are obtained from the full event selection, except for the \qq\ neural-network discrimination.
Linear corrections have been applied to the simulation. 
}
\label{fig:contcorrelrholnu}
\end{figure}

In addition to correcting the normalization, we also examine the shapes
of the \mES, \DeltaE, and $q^2$ distributions that are used to extract the signal yield.  Since the size of the off-resonance data set is small, we study samples with a looser selection, namely we bypass the \qq\ neural-network discrimination. The comparison of these \qq -enriched samples reveals small differences between data and simulation.  
We derive linear 
corrections from the bin-by-bin ratios and apply these corrections to the \mES, \DeltaE, and $q^2$ distributions.  
Figures~\ref{fig:contcorrelpilnu} and \ref{fig:contcorrelrholnu} 
show a comparison of the rescaled and corrected \qqbar\ MC samples with the 
off-resonance data for the \DeltaE, \mES, and $q^2$ distributions.
Within the relatively large statistical errors of the off-resonance data
the simulation agrees well with the data. The uncertainties in the 
shape of the simulated distributions will be assessed as a systematic 
uncertainty.

\subsection{{\boldmath \bclnu} Enhanced Sample}

The overall dominant background source in this analysis is \bclnu\ decays. 
Therefore it is important to verify that these decays are
correctly simulated. This has been done in two ways, a) by relaxing the
\bclnu\ suppression to obtain a charm-enhanced sample, and b)  
by reconstructing a specific decay mode, such as \Bzdslnu,  
in the same way we reconstruct the signal decays, and comparing the kinematic distributions  with MC simulations (see Section \ref{sec:controlsample}). 
 
\begin{figure}[htb]
  \twoFigOneCol
      {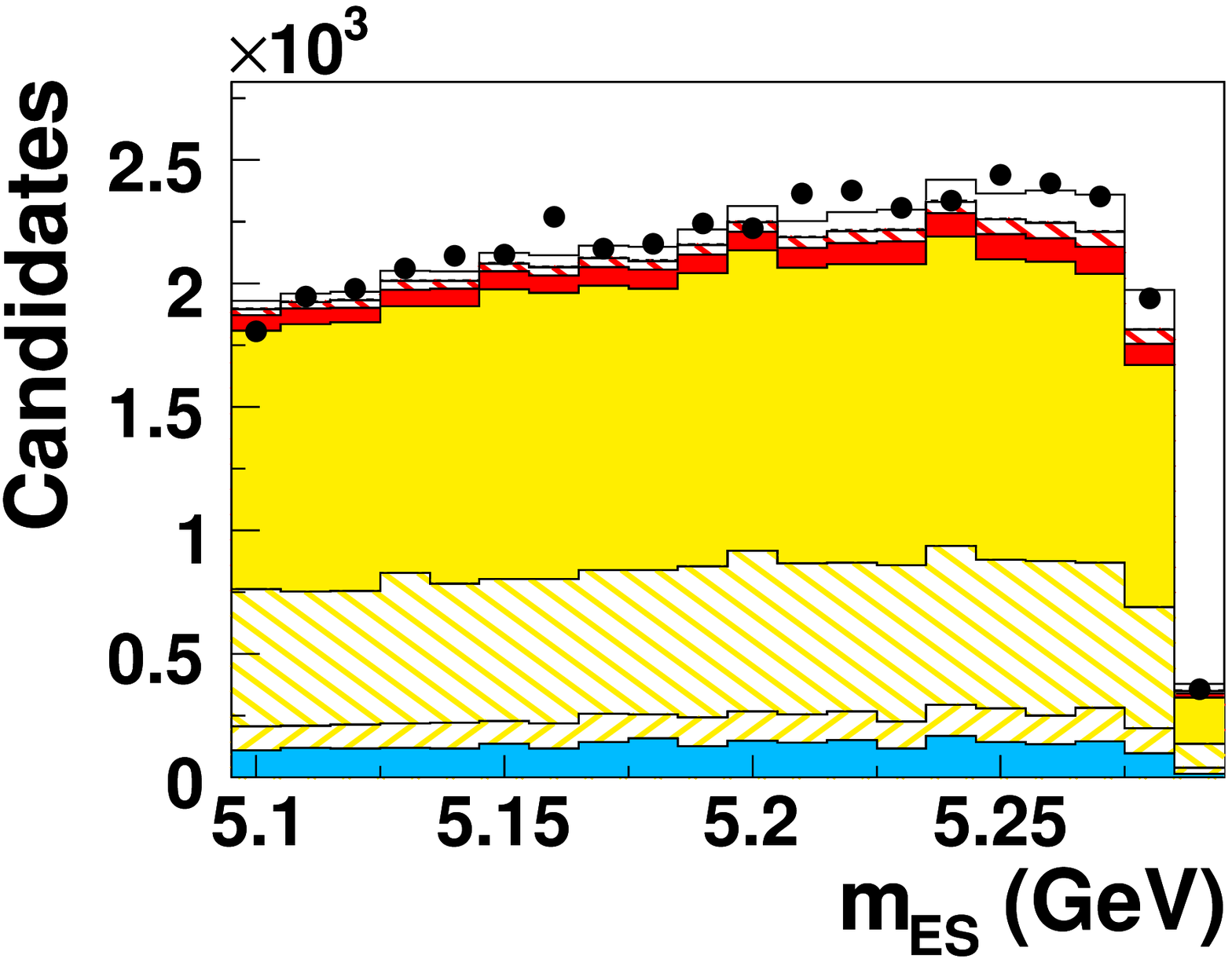}
      {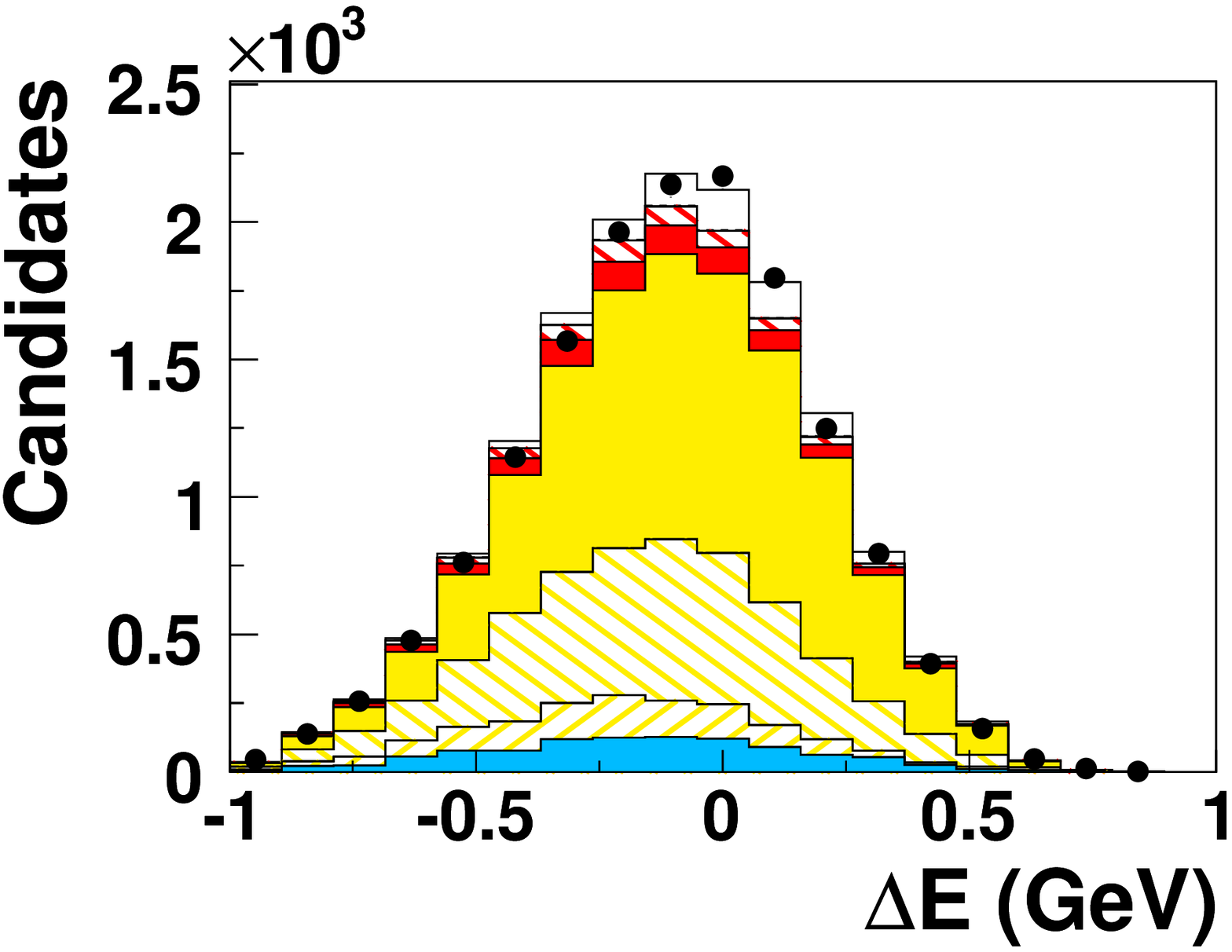}
  \twoFigOneCol
      {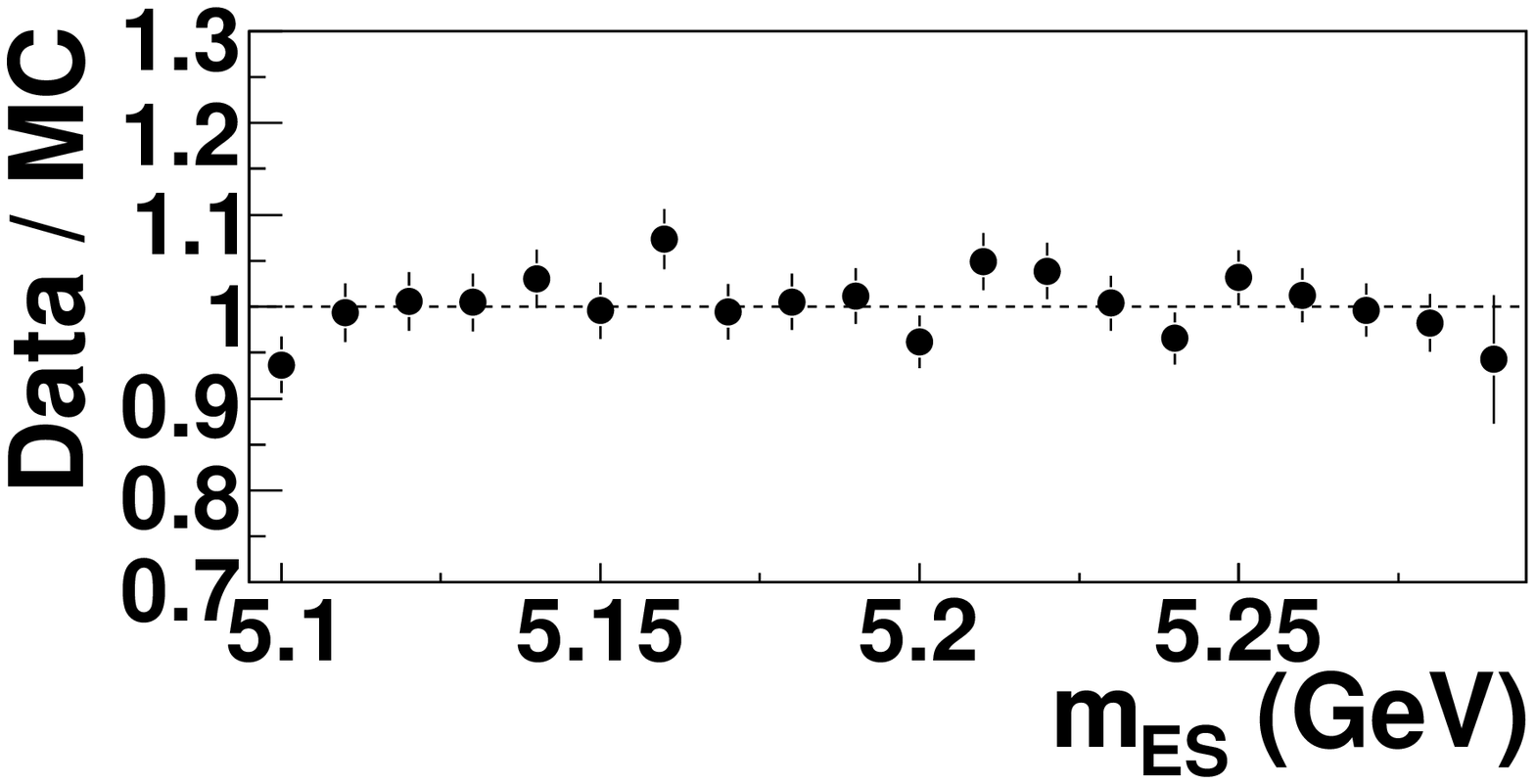}
      {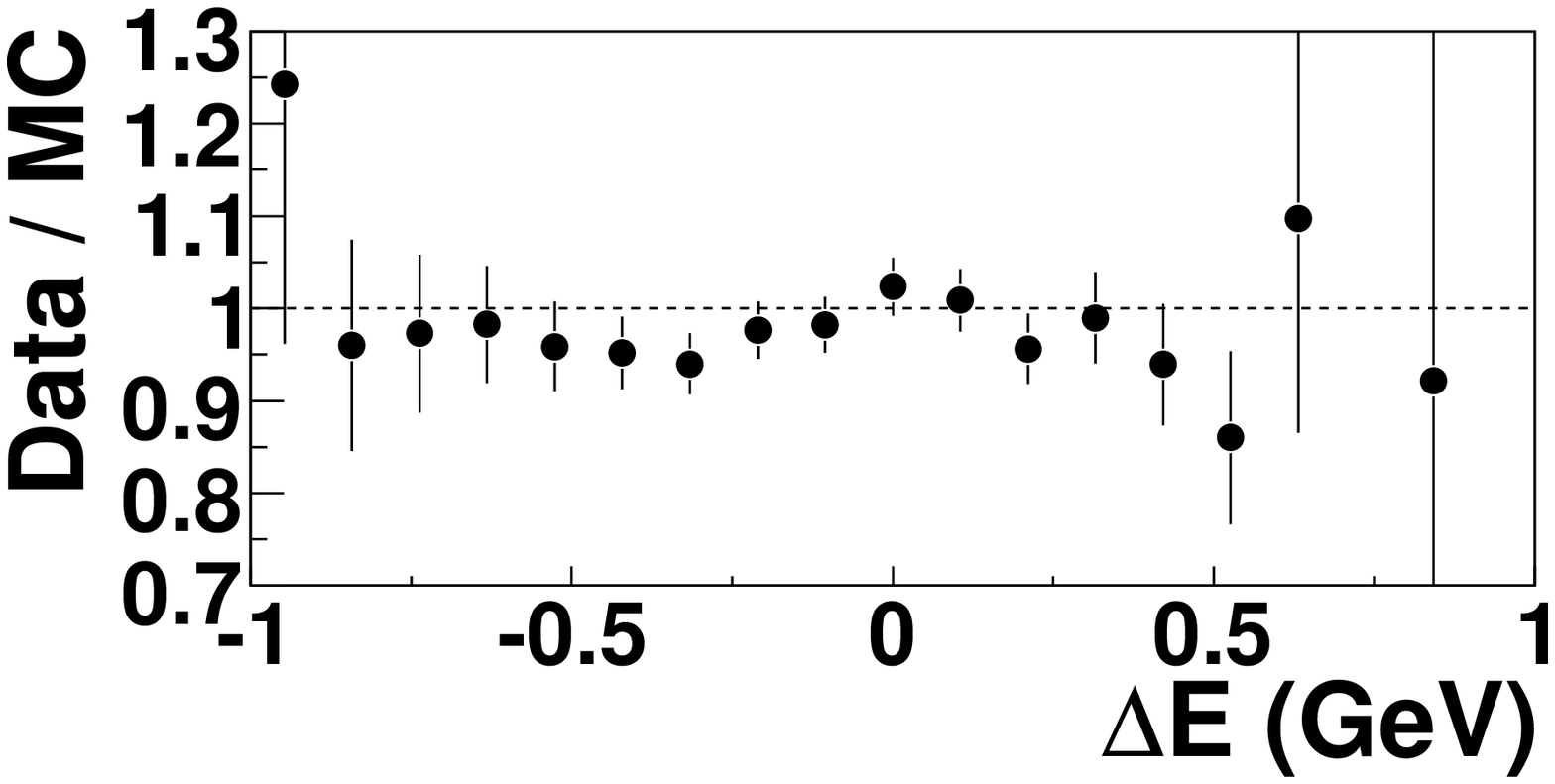}
  \twoFigOneCol
      {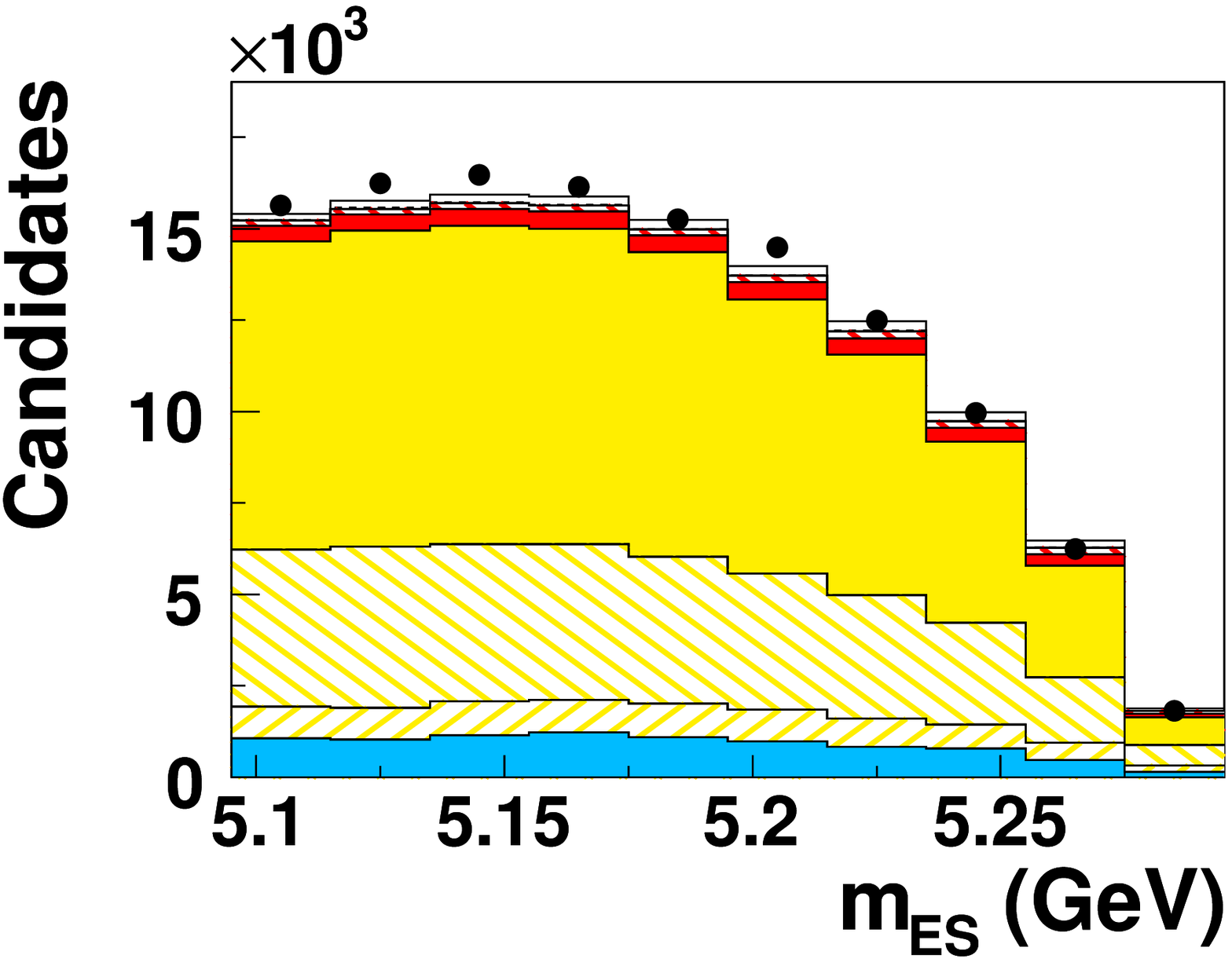}
      {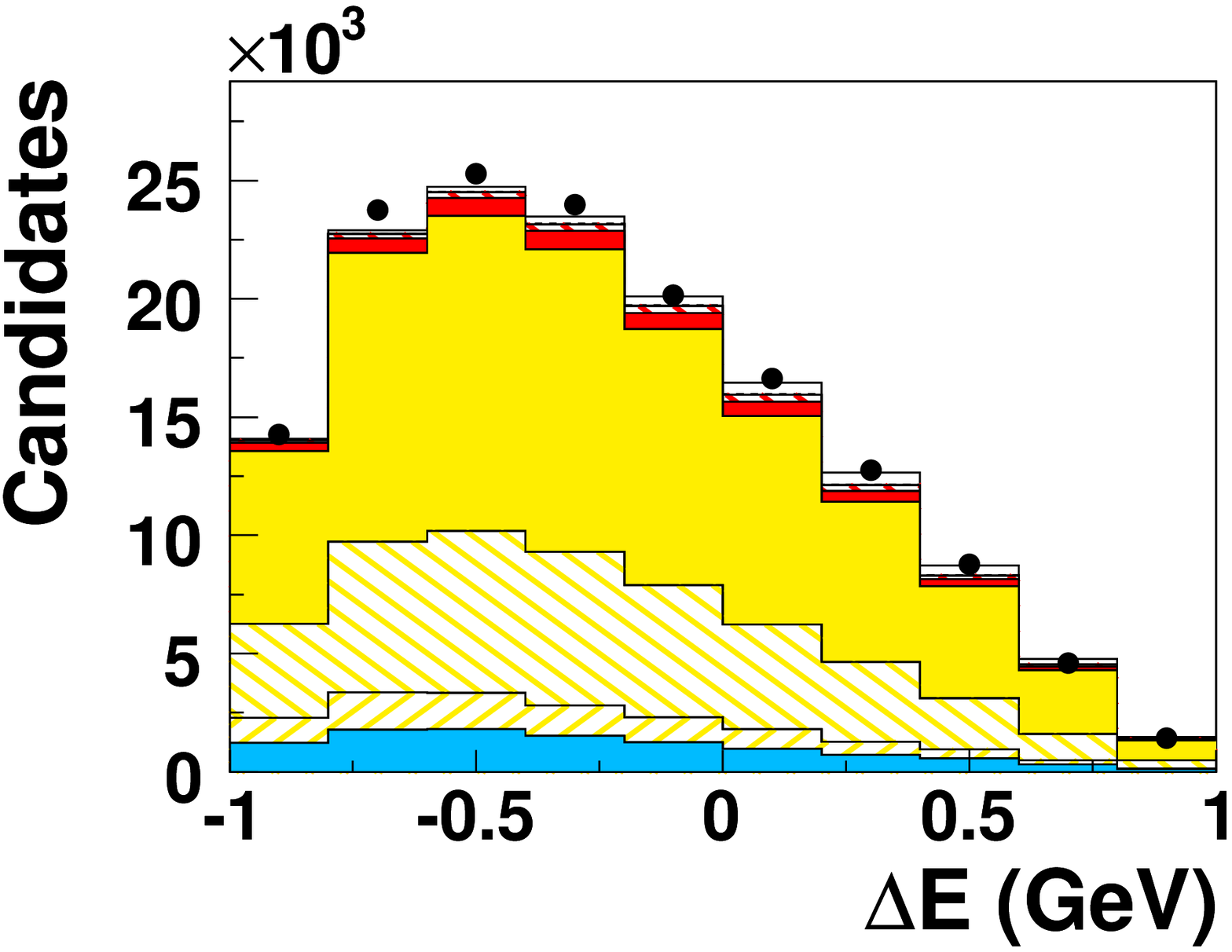}
  \twoFigOneCol
      {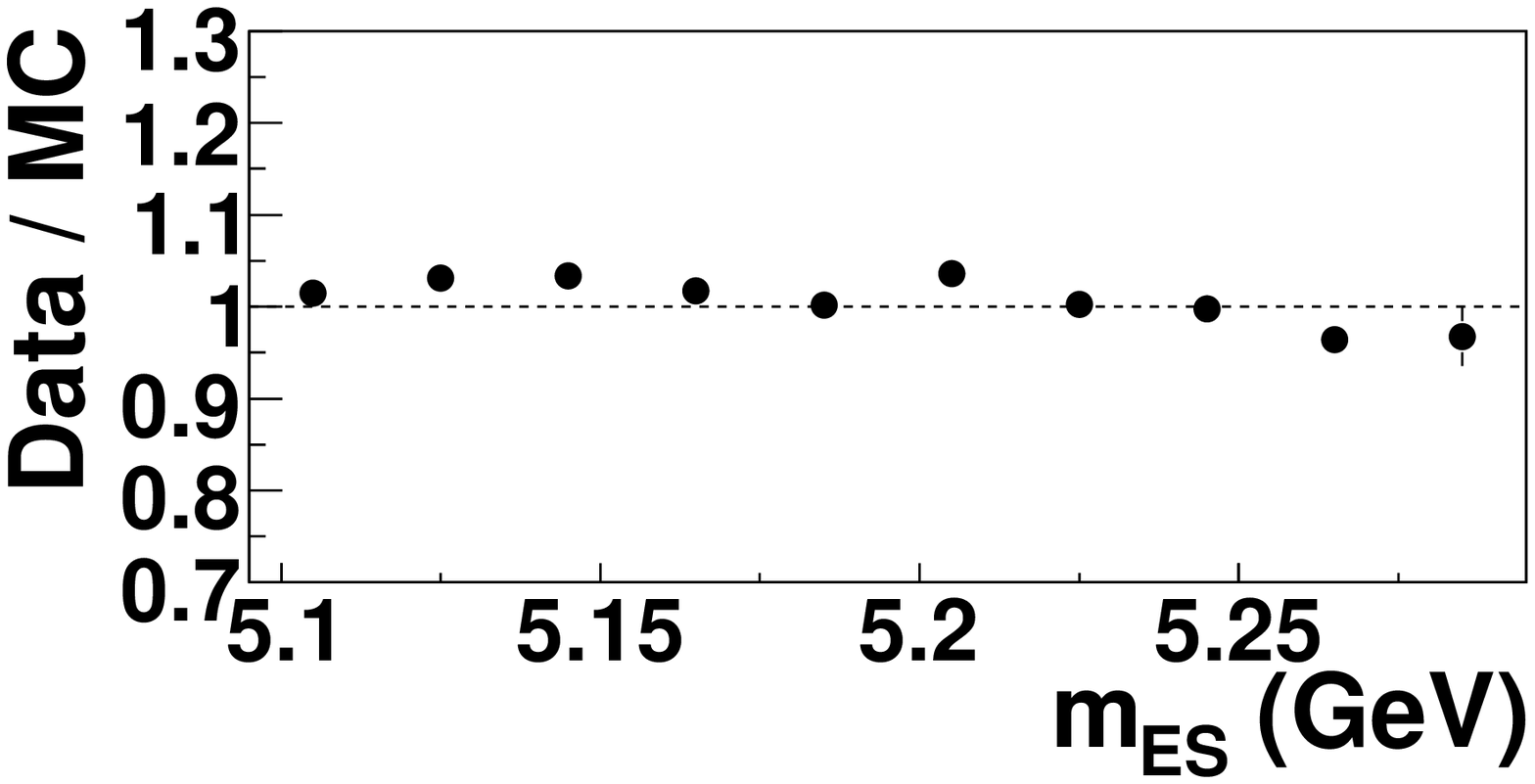}
      {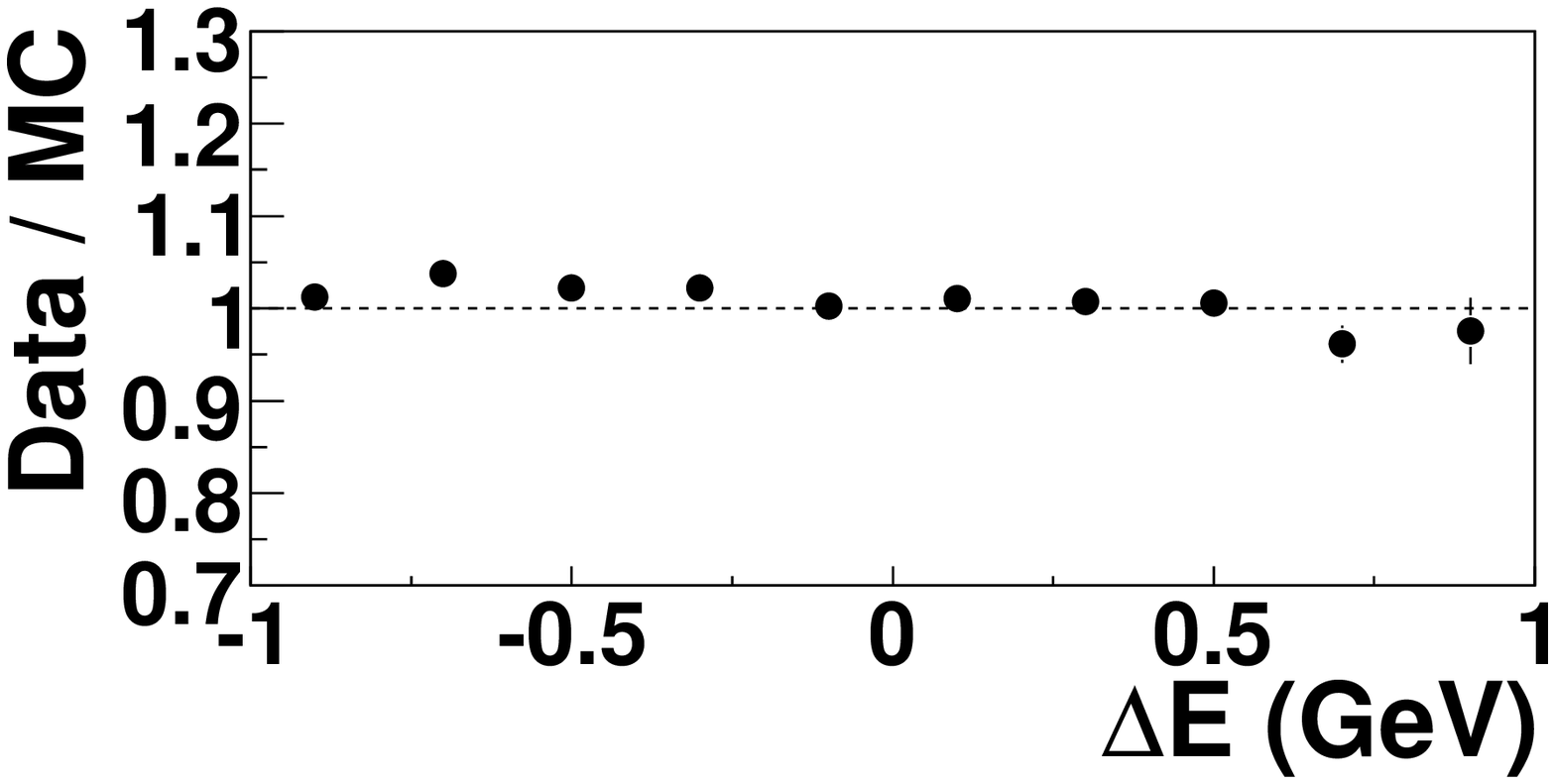}
      \includegraphics[width=.5\columnwidth]{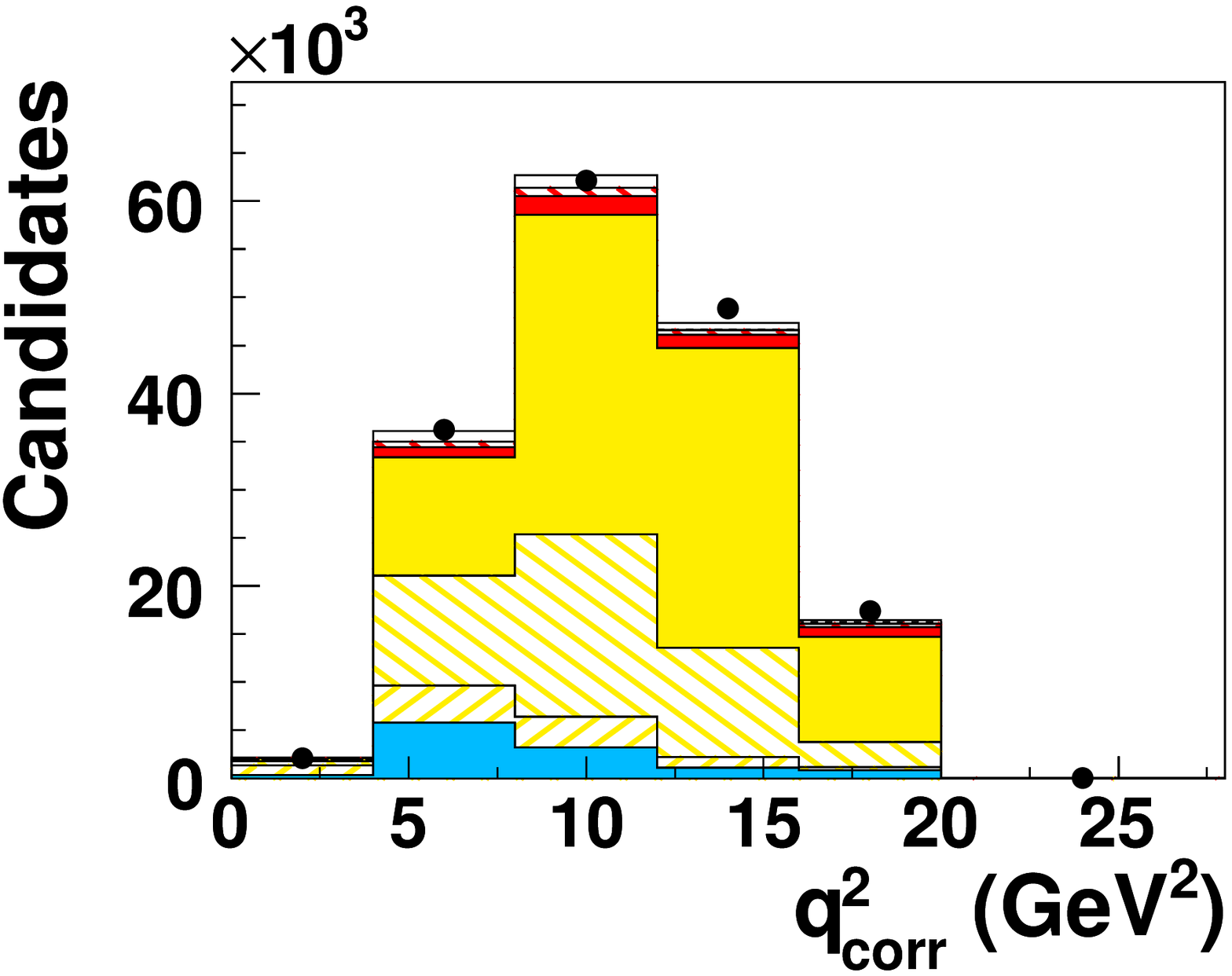}
      \includegraphics[width=.5\columnwidth]{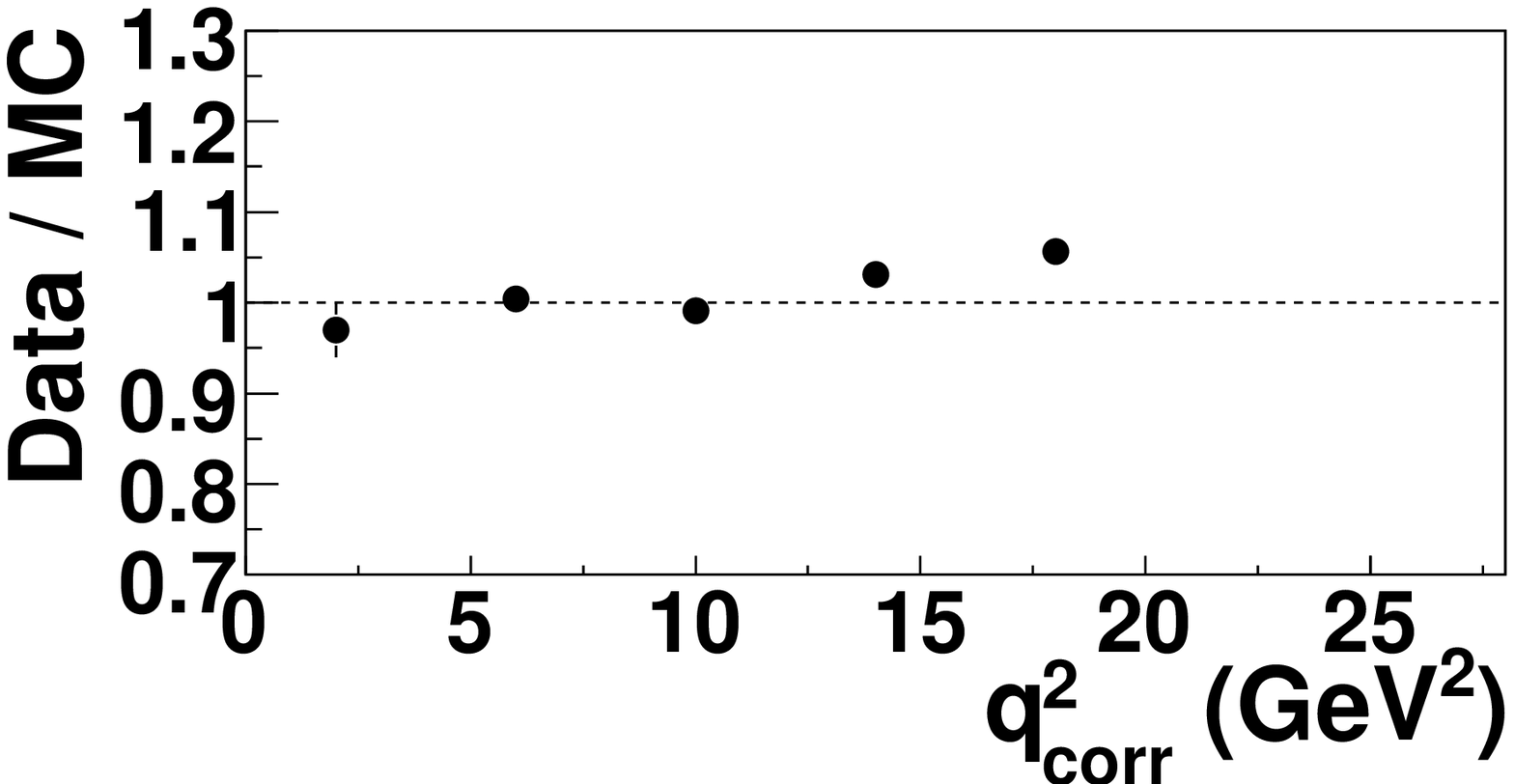}
      \caption{(color online)
	Comparison of data with MC distributions for the charm-enhanced selection 
      for the \Bzpilnu\ sample. 
       Top row:    \mES and \DeltaE for the signal bands, 
      center row:  \mES and \DeltaE  for the side bands, and
	bottom row:  $q^2_{\rm corr}$ for the whole fit region. 
	The bin-by-bin ratio of data over the sum of all MC contributions is given 
      in the plots below each histogram. 
             }
\label{fig:XclnuEnhanced_pilnu}
\end{figure}

\begin{figure}[htb]
  \twoFigOneCol
      {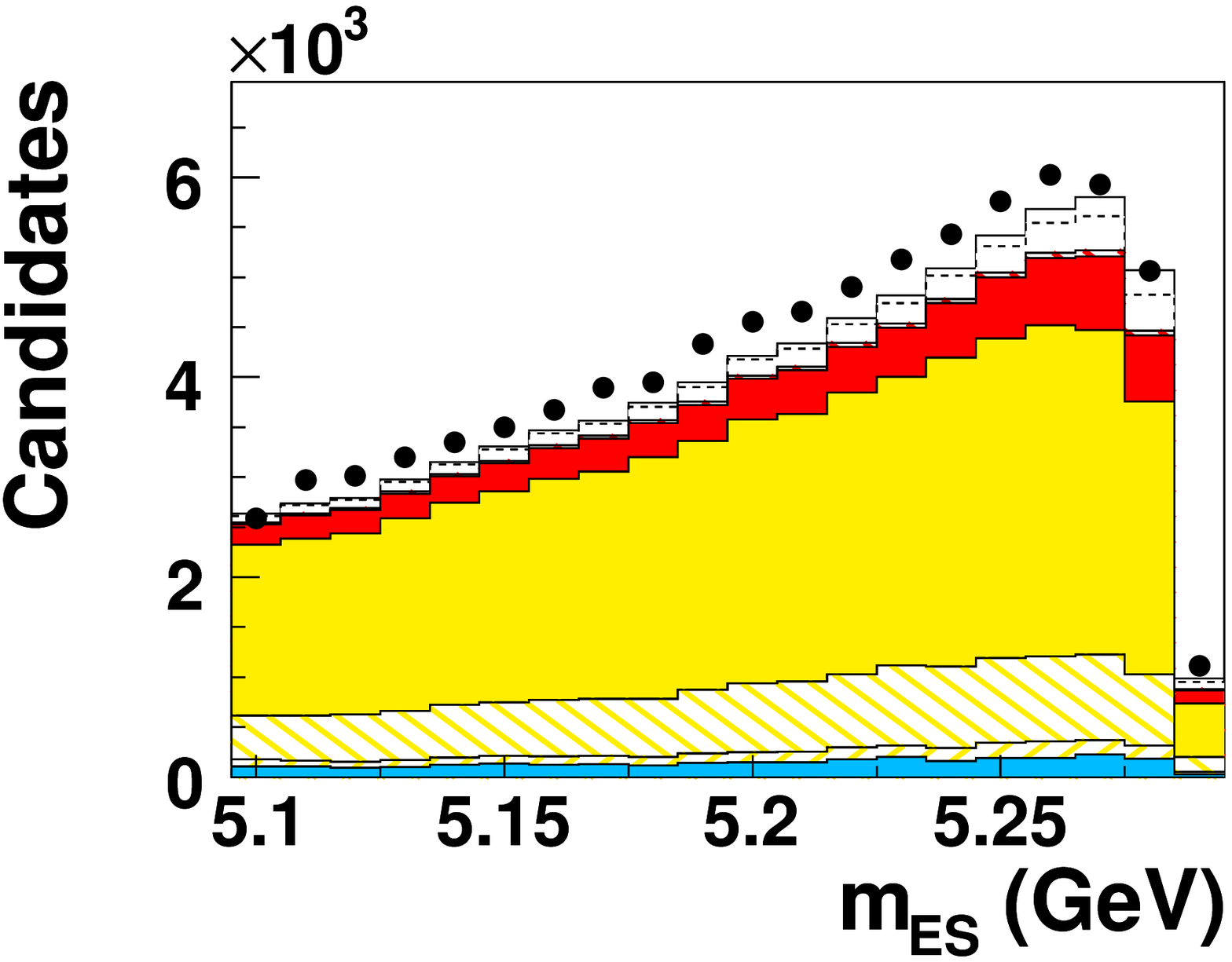}
      {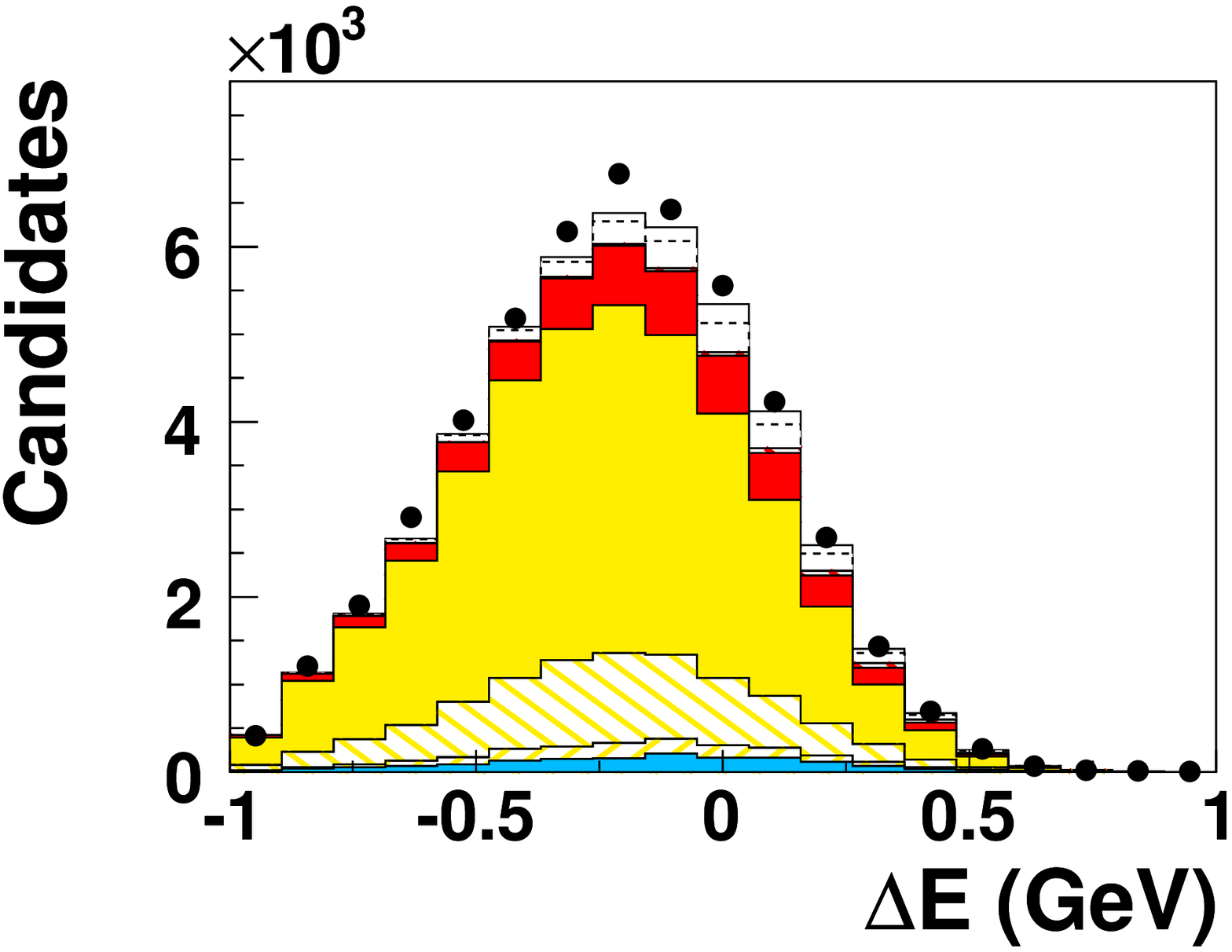}
  \twoFigOneCol
      {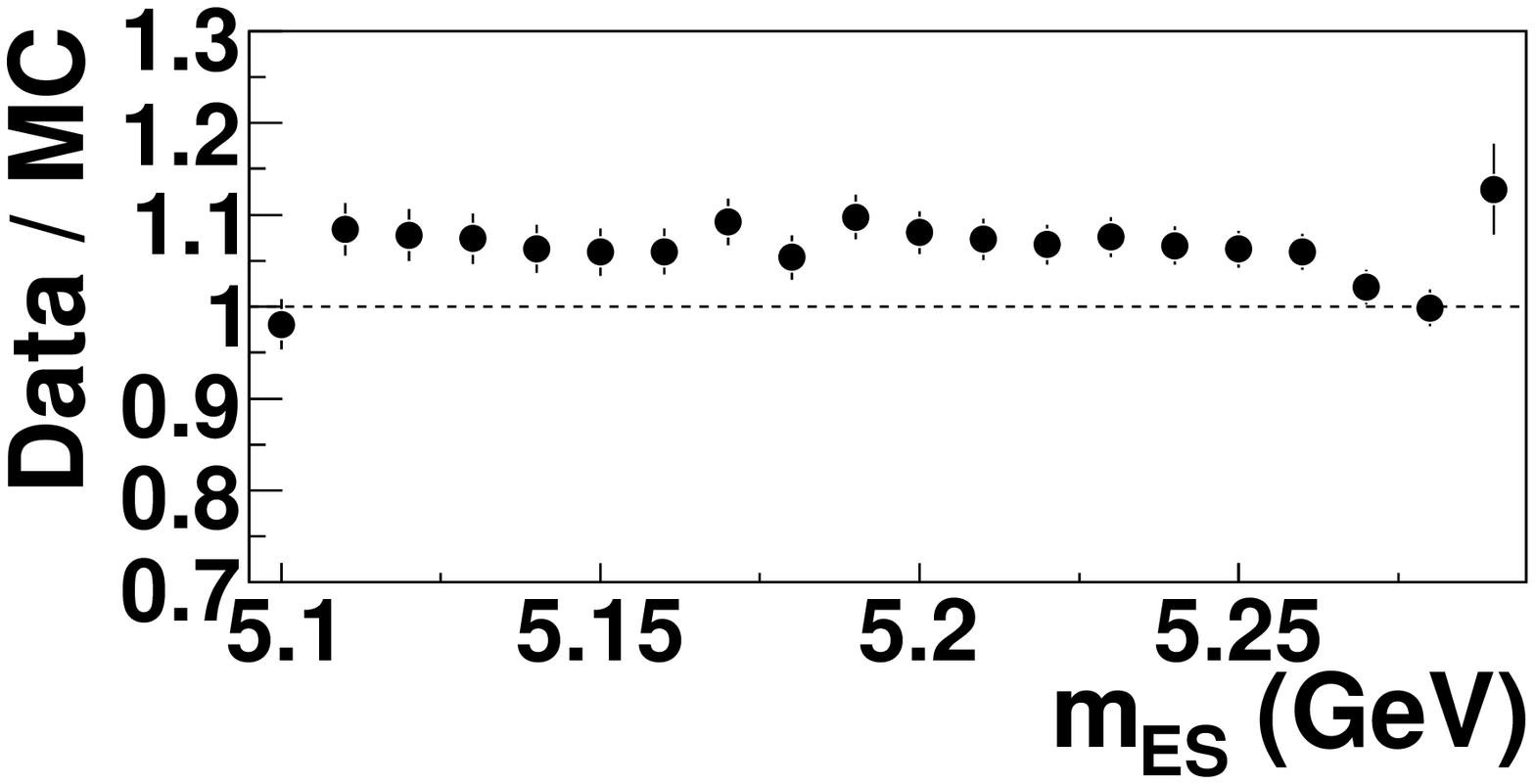}
      {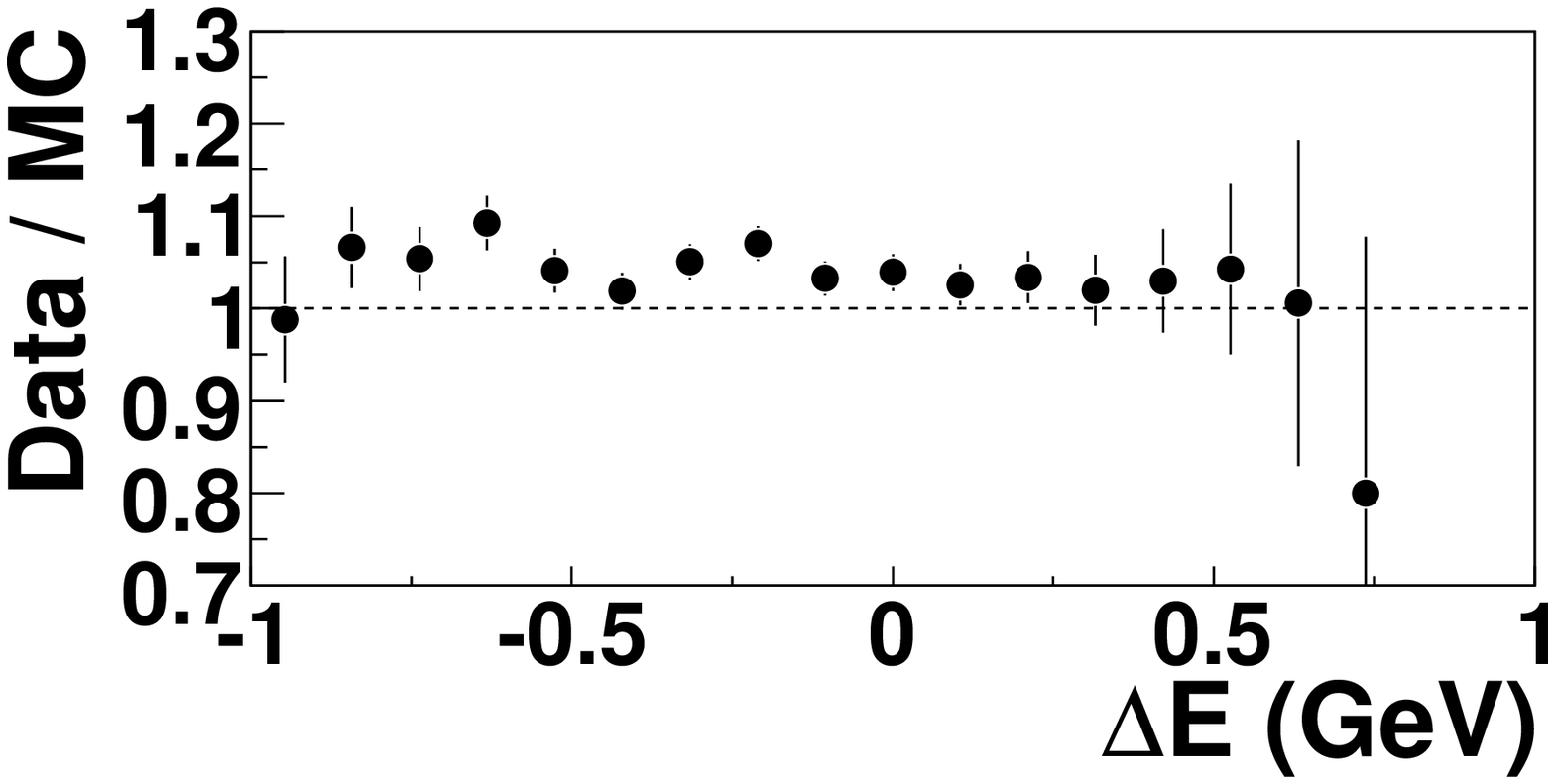}
  \twoFigOneCol
      {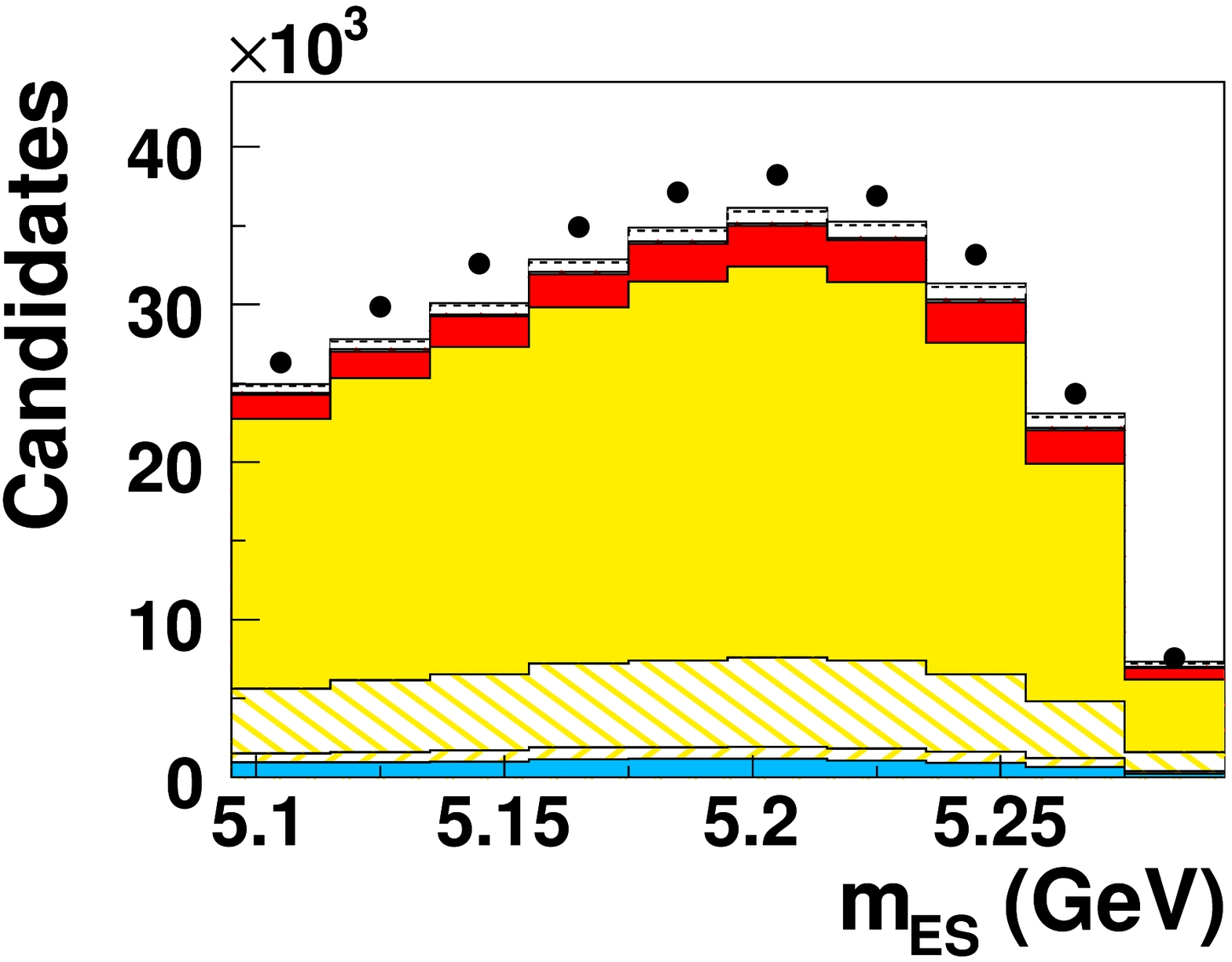}
      {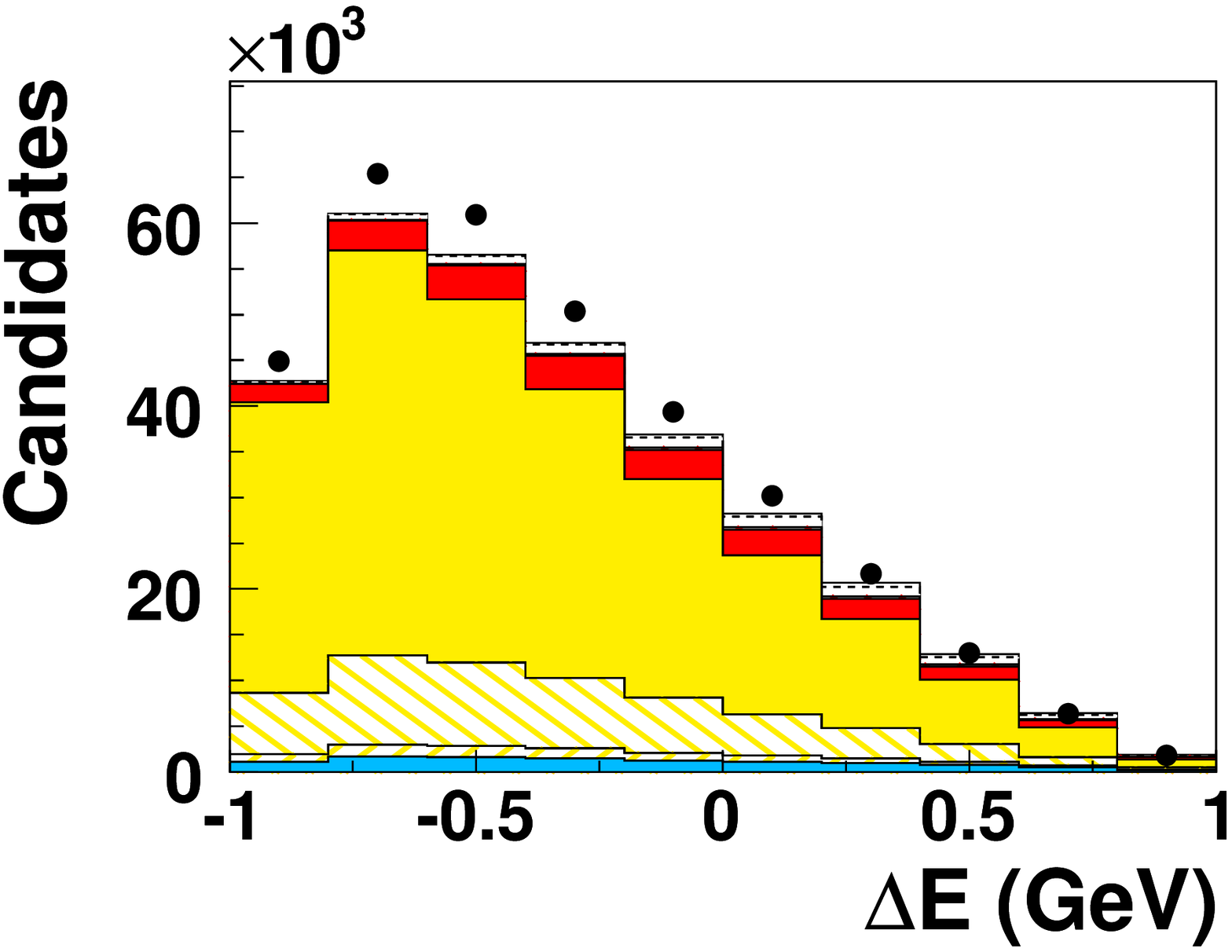}
  \twoFigOneCol
      {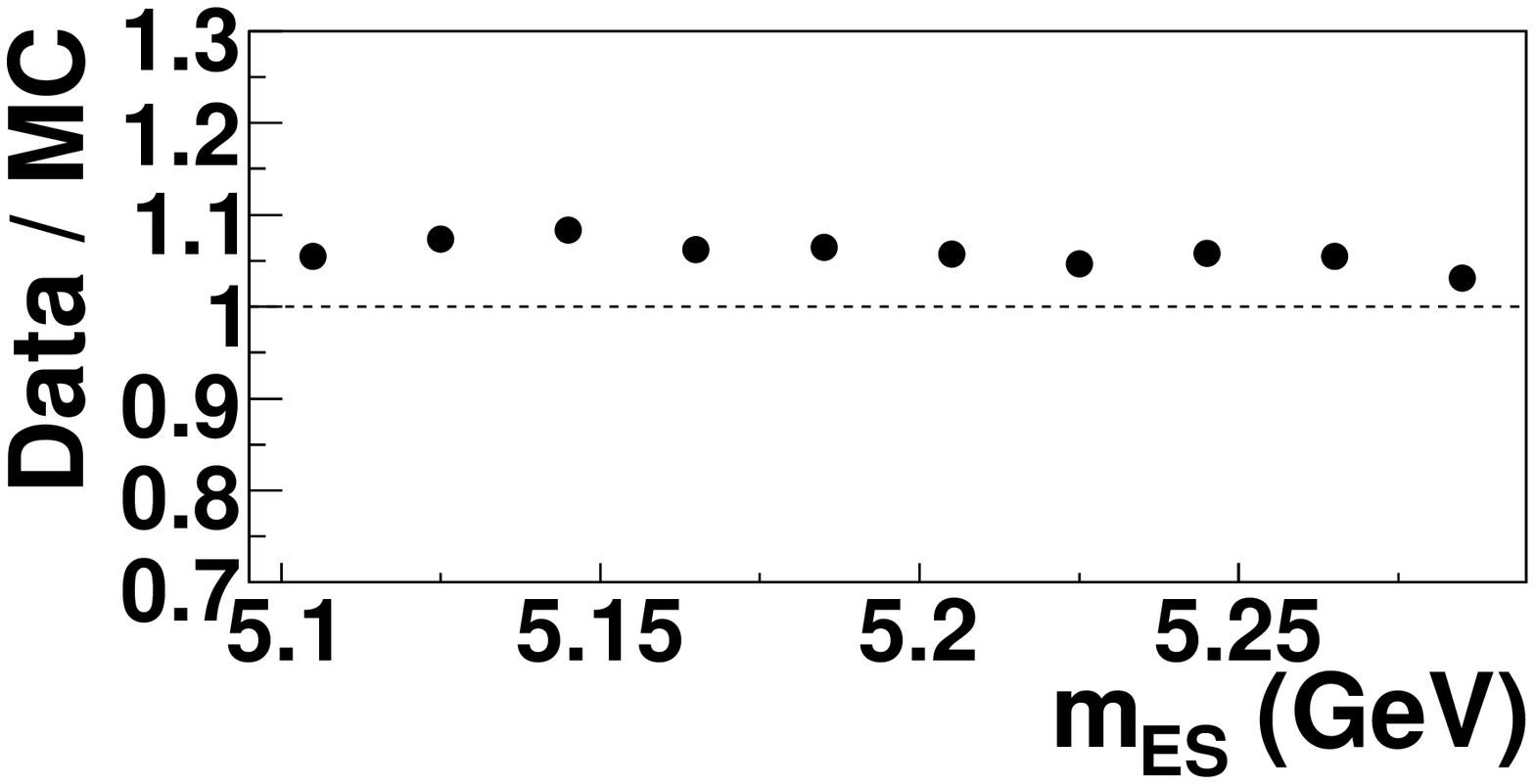}
      {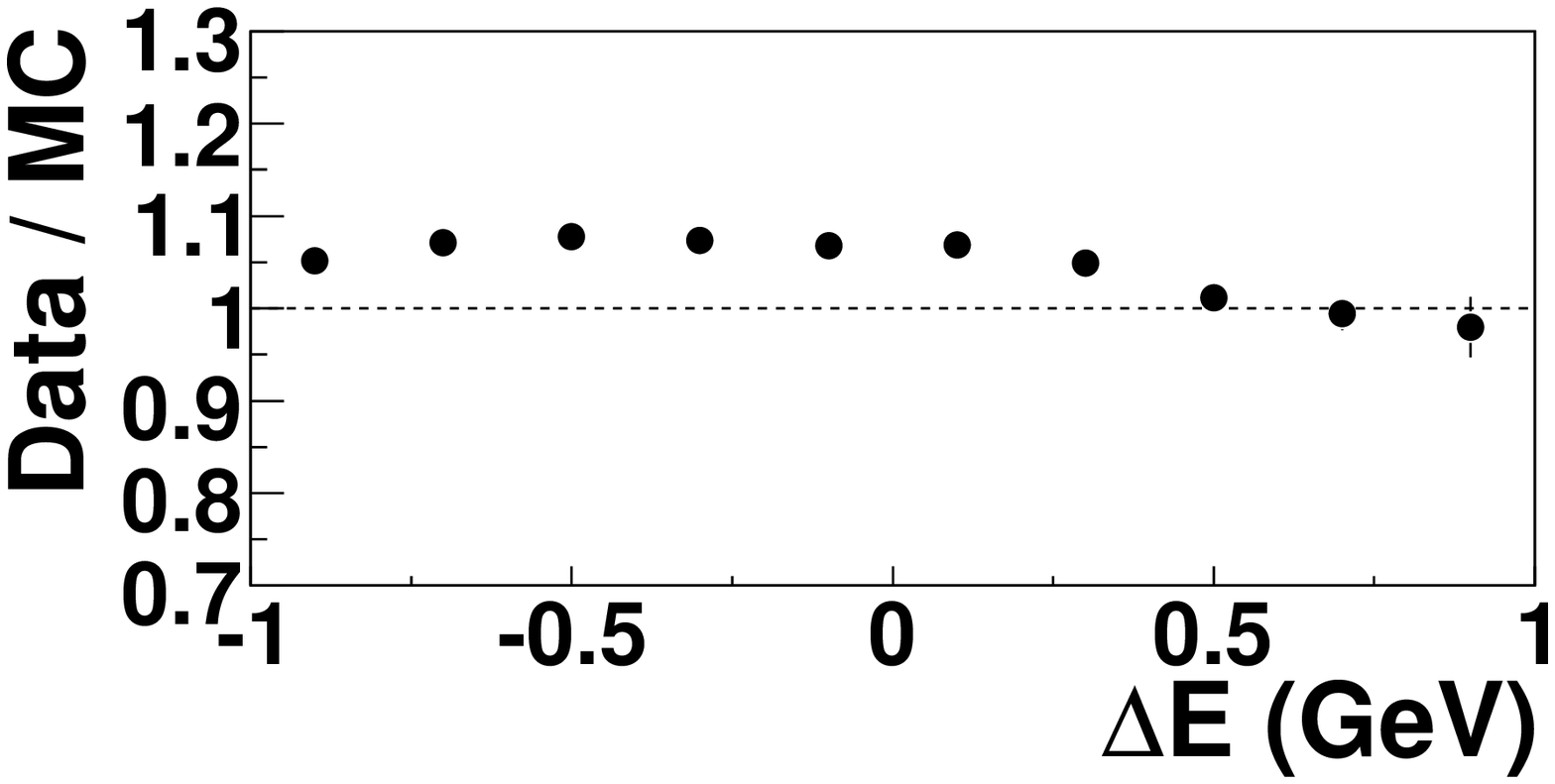}
      \includegraphics[width=.5\columnwidth]{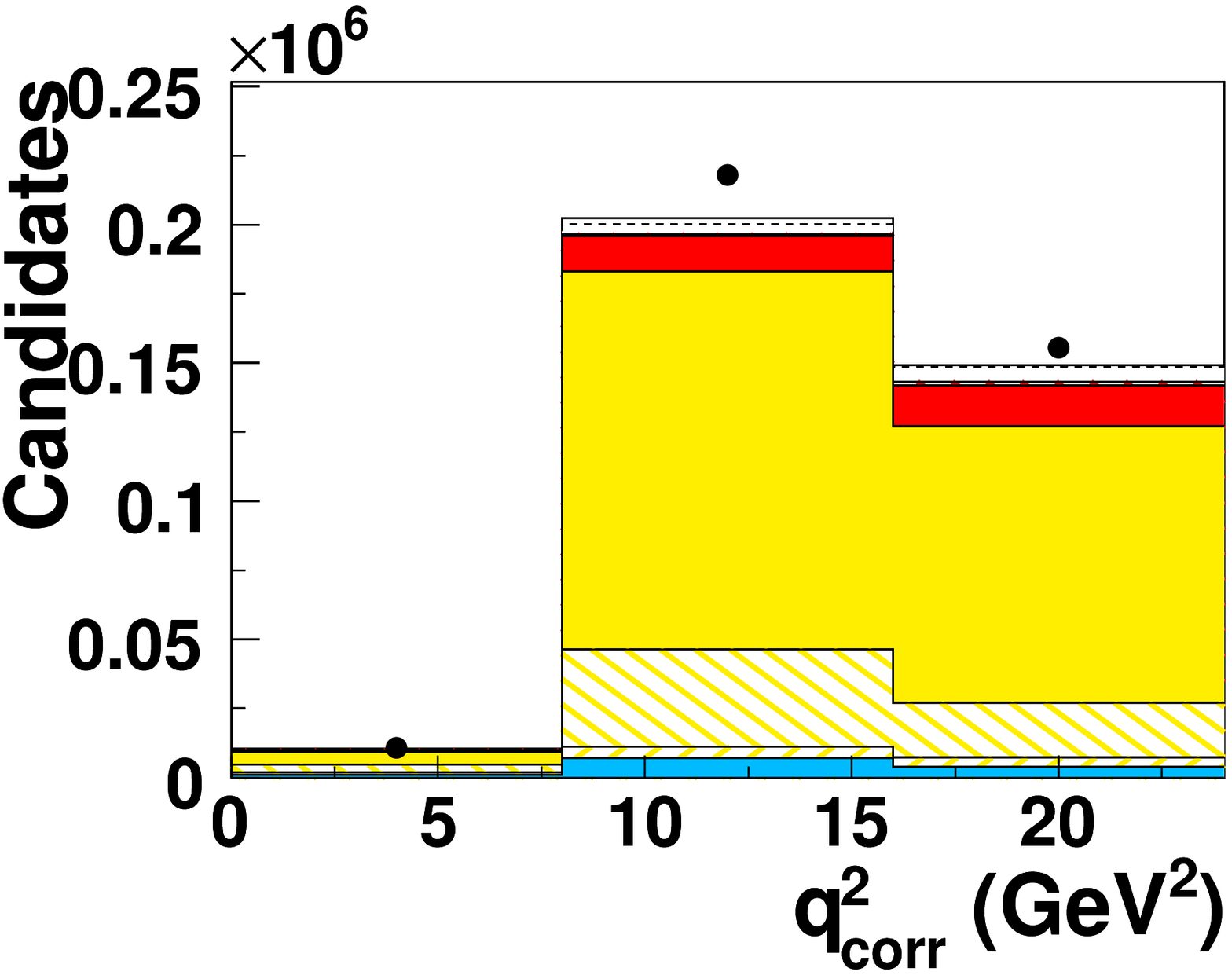}
      \includegraphics[width=.5\columnwidth]{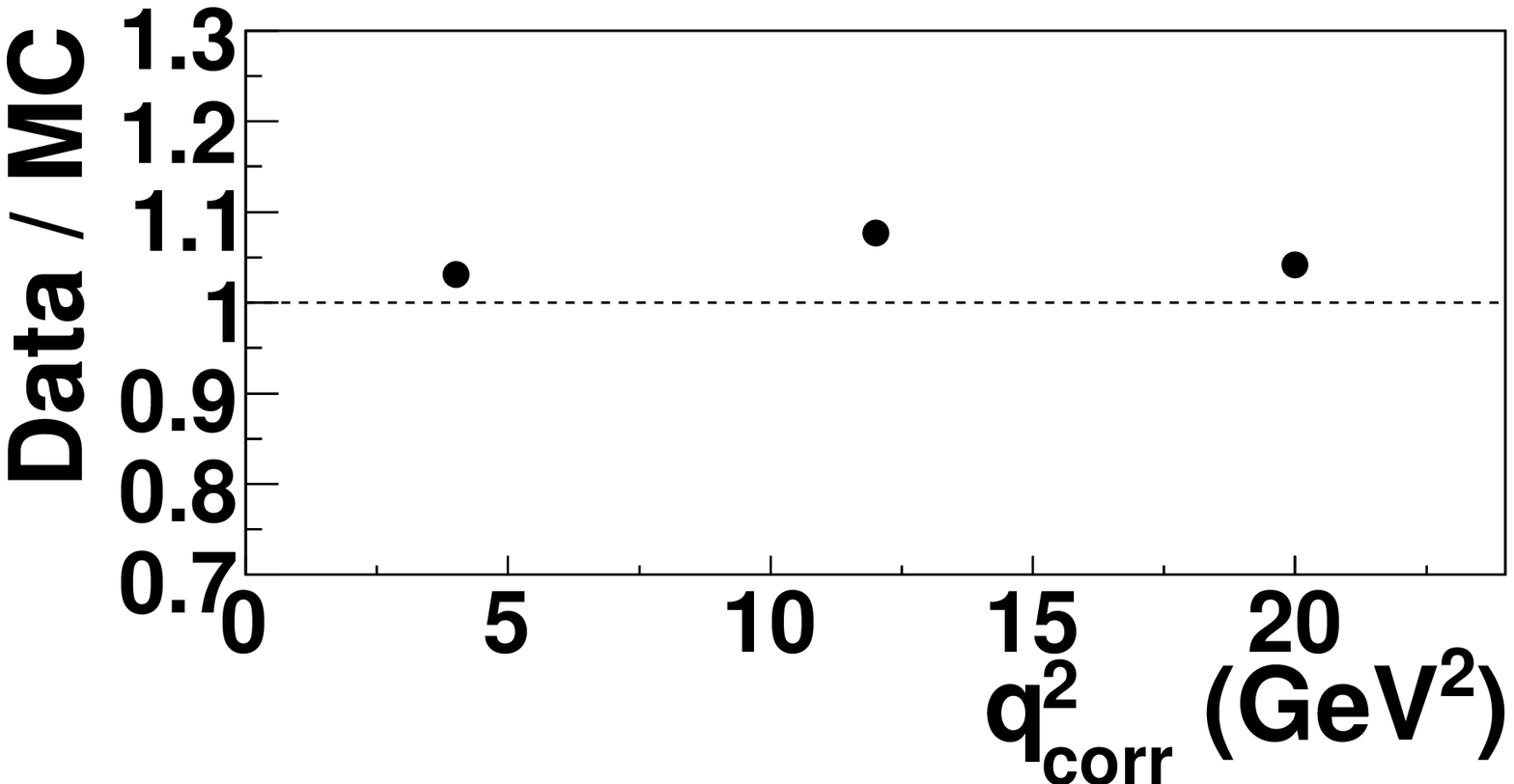}
      \caption{(color online)
	Comparison of data with MC distributions for the charm-enhanced selection 
      for the \Bzrholnu\ sample. 
      Top row:    \mES and \DeltaE for the signal bands, 
      center row:  \mES and \DeltaE  for the side bands, and
	bottom row:  $q^2_{\rm corr}$ for the whole fit region. 
	The bin-by-bin ratios of data over the sum of all MC contributions is given 
      in the plots below each histogram.
              }
\label{fig:XclnuEnhanced_rholnu}

\end{figure}

We select a charm-enhanced sample by inverting the cut on the \bclnu\ neural-network discriminator. 
Figures~\ref{fig:XclnuEnhanced_pilnu} and \ref{fig:XclnuEnhanced_rholnu}
show the \DeltaE\ and \mES\ distributions in the signal region and the side bands, as well 
as the $q^2$ distribution in the signal region. All distributions
show good agreement in shape; the absolute yields differ at a level that is expected,  considering that the MC distributions have not been adjusted.

\subsection{{\boldmath \Bzdslnu} Control Sample}
\label{sec:controlsample}

To study the Monte Carlo simulation of the neutrino reconstruction employed in this analysis, we use a control sample of
exclusively reconstructed \Bzdslnu\ decays with $D^{*-} \ra \Dzb \pi_s^-$ and $\Dzb \to K^+ \pi^-$.  
Since the \Bzdslnu\ decay rate exceeds the rate for $B^+ \to \rho^0 \ell^+ \nu$ by a factor of about 30 (including the $\Dzb$ branching fraction), this  control sample represents a high-statistics and high-purity sample of exclusive semileptonic decays.  Except for the low-momentum pion ($\pi_s^-$), this final state has the same number of tracks, and very similar kinematics, as the 
$B^+ \to \rho^0 \ell^+ \nu$ signal decay. Furthermore, since about $50 \%$ of the \bclnu\ background in the \Btopilnu\ and \Btorholnu\ samples comes from \Bzdslnu\ decays,  this \Bzdslnu\ sample can provide important tests of the shapes of the distributions that are used to discriminate the \bclnu\ background from signal.

Moreover, the distributions of  the primary background suppression variables, in particular $R2$, $L2$, \cosBY, $\theta_{\rm miss}$, and $\mmiss/(2\Emiss)$,
are relatively insensitive to the specific semileptonic decay mode.
Likewise, the resolutions for the fit variables \DeltaE, \mes, and $q^2$ are dominated by the resolution of the reconstructed neutrino, and thus depend little on the decay mode under study.

The reconstruction of the $D^{*-}$ from its decay products is straightforward. Except for the selection of the $D^{*-}$, we apply the same preselection as for the signal charmless decays.  We require the $K^+\pi^-$ invariant mass to be within 17 \mev\ of the nominal \Dzb\ mass, and restrict the mass difference, $\Delta m_{D^*} = m_{D\pi} - m_{D}$ to $0.1432 < \Delta m_{D^*} < 0.1478 \gev $.
The number of events in this data control sample exceeds the MC prediction by $3.8\pm 1.7\%$,
a result consistent with the uncertainties in the efficiency for the very-low-momentum charged pion from the $D^{*-} \ra \Dzb \pi^-$ decay.  We correct the MC yield and sequentially place requirements on the same seven variables we use in the neural networks to both the data and MC samples. We compare the step-by-step reduction in the number of events; the largest difference is $0.9 \pm 0.7\%$, for the cut on \costhetaWL.  For all other critical requirements the agreement is better than 0.5\% and one standard deviation. The remaining background is at the level of 10\%.

\begin{figure}[htb]
  \epsfig{file=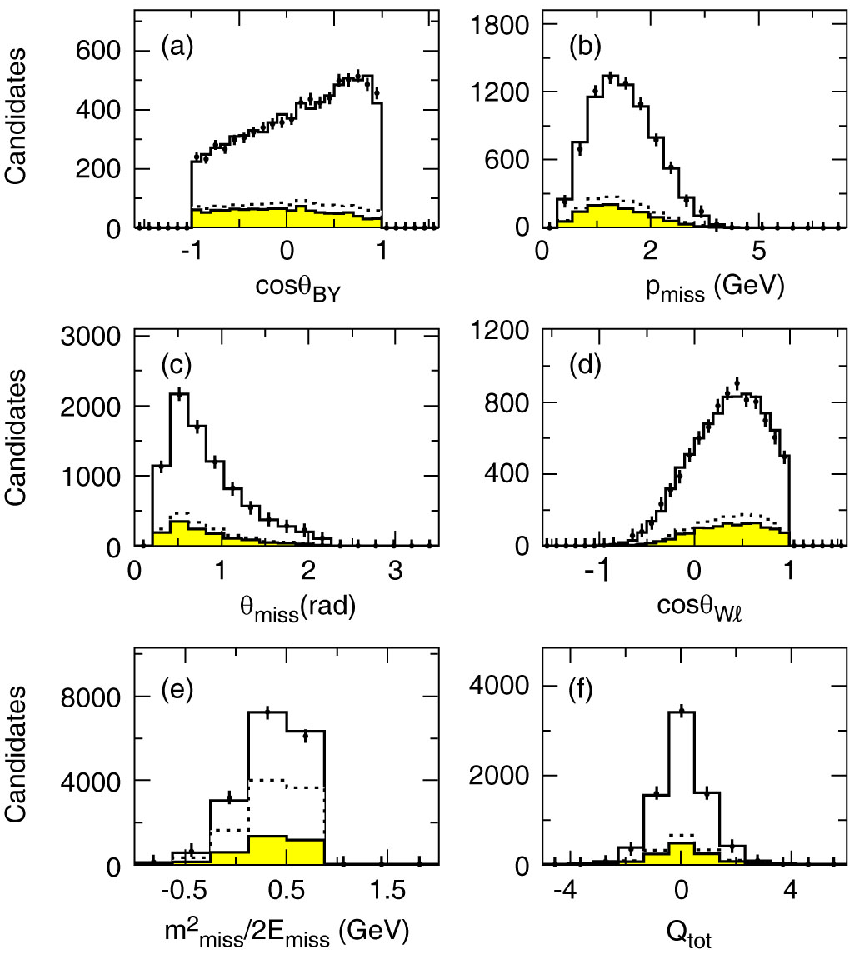,width=0.98\columnwidth}
  \caption{(color online)
Comparison of data and MC-simulated distributions for the \control sample, after selection criteria have been applied on all variables except the one presented.
a) $\cosBY$, b) $p_{\rm miss}$, c) $\theta_{\rm miss}$, d) $\cos\theta_{W\ell}$,
e) $\mmiss/2 \Emiss$, and f) the total charge per event $Q_{\rm tot}$.
The background to the sample is indicated as a shaded (yellow) histogram.
The combinatorial signal contribution is indicated as dashed histogram.
}   
  \label{fig:controlKin}
\end{figure}

We have compared the MC-generated distributions for the control sample
with the selected \Bzdslnu\ data sample and find very good agreement for
the basic event variables, {\it i.e.}, the multiplicity of charged particles and photons, and the total charge per event, indicating that the efficiency losses are well reproduced by the simulation. The distributions of the topological event variables $R2$ and $L2$ match well. 
Figure ~\ref{fig:controlKin} shows the distributions of the variables critical for the neutrino reconstruction, 
$p_{\rm miss}, \mmiss/(2 \Emiss)$, $\cosBY$, and $\theta_{\rm miss}$; they are also well reproduced. 

\begin{figure}[htb]
  \epsfig{file=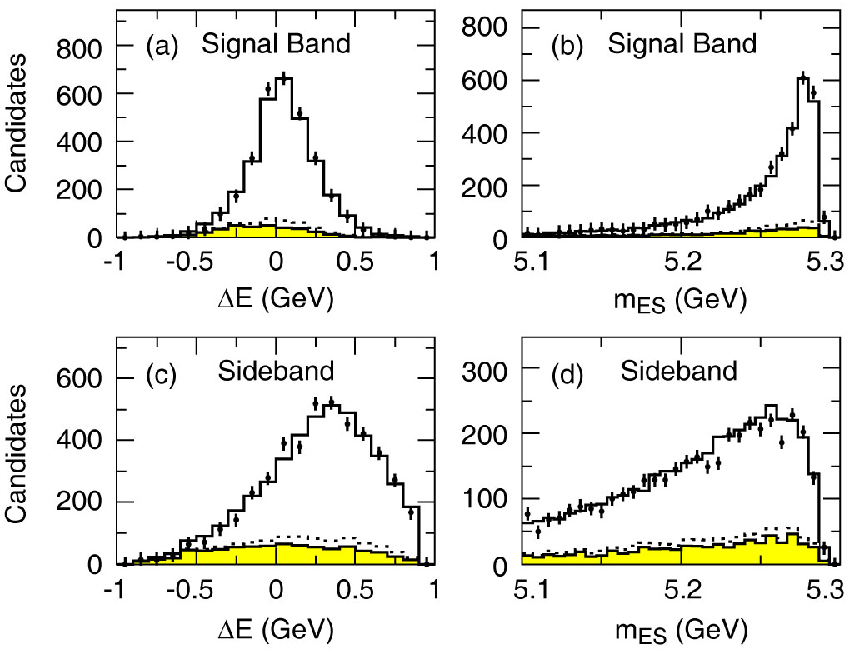,width=0.98\columnwidth}
\caption{(color online)
Comparison of data and MC-simulated distributions for the \control\ sample, after all selection cuts have been applied, a) \DeltaE\ for events in the \mes\ signal band, b) \mes\ for events in the  \DeltaE\ signal band, c) \DeltaE\ for events in the \mes\ side band, and d) \mes\ for events in the \DeltaE\ side bands.
The background to the sample is indicated as a shaded (yellow) histogram.
The combinatorial signal  contribution is indicated as dashed histogram.
}
\label{fig:DeltaEmesprojectionscontrolone}
\end{figure}

Figure~\ref{fig:DeltaEmesprojectionscontrolone} shows distributions of \DeltaE\ and \mes\ for events in the signal region and in the side bands.
Again, the agreement between data and the MC simulation is reasonable. 

\begin{figure}[htb]
  \epsfig{file=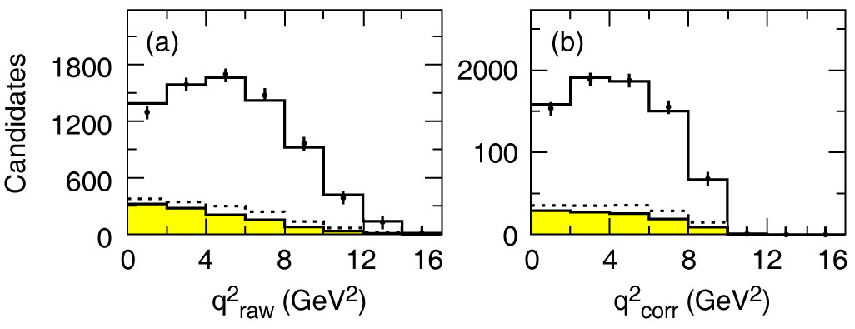,width=0.98\columnwidth}
\caption{(color online)
Comparison of the data and MC simulation of $q^2$ distributions for the \control\ sample after all selection criteria have been applied, a) the raw $q^2$, and b) the corrected $q^2$.
The background to the sample is indicated as a shaded (yellow) histogram.
The combinatorial signal contribution is indicated as dashed histogram.
}
\label{fig:datamcqsquaredcontrol}
\end{figure}

We have also compared the $q^2$ distributions of the simulation and the data control sample and find good agreement for both the raw and the corrected spectra, as illustrated in Figure~\ref{fig:datamcqsquaredcontrol}.  After corrections, no events appear above the kinematic limit of $10.7 \gev^2$.
The $q^2_{\rm corr}$ resolution function can be described by the sum of two 
Gaussian resolution functions, with widths of $0.27 \gev^2$ and $0.67 \gev^2$, close to the values obtained 
for events in the fit region for the signal \Btopilnu\ and \Btorholnu decays, respectively. 

\section{Maximum-Likelihood Fit}
\label{sec:fit}

\subsection{Overview}
\label{sec:fit_overview}

We determine the yields for the signal decay modes, \Bzpilnu, \Bppizlnu, \Bzrholnu, and \Bprhozlnu, by 
performing a maximum-likelihood fit to the three-dimensional $\DeltaE - \mES - q^2$ distributions for the four selected data samples corresponding to the four exclusive decay modes.  
The fit technique employed in this analysis is an extended binned maximum-likelihood fit that accounts for the statistical fluctuations not only of the data samples but also of the MC samples by allowing the MC-simulated
distributions to fluctuate in each bin according to the statistical uncertainty given by the number of events in the bin. This method was introduced by Barlow and Beeston~\cite{BarlowFit}. 

The parameters of the fit are scale factors for the signal and background 
yields of the four selected event samples. We use the following nomenclature for the fit parameters: $p_j^{source}$,
where the superscript $source$ denotes the fit source (signal or background type) and the subscript $j$ labels the $q^2_{\rm corr}$ bin (if no subscript is given, the same fit parameter is used across all $q^2$ bins).
Predictions for the shape of the $\DeltaE - \mES$ distributions are taken from simulation of both signal and the various background sources, separately for each bin in $q^2$.
The branching fractions for the four signal decays are obtained by multiplying the fitted values of the scale factors with the branching fractions that are implemented in the MC simulation.

\begin{figure}[htb]    
  \centering
  \begin{tabular}{cc}
    \begin{minipage}{0.53\linewidth}
      \epsfig{file=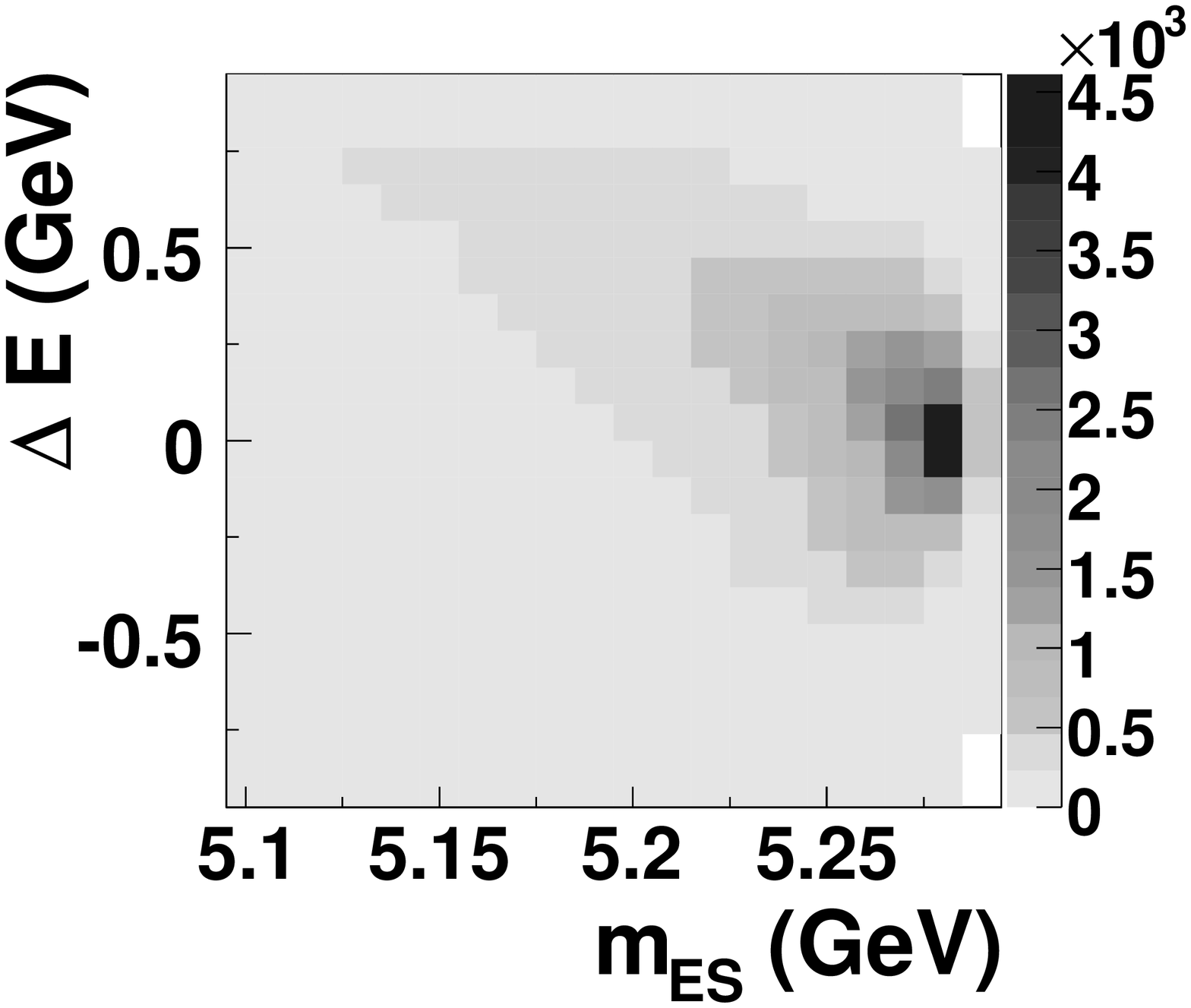, width = 4cm}
    \end{minipage}
    \begin{minipage}{0.5\linewidth}
      \epsfig{file=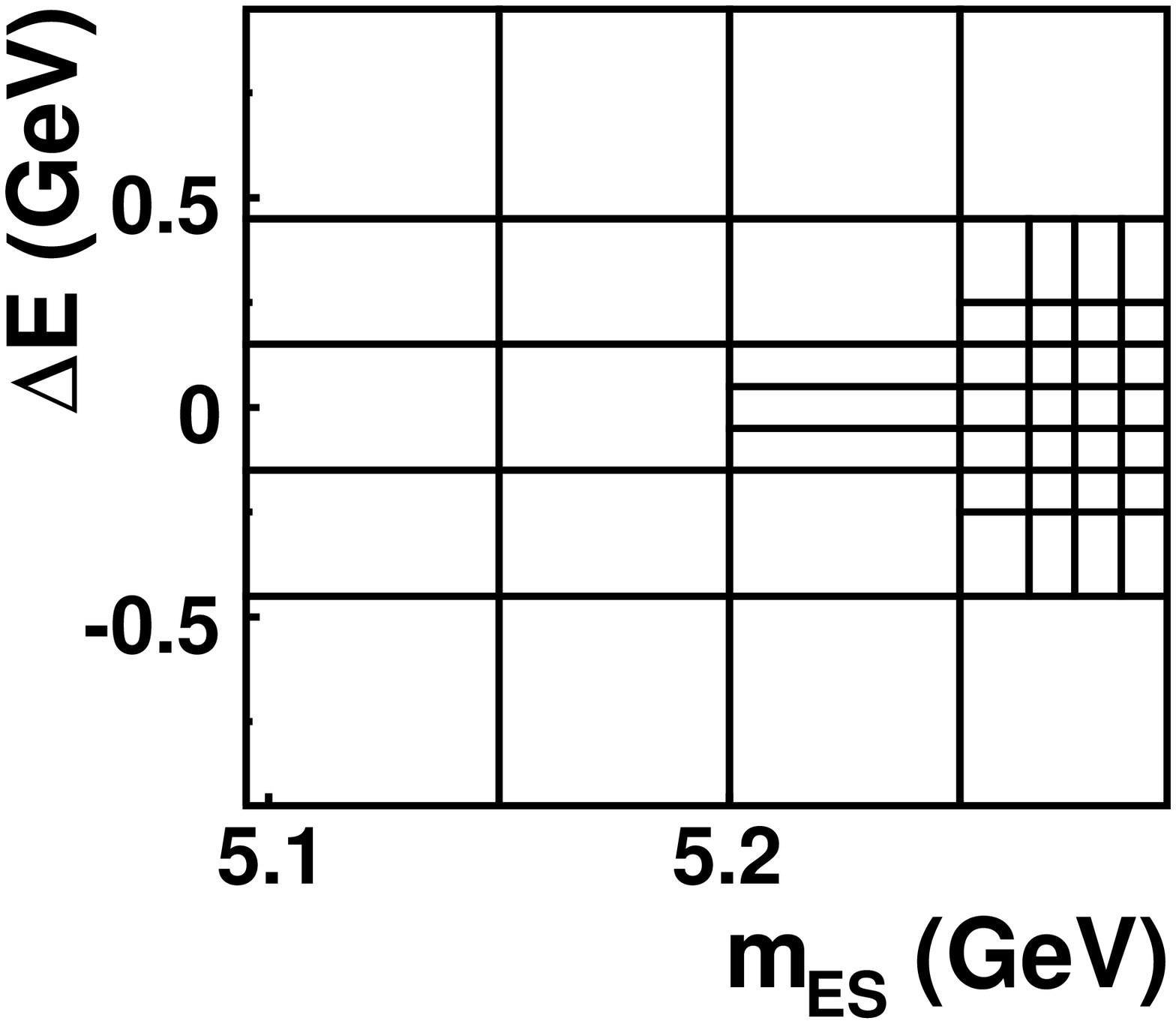, width = 4cm} 
    \end{minipage}
  \end{tabular}
  \caption{Distribution of \DeltaE\ vs \mES\ for true \Bzpilnu\ signal events (left) and 
    definition of bins in the $\DeltaE - \mES$ plane used in the fit for all samples (right).}
  \label{fig:binning}
\end{figure}

The choice of a two-dimensional distribution in \DeltaE\ and \mES\
is mandated because the two variables are correlated for both signal, in particular the combinatorial signal events, and for some of the background sources.
Since it would be difficult to determine reliable analytic expressions for these two-dimensional distributions, a binned maximum-likelihood method is used, with the bin sizes chosen to obtain a good signal and background separation while retaining adequate statistics in all bins. The bin sizes are small in the region where most of the signal is located and larger in the side bands. There are 47 $\DeltaE - \mES$ bins for each bin in $q^2_{\rm corr}$. 
Figure~\ref{fig:binning} shows the $\DeltaE - \mES$ distribution for
signal events and the binning used in the fit. 
As mentioned in Section~\ref{sec:Neuralnets}, for the two \Bpilnu\ samples 
the $q^2$ range $0 < q^2_{\rm corr}< 26.4 \gev^2$ is divided into six bins, and for the two \Brholnu\ samples the range $0 < q^2_{\rm corr}< 20.3 \gev^2$  is divided into three bins.

\subsection{Fit Method}
\label{sec:fitmethod}

Since the MC samples available to create the probability density functions (PDFs) for the individual sources that are input to the fits are rather limited in size, it is necessary to take into account the statistical uncertainties, given by the number of events generated for each bin. 
For this reason we have adopted a generalized binned maximum-likelihood fit method.
The MC samples that are used to define the PDFs are to a good approximation statistically independent
of those used to train the neural networks for background suppression, since for the latter 
relatively small subsamples of the full MC samples have been used.

As mentioned above, the data are divided into $n$ bins 
in a three-dimensional array in $\DeltaE-\mES-q^2_{\rm corr}$.  

If $d_i$ is the number of selected events in bin $i$ 
for a given single data sample corresponding to candidates for a specific decay mode, 
and $a_{ji}$ is the number of MC events from source $j$ in this bin, then 
\begin{equation}
N_D = \sum_{i=1}^n d_i, \hspace{1cm}N_j = \sum_{i=1}^{n} a_{ji},
\end{equation}
where $N_D$ is the total number of events in the data sample, and $N_j$ is the total number in the MC sample for source $j$. We assume that there are $m$ different MC-generated source distributions that add up to describe the data.
The predicted number of events in each bin $f_i(P_j)$ can be written in terms of the strength of the individual contributions $P_j$ ($j=1,..,m$) as
\begin{equation}
f_i = N_D \sum_{j=1}^{m}P_j w_{ji} a_{ji}/N_j = \sum_{j=1}^{m}p_j w_{ji} a_{ji},
\end{equation}
with $p_j=N_D P_j/N_j$.  In each bin, the weights $w_{ji}$ account for the relative normalization of the samples and various other corrections. 

Since the MC samples are limited in size, the generated numbers of events $a_{ji}$ have statistical fluctuations relative to the value $A_{ji}$ expected for infinite statistics, and thus the more correct prediction for each bin is
\begin{equation}
f_i = \sum_{j=1}^{m} p_j w_{ji} A_{ji}.
\label{eq:finew}
\end{equation}
If we assume Poisson statistics for both the data and MC samples, the total likelihood function ${\cal L}$ is the combined probability for the observed $d_i$ and  $a_{ji}$~\cite{BarlowFit},
\begin{equation}
\ln {\cal L} =
 \sum_{i=1}^{n}(d_i\ln f_i - f_i)
+ \sum_{i=1}^n
    \sum_{j=1}^{m}
    (a_{ji} \ln A_{ji} -A_{ji}).
\label{eq:logl}
\end{equation}
The first sum has the usual form associated with the uncertainty of the data 
and the second term refers to the MC statistics and is not dependent on data.  
There are $(n+1)\times m$ unknown parameters that need to be determined: the $m$ relative normalization factors $p_j$,
 which are of interest to the signal extraction, and $n\times m$ values $A_{ji}$. 

The problem can be significantly simplified. The $n\times m$ quantities $A_{ji}$ can be 
determined by solving $n$ simultaneous equations for $A_{ji}$ of the form 
\begin{equation}
f_i = \sum_{j=1}^m p_j w_{ji} A_{ji} =  \sum_{j=1}^m \frac{p_j w_{ji} a_{ji}}{1+p_j w_{ji} t_i},
\label{eq:tidef}
\end{equation}
with $A_{ji}=a_{ji}/(1+  p_j w_{ji} t_i)$ and  $t_i=1-d_i/f_i$
(for $d_i=0$ we define $t_i=1$). At every step in the minimization of $-2$ln${\cal L}$ these $n$ independent equations need to be solved.  This procedure results not only in the determination of the parameters $p_j$, but also in improved estimates for the various contributions $A_{ji}$ in each bin. 

For fits to the individual data samples corresponding to the four signal decay modes, there is a specific likelihood function (Eq. \ref{eq:logl}).  To perform a simultaneous fit to all four data samples the log-likelihood function is the sum of the individual ones. 
Some of the parameters $p_j$ may be shared among the four likelihood functions,
\begin{eqnarray}
 \ln {\cal L} = \sum_{h=1}^{4} \ln {\cal L}_h &=& \sum_{h=1}^{4} \sum_{i=1}^{n}{(  d_{hi}  \ln{f_{hi}} - f_{hi})} \\ \nonumber
&+& \sum_{h=1}^{4}{\sum_{i=1}^{n} \sum_{j=1}^{m}{( a_{hji} \ln{ A_{hji}} %
-  A_{hji})}}.
\end{eqnarray}

\subsection{Fit Parameters and Inputs}
\label{sec:fitparam}

The fits can be performed separately for each of the four data samples or combined for all four data samples, and where possible, with common fit parameters. The nominal fit in this analysis is a simultaneous fit of all four data samples: \Bzpilnu , \Bppizlnu , \Bzrholnu, and \Bprhozlnu.  
A signal decay in one data sample may contribute to the background in another sample, and therefore these sources share a common
fit parameter. For example, the scale factor for the $\Bzrholnu$ signal in the $\Bzrholnu$ sample is also applied in the $\Bzpilnu$ sample,  
where it represents cross-feed background. 

We impose isospin invariance for the signal decay modes,
\begin{eqnarray} 
  \Gamma(\B^0 \ra \pi^- \ell^+ \nu)  & = & 2 \Gamma(\B^+ \ra \pi^0 \ell^+ \nu), \nonumber \\ 
  \Gamma(\B^0 \ra \rho^- \ell^+ \nu) & = & 2 \Gamma(\B^+ \ra \rho^0 \ell^+ \nu) \ .
  \label{eq:isospin}
\end{eqnarray}
\noindent
The yields of the true and combinatorial signal decays as well as isospin-conjugate decays are related to the 
same branching fraction and therefore share the same fit parameter. 
The $B \to X_u\ell\nu$ background, which contains exclusive
and non-resonant decays, is scaled by two parameters, one for low and intermediate $q^2$  ($q^2<20\gev^2$) 
and one for high $q^2$ ($q^2>20\gev^2$), for the fits to the \Bpilnu\ samples.
Because of the large correlation between $B \to X_u \ell\nu$ background and $B\to\rho\ell\nu$ signal
($>90\%$ for both \Brholnu\ modes), we rely on MC simulation for the $B \to X_u \ell\nu$ background
and keep it fixed in the fits to the \Brholnu\ samples.
The \BB\ background is split into two sources.
Among the \bclnu\ decays, we treat the dominant decay mode,
$B \to D^* \ell\nu$, as a separate source and combine the other semileptonic decays 
($B \to D\ell\nu$, $B \to D^{(*)}(n\pi) \ell\nu$) 
and the remaining (or ``other'') \BB\ background (secondary leptons and fake leptons)
into a single source.  The continuum \qqbar\ background sources containing true and fake leptons 
are combined into one fit source and scaled by a single fit parameter.

The complete list of fit sources and corresponding fit parameters is given in Table~\ref{tab:fitSources}. 
The $\pi\leftrightarrow\rho$ cross feed is a free fit parameter in the four-mode fit;  
for one-mode fits, it is fixed to the value obtained from the four-mode fit.
In the four-mode fit, all background sources that are not fixed 
are fit separately for each signal mode,
since the different hadrons of the signal decays lead to different
combinatorial backgrounds. 
\begin{table}[hbt]
\centering
\caption{
List of fit parameters representing scale factors for the different signal samples and background sources. 
Parameters with index $j$ are free parameters in the fit, one for each $q^2$ bin $j$. The $\pi\leftrightarrow\rho$ crossfeed parameter is free 
only in the four-mode fit; for one-mode fits, it is fixed to the values obtained from 
the four-mode fit. There are independent scale factors for \qqbar\ background, $B \to D^* \ell \nu$ decays 
and for all other background sources from \BB\ events for all four signal modes (subscripts $\pi^{\pm}, \pi^0, \rho^{\pm}, \rho^0$). 
For the \Bpilnu\ decays, the \BXulnu background is fit in two $q^2$ intervals (index $k=1,2$); for the \Brholnu\ decays it is fixed.}
\begin{tabular}{lcc} 
\hline \hline
Source / sample                   	&  $B \to \pi\ell\nu$                                   & 
$B \to \rho\ell\nu$ \\ \hline

Signal                     		&  $p_j^{\pi\ell\nu}$                                   & $p_j^{\rho\ell\nu}$      \\
Combinatorial signal       		&  $p_j^{\pi\ell\nu}$                                   & $p_j^{\rho\ell\nu}$      \\
Isospin-conjugate signal   		&  $p_j^{\pi\ell\nu}$                                   & $p_j^{\rho\ell\nu}$      \\
Cross feed $\pi\leftrightarrow\rho$     
                           		&  $p_j^{\rho\ell\nu}$                                  & $p_j^{\pi\ell\nu}$       \\

\bulnu\ background                	&  $p^{u\ell\nu}_{\pi^\pm,k}$, $p^{u\ell\nu}_{\pi^0,k}$ &  fixed                   \\
$B \to D^* \ell\nu $ background         &  $p_{\pi^\pm}^{D^*\ell\nu}$, $p_{\pi^0}^{D^*\ell\nu}$ & $p_{\rho^\pm}^{D^*\ell\nu}$, $p_{\rho^0}^{D^*\ell\nu}$  \\
Other \BB\ background                   &  $p_{\pi^\pm}^{other\BB}$, $p_{\pi^0}^{other\BB}$     & $p_{\rho^\pm}^{other\BB}$, $p_{\rho^0}^{other\BB}$  \\
\qqbar\ background                      &  $p_{\pi^\pm}^{\qqbar}$, $p_{\pi^0}^{\qqbar}$         & $p_{\rho^\pm}^{\qqbar}$, $p_{\rho^0}^{\qqbar}$    \\ 
\hline\hline
\end{tabular}
\label{tab:fitSources}
\end{table} 

\subsection{Fit Results}
\label{sec:fitresults}

\begin{figure*}    
\centering
  \begin{tabular}{ccc}
    \begin{minipage}{0.33\linewidth}
      \epsfig{file=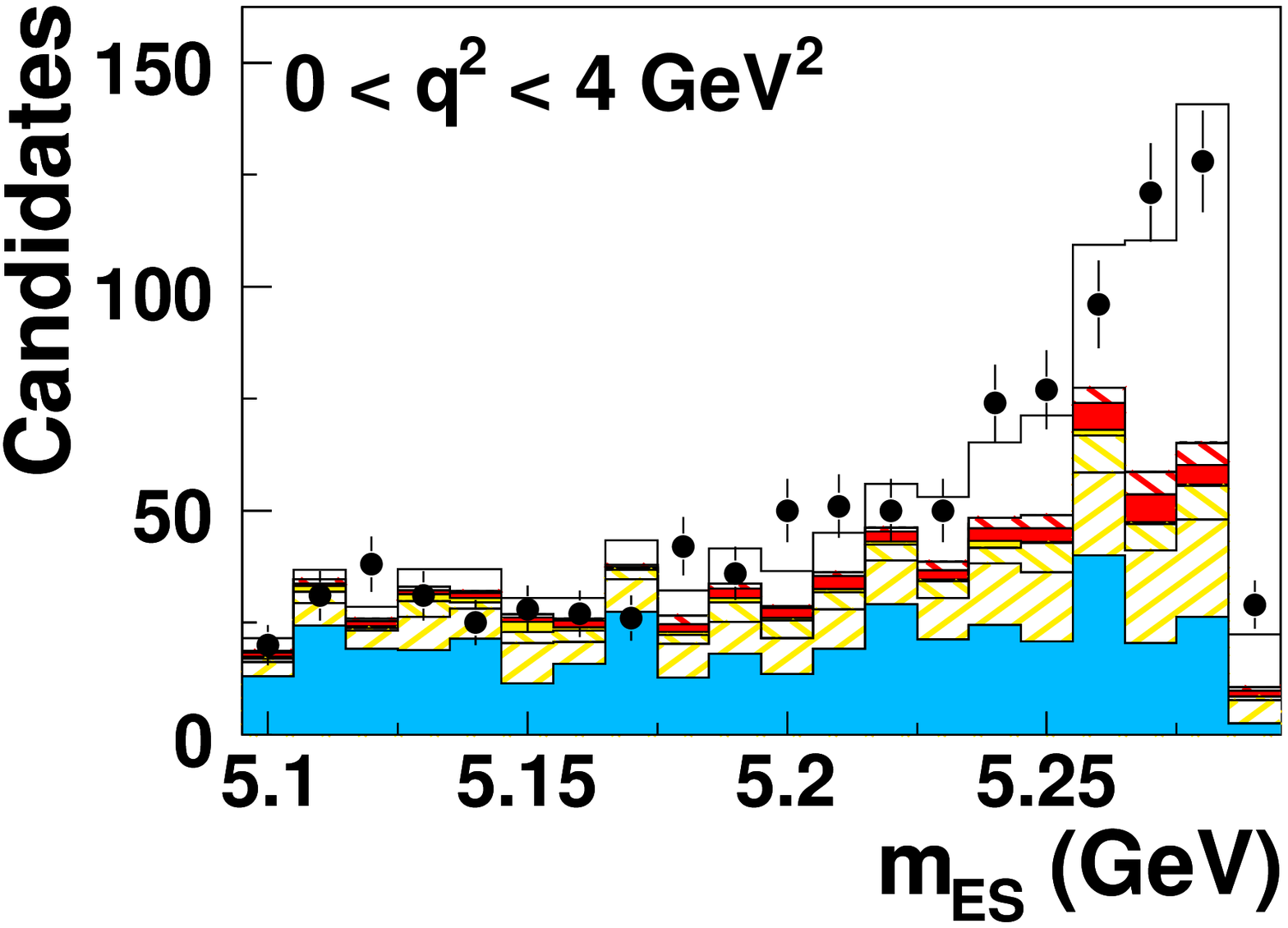, width = 6cm}\\
      \epsfig{file=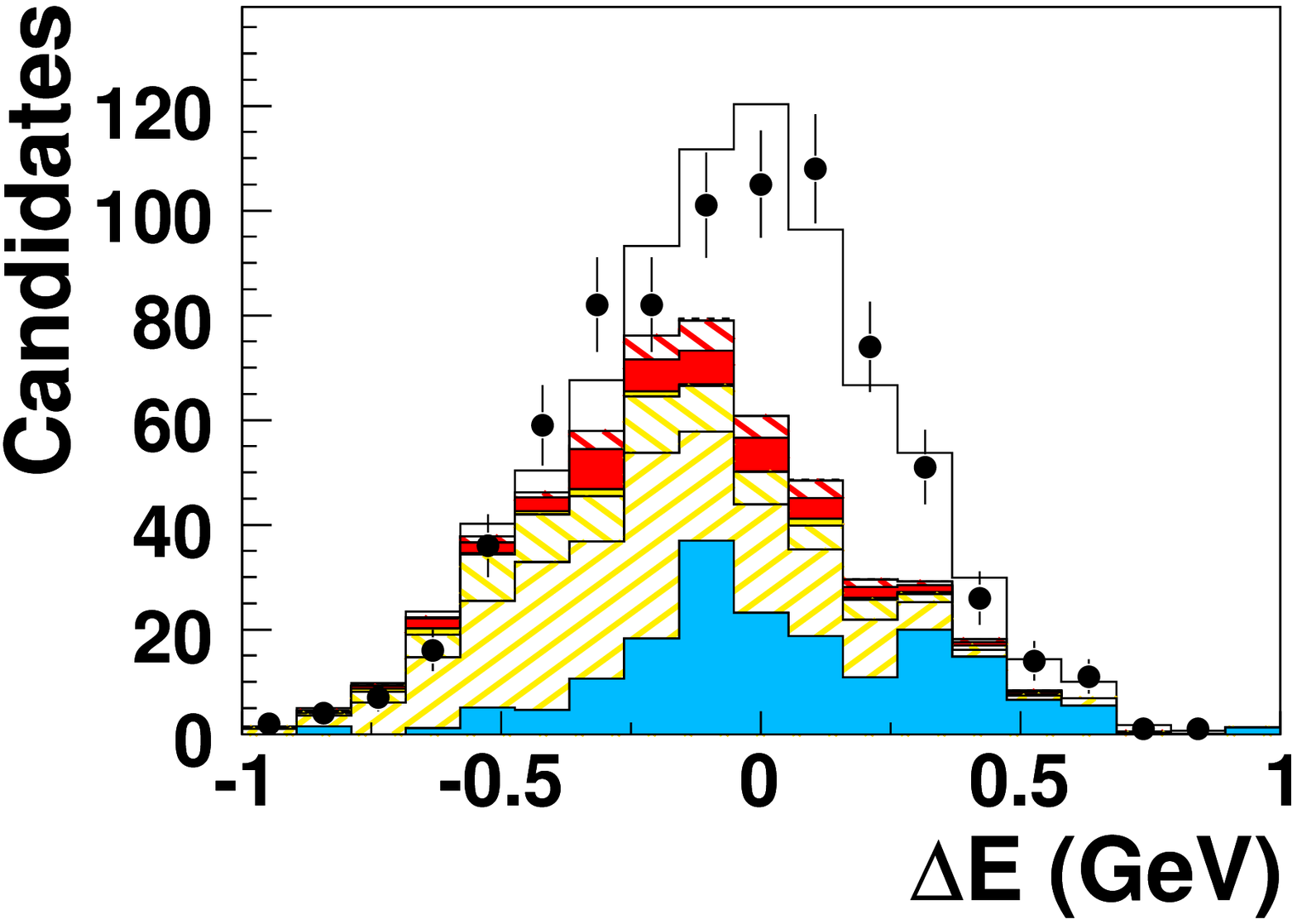, width = 6cm}\\
    \end{minipage}
    \begin{minipage}{0.33\linewidth}
      \epsfig{file=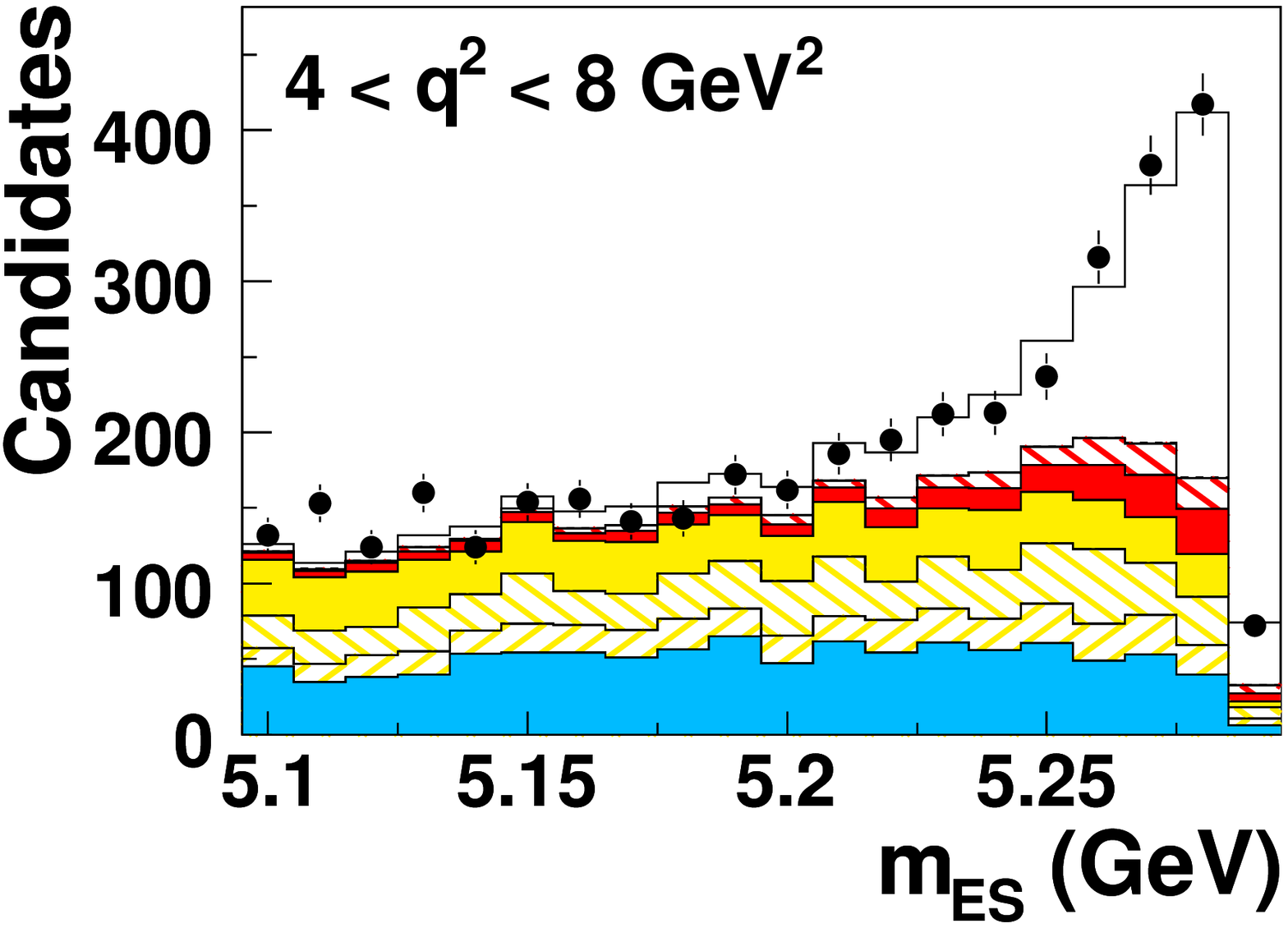, width = 6cm}\\
      \epsfig{file=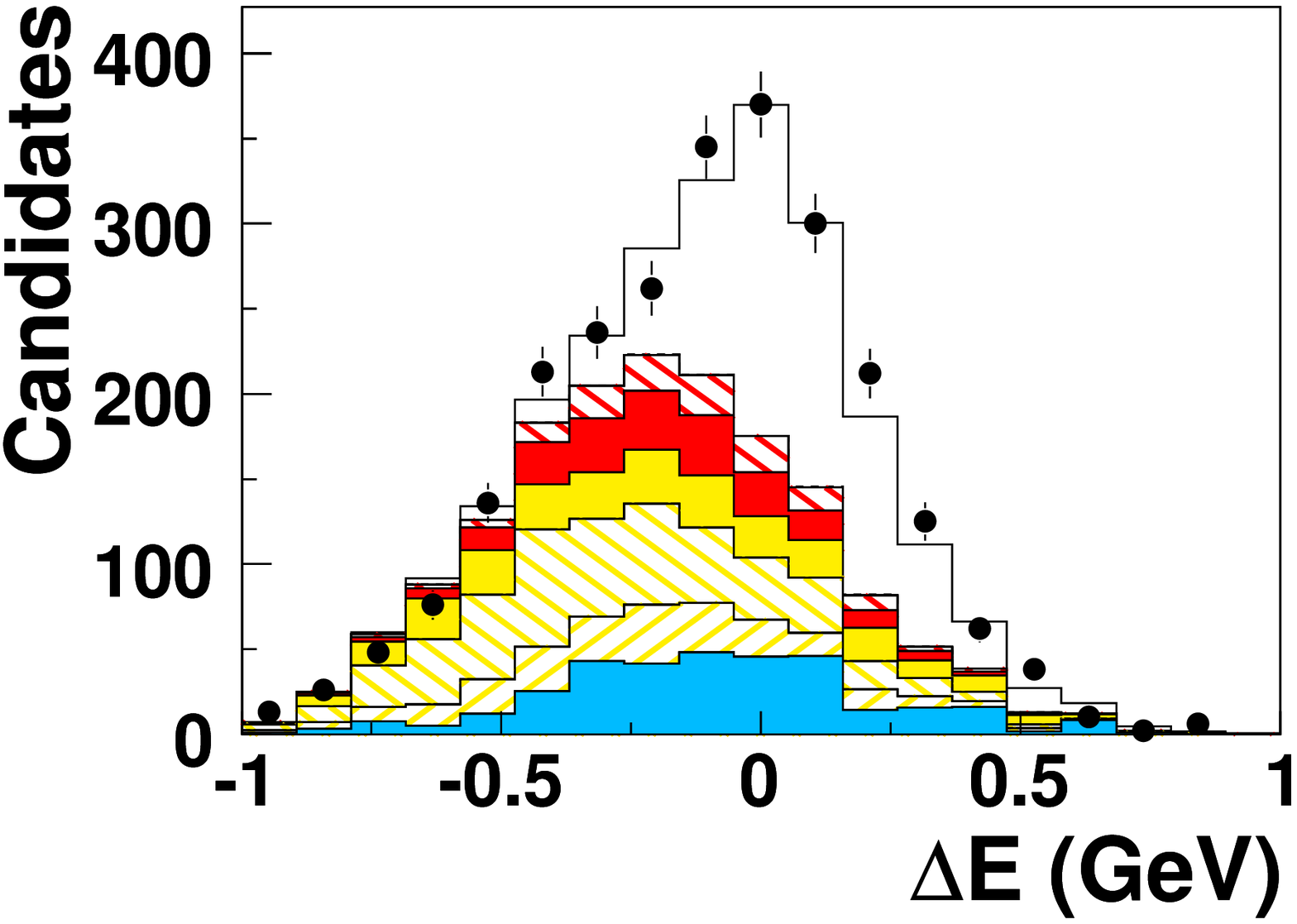, width = 6cm}\\
    \end{minipage}
    \begin{minipage}{0.33\linewidth}
      \epsfig{file=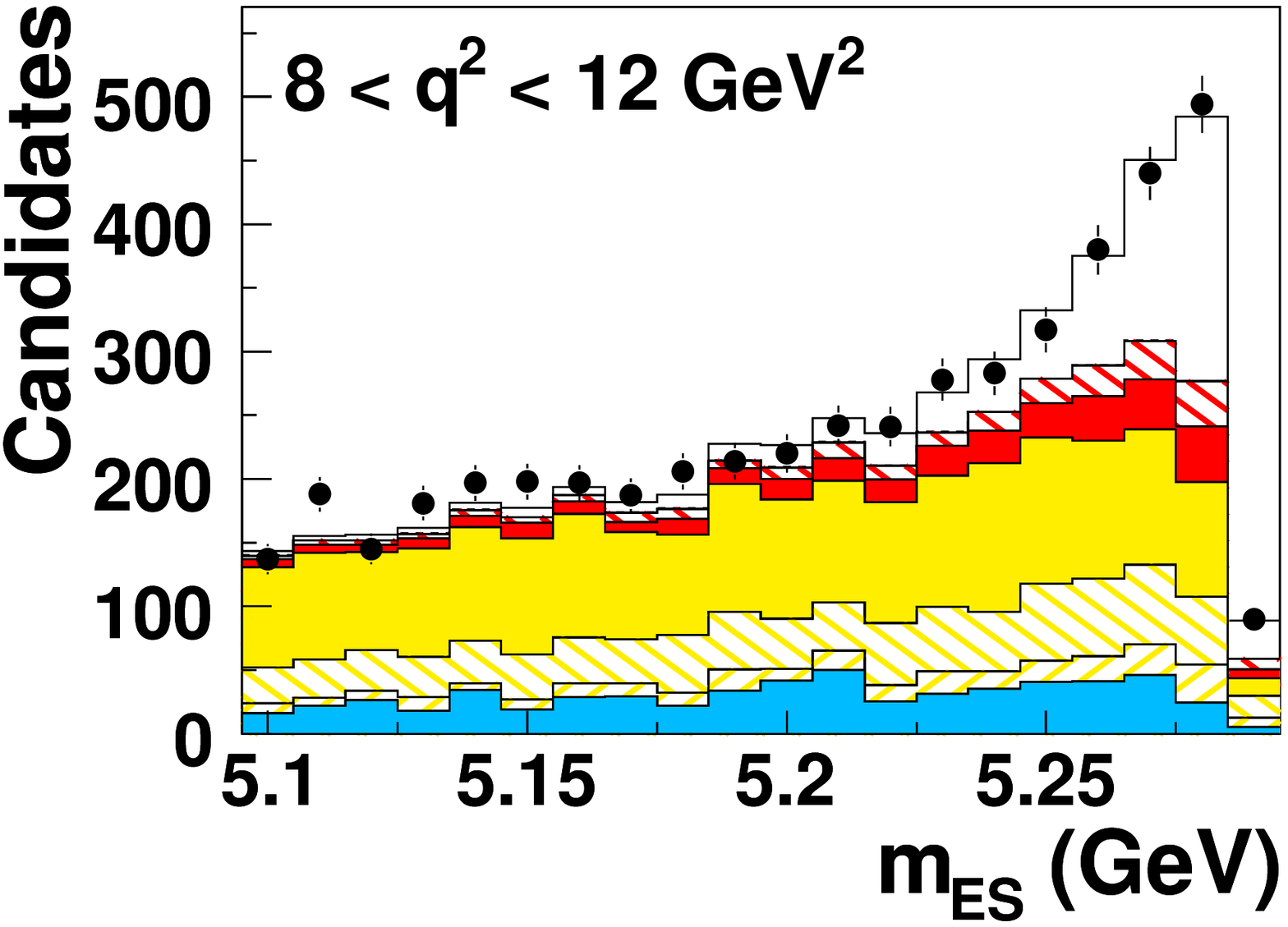, width = 6cm}\\
      \epsfig{file=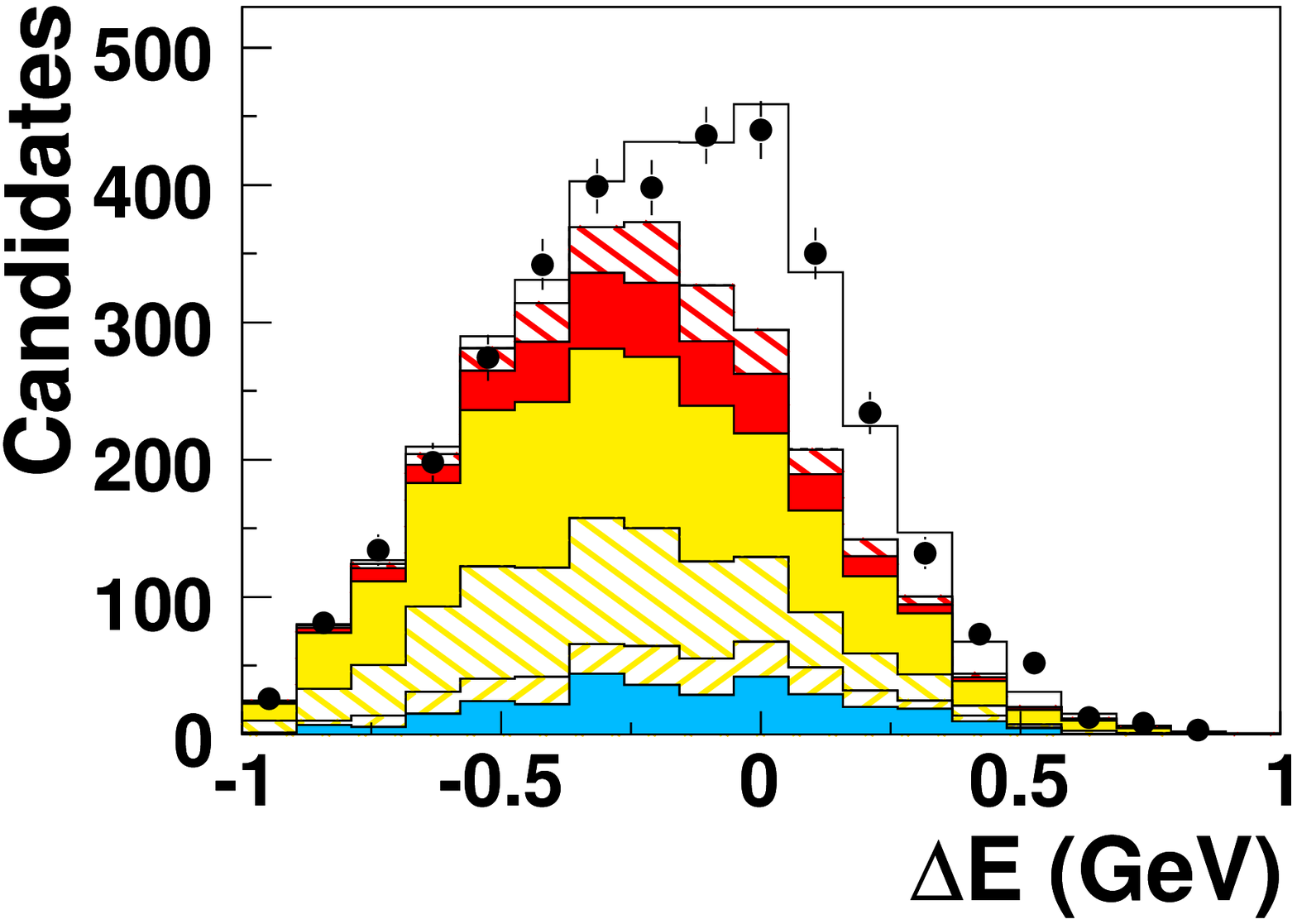, width = 6cm}\\
    \end{minipage}\\
    \begin{minipage}{0.33\linewidth}
       \epsfig{file=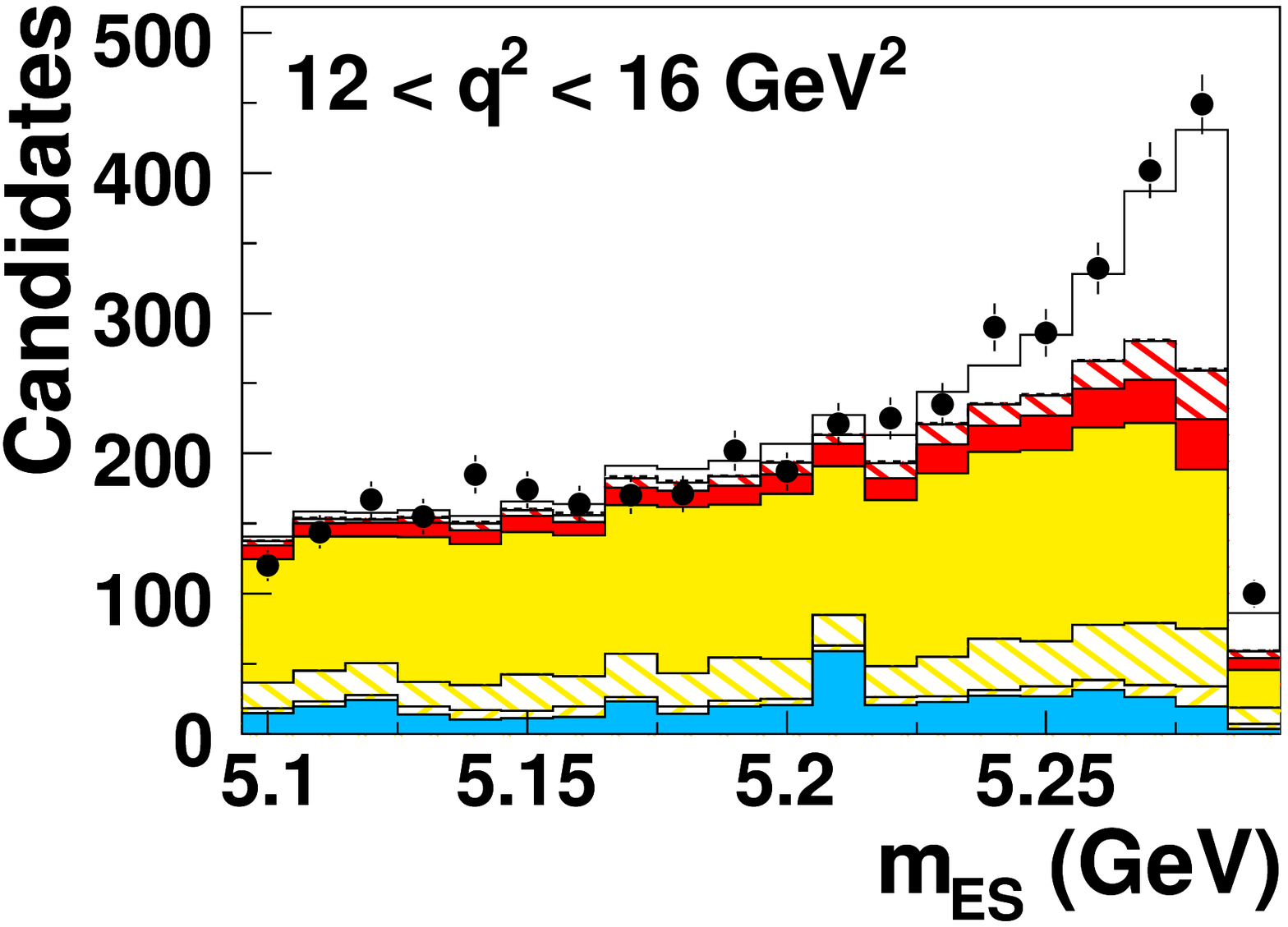, width = 6cm}\\
      \epsfig{file=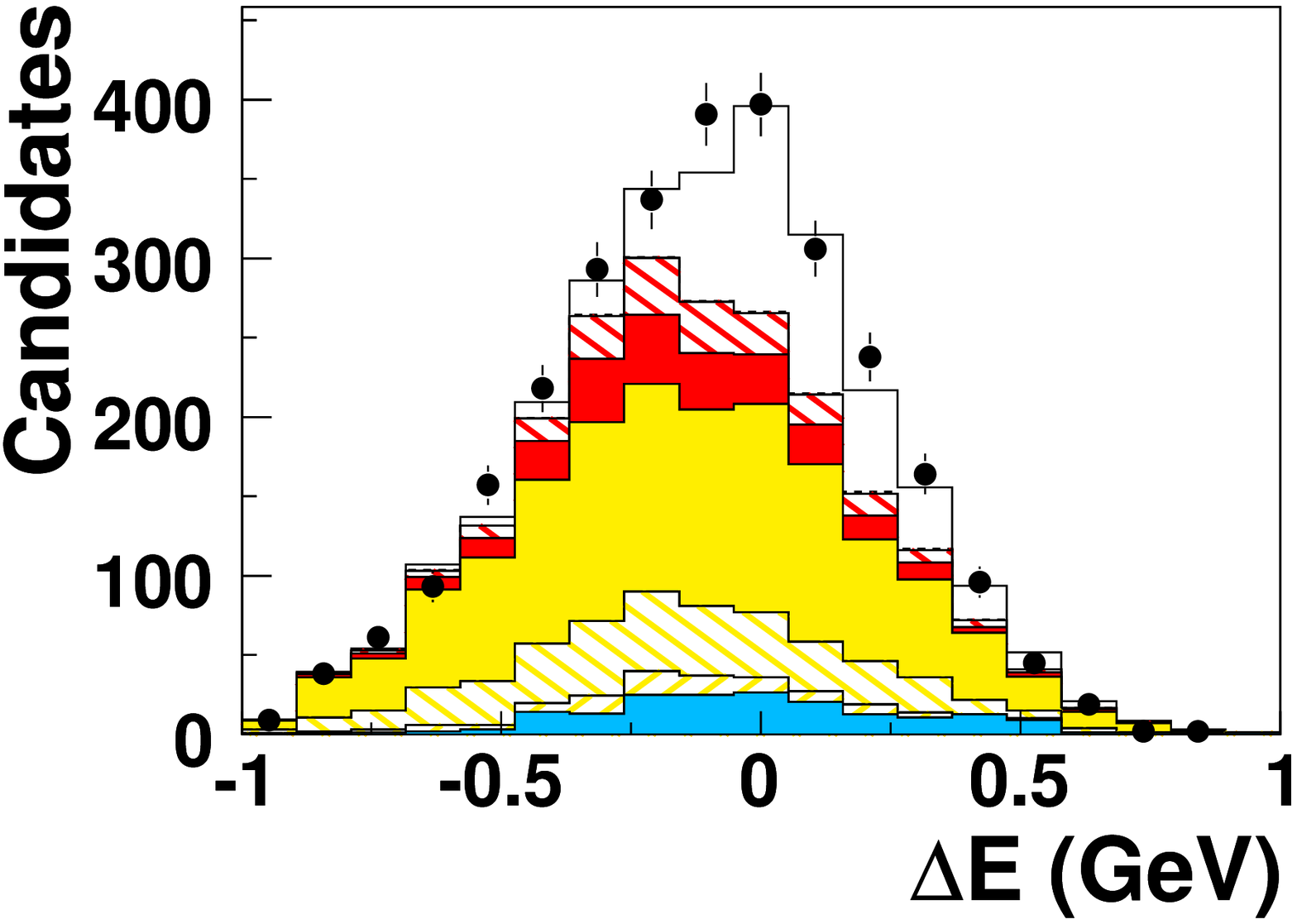, width = 6cm}\\
    \end{minipage}
    \begin{minipage}{0.33\linewidth}
      \epsfig{file=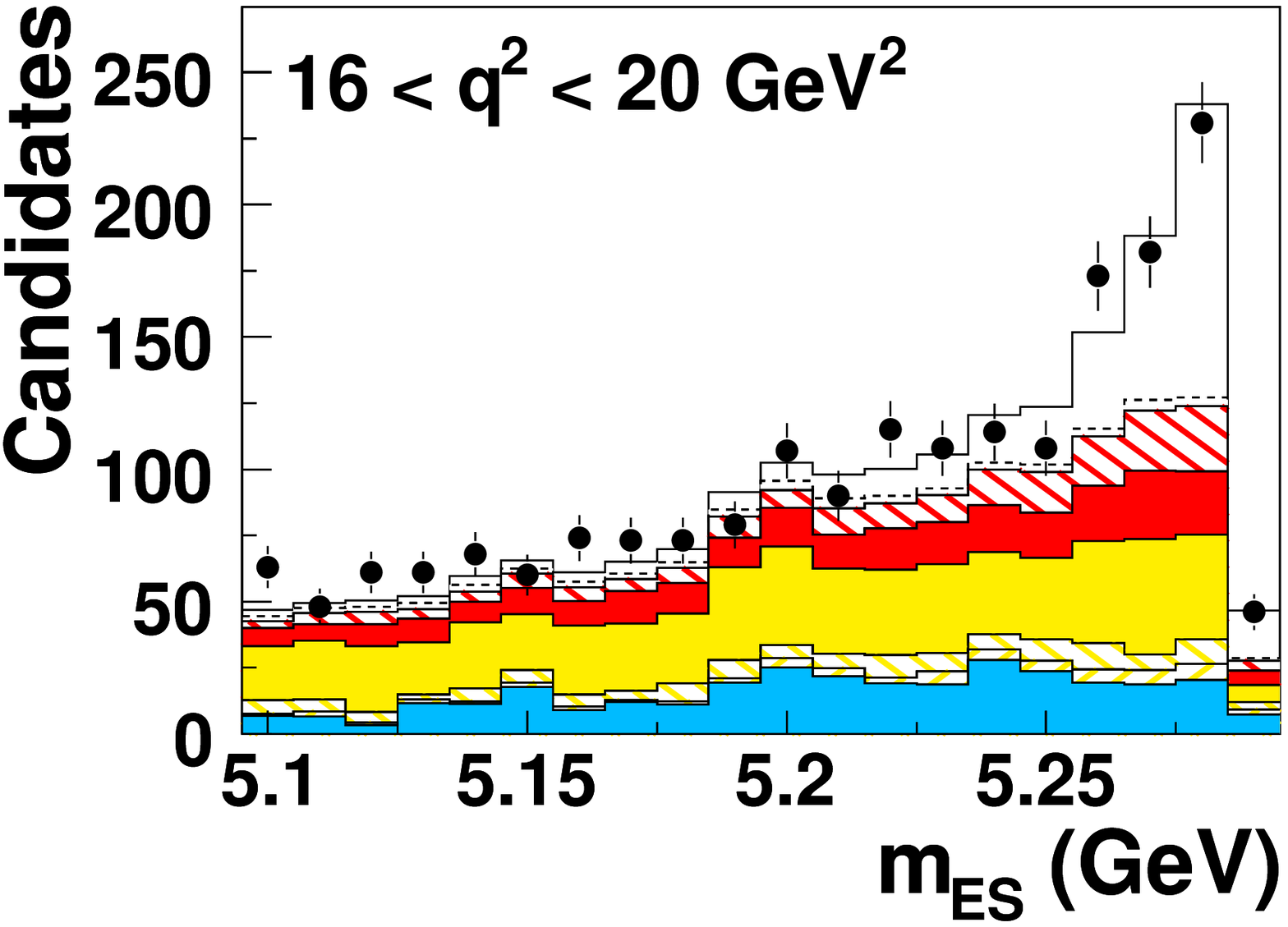, width = 6cm}\\
      \epsfig{file=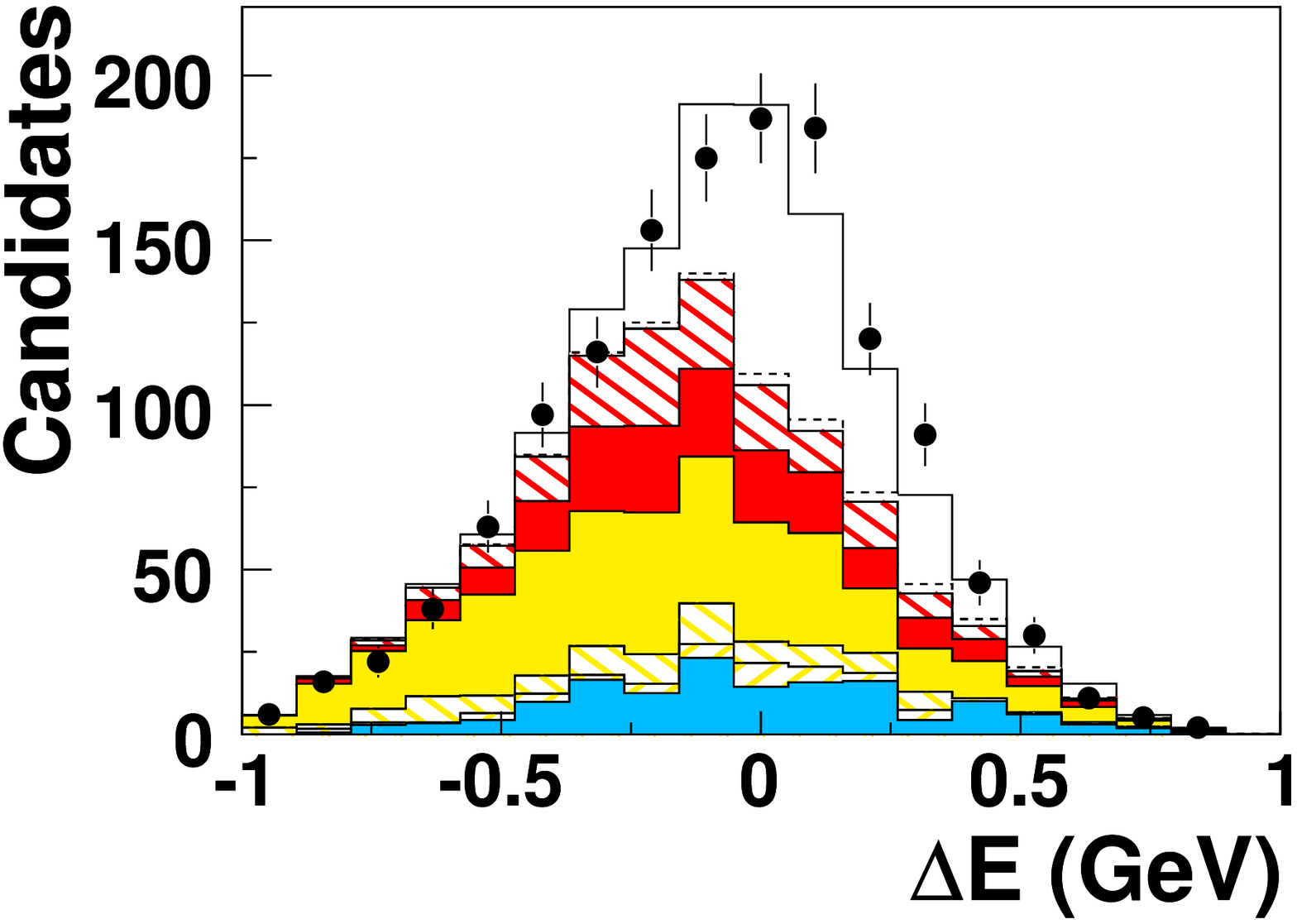, width = 6cm}\\
    \end{minipage}
    \begin{minipage}{0.33\linewidth}
      \epsfig{file=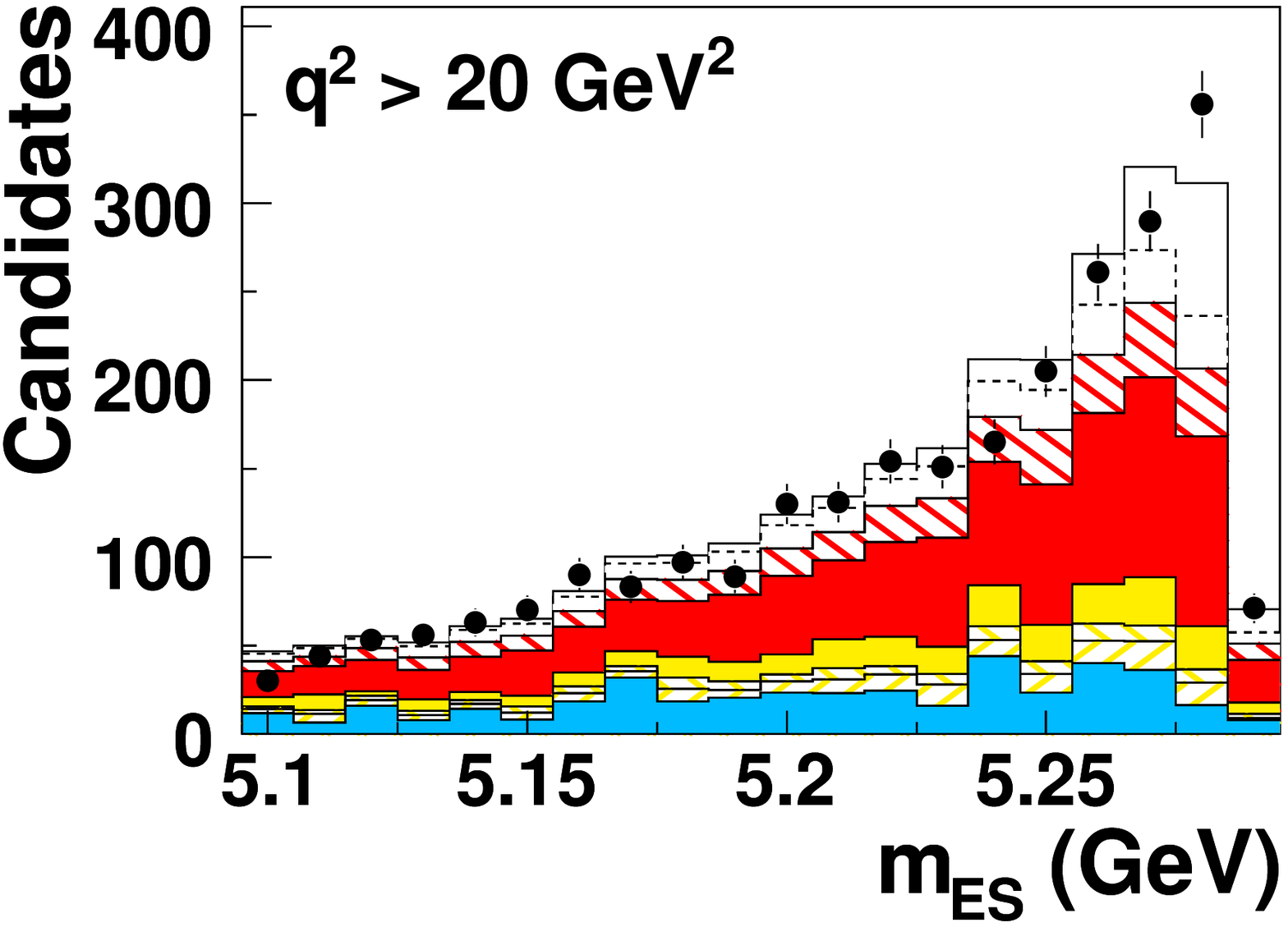, width = 6cm}\\
      \epsfig{file=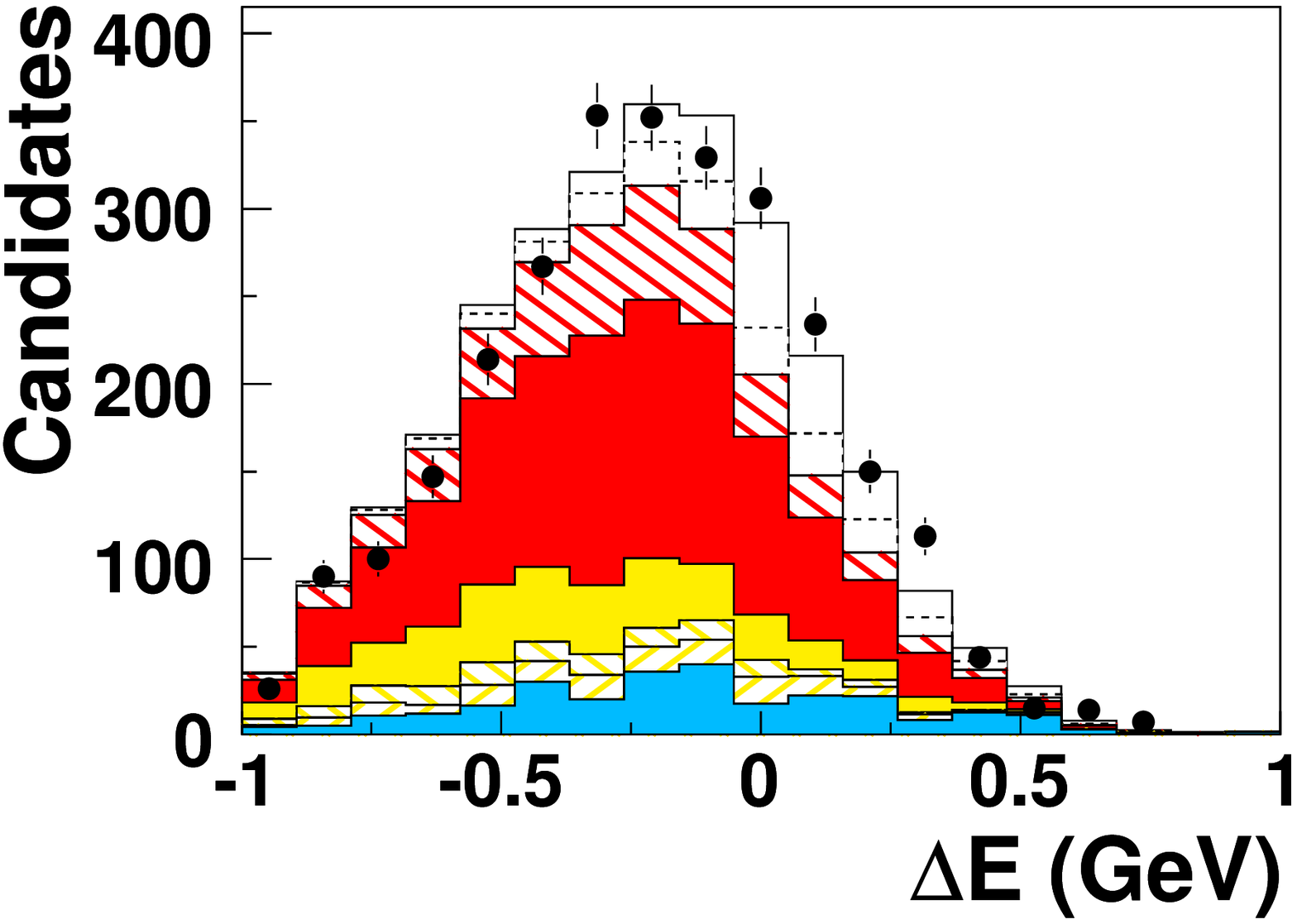, width = 6cm}\\
    \end{minipage}
  \end{tabular}
  \caption{(color online)
  \mES and \DeltaE distributions in each $q^2$~bin for 
           \Bzpilnu after the fit.
           The distributions are shown in the \DeltaE and \mES signal bands,
           respectively.
	   Legend: see Figure~\ref{fig:legend}.}
  \label{fig:DeltaE_pilnu_fit}
\end{figure*}

\begin{figure*}    
\centering
  \begin{tabular}{ccc}
    \begin{minipage}{0.33\linewidth}
      \epsfig{file=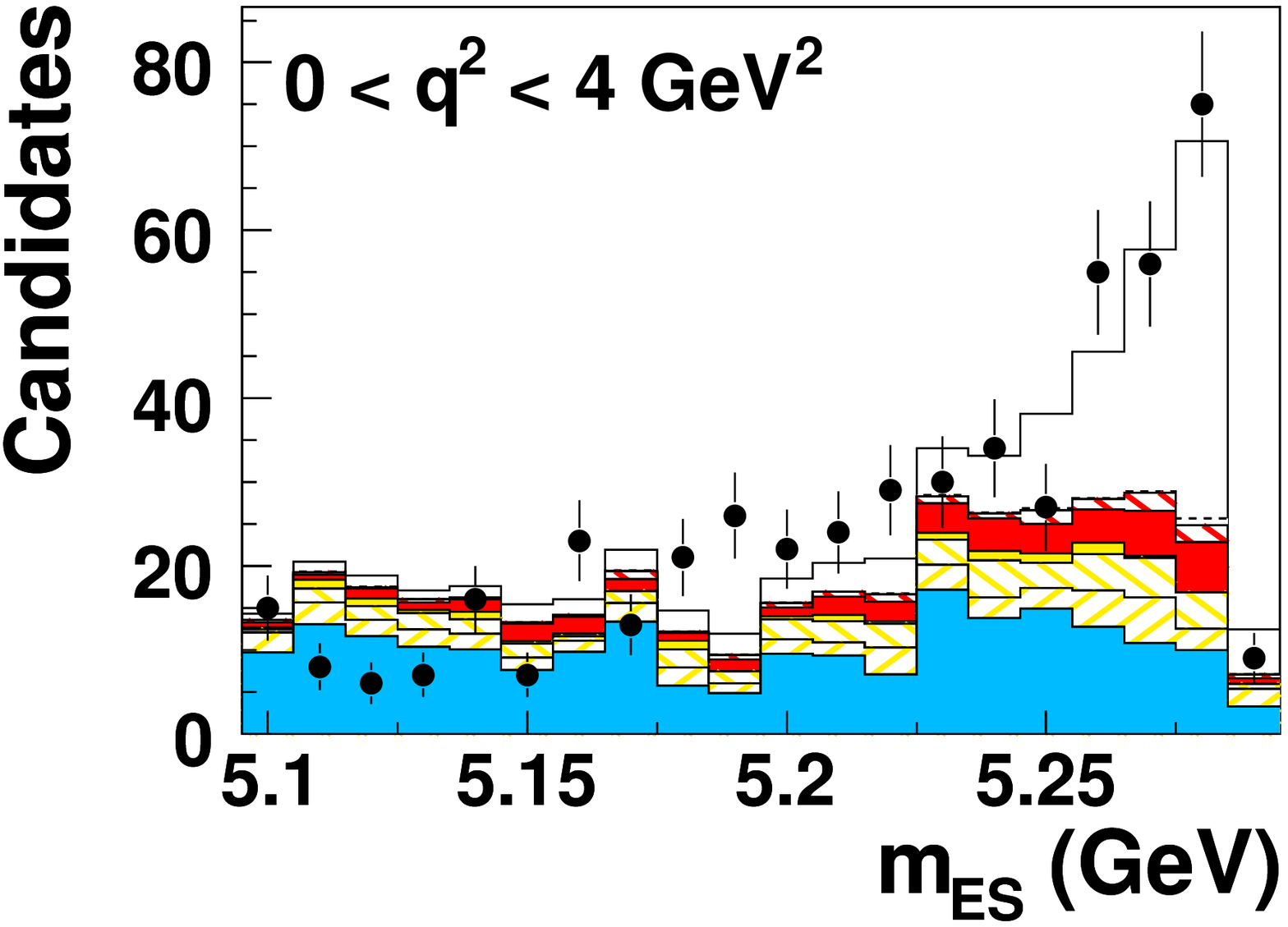, width = 6cm}\\
      \epsfig{file=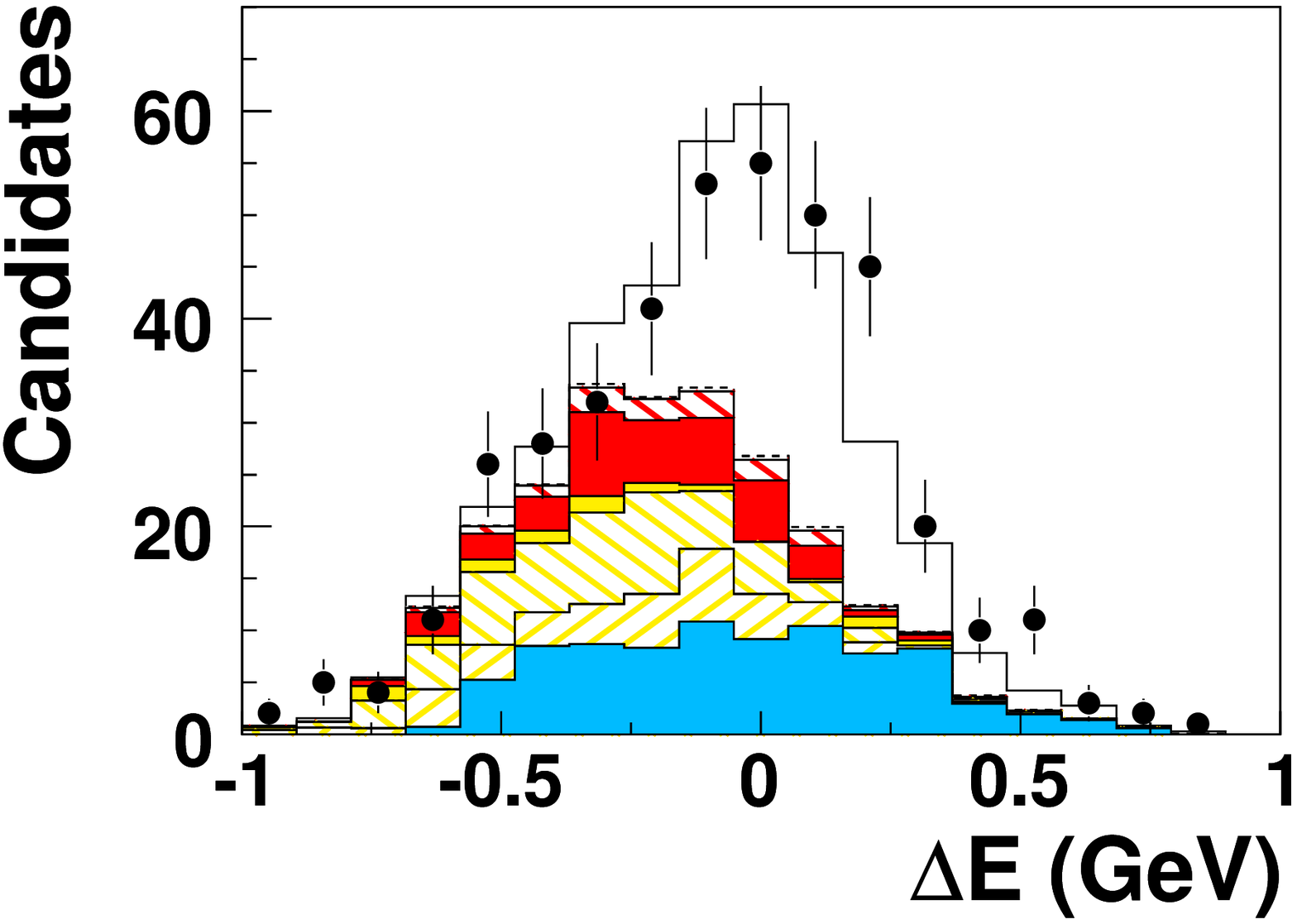, width = 6cm}\\
    \end{minipage}
    \begin{minipage}{0.33\linewidth}
      \epsfig{file=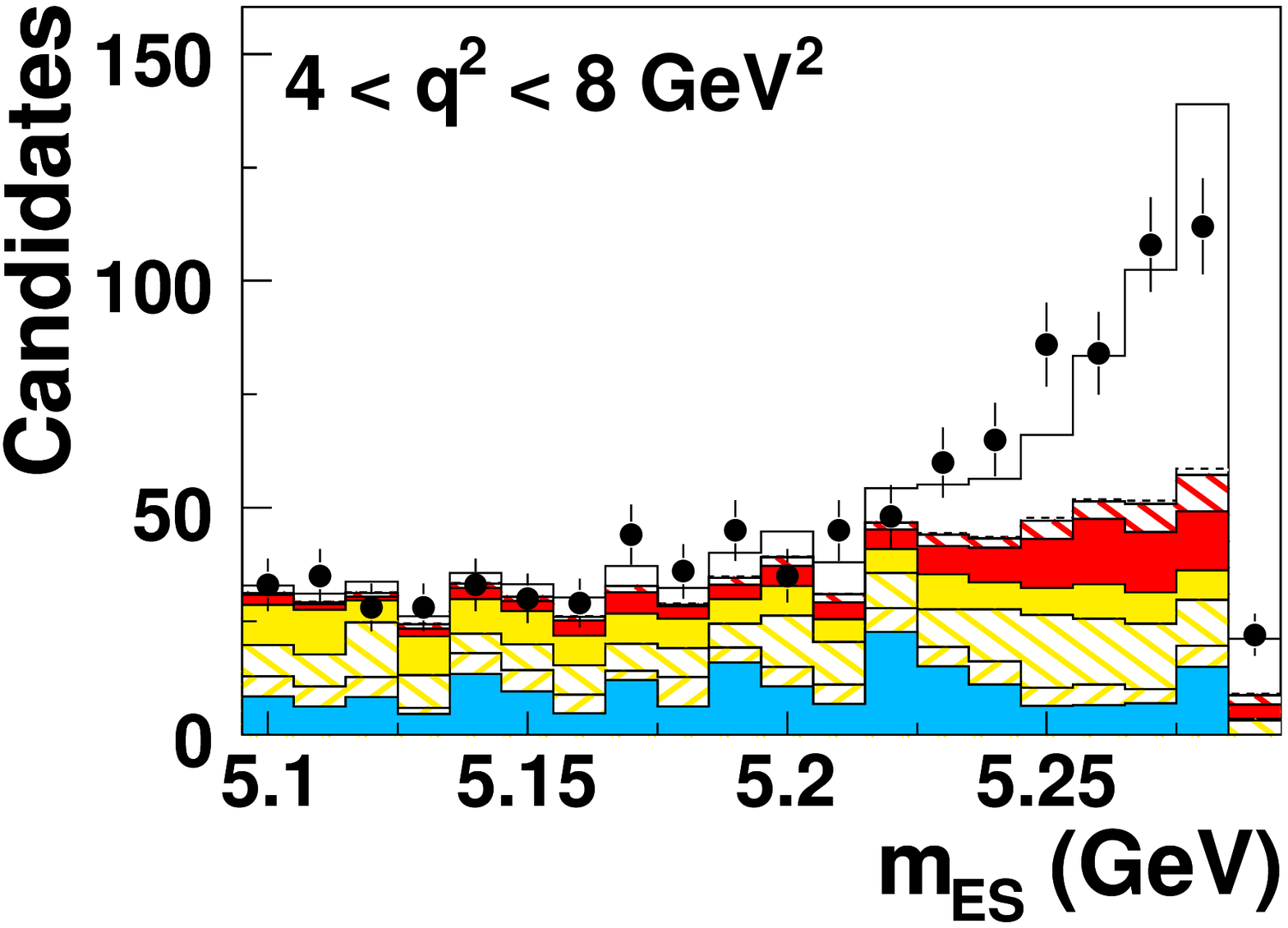, width = 6cm}\\
      \epsfig{file=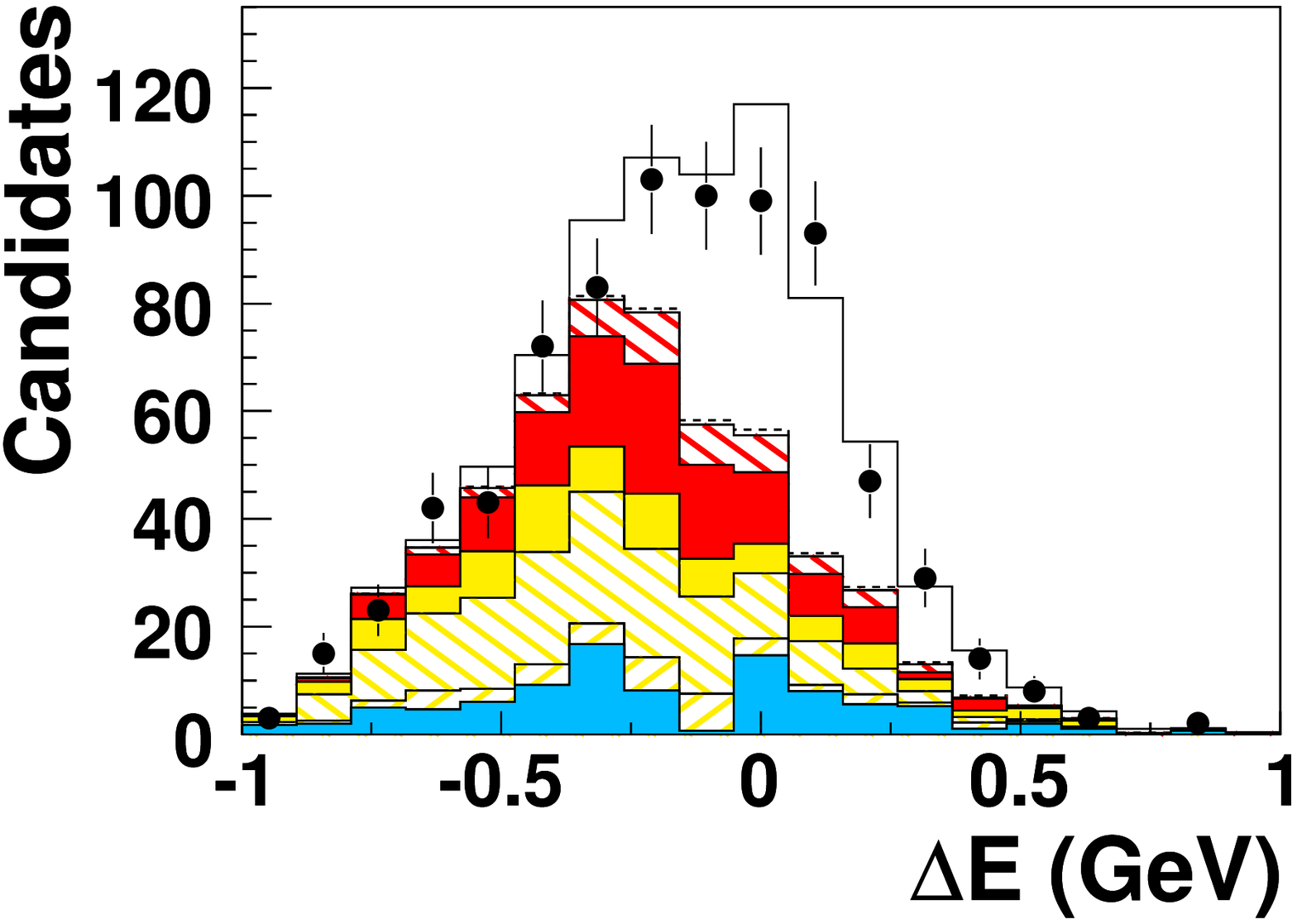, width = 6cm}\\
    \end{minipage}
    \begin{minipage}{0.33\linewidth}
      \epsfig{file=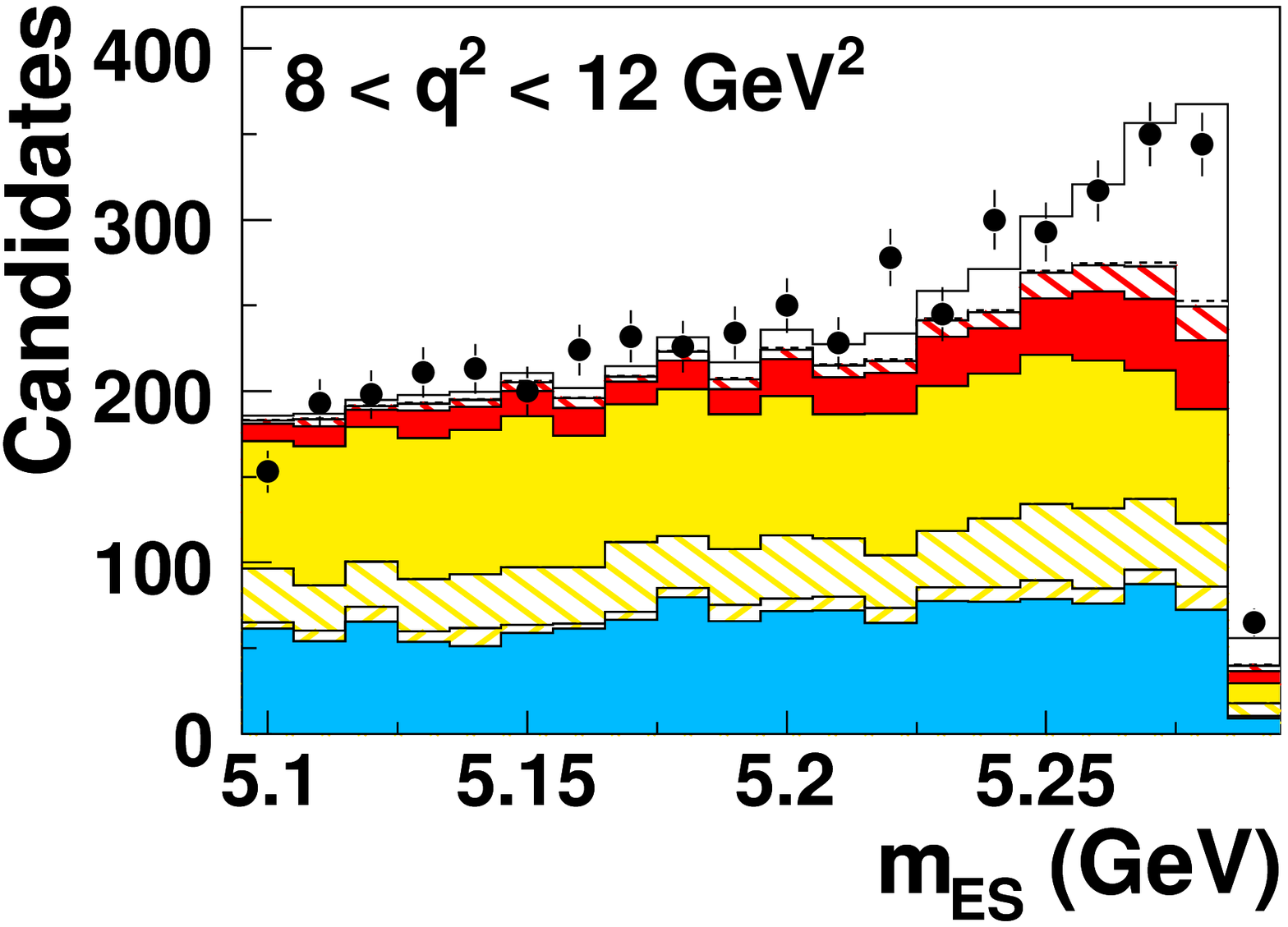, width = 6cm}\\
      \epsfig{file=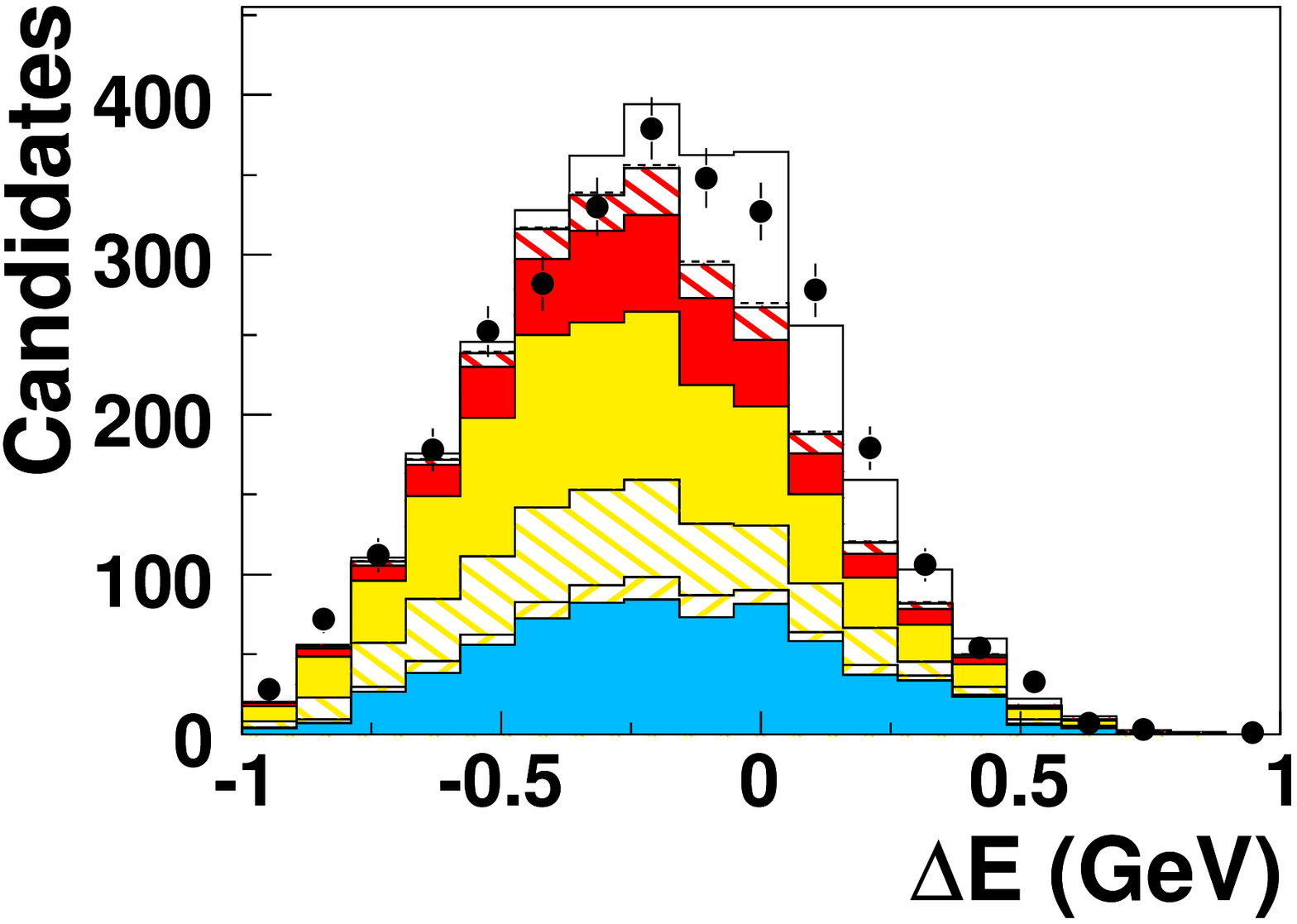, width = 6cm}\\
    \end{minipage}\\
    \begin{minipage}{0.33\linewidth}
       \epsfig{file=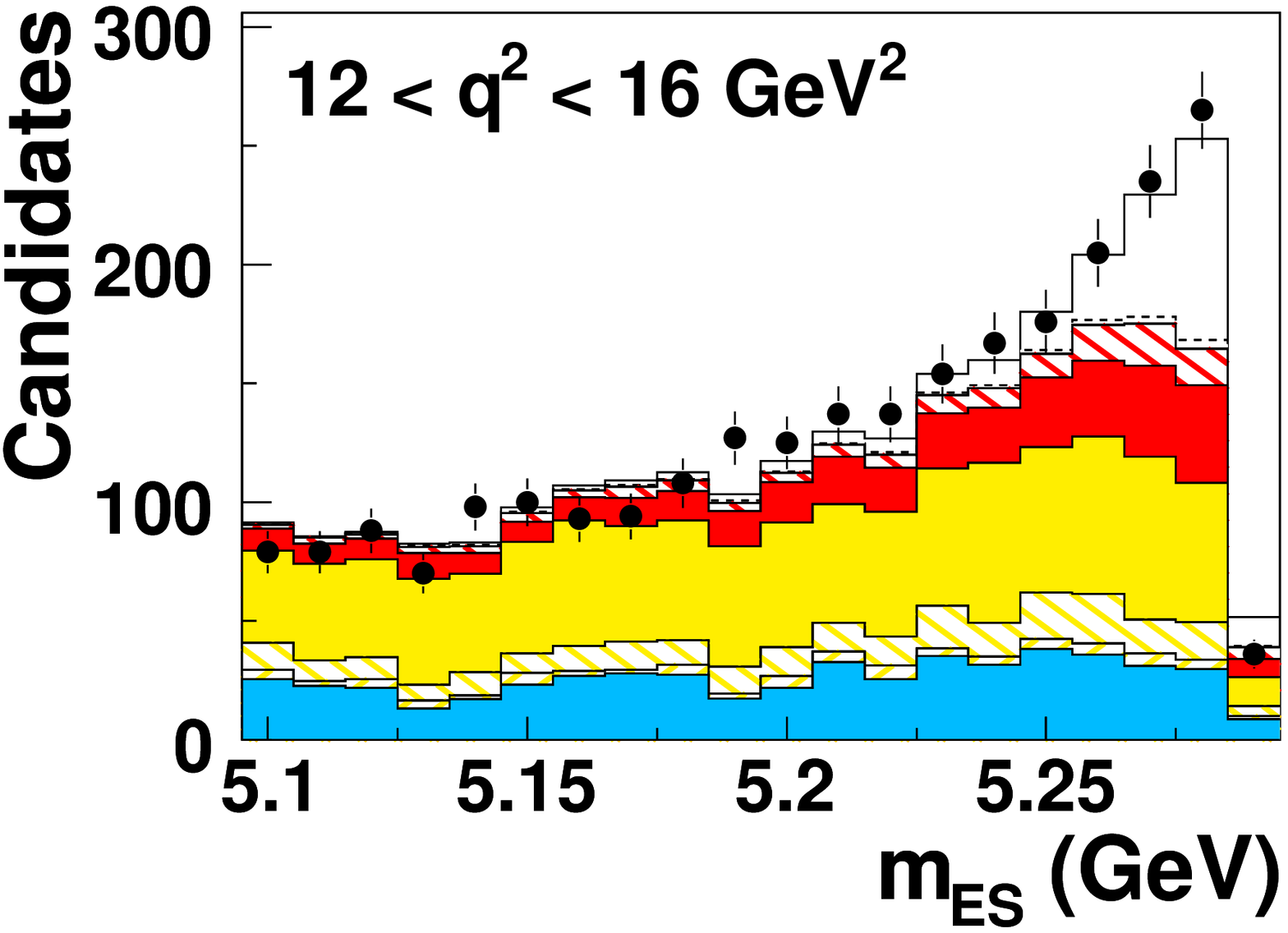, width = 6cm}\\
      \epsfig{file=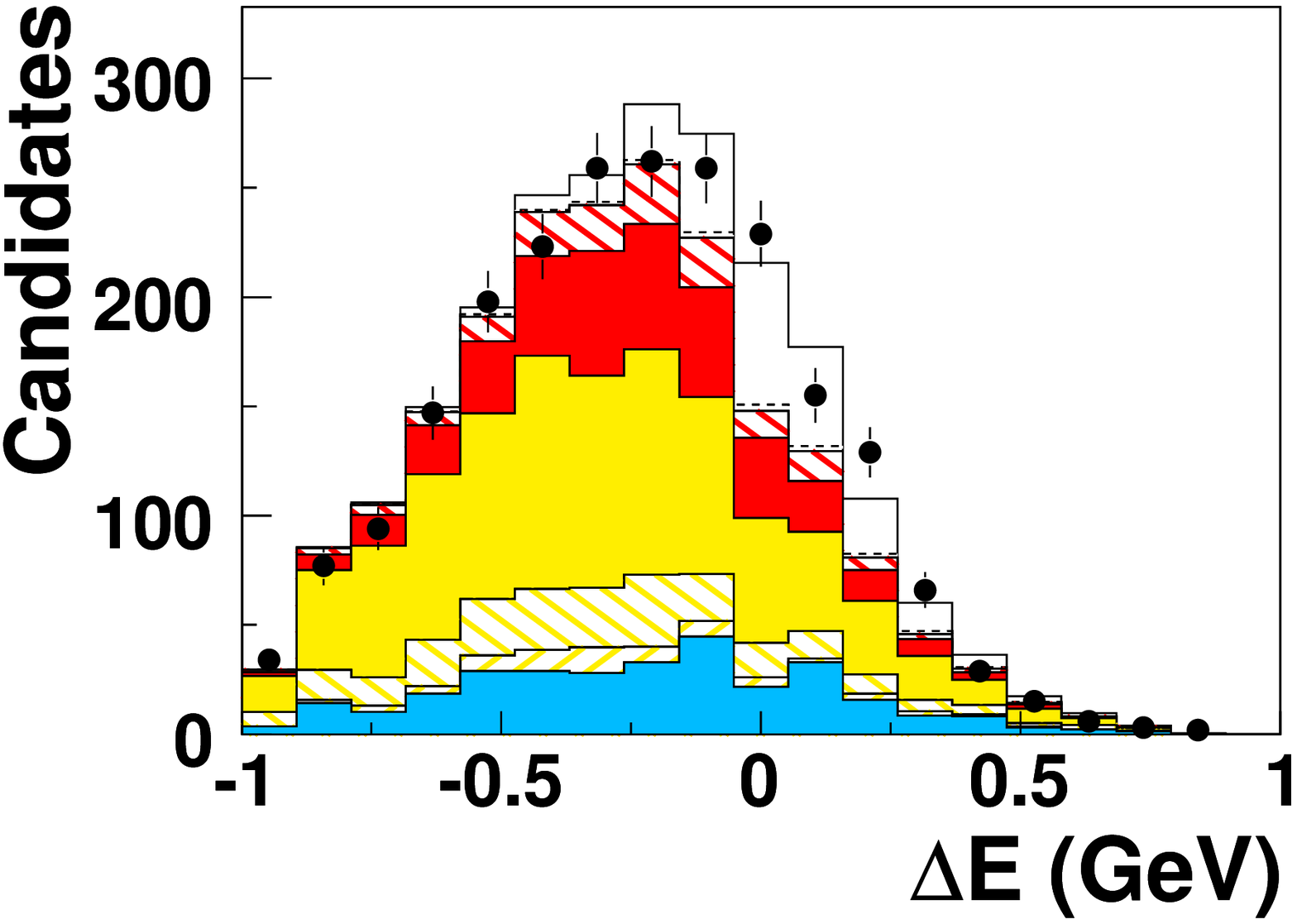, width = 6cm}\\
    \end{minipage}
    \begin{minipage}{0.33\linewidth}
      \epsfig{file=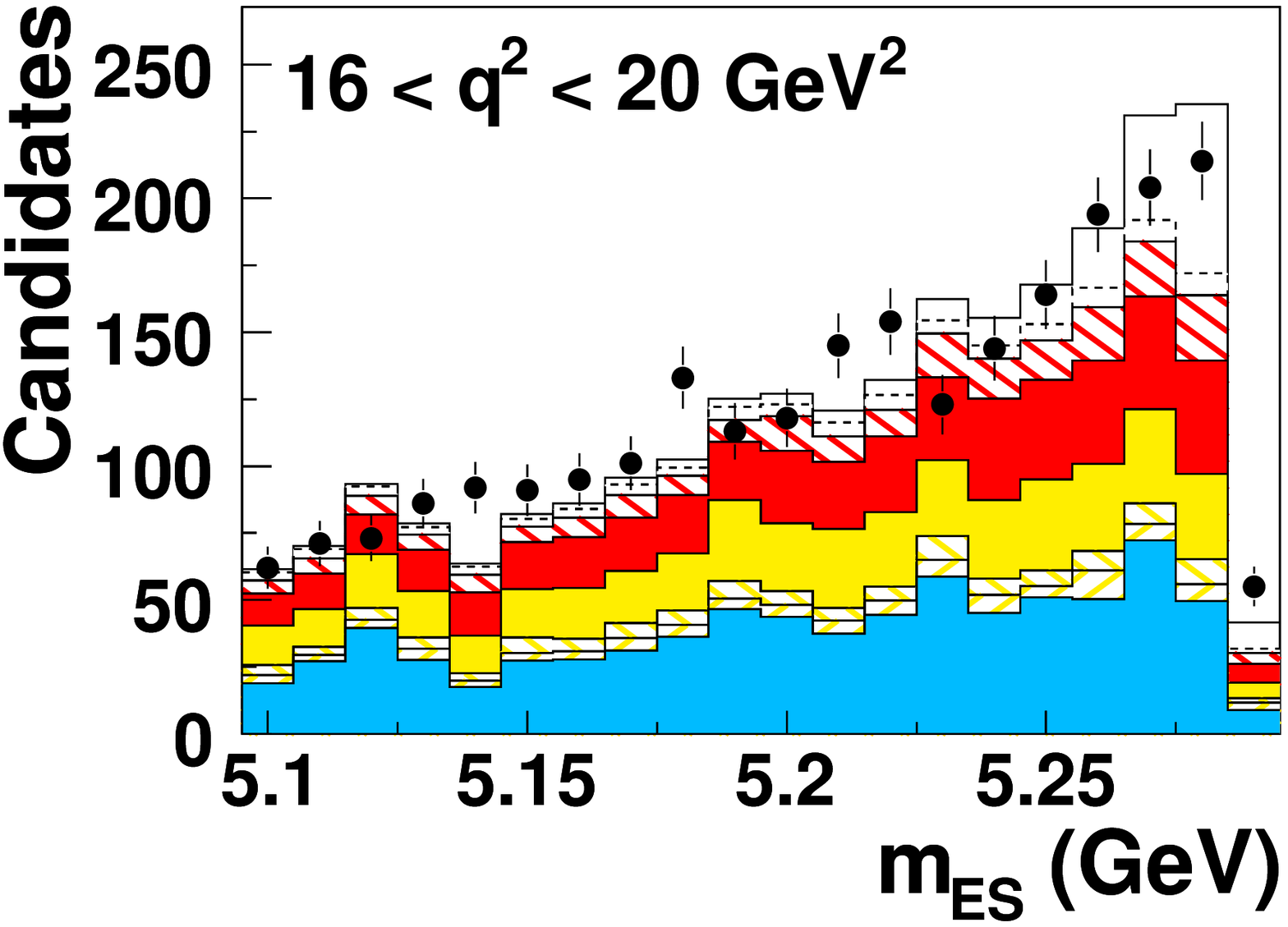, width = 6cm}\\
      \epsfig{file=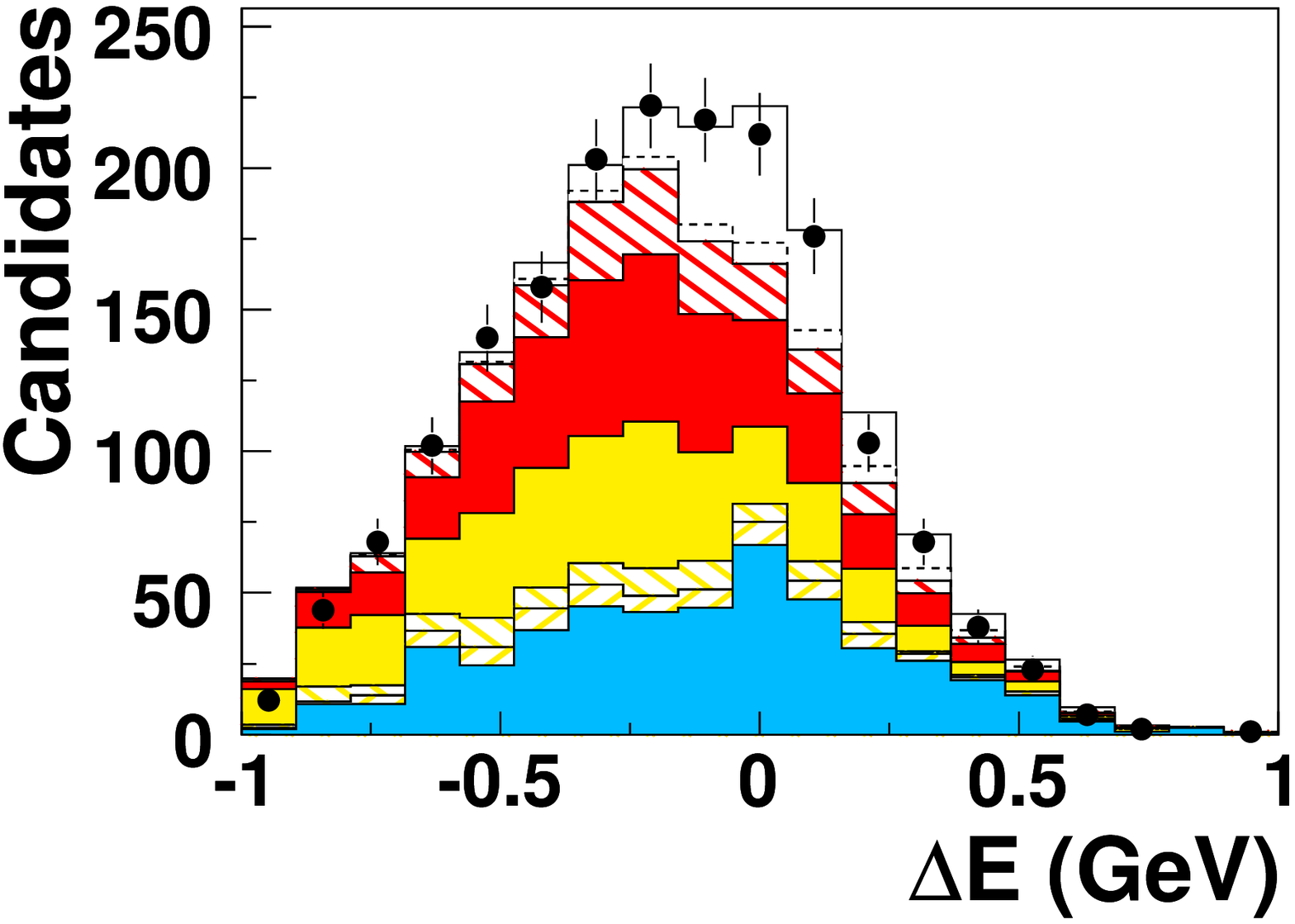, width = 6cm}\\
    \end{minipage}
    \begin{minipage}{0.33\linewidth}
      \epsfig{file=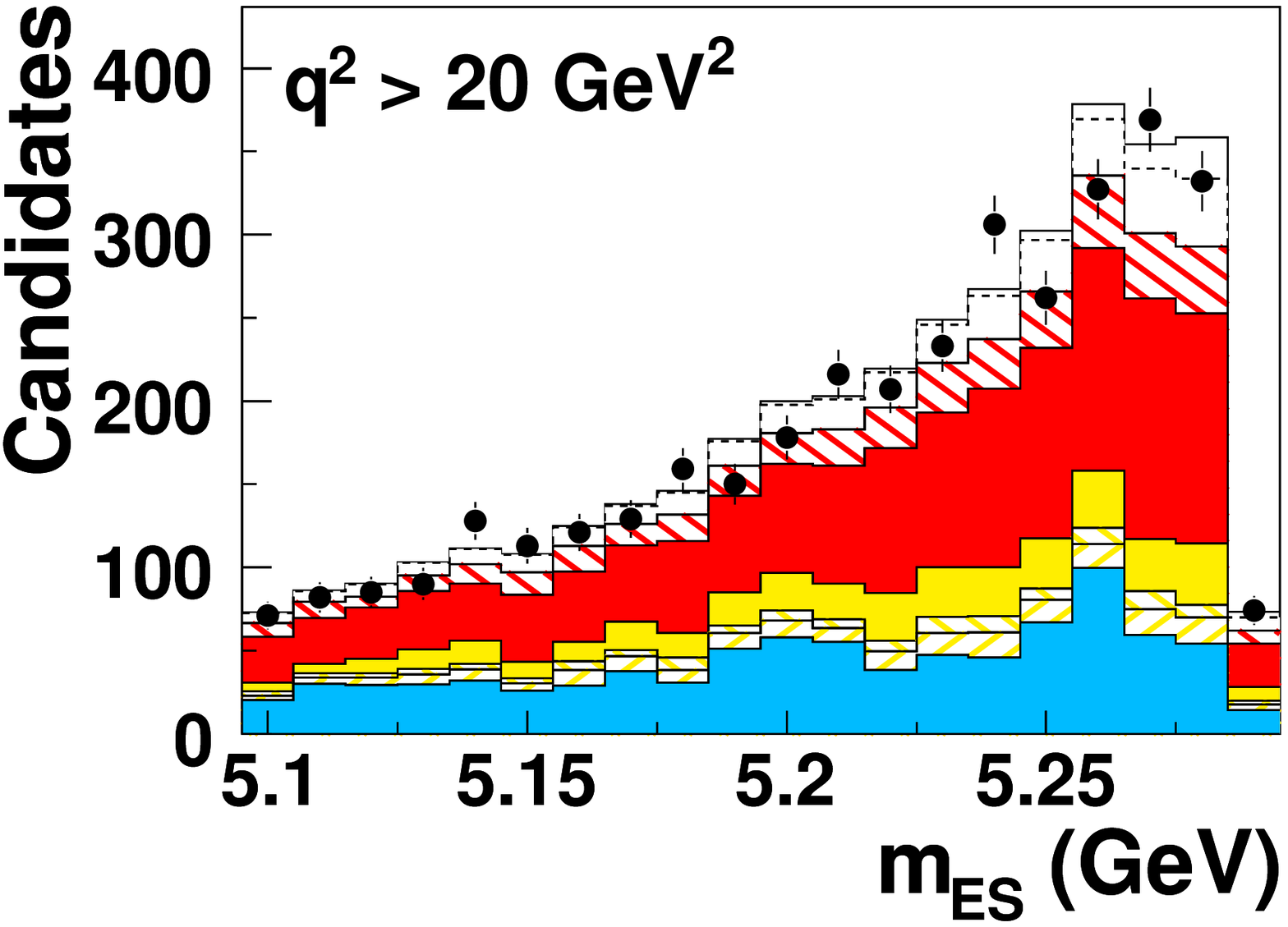, width = 6cm}\\
      \epsfig{file=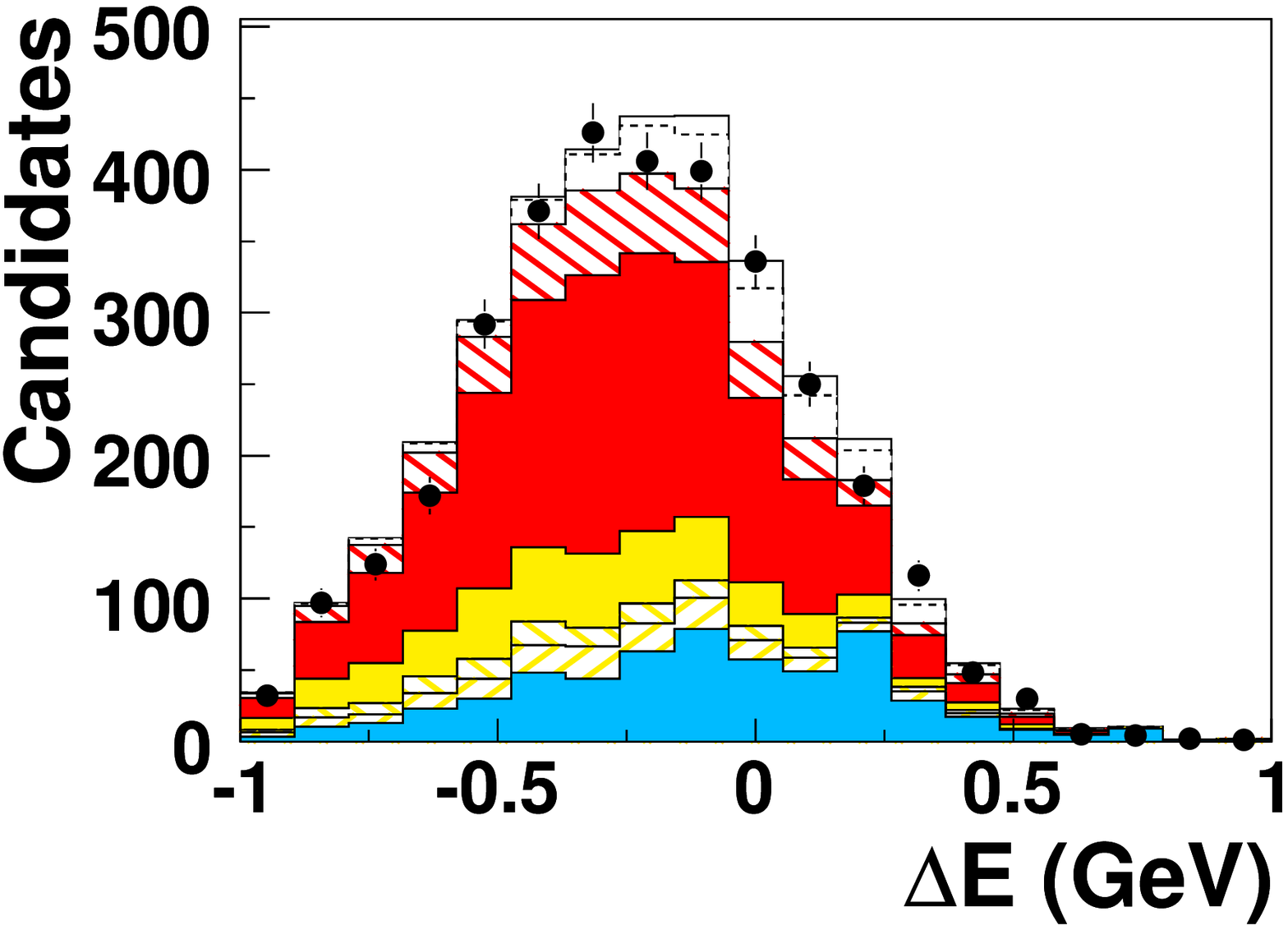, width = 6cm}\\
    \end{minipage}

  \end{tabular}
  \caption{(color online)
\mES and \DeltaE distributions in each $q^2$~bin for 
           \Bppizlnu after the fit.
           The distributions are shown in the \DeltaE and \mES signal bands,
           respectively.
	   Legend: see Figure~\ref{fig:legend}.}
  \label{fig:DeltaE_pi0lnu_fit}
\end{figure*}

\begin{figure*}    
\centering
  \begin{tabular}{ccc}
    \begin{minipage}{0.33\linewidth}
      \epsfig{file=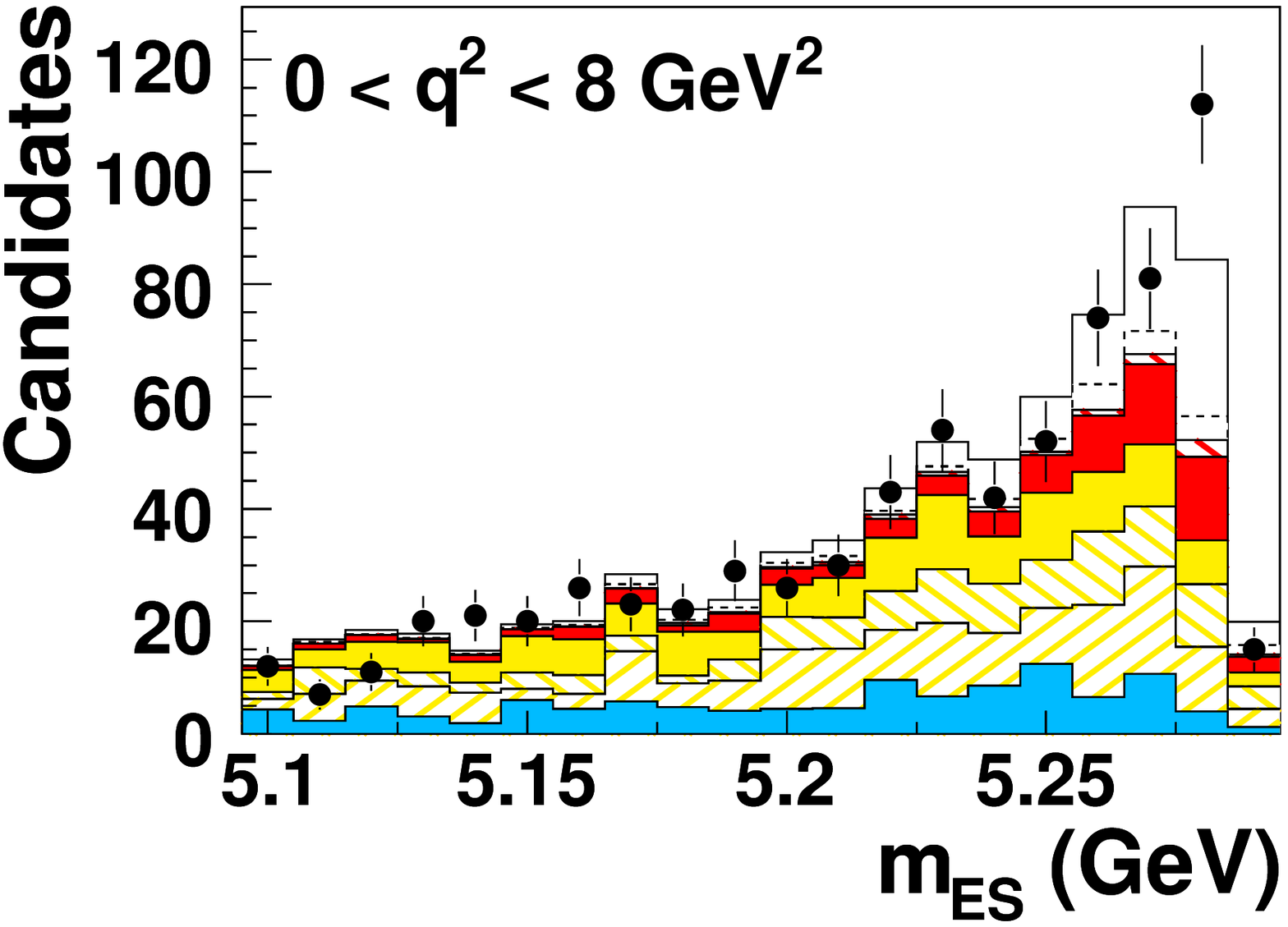, width = 6cm}
      \epsfig{file=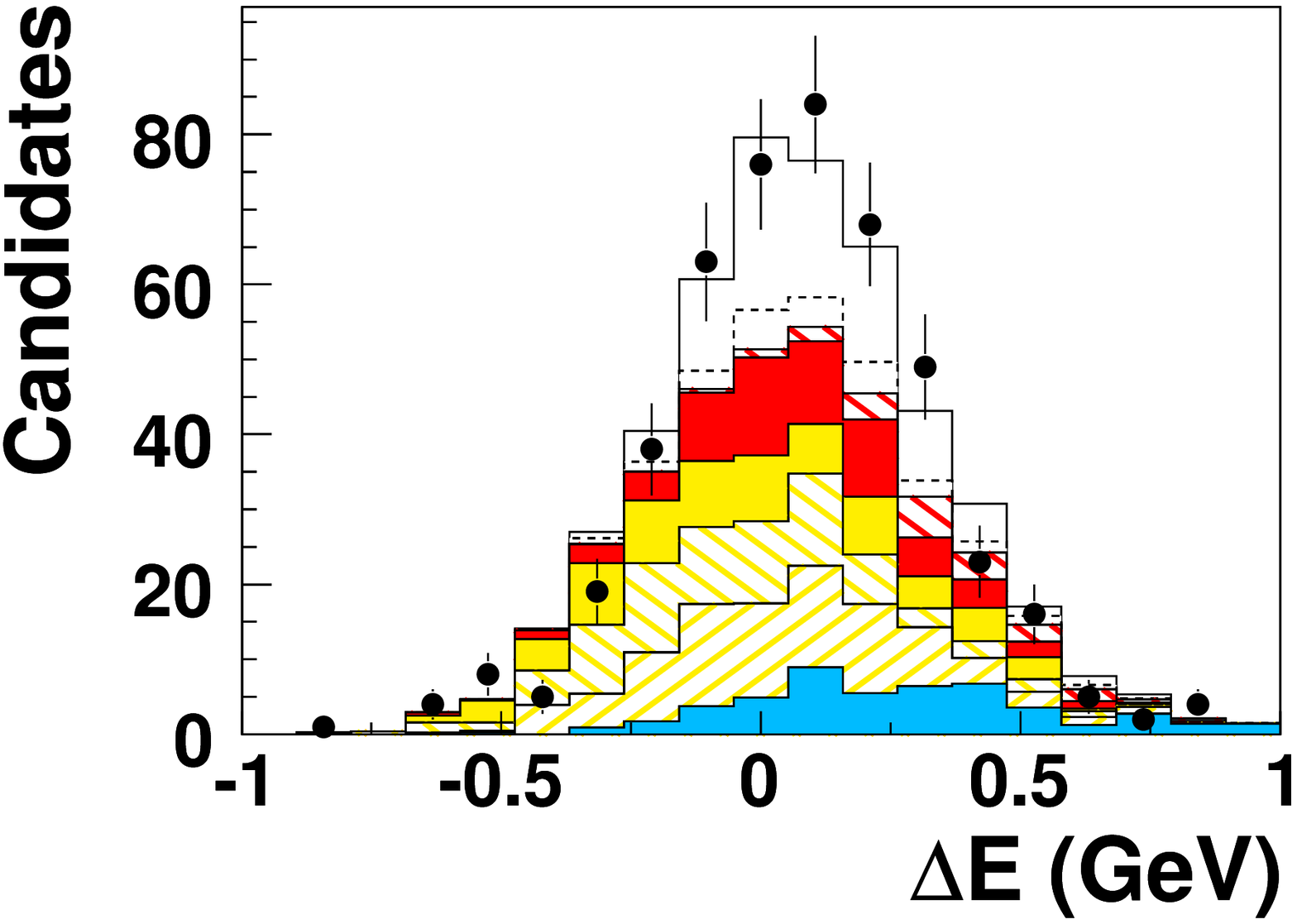, width = 6cm}
    \end{minipage}
    \begin{minipage}{0.33\linewidth}
      \epsfig{file=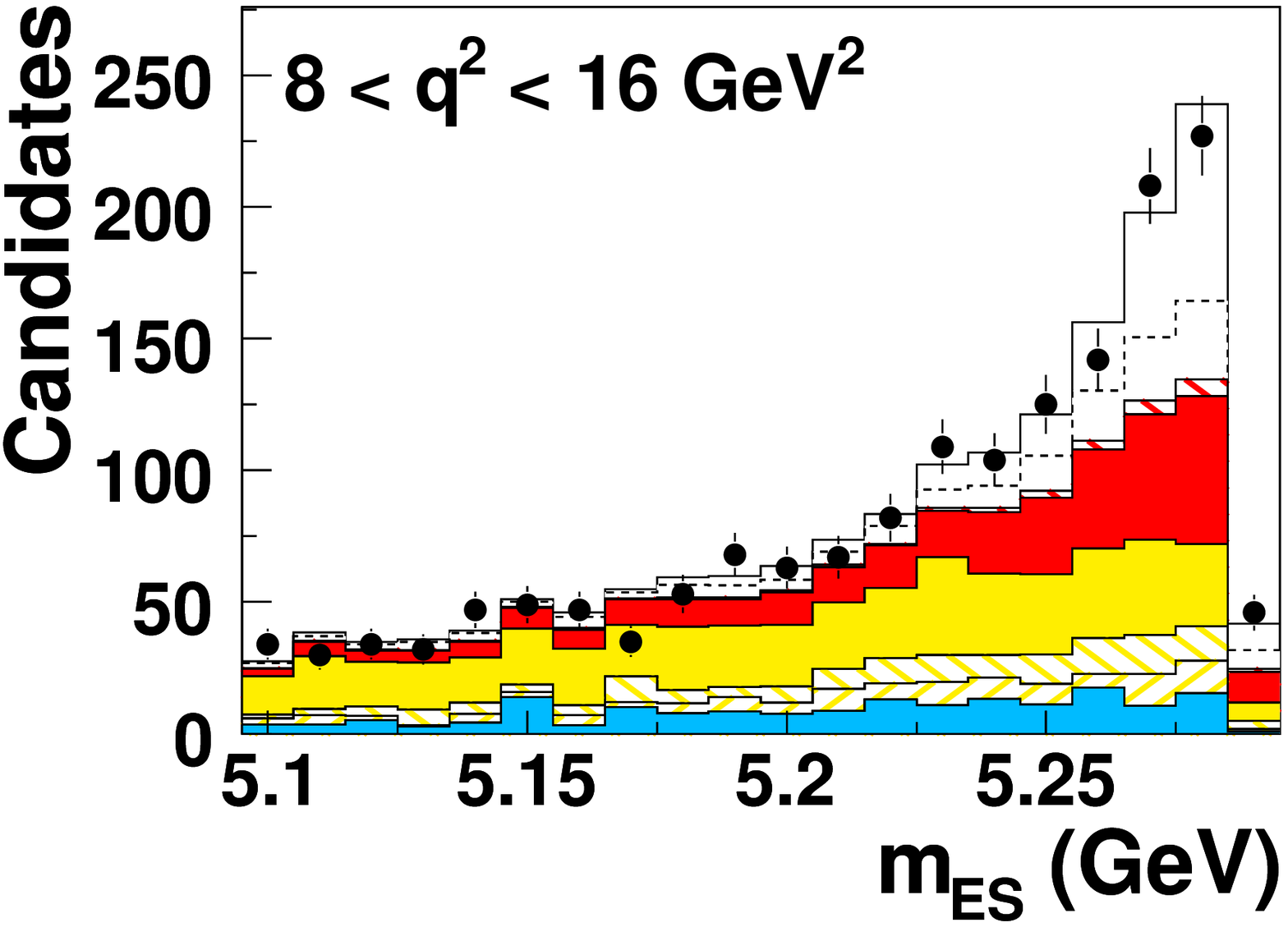, width = 6cm}
      \epsfig{file=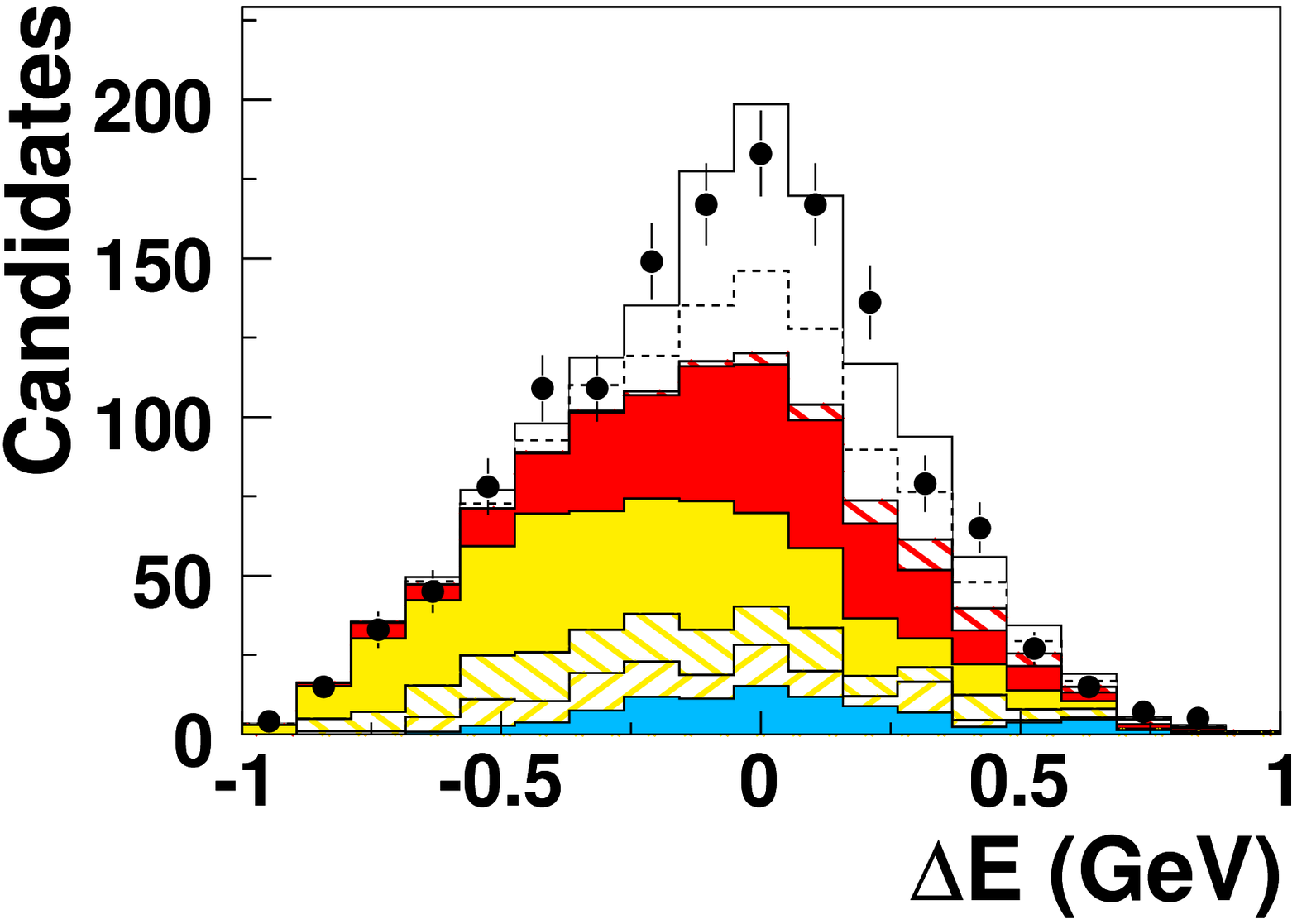, width = 6cm}
    \end{minipage}
    \begin{minipage}{0.33\linewidth}
      \epsfig{file=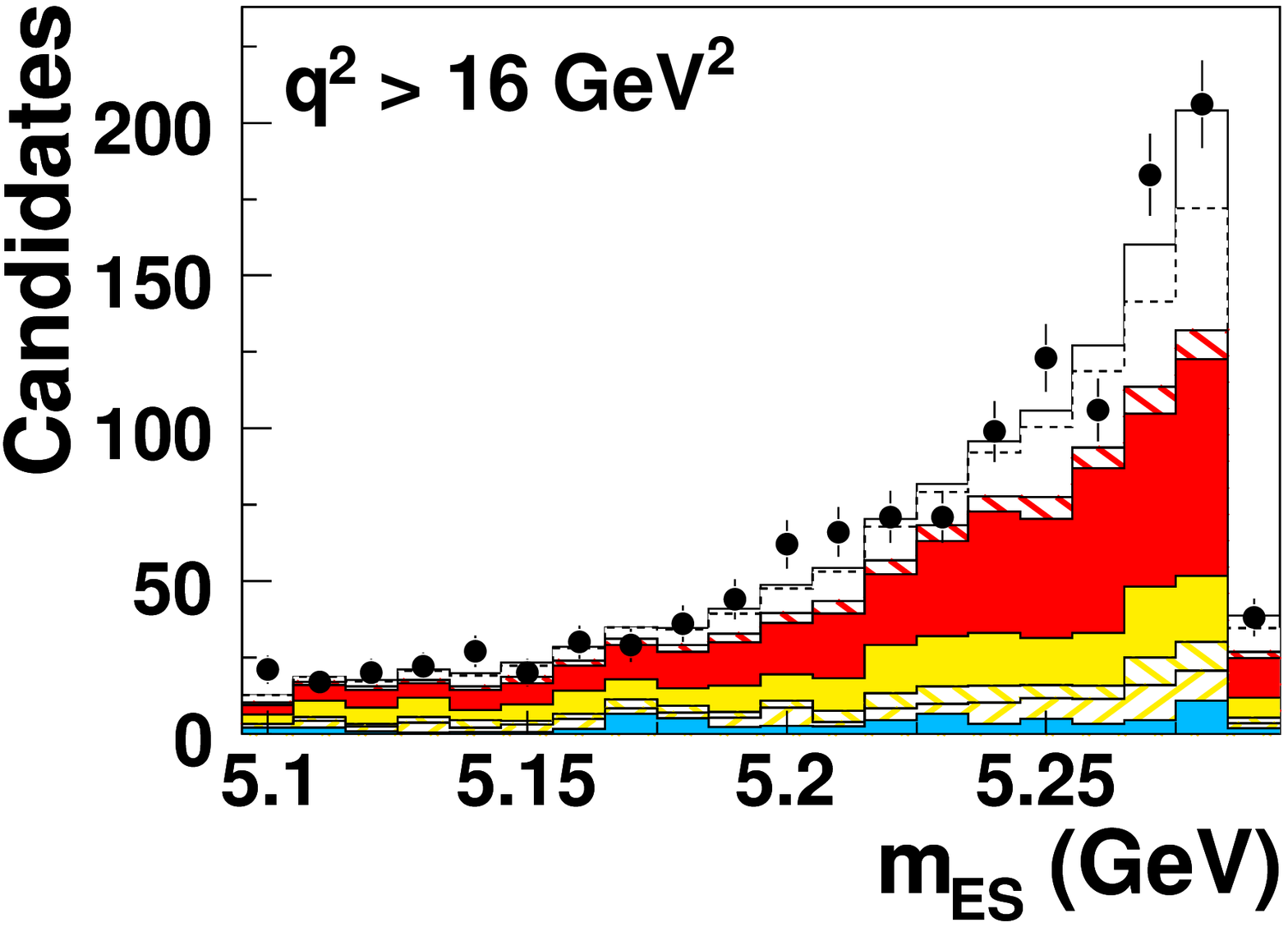, width = 6cm}
      \epsfig{file=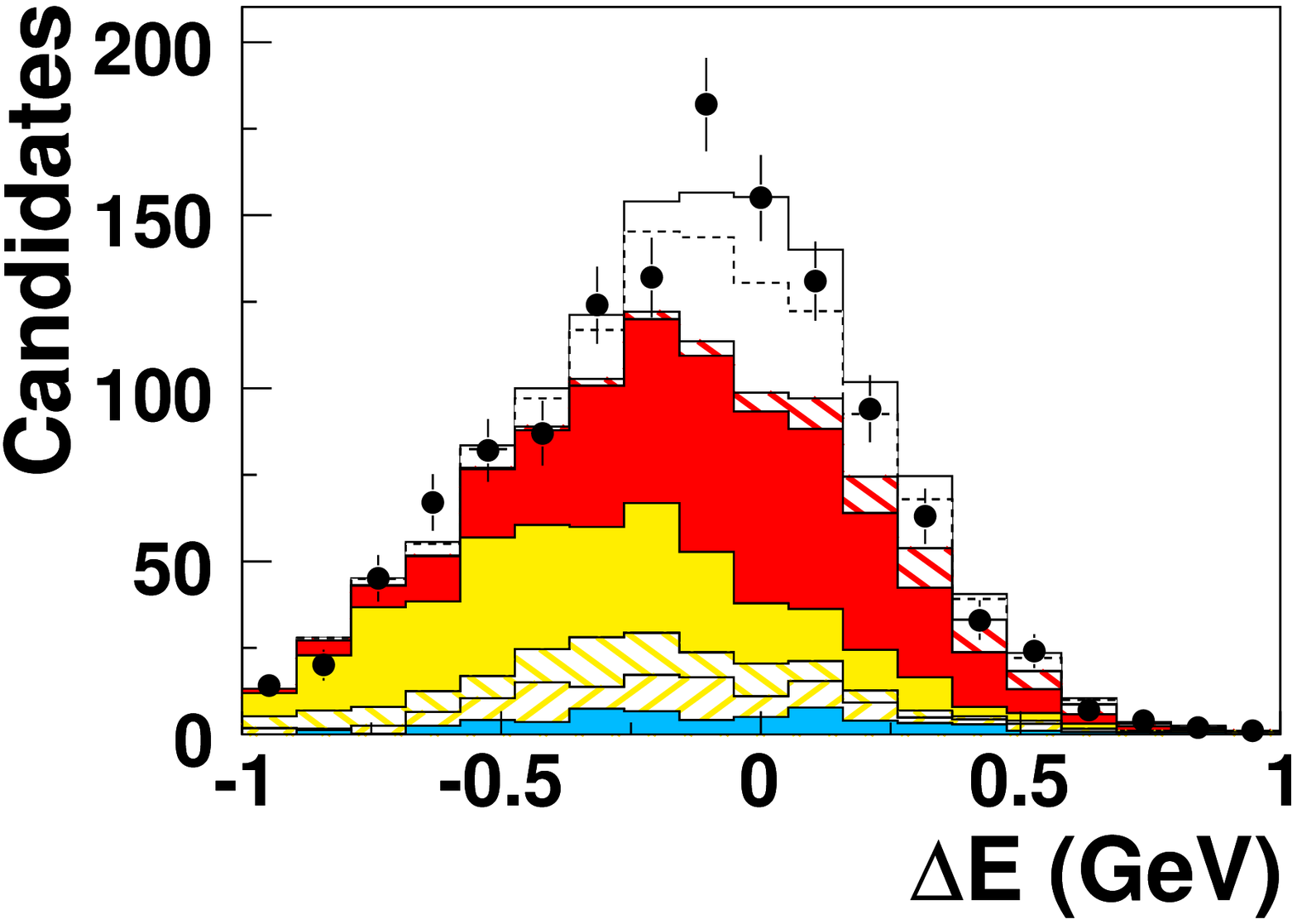, width = 6cm}
    \end{minipage}
  \end{tabular}
  \caption{(color online)
\mES and \DeltaE distributions in each $q^2$~bin for 
           \Bzrholnu after the fit.
           The distributions are shown in the \DeltaE and \mES signal bands,
           respectively.
	   Legend: see Figure~\ref{fig:legend}.}
  \label{fig:DeltaE_rholnu_fit}
\end{figure*}

\begin{figure*}    
\centering
  \begin{tabular}{ccc}
    \begin{minipage}{0.33\linewidth}
      \epsfig{file=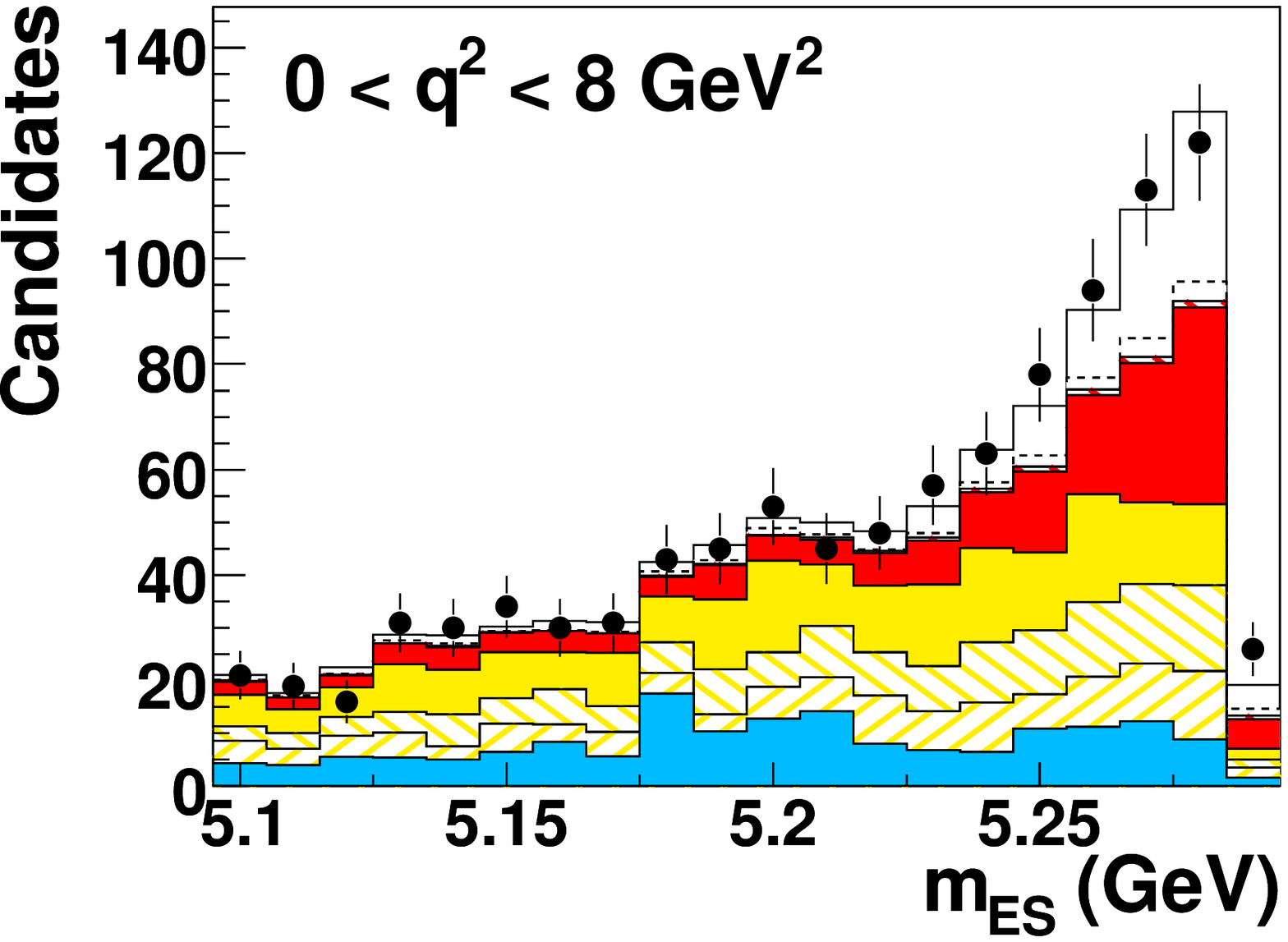, width = 6cm}
      \epsfig{file=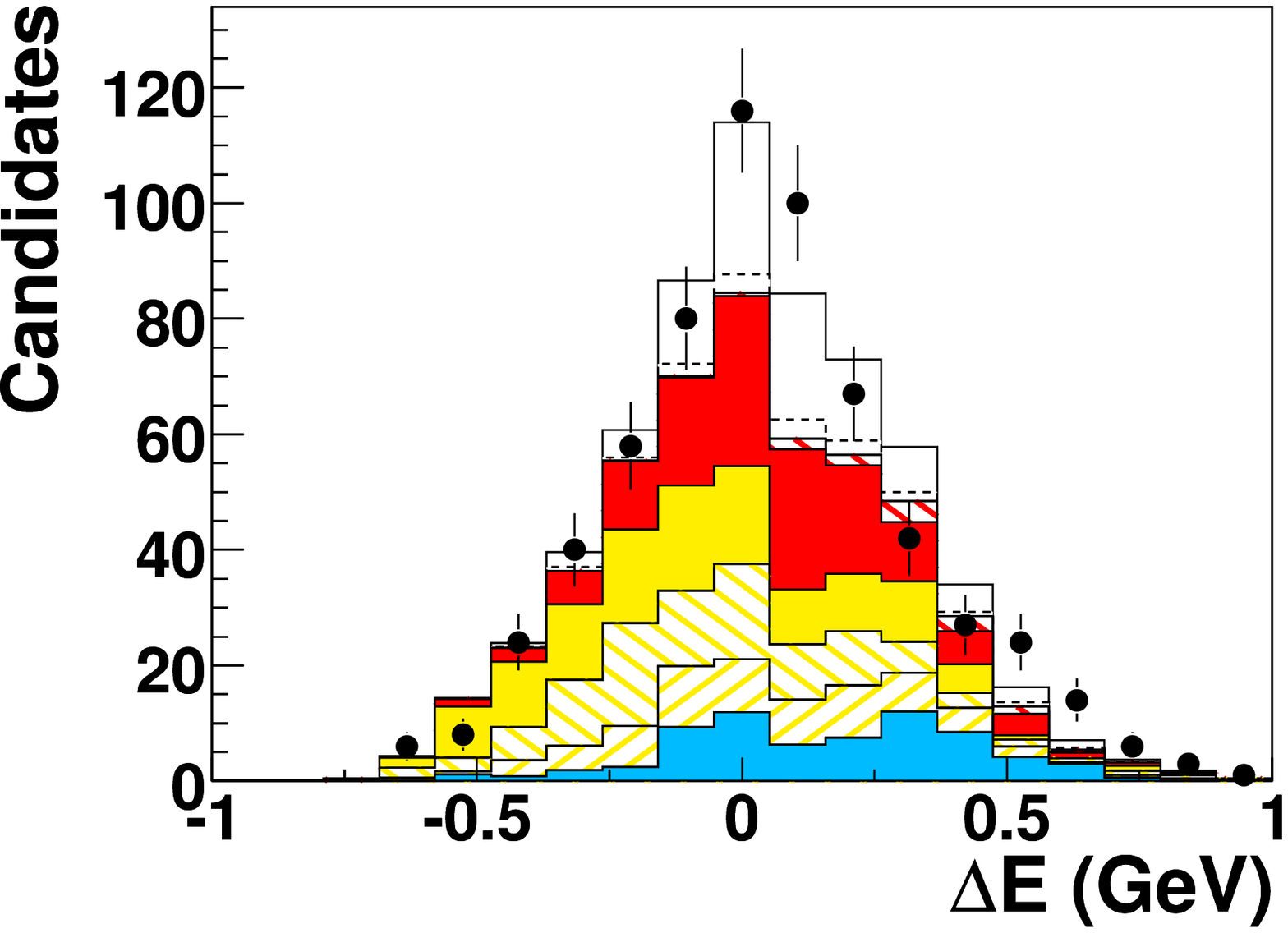, width = 6cm}
    \end{minipage}
    \begin{minipage}{0.33\linewidth}
      \epsfig{file=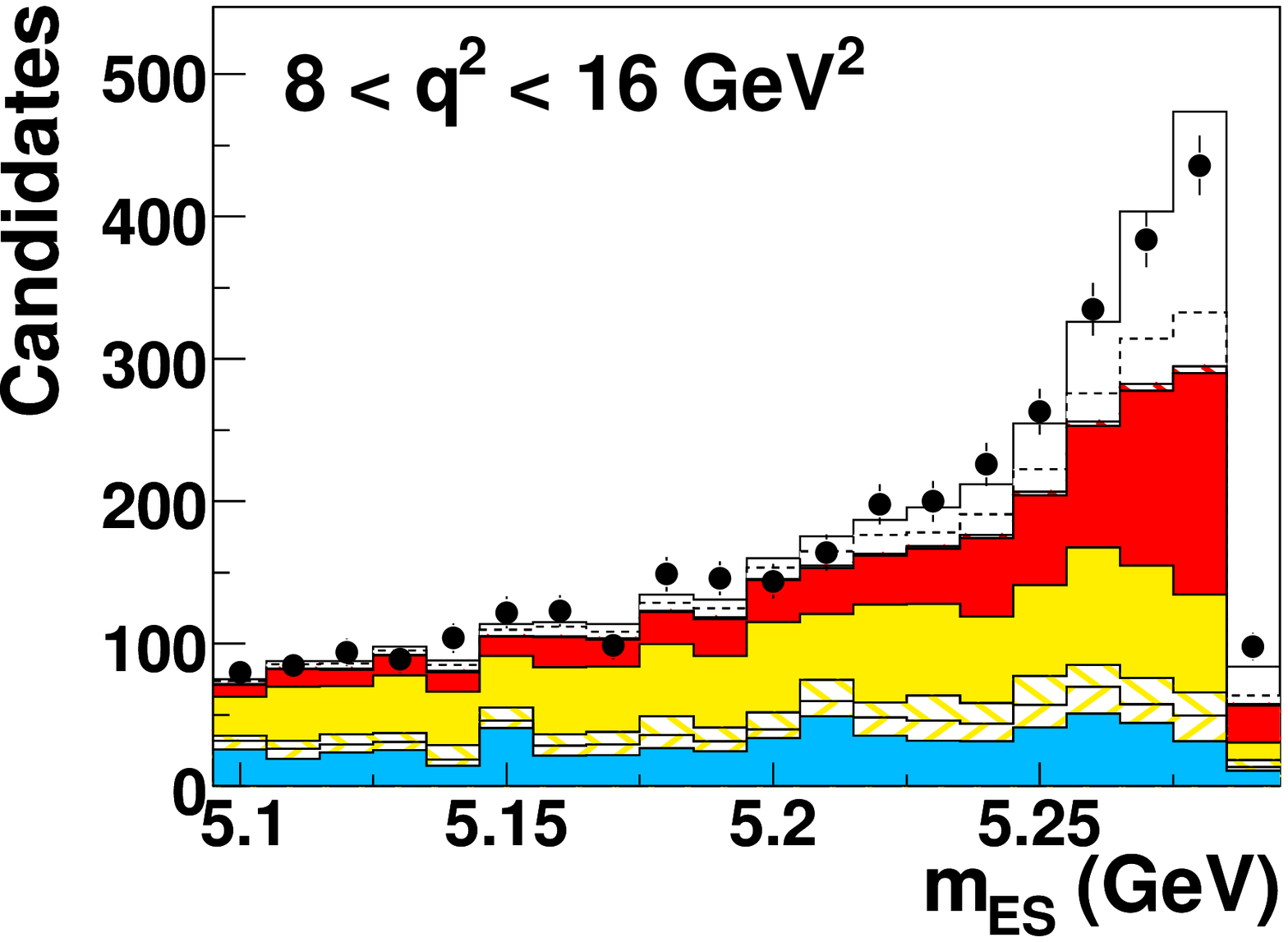, width = 6cm}
      \epsfig{file=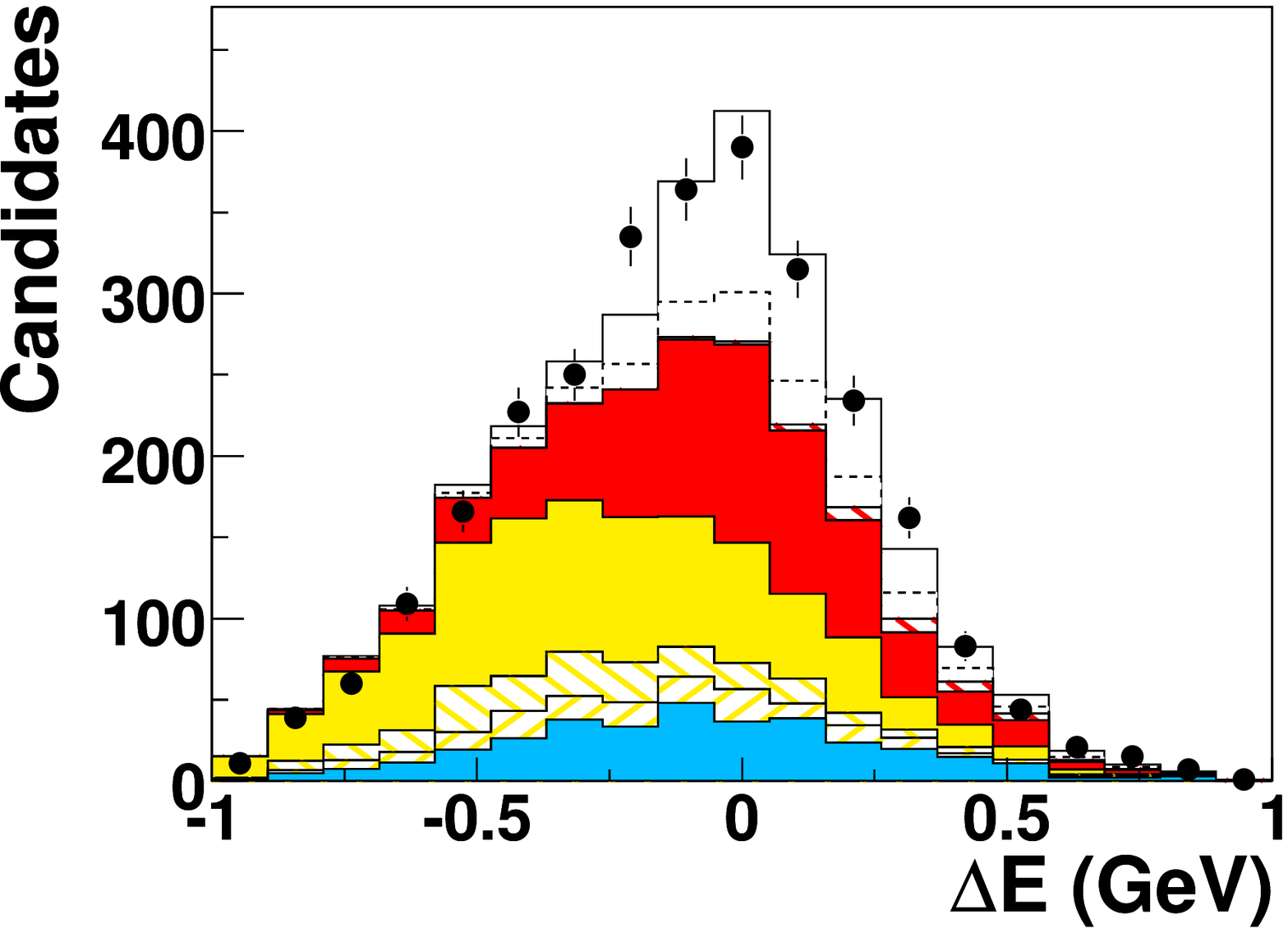, width = 6cm}
    \end{minipage}
    \begin{minipage}{0.33\linewidth}
      \epsfig{file=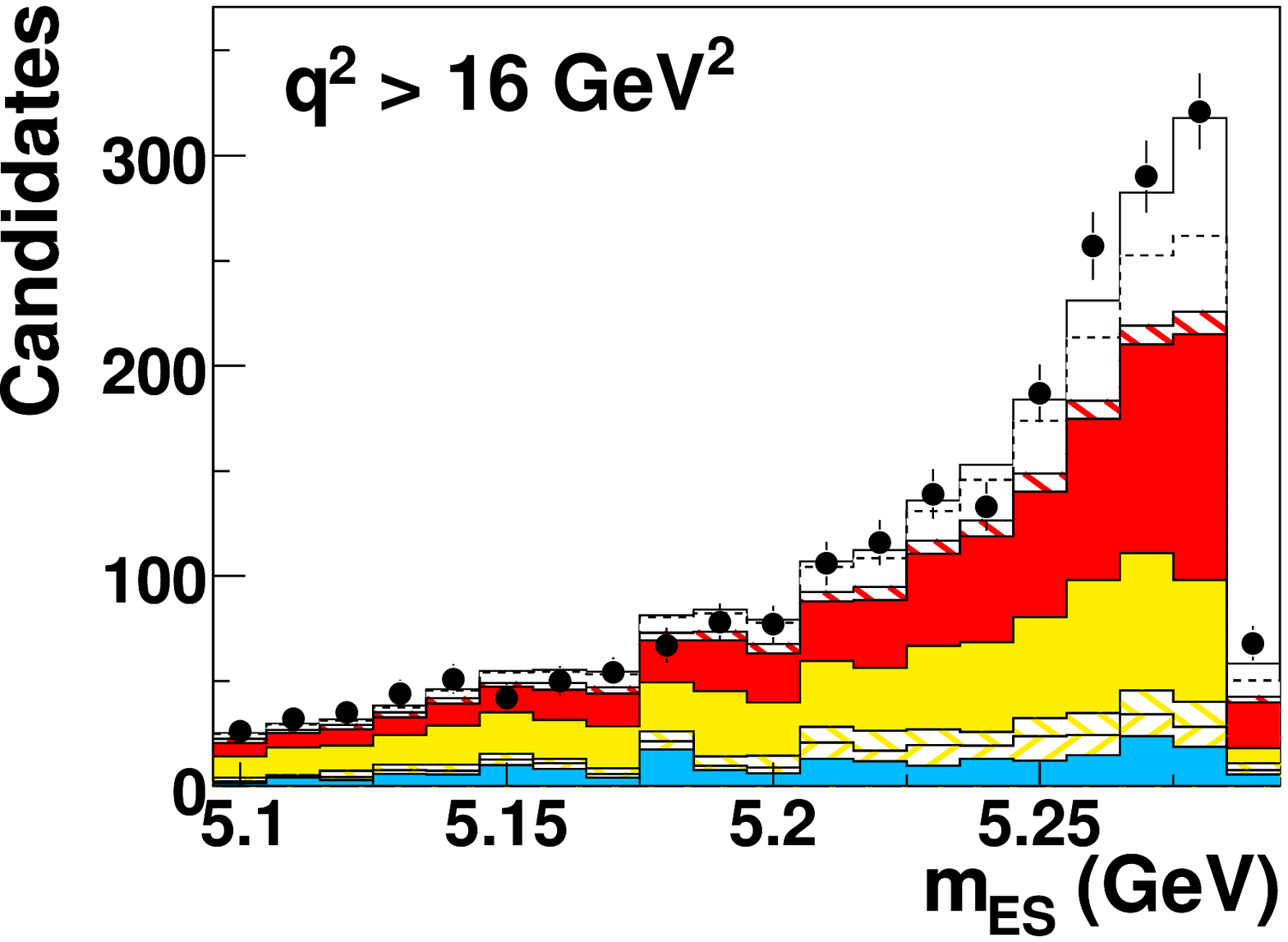, width = 6cm}
      \epsfig{file=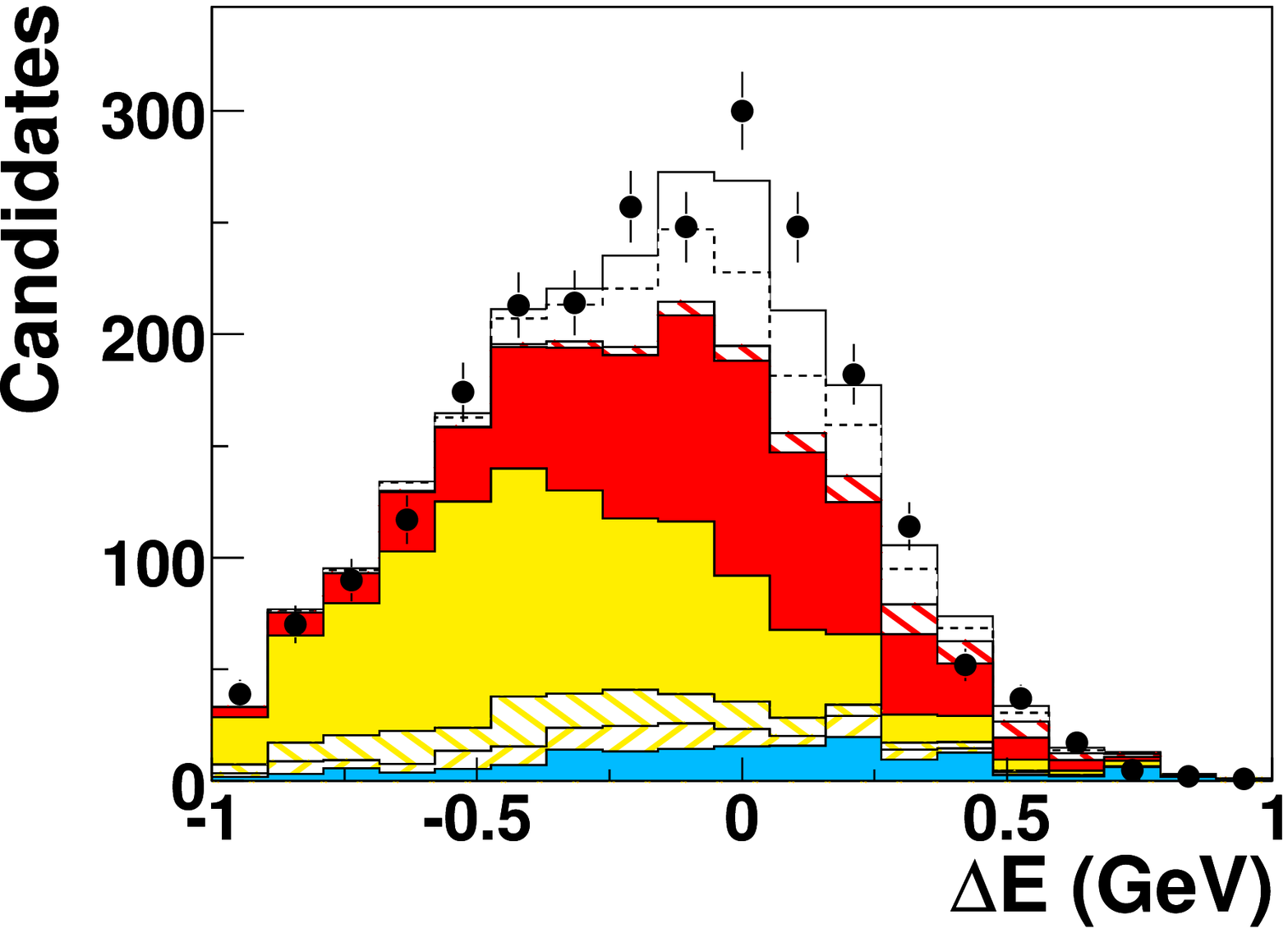, width = 6cm}
    \end{minipage}
  \end{tabular}
  \caption{(color online)
\mES and \DeltaE distributions in each $q^2$~bin for 
           \Bprhozlnu after the fit.
           The distributions are shown in the \DeltaE and \mES signal bands,
           respectively.
	   Legend: see Figure~\ref{fig:legend}.}
  \label{fig:DeltaE_rho0lnu_fit}
\end{figure*}

The fits are performed both separately and simultaneously for the four signal decay modes, \Bzpilnu, \Bppizlnu, \Bzrholnu, and \Bprhozlnu. 
Figures~\ref{fig:DeltaE_pilnu_fit}--\ref{fig:DeltaE_rho0lnu_fit} show projections of the
fitted $\DeltaE - \mES$ distributions in the signal bands for these decays, separately for each bin in $\q^2_{\rm corr}$. As a measure of the goodness-of-fit we use $\chi^2$ per degree of freedom; all fits have values in the range $1.05 - 1.11$ (for details see Table \ref{tab:signalYields}).

The scale factors for the signal contributions, which are determined by the fits, can be translated to numbers of background-subtracted signal events for the four signal decays.
These signal yields are listed in Table \ref{tab:signalYields} with errors
that are a combination of the statistical uncertainties of the data and MC samples and the uncertainties of the fitted yields of the various backgrounds.
For each signal decay mode, the table specifies the number of true and combinatorial signal decays. Their relative fraction is taken from simulation. This fraction is larger for decays with a \piz\ in the final state.   For all signal
modes, the fraction of combinatorial signal events is small at low $q^2$, 
increases with $q^2$, and at the highest $q^2$  it is similar to or exceeds the one of true signal decays. This leads to 
larger errors in the measurement of $q^2$, $\mES$ and $\DeltaE$.

In Table~\ref{tab:fullCorrMat} in Appendix~\ref{sec:appendix_corrMat} the correlation matrix of the four-mode fit is presented.
We observe correlations of about $40-60\%$ between
the \qqbar\ and the other \BB backgrounds and 
between the $B \to D^* \ell \nu$ and the other \BB backgrounds
for all signal modes.  For \Bpilnu, the correlation between the  
\bulnu\ background and the signal at high $q^2$ is also sizable ($\simeq 60\%$).
For \Brholnu, this correlation is larger than $90\%$, which is 
why we choose to fix the \bulnu\ background normalization
for these two samples.
As a test, we let the $B \to X_u \ell\nu$ background normalization in the $B\to\rho\ell\nu$ modes vary as free
parameter in the four-mode fit.
This results in a $B \to X_u \ell\nu$ contribution that is lower by a factor of $0.85 \pm 0.15$ for \Bzrholnu\
and $0.90 \pm 0.14$ for \Bprhozlnu\ and an increase of the $B \to \rho \ell \nu$ signal yields 
by $10\%$ in the first two $q^2$ bins and by $15\%$ in the last $q^2$ bin.
These changes are covered by the systematic uncertainties due to the \bulnu\ background 
stated in Section~\ref{sec:systematics}.

To cross-check the results of the nominal four-mode fit, we also perform 
fits for each signal mode separately. 
The contributions from the 
other signal decay modes are fixed to the result obtained from the four-mode fit. 
Since the shape of the $\pi \leftrightarrow \rho$ cross-feed contribution is very similar to the other \bulnu\ background, 
we fix its normalization to the one obtained from the four-mode fit. 
A comparison of the results of the one-mode fits with the combined four-mode fit
shows agreement within the fit errors of the \Bzpilnu\ and \Bppizlnu\ modes and 
the \Bzrholnu\ and \Bprhozlnu\ modes in all $q^2$ bins.

The partial branching fractions for the different  $\q^2_{\rm corr}$ bins
are derived as the products of the fitted signal scale factors and 
the signal decay branching fractions used in the simulation. The total branching
fraction integrated over the entire $q^2$ range and its error are calculated 
as the sum of all partial branching fractions,
taking into account the correlations of the fitted yields in different $\q^2$ bins.
The branching fraction for $\Bz$ decays, ${\cal B}^0_{\rm signal}$, is related to the fitted signal yields, $N^0_{\rm signal}$, in the following way,
\begin{equation}
\label{eq:BFformula}
{\cal B}^0_{\rm signal} = \frac{N^0_{\rm signal}}{4 \times \epsilon^0_{\rm signal} f_{00}  N_{\BB}} \ ,
\end{equation}
where $f_{00}=0.484 \pm 0.006$~\cite{HFAG2009} denotes the fraction of $B^0\bar{B^0}$ events
produced in \FourS\ decays and $\epsilon^0_{\rm signal}$ is the total signal efficiency (averaged over the electron and muon samples) as predicted by the MC simulation. The factor of four accounts for the 
fact that each event contains two $B$ mesons, and that the branching fraction is quoted 
for a single charged lepton, not for the sum of the decays to electrons or muons. 
The branching fraction results are presented in Section~\ref{sec:results}.

\begin{table*}
\centering
\caption{Results of fits performed separately for each of the four signal decay modes and simultaneously for all four decay modes in bins of $q^2$: $\chi^2$ per degree of freedom, 
signal yields for true signal decays, $N^{\rm sig}$, and combinatorial signal, $N^{\rm comb}$.  The stated errors are the fit errors.}

\renewcommand{\arraystretch}{1.6}
\begin{tabular}{lcccccccc} 
\hline \hline

$q^2$ range ($\gev^2$) & $\chi^2/ndf$  & 0-4          & 4-8          & 8-12         & 12-16        & 16-20        & $>$ 20       & $0<q^2<26.4$    \
       \\ \hline
$N^{sig}_{\pi^\pm}$           &259/268& $ 701\pm 64$ &$1950\pm  104$ &$1552\pm  113 $ &$1184\pm   104 $&$ 732\pm   80$&$ 541\pm   68$&$ 6660 \pm  2\
78$               \\
$N^{comb}_{\pi^\pm}$           &      & $   1\pm 0.1 $&$   4\pm 0.2 $&$   9\pm    1$&$  30\pm    3$&$  77\pm    8$&$ 401\pm   51$&$  521 \pm   22$ \
              \\
$N_{\pi^\pm}=N^{sig}_{\pi^\pm}+N^{comb}_{\pi^\pm}$
                               &      & $702\pm 64$&$1954\pm  104 $&$1561\pm 113$&$1214\pm 104$&$809\pm   80$&$  942\pm    85$&$ 7181 \pm  279$    \
           \\  \hline
$N^{sig}_{\pi^0}$             &237/268& $ 315\pm   42$&$ 576\pm   54$&$ 904\pm  107$&$ 471\pm   68$&$ 414\pm   83$&$ 159\pm   55$&$ 2840 \pm  203$ \
              \\
$N^{comb}_{\pi^0}$             &      & $   5\pm 0.7 $&$  13\pm 1   $&$  36\pm    4$&$  36\pm    5$&$ 119\pm   24$&$ 397\pm  137$&$  606 \pm   43$ \
              \\
$N_{\pi^0}=N^{sig}_{\pi^0}+N^{comb}_{\pi^0}$
                               &    & $320\pm   42$&$589\pm  54$&$940\pm 107$&$507\pm  68$&$533\pm  86$&$  556\pm  148$&$ 3446 \pm  208$           \
    \\  \hline
$N_{\pi^\pm}+N_{\pi^0}$ 4-mode &799/819& $1012\pm   76$&$2535\pm  128$&$2485\pm 157$&$1729\pm  130$&$1291\pm  125$&$ 1552\pm  180$&$ 10604 \pm  376\
$               \\  \hline\hline

$q^2$ range ($\gev^2$)          &       &\multicolumn{2}{c}{0-8}           &\multicolumn{2}{c}{8-16}           &\multicolumn{2}{c}{$>$16}          \
& $0<q^2<20.3$      \\ \hline
$N^{sig}_{\rho^\pm}$            &147/131&\multicolumn{2}{c}{$ 237\pm   56$}&\multicolumn{2}{c}{$ 459\pm   44$} &\multicolumn{2}{c}{$ 170\pm   17$} \
&$  866\pm  101$ \\
$N^{comb}_{\rho^\pm}$           &       &\multicolumn{2}{c}{$  56\pm   13$}&\multicolumn{2}{c}{$ 287\pm   27$} &\multicolumn{2}{c}{$ 368\pm   38$} \
&$  711\pm   82$ \\
$N_{\rho^\pm}=N^{sig}_{\rho^\pm}+N^{comb}_{\rho^\pm}$
                                &       &\multicolumn{2}{c}{$ 293\pm   57$}&\multicolumn{2}{c}{$746\pm  52$}  &\multicolumn{2}{c}{$ 538\pm   42$} &\
$ 1577\pm  130$ \\ \hline
$N^{sig}_{\rho^0}$             &162/131&\multicolumn{2}{c}{$ 253\pm   63$}&\multicolumn{2}{c}{$ 856\pm  74$} &\multicolumn{2}{c}{$ 294\pm   29$} &$\
 1403\pm  143$ \\
$N^{comb}_{\rho^0}$            &       &\multicolumn{2}{c}{$  31\pm    8$}&\multicolumn{2}{c}{$ 267\pm   23$}&\multicolumn{2}{c}{$ 270\pm   27$} &$\
  567\pm   58$ \\
$N_{\rho^0}=N^{sig}_{\rho^0}+N^{comb}_{\rho^0}$
                               &       &\multicolumn{2}{c}{$ 284\pm   64$}&\multicolumn{2}{c}{$1123\pm 77$}  &\multicolumn{2}{c}{$  564\pm  40$} &$\
 1970\pm  154$ \\ \hline
$N_{\rho^\pm}+N_{\rho^0}$  4-mode &799/819&\multicolumn{2}{c}{$ 471\pm  101$}&\multicolumn{2}{c}{$ 1754\pm  120$} &\multicolumn{2}{c}{$1105\pm   86\
$} &$ 3332\pm  286$ \\ 

\hline\hline
\end{tabular}
\label{tab:signalYields}
\end{table*}

\subsection{Fit Validation and Consistency}
\label{sec:fitvalidation}

The fit procedure is validated several ways.
First of all, the implementation of the Barlow-Beeston fit technique allowing statistical fluctuations of the MC distributions to be incorporated is checked by verifying the consistency of the fit variations with the statistical error of the input distributions.
Secondly, a large number of simulated experiments are generated based on random samples drawn from the three-dimensional histograms used in the standard fit. 
Specifically, we  create 500 sets of distributions 
by fluctuating each simulated source distribution bin-by-bin using Poisson statistics. For each of the sets, we add the source distributions to make up to total distribution that corresponds to the data distribution (``toy data''), which are then fitted by the standard procedure.  In addition, we create independent fluctuations for the distributions that make up the source PDFs for the fit, in the same way as for the toy data described above. 
For a compilation of these 500 ``toy experiments'', we study the distributions of the deviation of the fit result
from the input value divided by the fit error. These distributions show no significant bias
for any of the free parameters and confirm that the errors are correctly estimated.

Additional fits are performed to check the consistency of the data. For instance, the data samples are divided into subsamples, {\it i.e.}, the electron sample separated from the muon sample or the data separated into different run periods.  These subsamples are fitted separately; the results agree within the statistical uncertainties. 
\section{Systematic Uncertainties}
\label{sec:systematics}

Many sources of systematic uncertainties have been assessed for the measurement of the exclusive branching fractions as a function of $q^2$.
Since this analysis does not depend only on the reconstruction of the charged lepton and hadron from the signal decay mode, but also on the measurement of all remaining tracks and photons in the event, 
the uncertainties in the detection efficiencies of all particles as well as 
the uncertainties in the background yields and shapes enter into
the systematic errors.

Tables \ref{tab:syst_pi_allModes} and \ref{tab:syst_rho_allModes}
summarize the systematic uncertainties for \Bpilnu\ and \Brholnu for the four-mode fit.
In Appendix~\ref{sec:appendix_systTable} the systematic error tables for the one-mode fits are presented.
The individual sources are, to a good approximation, uncorrelated and can therefore be added in quadrature to obtain the total systematic errors
for each decay mode.
In the following, we discuss the assessment of the systematic uncertainties in detail.

\begin{table}[hbt]
\centering
\caption{
Systematic errors in \%  for \BRBzpilnu\ from the four-mode fit for bins in $q^2$ and the total $q^2$ range.  The total errors are derived from the individual contributions taking into account the complete covariance matrix.
}
\begin{tabular}{lccccccc} \hline\hline

\multicolumn{8}{c}{\boldmath \Bpilnu}\\ \hline
$q^2$ range ($\gev^2$)  & 0-4 & 4-8 & 8-12 &12-16&16-20&$>$20& 0-26.4  \\ \hline  

Track efficiency        & 3.4  &  1.5  &  2.3  &  0.1  &  1.5  &  2.8  & 1.9   \\
Photon efficiency       & 0.1  &  1.4  &  1.0  &  4.6  &  2.8  &  0.3  & 1.8   \\ \hline

Lepton identification   & 3.8  &  1.6  &  1.9  &  1.8  &  1.9  &  3.0  & 1.8   \\ \hline

$K_L$ efficiency        & 1.0  &  0.1  &  0.5  &  4.5  &  0.4  &  2.0  & 1.4  \\
$K_L$ shower energy     & 0.1  &  0.1  &  0.1  &  0.8  &  0.9  &  3.8  & 0.7  \\
$K_L$ spectrum          & 1.6  &  1.9  &  2.2  &  3.1  &  4.4  &  2.3  & 2.5   \\ \hline

$\Bpilnu FF$ $f_+$      & 0.5  &  0.5  &  0.5  &  0.6  &  1.0  &  1.0  & 0.6  \\
$\Brholnu FF A_1$       & 1.7  &  1.2  &  3.4  &  2.0  &  0.1  &  1.6  & 1.7  \\
$\Brholnu FF A_2$       & 1.3  &  0.8  &  2.6  &  1.0  &  0.1  &  0.4  & 1.1  \\
$\Brholnu FF V$         & 0.2  &  0.3  &  0.9  &  0.7  &  0.1  &  0.5  & 0.5   \\ \hline

$\BRBpomlnu$            & 0.1  &  0.1  &  0.1  &  0.2  &  0.3  &  1.5  & 0.2  \\
$\BRBpetalnu$           & 0.1  &  0.1  &  0.2  &  0.2  &  0.2  &  0.5  & 0.2   \\
$\BRBpetaplnu$          & 0.1  &  0.1  &  0.1  &  0.1  &  0.1  &  0.3  & 0.1  \\
$\BRXulnu$              & 0.2  &  0.1  &  0.1  &  0.1  &  1.1  &  1.6  & 0.4  \\
$\BXulnu$ SF param.     & 0.4  &  0.1  &  0.2  &  0.2  &  0.5  &  4.2  & 0.7  \\ \hline 

$\B \to D\ell\nu$ FF $\rho^{2}_{D}$    & 0.2  &  0.1  &  0.5  &  0.3  &  0.2  &  0.7  & 0.3      \\
$\B \to D^*\ell\nu$ FF $R_1$           & 0.1  &  0.4  &  0.8  &  0.6  &  0.3  &  0.6  & 0.5    \\
$\B \to D^*\ell\nu$ FF $R_2$           & 0.5  &  0.2  &  0.1  &  0.2  &  0.1  &  0.4  & 0.2    \\
$\B \to D^*\ell\nu$ FF $\rho^{2}_{D^*}$& 0.7  &  0.2  &  0.6  &  0.8  &  0.4  &  1.1  & 0.6     \\

$\BRDlnu $              & 0.2  &  0.2  &  0.3  &  0.4  &  0.5  &  0.5  & 0.3   \\
$\BRDslnu $             & 0.4  &  0.1  &  0.3  &  0.3  &  0.3  &  0.7  & 0.3   \\
$\BRDssnlnu $           & 0.4  &  0.1  &  0.1  &  0.3  &  0.1  &  0.5  & 0.2    \\
$\BRDssblnu $           & 0.1  &  0.1  &  0.1  &  0.5  &  0.1  &  0.2  & 0.2   \\
Secondary leptons       & 0.5  &  0.2  &  0.3  &  0.2  &  0.2  &  0.7  & 0.3   \\ \hline

Continuum               & 5.3  &  1.0  &  2.6  &  1.8  &  3.1  &  6.1  & 2.0 \\ \hline

Bremsstrahlung          & 0.3  &  0.1  &  0.1  &  0.1  &  0.1  &  0.4  & 0.2   \\
Radiative corrections   & 0.5  &  0.1  &  0.1  &  0.2  &  0.2  &  0.6  & 0.3   \\ \hline

$N_{\BB}$               & 1.2  &  1.0  &  1.2  &  1.2  &  1.1  &  1.6  & 1.2  \\
$B$ lifetimes           & 0.3  &  0.3  &  0.3  &  0.3  &  0.3  &  0.7  & 0.3  \\
$f_\pm / f_{00}$        & 1.0  &  0.4  &  0.8  &  0.8  &  0.5  &  1.3  & 0.8  \\  \hline

Total syst. error       & 8.2  &  3.9  &  6.7  &  8.3  &  6.9  &  10.6 & 5.0  \\ \hline\hline

\end{tabular}
\label{tab:syst_pi_allModes}
\end{table}
\begin{table}[hbt]
\centering
\caption{
Systematic errors in \%  for \BRBzrholnu\ from the four-mode fit for three bins in $q^2$ and the total $q^2$ range.  The total errors are derived from the individual contributions taking into account the complete covariance matrix.
}
\begin{tabular}{lcccc} \hline\hline

\multicolumn{5}{c}{\boldmath \Brholnu} \\ \hline
$q^2$ range ($\gev^2$)  & 0-8 & 8-16 & $>$16   & 0-20.3  \\ \hline  

Track efficiency        & 3.2  &  2.9  &  0.3  & 2.5   \\
Photon efficiency       & 2.6  &  2.0  &  2.6  & 2.4   \\ \hline
Lepton Identification   & 5.7  &  3.0  &  4.0  & 3.4   \\ \hline 

$K_L$ efficiency	& 10.3 &  1.2  &  4.9  & 4.8   \\
$K_L$ shower energy     & 1.6  &  0.8  &  1.0  & 1.1   \\
$K_L$ spectrum          & 4.2  &  6.1  &  7.0  & 5.7    \\ \hline

$\Bpilnu$  FF $f_+$     & 0.1  &  0.1  &  0.7  & 0.2   \\ 
$\Brholnu$ FF $A_1$      & 10.7 &  6.6  &  4.5  & 7.5   \\ 
$\Brholnu$ FF $A_2$      & 8.5  &  3.8  &  0.8  & 4.7   \\ 
$\Brholnu$ FF $V$       & 3.4  &  3.0  &  3.6  & 3.2   \\ \hline

$\BRBpomlnu$            & 0.7  &  0.7  &  3.4  & 1.2             \\
$\BRBpetalnu$           & 0.8  &  0.1  &  0.6  & 0.4   \\
$\BRBpetaplnu$          & 0.8  &  0.5  &  1.2  & 0.7   \\
$\BRXulnu$              & 7.4  &  7.3  &  10.6  & 8.0   \\ 
$\BXulnu$ SF param.     & 11.9 &  7.6  &  12.8  & 10.0  \\ \hline

$B\to D \ell\nu$ FF $\rho^2_{D}$    & 0.9  &  0.2  &  0.1  & 0.4   \\
$B\to D^* \ell\nu $FF $R_1$         & 0.7  &  0.1  &  0.3  & 0.3    \\
$B\to D^* \ell\nu $FF $R_2$         & 1.7  &  0.1  &  0.2  & 0.6   \\
$B\to D^* \ell\nu $FF $\rho^2_{D^*}$& 2.0  &  0.2  &  0.1  & 0.7     \\ 

$\BRDlnu $              & 1.6  &  0.3  &  0.1  & 0.7   \\
$\BRDslnu $             & 0.5  &  0.1  &  0.3  & 0.3  \\
$\BRDssnlnu $           & 1.3  &  0.1  &  0.1  & 0.5   \\
$\BRDssblnu $           & 0.7  &  0.1  &  0.1  & 0.3   \\ 

Secondary leptons       & 1.5  &  0.1  &  0.1  & 0.5   \\ \hline
Continuum               & 8.9  &  3.8  &  5.0  & 4.0   \\ \hline

Bremsstrahlung          & 0.9  &  0.1  &  0.2  & 0.4   \\
Radiative corrections   & 1.3  &  0.1  &  0.7  & 0.6   \\ \hline

$N_{\BB}$               & 2.7  &  2.0  &  2.5  & 2.3    \\
$B$ lifetimes           & 1.5  &  0.4  &  0.4  & 0.7    \\
$f_\pm / f_{00}$        & 1.2  &  0.1  &  0.1  & 0.4   \\ \hline 

Total syst. error       & 26.1  &  16.1  &  21.3  & 15.7  \\ \hline\hline

\end{tabular}
\label{tab:syst_rho_allModes}
\end{table}

For the estimation of the systematic errors of the fitted branching fractions,
we compare the differential branching fractions obtained from the nominal fit with results obtained after changes to the MC simulation that reflect the uncertainty in the parameters that impact the detector efficiency and resolution or the simulation of signal and background processes. 
For instance, we vary the tracking efficiency, reprocess the MC samples, 
reapply the fit to the data, and take the difference compared to the results obtained with the nominal MC simulation as an estimate of the systematic error.
The sources of systematic errors are not identical for all four signal decay modes,
and the size of their impact on the event yields depends on the sample composition and $q^2$.

\subsection{Detector Effects}

Uncertainties in the reconstruction efficiencies for charged and neutral particles and
in the rate of tracks and photons from beam background, fake tracks, failures in the matching of EMC clusters to charged tracks, and showers split off from hadronic interactions, undetected $K_L$, and additional neutrinos, all contribute to the quality of the neutrino reconstruction and impact the variables that are used in the preselection and the neural networks.   For all these effects the uncertainties in the efficiencies and resolution have been derived independently from comparisons of data and MC simulation for selected control samples.

\subsubsection{Track, Photon, and Neutral-Pion Reconstruction}

We evaluate the impact of uncertainties in the tracking efficiency by randomly eliminating tracks
with a probability that is given by the uncertainty ranging from 0.25\% to 0.5\% per track,  as measured with data control samples.

Similarly, we evaluate the uncertainty due to photon efficiency by eliminating photons at random with an energy-dependent probability, ranging from 0.7\% per photon above 1\gev to 1.8\% at lower energies. This estimate includes the uncertainty
in the $\pi^0$ efficiency for signal decays with a $\pi^0$, since photons originating from the signal hadron are also eliminated.

\subsubsection{Lepton Identification}

The average uncertainties in the identification of electrons and muons 
have been assessed to be 1.4\% and 3\%, respectively. The
uncertainty in the misidentification of hadrons as electrons or muons is
about 15\%.

\subsubsection{$K^0_L$ Production and Interactions}
\label{sec:klong}

Events containing a $K^0_L$ have a significant impact on the neutrino reconstruction, 
because only a small fraction of the $K^0_L$ energy is deposited in the 
electromagnetic calorimeter. Based on  detailed studies of data control 
samples of $D^0 \to K^0 \pi^+\pi^-$ decays and inclusive $K^0_S$ samples 
in data and MC, corrections to the efficiency, shower deposition and 
the production rates have been derived and applied to the simulation
as a function of the $K^0_L$ momentum and angles (see Section \ref{sec:Detector}).  
To determine the systematic uncertainties in the MC simulations we vary 
the scale factors within their statistical and systematic uncertainties. 
The average uncertainty of the 
energy deposition in the EMC due to $K^0_L$ interactions is estimated 
to be 7.5\%. Above 0.7\gev, the $K^0_L$ detection efficiency is well 
reproduced by the simulation, with an estimated average uncertainty of 2\%.  
At lower momenta, the simulation is corrected to match the data, and 
the uncertainty increases to 25\% below 0.4\gev. 

The production rates for $K^0_S$ in data and MC agree within errors, except for momenta
below 0.4\gev where the data spectrum is low by $22\pm 7\%$ compared to the MC simulation and a correction is applied. To assess the impact of the uncertainty of the correction procedure, the size of the correction is varied by its estimated uncertainty.

\subsection{Simulation of Signal and Background}

\subsubsection{Signal Form Factors}

To assess the impact of the form-factor (FF) uncertainty on the shape of the 
simulated signal distributions, we  vary the $B\to\pi$ form factor within the uncertainty of the previous 
\babar\ measurement~\cite{pilnu_cote} and the $B\to\rho$ form factors within the uncertainties of the 
LCSR calculation assessed by Ball and Zwicky~\cite{ball_rho}. 
For the latter we assume uncertainties on the form factors $A_1$, $A_2$ and $V$ of $10\%$ at $q^2=0$.
They rise linearly to $13\%$ at $q^2=14\gev^2$ and are extrapolated up to the kinematic endpoint.
We add the uncertainties due to the three form factors in quadrature.
For \Btopilnu, the form-factor uncertainty is small, since we extract the signal in six bins of $q^2$.
In contrast, for \Btorholnu the form-factor uncertainty is one of the dominant sources of systematic error.
This is partly due to the stricter requirement on the lepton momentum, $p^*_\ell>1.8\gev$, 
which is imposed to suppress the large \bclnu\ background. 
We refrain from using the difference between LCSR and ISGW2 as systematic
uncertainty, but this difference is comparable to the estimate we obtain 
from the uncertainties in the LCSR calculation. 

\subsubsection{\bulnu\ Background}

The \bulnu\ background contribution is composed of the sum of exclusive decays,
\Bpomegalnu,
\Bpetalnu, and \Bpetaplnu\ decays, and the remaining resonant and non-resonant \bulnu\ 
decays that make up the total \bulnu\ branching fraction.  
We estimate the total error of the \bulnu\ background composition by repeating the fit with branching fractions for various exclusive and non-resonant decays varied independently within their current measurement errors.
The uncertainty of the branching fraction for non-resonant decays is dominant; it is equal to the error on the total \bulnu branching fraction, $\BRXulnu = (2.33 \pm 0.22) \times 10^{-3}$~\cite{HFAG2009}.

In addition, the analysis is sensitive to the mass and composition of the charmless hadronic states.  We assess the uncertainty of the predictions by varying the QCD parameters that define the mass, the lepton spectrum, and the $q^2$ distributions predicted by calculations~\cite{dFN}
based on HQE. We vary the shape-function (SF) parameters $m_b$ and $\mu_\pi^2$ within the uncertainties (error ellipse) given in Ref.~\cite{moments}.

For the two \Btorholnu\ samples, the \bulnu\ background is large compared to the signal
and very difficult to separate.  Consequently, the fit shows very high correlations between the fitted yields for signal and this background.  We therefore choose to fix the background yields and shapes to those provided by the simulation, and account for the uncertainty by assessing the sensitivity of the fitted signal yield to variations of the \bulnu\ branching fraction and the shapes of the background distributions, corresponding to the estimated error of the shape-function parameters. The resulting estimated errors are the two dominant contributions to the systematic errors of the \Btorholnu\ partial and total branching fractions.

\subsubsection{\bclnu Background}

The systematic error related to the shapes of the \bclnu\ background distributions is dominated by the uncertainties in the branching fractions and form factors for the various semileptonic decays.
We vary the composition of the \bclnu\ background based on a compilation  of the individual branching fractions of $B \to D \ell \nu$,  
$B \to D^* \ell\nu$ and $B \to D^{**} \ell \nu$ (narrow and broad $D^{**}$ states) 
decays within the ranges given by their errors, see Table~\ref{tab:BF}.  
Since we scaled up the four $B \to D^{**} \ell \nu$ branching fractions to take into account the unknown $D^{**}$ partial branching fractions, the errors were increased by a factor of three relative to the published values.

To evaluate the effect of uncertainties in the form-factor parameters for the  dominant $B \to D^* \ell\nu$ component, we
repeat the fit with $\pm 1\sigma$ variations in each of the three 
form-factor parameters, $\rho^2_{D^*}, R_1$ and $R_2$.  
The impact of the form factor for the $B \to D \ell\nu$ background is evaluated
by varying the parameter $\rho^2_{D}$ within its uncertainty.

\subsubsection{Continuum Background}

In Section~\ref{sec:offres}, we have described the correction of  the simulated shapes of the \mES, \DeltaE, and $q^2$
distributions for the continuum using linear functions derived from 
comparison with off-resonance data. 
The uncertainties of the fitted slopes of these correction functions are
used to evaluate the errors due to modeling of the shape of the continuum 
background distributions. 
They represent a sizable contribution to the systematic error, which is mainly due to the low statistics of the off-resonance data sample.

\subsection{Other Systematic Uncertainties}

\subsubsection{Final-State Radiation and Bremsstrahlung}

The kinematics of the signal decays are corrected for radiative effects such as final-state radiation and bremsstrahlung in detector material. 

In the MC simulation, final-state radiation (FSR) is modeled using PHOTOS~\cite{photos}, which is based on ${\cal O}(\alpha)$ calculations but includes multiple-photon emission from the electron.
We have studied the effects of FSR on the $q^2$ dependence of the measured signal and background yields by comparing events generated with and without PHOTOS. 
The observed change is largest, up to 5\%, for electron momenta of about 0.6\gev (i.e. well below our cut-off at 1\gev for \Bpilnu\ 
and 1.8\gev for \Brholnu). Comparisons of 
the PHOTOS simulation with semi-analytical calculations~\cite{richter-was} show excellent agreement.  Allowing for the fact that non-leading terms from possible electromagnetic corrections to the strong interactions of the quarks in the initial and final state have not been calculated to any precision~\cite{ginsberg}, we adopt an uncertainty in the PHOTOS calculations of 20\%. 

The uncertainty of the bremsstrahlung correction is determined by the uncertainty
of the amount of detector material in the inner detector. 
We have adopted as the systematic uncertainty due to bremsstrahlung the impact of a change in the thickness of the detector material by $0.14\%$ radiation lengths, the estimated uncertainty in the thickness of inner detector and the beam vacuum pipe. As for final-state radiation, the uncertainty in the effective radiator thickness impacts primarily the electron spectrum. 

The uncertainties due to final-state radiation and bremsstrahlung combined amount to far less than $1\%$ for most of the $q^2$ range.

\subsubsection{Number of \BB\ Events}

The determination of the on-resonance luminosity and the number of $\BB$ events is described in detail elsewhere~\cite{bcount}.  The uncertainty of the 
total number of \BB\ pairs is estimated to be 1.1\%. 

At the \FourS\ resonance, the fraction of  $B^0\bar{B^0}$ events
is measured to be $f_{00}=0.484 \pm 0.006$, 
with the ratio $f_{+-}/f_{00}=1.065 \pm 0.026$~\cite{HFAG2009}.
This error impacts the branching-ratio measurements by 0.8\%.

\subsubsection{$B^0$ and $B^+$ Lifetimes}

Since we combine fits to decays of charged and neutral $B$ mesons and make
use of isospin relations, the $B$-meson lifetimes enter into the four-mode fit.
We use the PDG \cite{PDG2008} value for the $B$ lifetime,
$\tau_0=1.530\pm 0.009$~ps , 
and the lifetime ratio, $\tau_+/\tau_0= 1.071 \pm 0.009$.
These uncertainties lead to a systematic error of $0.3\%$ for \Btopilnu\ and 0.7\% for \Btorholnu decays.
\section{Results}
\label{sec:results}

Based on the signal yields obtained in the four-mode fit, 
integrated over the full $q^2$ range (see Table~\ref{tab:signalYields}),
we derive the following total branching fractions, constrained by the 
isospin relations stated in Eqs.~\ref{eq:isospin},  
\begin{eqnarray}
  \BRBzpilnu  &=& (1.41 \pm 0.05 \pm 0.07) \times 10^{-4} \ , \nonumber \\
  \BRBzrholnu &=& (1.75 \pm 0.15 \pm 0.27) \times 10^{-4} \ . \nonumber
\end{eqnarray}
Here and in the following, the first error reflects the statistical (fit) error and the second the estimated systematic error.
The total branching fractions obtained from the single-mode fits
for the charged and neutral \Bpilnu\ samples are 
\begin{eqnarray}
\BRBzpilnu     &=& (1.44 \pm 0.06 \pm 0.07)\times 10^{-4} \ , \nonumber \\
\BRBppizlnu \times 2 \frac{\tau_0}{\tau_+}   
               &=& (1.40 \pm 0.10 \pm 0.11)\times 10^{-4} \ .  \nonumber
\end{eqnarray}
For the charged and neutral \Brholnu\ samples, we obtain
\begin{eqnarray}
\BRBzrholnu     &=& (1.98 \pm 0.21 \pm 0.38)\times 10^{-4}\ , \nonumber \\
\BRBprhozlnu \times 2 \frac{\tau_0}{\tau_+}   
                &=& (1.87 \pm 0.19 \pm 0.32)\times 10^{-4}\ . \nonumber
\end{eqnarray}
The single-mode fits result in higher values for  
\BRBzrholnu\ and \BRBprhozlnu\ than the average branching fraction obtained 
from the four-mode fit. 
This may be explained by different treatments of the isospin-conjugate signal and the 
$\pi \leftrightarrow \rho$ cross feed in the single- and four-mode fits. 
In contrast to the four-mode fit, the isospin-conjugate signal contribution in the
single-mode fits is not constrained by the isospin-conjugate mode.
In addition, the four-mode fit uses the same fit parameter for the signal and the cross feed
from the signal mode into other modes, which leads to a slight decrease in the $B\to\rho\ell\nu$
branching fraction compared to the single-mode fits. Since the $\rho \to \pi$ cross feed is
significantly larger than the $\pi \to \rho$ cross feed, the effect on the 
$B\to\rho\ell\nu$ results is larger than for $B\to\pi\ell\nu$.

Both the \Bpilnu\ and the \Brholnu\ results are consistent within errors
with the isospin relations, 
\begin{eqnarray}
\label{eq:bfratio_pi}
\frac{\BRBzpilnu}{\BRBppizlnu \times 2 \frac{\tau_0}{\tau_+}}
  &=& 1.03 \pm 0.09 \pm 0.06 \ , \nonumber \\
\label{eq:bfratio_rho}
\frac{\BRBzrholnu}{\BRBprhozlnu \times 2 \frac{\tau_0}{\tau_+}}   
  &=& 1.06 \pm 0.16 \pm 0.08 \ .  \nonumber
\end{eqnarray}

By extracting the signal in several $q^2$ bins 
we also measure the $q^2$ spectra of \Bpilnu\ and \Brholnu\ decays. 
These spectra need to be corrected for effects such as
detector resolution, bremsstrahlung, and final-state
radiation.

\subsection{Partial Branching Fractions}
\label{sec:unfolding}

We correct the measured $q^2$ spectra for 
resolution, radiative effects and bremsstrahlung 
by applying an unfolding technique that is based on singular-value decomposition
of the detector response matrix~\cite{unfolding_SVD}. The detector response matrix
in the form of a two-dimensional histogram of the reconstructed versus the true $q^2$ 
values (see Figure~\ref{fig:q2resolutions}) is used as input to the unfolding algorithm.
This algorithm contains a regularization term to suppress
spurious oscillations originating from statistical fluctuations. To find the best
choice of the regularization parameter $\kappa$ we have studied the
systematic bias on the partial branching fractions compared to the
statistical uncertainty as a function of $\kappa$ using a set of simulated distributions.
The data samples in this analysis are large enough that no
severe distortions due to statistical fluctuations are expected. 
We choose the largest possible value of $\kappa$, {\it i.e.}, we set $\kappa$ 
equal to the number of $q^2$ bins, to minimize a potential bias.

The $\Delta{\cal B}/\Delta q^2$ distributions resulting from the unfolding procedure 
are presented in 
Figure~\ref{fig:FFfit} for \Bpilnu\ and in Figure~\ref{fig:q2Spectrum_rho} for \Brholnu. 
Tables~\ref{tab:BF_pi} and~\ref{tab:BF_rho} list the partial branching fractions $\Delta {\cal B}$ 
for \Bpilnu\ and \Brholnu, respectively. 


\begin{table*}[hbt]
\centering
\caption{
Partial and total branching fractions (corrected for radiative effects) 
for $\Bz \to \pim \ell^+ \nu$ and $\Bp \to \piz \ell^+ \nu$  decays obtained from the
single-mode fits and \Bpilnu\ decays from the four-mode fit with
statistical (fit), systematic and total errors. The branching fraction for \Bppizlnu\ 
has been scaled by twice the lifetime ratio of neutral and charged $B$ mesons.
All branching fractions and associated errors are given in units of $10^{-4}$. 
}

\renewcommand{\arraystretch}{1.6}

\def\sc{\ensuremath{\times 2\tau_0/\tau_+}\xspace}

\begin{tabular}{l|cccccc|c|cc} 
\hline \hline

$q^2$ range ($\gev^2$)         & 0-4 & 4-8 & 8-12 & 12-16 & 16-20 & 20-26.4 & Total & $<$16 & $>$16  \\ \hline  
$\Delta\BRBzpilnu$             & 0.313 & 0.329 & 0.241 & 0.222 & 0.206 & 0.124 & 1.435 & 1.105 & 0.330 \\ 
Fit error   	               & 0.030 & 0.018 & 0.018 & 0.020 & 0.020 & 0.018 & 0.061 & 0.049 & 0.027 \\
Syst. error       	       & 0.025 & 0.016 & 0.015 & 0.015 & 0.013 & 0.010 & 0.068 & 0.059 & 0.019 \\
Total error         	       & 0.039 & 0.024 & 0.023 & 0.025 & 0.024 & 0.021 & 0.092 & 0.077 & 0.033 \\ \hline

$\Delta\BRBppizlnu\sc$         & 0.357 & 0.294 & 0.234 & 0.210 & 0.206 & 0.099 & 1.401 & 1.096 & 0.305 \\
Fit error 	               & 0.049 & 0.031 & 0.031 & 0.033 & 0.039 & 0.043 & 0.102 & 0.075 & 0.062 \\
Syst. error 	       	       & 0.050 & 0.015 & 0.028 & 0.019 & 0.024 & 0.028 & 0.106 & 0.089 & 0.037 \\
Total error 	               & 0.070 & 0.035 & 0.041 & 0.038 & 0.046 & 0.051 & 0.147 & 0.117 & 0.072 \\ \hline

$\Delta\BRBzpilnu$  4-mode     & 0.320 & 0.321 & 0.235 & 0.220 & 0.201 & 0.118 & 1.414 & 1.095 & 0.319 \\
Fit error 		       & 0.025 & 0.017 & 0.015 & 0.017 & 0.018 & 0.016 & 0.050 & 0.041 & 0.024 \\
Syst. error 	       	       & 0.027 & 0.012 & 0.016 & 0.018 & 0.014 & 0.014 & 0.074 & 0.061 & 0.024 \\
Total error 	               & 0.037 & 0.021 & 0.022 & 0.025 & 0.023 & 0.022 & 0.089 & 0.074 & 0.034 \\ \hline \hline

\end{tabular}
\label{tab:BF_pi}
\end{table*}
\def\sc{\ensuremath{\times 2\tau_0/\tau_+}\xspace}

\begin{table}[hbt]
\centering
\caption{
Partial and total branching fractions (corrected for radiative effects) 
for $\Bz \to \rho^- \ell^+ \nu$  and $\Bp \to \rho^0 \ell^+ \nu$ decays obtained from the
single-mode fits and for \Brholnu\ decays from the four-mode fit with
statistical (fit), systematic and total errors.  The branching fractions for \Bprhozlnu
have been scaled by twice the ratio of the lifetimes of neutral and charged $B$ mesons.
All branching fractions and associated errors are given in units of $10^{-4}$. 
}
\renewcommand{\arraystretch}{1.6}

\begin{tabular}{l|ccc|c} 
\hline \hline

$q^2$ range ($\gev^2$)      & 0-8   & 8-16  & 16-20.3 & Total     \\ \hline  
$\Delta\BRBzrholnu$         & 0.747 & 0.980 & 0.256   & 1.984    \\ 
Fit error                   & 0.151 & 0.087 & 0.030   & 0.214    \\ 
Syst. error                 & 0.178 & 0.165 & 0.066   & 0.379    \\ 
Total error                 & 0.234 & 0.187 & 0.072   & 0.435    \\ \hline           

$\Delta\BRBprhozlnu\sc$     & 0.627 & 0.977 & 0.265 & 1.871    \\ 
Fit error                   & 0.136 & 0.079 & 0.028 & 0.190    \\ 
Syst. error                 & 0.152 & 0.161 & 0.061 & 0.320    \\ 
Total error                 & 0.204 & 0.179 & 0.068 & 0.373    \\ \hline           

$\Delta\BRBzrholnu$ 4-mode  & 0.564 & 0.912 & 0.268 & 1.745    \\ 
Fit error   	            & 0.107 & 0.059 & 0.022 & 0.149    \\ 
Syst. error  	            & 0.126 & 0.135 & 0.058 & 0.272    \\ 
Total error                 & 0.166 & 0.147 & 0.062 & 0.310    \\ \hline \hline          

\end{tabular}
\label{tab:BF_rho}
\end{table}

\subsection{Form-factor Shape}

For \Bpilnu\ decays, we extract the shape of the form factor $f_+(q^2)$ directly from data. 
For \Brholnu\ decays,  we restrict ourselves to the measurement of the $q^2$ dependence, 
since the current experimental precision is not adequate to extract the three different form factors involved.

Several parameterizations of $f_+(q^2)$
are used to interpolate between results of various form-factor calculations
or to extrapolate these calculations from a partial to the whole $q^2$ range. 
The four most common parameterizations, the BK~\cite{BK}, BZ~\cite{ball_pi}, 
BGL~\cite{BGL,BHill} and BCL~\cite{BCL} parameterizations, have been introduced 
in Section~\ref{sec:theory}.  
For the BGL and BCL parameterizations, we consider a linear ($k_{max}=2$) and a quadratic 
($k_{max} = 3$) ansatz.

We perform  $\chi^2$ fits to the measured $q^2$ spectrum to determine the free 
parameters for each of these parameterizations.
The fit employs the following $\chi^2$ definition, with integration of the fit function 
over the $q^2$ bins, 
\begin{equation}
  \label{eq:chi2}
  \chi^2 = \sum\limits_{i,j=1}^{N_{\rm bins}} \Delta_i \ V_{ij}^{-1} \Delta_j \ ,
\end{equation}
where $V_{i,j}^{-1}$ is the inverse covariance matrix of the
partial-branching-fraction measurements. $\Delta_k$ for $q^2$ bin~$k$ is defined as
\begin{eqnarray}
  \label{eq:delta}
  \Delta_k &=& \left({\frac{\Delta B}{\Delta q^2}}\right)^{\rm data}_{k} 
           - \frac{C}{\Delta q^2_k} 
               \int\limits_{\Delta q^2_k} p_{\pi}^3 |f_+(q^2; {\alpha})|^2  dq^2 , \ \ \ 
\end{eqnarray}
where $\alpha$ denotes the 
set of parameters for a chosen parameterization of $f_+(q^2)$, and
$C = \Vub^2 \tau_0 G^2_F / (24 \pi^3)$ is an overall normalization factor
whose value is irrelevant for these fits since the data can only constrain the shape
of the form factor, but not its normalization.
\begin{table*}[hbt]
\renewcommand{\arraystretch}{1.2}
\caption{
Results of fits to the measured $\Delta{\cal B}/\Delta q^2$ for \Bpilnu\ decays, based on different form-factor parameterizations. 
}

\begin{tabular}{lcclc} \hline\hline
Parametrization &  $\chi^2$/ndf  & Prob($\chi^2$/ndf) & Fit parameters                    
& $f_+(0)|V_{ub}|~[10^{-3}]$    \\ \hline
BK              & $6.8/4$  & 0.148    & $\alpha_{BK} =+0.310\pm 0.085$ & $1.052 \pm 0.042$ \\ 
BZ              & $6.0/3$  & 0.112    & $     r_{BZ} =+0.170\pm 0.124$ & $1.079 \pm 0.046$ \\
                &          &          & $\alpha_{BZ} =+0.761\pm 0.337$ &                   \\
 
BCL (2 par.)    & $6.3/4$  & 0.179    & $b_1/b_0 = -0.67 \pm 0.18$     & $1.065 \pm 0.042$  \\
BCL (3 par.)    & $6.0/3$  & 0.112    & $b_1/b_0 = -0.90 \pm 0.46$     & $1.086 \pm 0.055$  \\
                &          &          & $b_2/b_0 = +0.47 \pm 1.49$     &                   \\

BGL (2 par.)    & $6.6/4$  & 0.156    & $a_1/a_0=-0.94 \pm 0.20 $      & $1.103 \pm 0.042$ \\
BGL (3 par.)    & $6.3/3$  & 0.100    & $a_1/a_0=-0.82 \pm 0.29 $      & $1.080 \pm 0.056$ \\
                &          &          & $a_2/a_0=-1.14 \pm 1.81 $      &             \\  \hline \hline                
\end{tabular}
\label{tab:FFfits}
\end{table*}

\begin{figure*}[hbt]
  \begin{minipage}{0.45\linewidth}
    \epsfig{file=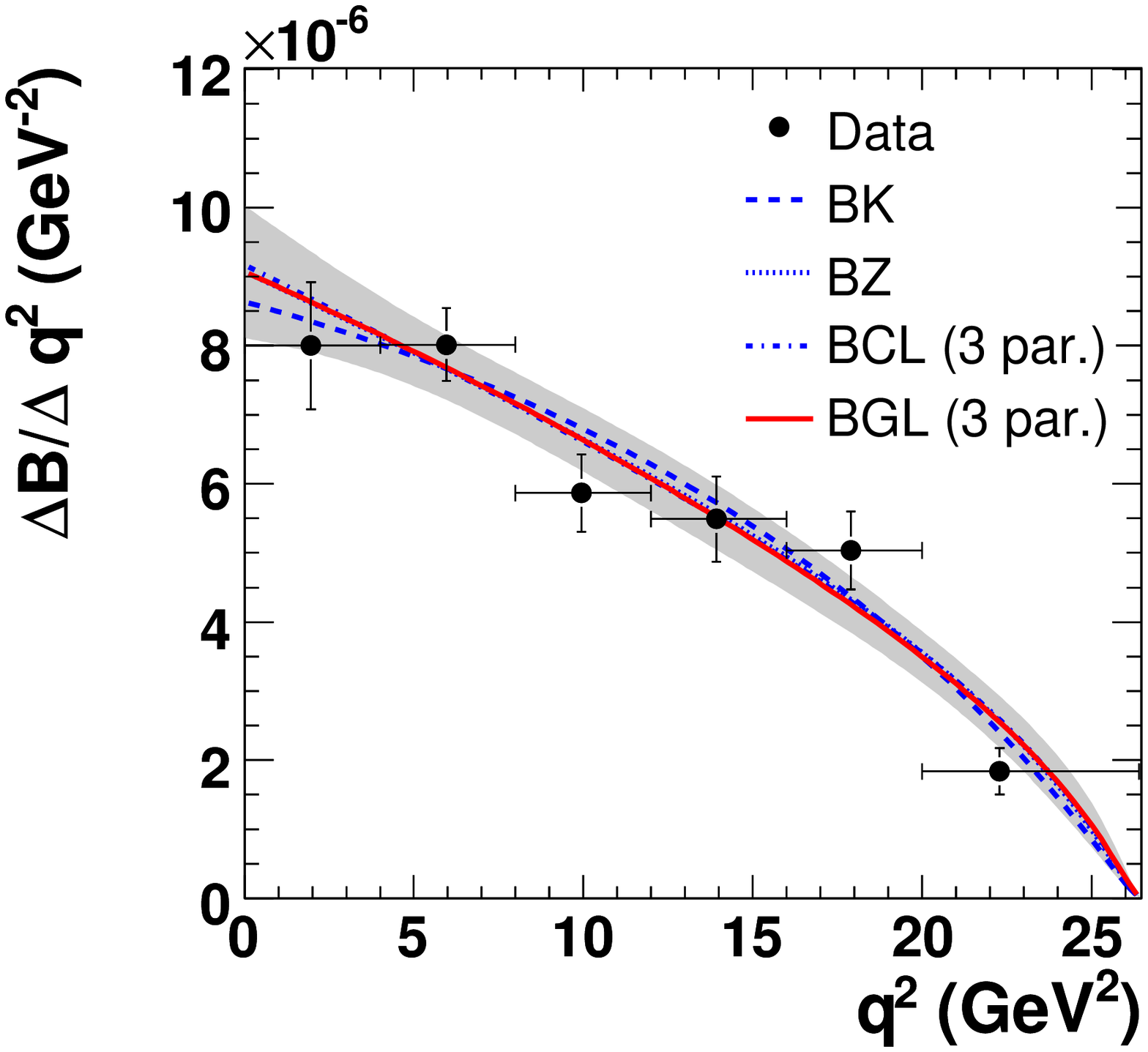, width = 7cm} 
  \end{minipage}
  \begin{minipage}{0.45\linewidth}
    \epsfig{file=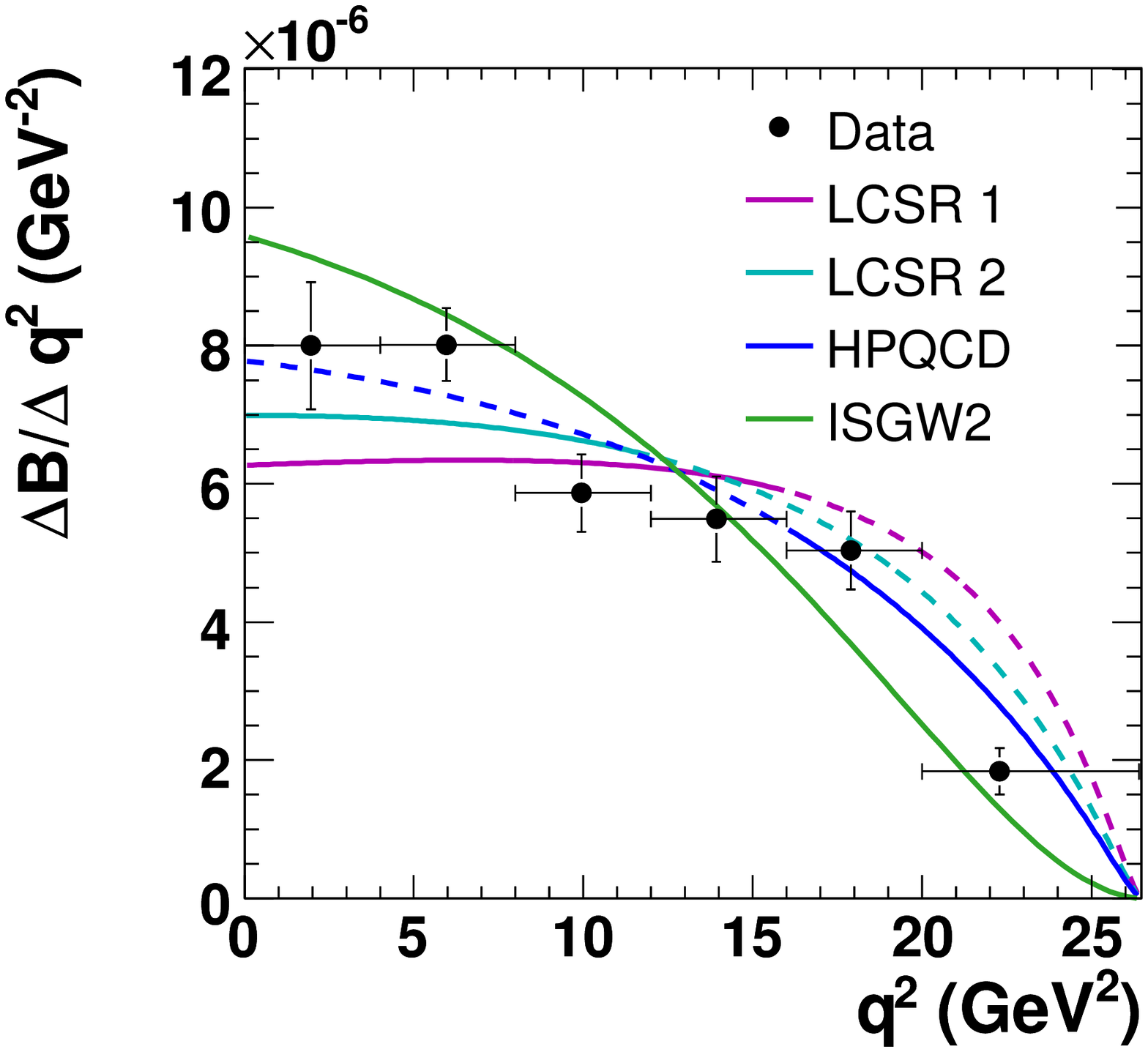, width = 7cm} 
  \end{minipage}
  \caption{
    Measured $\Delta{\cal B}/\Delta q^2$ distribution 
    for \Bpilnu. 
    The vertical error bars correspond to the combined statistical and systematic 
    uncertainties.  The positions of the data points have been adjusted to 
    correspond to the mean $q^2$ value in each bin, based on the quadratic
    BGL ansatz. 
    Left: fits of four different form-factor parameterizations
    to the $\Delta{\cal B}/\Delta q^2$ data spectrum. 
    The fit result for the BZ and BCL parameterizations are barely visible, 
    since they overlap almost completely with the BGL result. 
    The shaded band illustrates the uncertainty of the quadratic BGL fit to data.
    Right: shape comparisons of the data to various \Bpilnu\ form-factor predictions
    (LCSR~1~\cite{ball_pi}, LCSR~2~\cite{Siegen}, HPQCD~\cite{hpqcd04}, ISGW2~\cite{isgw2}), 
    which have been normalized to the measured total branching fraction.   
    The extrapolations of the QCD predictions to the full $q^2$ range are marked as dashed lines.} 
  \label{fig:FFfit}
\end{figure*}

In Table~\ref{tab:FFfits} and Figure~\ref{fig:FFfit} we present the results of these fits to the \Bpilnu\ samples.
All parameterizations describe the data well, with $\chi^2$ probabilities ranging from 10\% to 18\%.  
Thus, within the current
experimental precision,  all parameterizations are valid choices, and the central values 
for $\Vub f_+(0)$ agree with each other.
We choose the quadratic BGL parameterization as the default, though even a linear parameterization results in a very good fit to the data. The error band represents the uncertainties of the fit to data, based on the quadratic BGL parameterization (solid line in Figure~\ref{fig:FFfit}). It has been computed using standard error propagation,
taking the correlation between the fit parameters into account. 

\begin{figure}[hbt]
    \epsfig{file=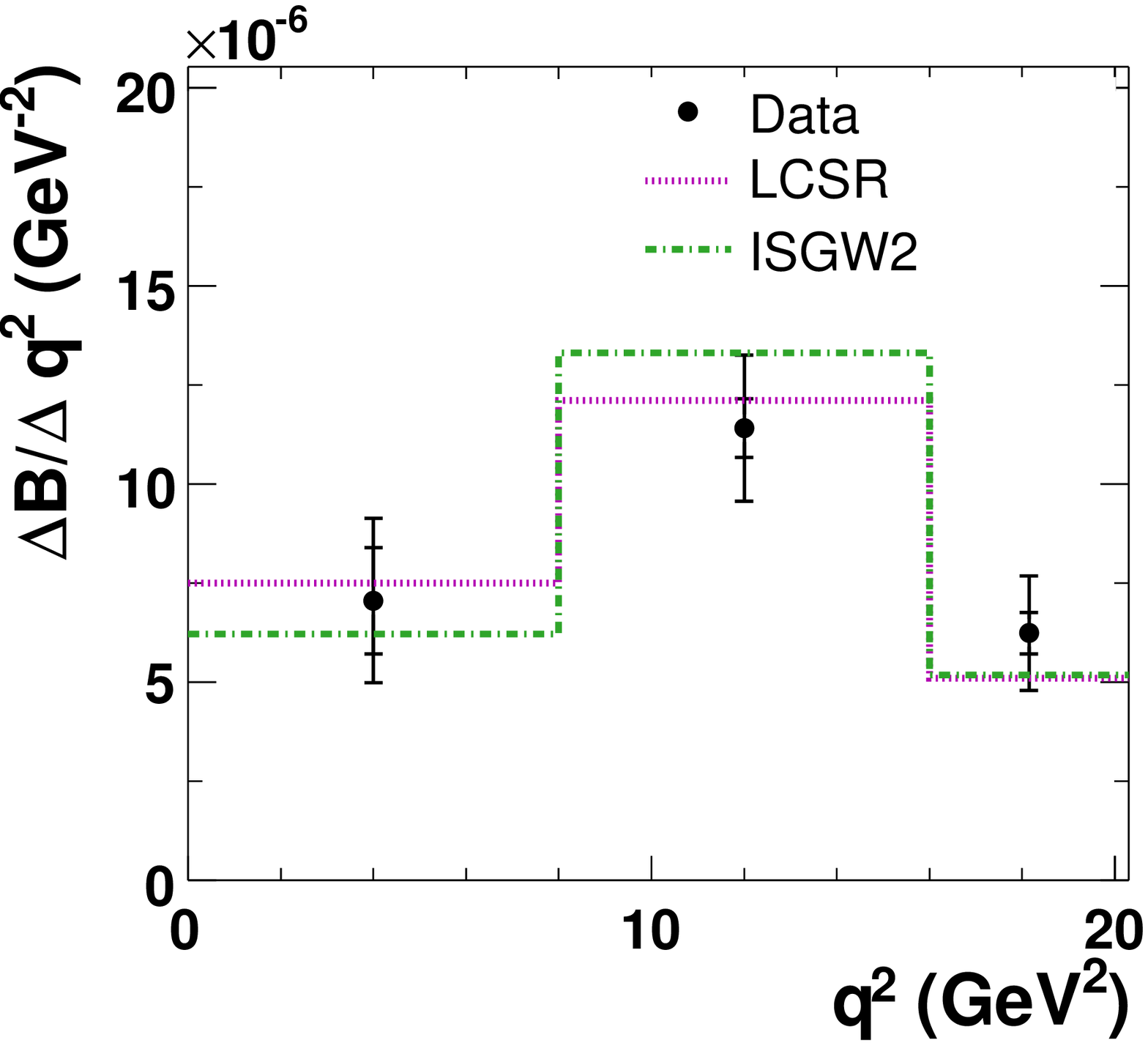, width = 7cm}
  \caption{(color online)
    Measured $\Delta{\cal B}/\Delta q^2$ distribution 
    for \Brholnu. 
    The inner and outer error bars correspond to the statistical uncertainty
    and the combined statistical and systematic uncertainty, respectively.
    The data are compared with the \Brholnu form-factor predictions from LCSR~\cite{ball_rho}
    and from the ISGW2 quark model~\cite{isgw2}.}
  \label{fig:q2Spectrum_rho}
\end{figure}

We compare the measured $q^2$ spectra with the shapes
predicted by form-factor calculations based on lattice QCD~\cite{hpqcd04}, 
light-cone sum rules~\cite{ball_pi,Siegen}, and the ISGW2~\cite{isgw2} relativistic quark model.
Among the available calculations for \Bpilnu decays, the HPQCD lattice calculation agrees best with the data.
It should be noted that the LQCD predictions are only valid for $q^2 > 16 \gev^2$, the earlier LCSR calculation
(LCSR~1) for $q^2 < 16 \gev^2$, and the more recent LCSR calculation (LCSR~2) for $q^2 < 12 \gev^2$; 
their extrapolation is impacted by sizable uncertainties. 

In Table~\ref{tab:BF_rho} and Figure~\ref{fig:q2Spectrum_rho} we present the results of the fits to the \Brholnu\ samples.
The LCSR calculation  and the ISGW2 model are in good agreement with the data. However, the errors of the measured \Brholnu\ 
partial branching fractions are relatively large, at the level of 15-30\%, depending
on the $q^2$ interval.

It should be noted that the theoretical calculations differ most for low and high $q^2$.
In these regions of phase space, the measurements are impacted significantly by higher levels of backgrounds, specifically continuum events at low $q^2$ and other $\bulnu$ decays that are difficult to separate from the signal modes at higher $q^2$.
These two background sources have been examined in detail, and the uncertainties in their normalization and shape are included in the systematic uncertainties.
For the inclusive $\bulnu$ background, the $q^2$ and the hadronic mass spectra are derived from theoretical predictions that depend on non-perturbative parameters that
are not well measured~\cite{moments}.  
For \Brholnu\ the correlation between the signal and the $\bulnu$ background is so large that they cannot both be fitted simultaneously. Thus the 
$\bulnu$ background scale factor and shape are fixed to the MC predictions, which have large uncertainties.  
MC studies indicate that this may introduce a bias affecting the signal yield.  The stated errors account for this potential bias.

\subsection{\boldmath Determination of \Vub}

We choose two different approaches to determine the magnitude of the 
CKM matrix element $V_{ub}$. 

First, we use the traditional method to derive \Vub. 
As in previous publications~\cite{pilnu_jochen, pilnu_cleo, pilnu_cote, pilnu_babar_tag, pilnu_belle_tag},
we combine the measured partial branching fractions with integrals of 
the form-factor calculations over a certain $q^2$ range using the relation
\begin{equation}
  \Vub = \sqrt{\frac{\Delta{\cal B}(q^2_{min},q^2_{max})}{\tau_0 \                              \Delta\zeta(q^2_{min},q^2_{max})}} \ ,
\end{equation}
where $\tau_0 = (1.530 \pm 0.009)$~ps is the $B^0$~lifetime and $\Delta \zeta$ 
is defined as
\begin{equation}
\label{eq:GammaThy}
\Delta \zeta(q^2_{min},q^2_{max})  = 
\frac{G^2_F} {24 \pi^3} 
\int\limits_{q^2_{min}}^{q^2_{max}} 
p_{\pi}^3 |f_+(q^2)|^2 dq^2 \ .
\end{equation} 
The values of $\Delta\zeta$ are derived from theoretical form-factor calculations 
for different $q^2$ ranges.
Table~\ref{tab:vub} summarizes the $\Delta\zeta$ values, the partial branching fractions and the \Vub results.

\begin{table}[hbt] 
  \renewcommand{\arraystretch}{1.4}
  \centering
  \caption{$|V_{ub}|$ derived from \Btopilnu and \Btorholnu decays for 
  various $q^2$ regions and form-factor calculations.
  Quoted errors are
  experimental uncertainties and theoretical uncertainties of the
  form-factor integral~$\Delta\zeta$. 
  No uncertainties on~$\Delta\zeta$ for \Brholnu\ are given in Refs.~\cite{ball_rho} and~\cite{isgw2}.} 

  \begin{tabular}{llccl} \hline\hline
  & $q^2$ Range & $\Delta {\cal B}$ & $\Delta\zeta$ &
    \multicolumn{1}{c}{$|V_{ub}|$} \\
  & ($\gev^2$) & (10$^{-4}$) & (ps$^{-1}$) &
    \multicolumn{1}{c}{(10$^{-3}$)} \\
\hline
  \Btopilnu \\
  LCSR~1~\cite{ball_pi}  & $0-16$    & $1.10 \pm 0.07$ & $5.44{\pm}1.43$        & $3.63 \pm 0.12^{+0.59}_{-0.40}$ \\
  LCSR~2~\cite{Siegen}   & $0-12$    & $0.88 \pm 0.06$ & $4.00^{+1.01}_{-0.95}$ & $3.78 \pm 0.13^{+0.55}_{-0.40}$ \\
  HPQCD~\cite{hpqcd04}   & $16-26.4$ & $0.32 \pm 0.03$ & $2.02{\pm}0.55$        & $3.21 \pm 0.17^{+0.55}_{-0.36}$ \\ 
\hline
  \Btorholnu \\
  LCSR~\cite{ball_rho}   & $0-16.0$  & $1.48 \pm 0.28$ & $13.79$  & $2.75 \pm 0.24$ \\ 
  ISGW2~\cite{isgw2}     & $0-20.3$  & $1.75 \pm 0.31$ & $14.20$  & $2.83 \pm 0.24$ \\
\hline\hline
  \end{tabular}
  \label{tab:vub}
\end{table}

For \Brholnu, values of $\Delta\zeta$ are taken from the LCSR calculation in the range $q^2<16\gev$ and the quark model predictions of ISGW2 over the full $q^2$ range.  The results are also presented in Table~\ref{tab:vub}.  Estimates of the uncertainties for $\Delta\zeta$ 
are not given in Refs.~\cite{ball_rho} and~\cite{isgw2} .

Second, we perform a simultaneous fit to the most recent lattice results and \babar\ data
to make best use of the available information on the form factor
from data (shape) and theory (shape and normalization).
A fit of this kind was first presented by the FNAL/MILC Collaboration~\cite{fnal09}
using the earlier \babar\ results on \Bzpilnu decays~\cite{pilnu_cote}. 

To perform this fit, we translate the $f_+(q^2)$ predictions from LQCD to 
$1/(\tau_0 |V_{ub})|^2) \Delta{\cal B}/\Delta{q^2}$.
We simultaneously fit this distribution  
and the $\Delta{\cal B}/\Delta{q^2}$ distribution from data as a function of $q^2$. 
We use the BGL form-factor parameterization as the fit function, with the additional 
normalization parameter $a_{\rm norm}=\tau_0 |V_{ub}|^2$, which allows us to determine
$|V_{ub}|$ from the relative normalization of data and LQCD predictions.

The $\chi^2$ for this fit is given by
\begin{eqnarray}
  \label{eq:chi2_Vubfit}
  \chi^2 &=& \chi^2(\rm data) + \chi^2(\rm lattice) \nonumber \\
         &=& \sum\limits_{i,j=1}^{N_{\rm bins}} \Delta_i^{\rm data} \ (V_{ij}^{\rm data})^{-1} \Delta_j^{\rm data} \nonumber \\ 
         &+& \sum\limits_{l,m=1}^{N_{\rm points}} \Delta_l^{\rm lat} \ (V_{lm}^{\rm lat})^{-1} \Delta_m^{\rm lat}
\end{eqnarray}
where 
\begin{eqnarray}
  \Delta_i^{\rm data} &=& \left(\frac{\Delta B}{\Delta q^2}\right)^{\rm data}_{i} 
   - \frac{1}{\Delta q^2_i} \int\limits_{\Delta q^2_i} g(q^2; \alpha) dq^2 \ , \\ 
  \Delta_l^{\rm lat} &=& \frac{G^2_F} {24 \pi^3} p_\pi^3(q^2_l) |f_+^{\rm lat}(q^2_l)|^2- g(q^2_l; \alpha)
\end{eqnarray}
and 
\begin{eqnarray}
  g(q^2; \alpha) &=& \frac{G^2_F} {24 \pi^3} p_\pi^3(q^2) |f_+(q^2)|^2 \nonumber \\
  &\times& \left\{\begin{array}{cl} a_{\rm norm} & \mbox{for data}\\ 1 & \mbox{for LQCD} \end{array}\right. , \\ 
f_+(q^2) &=& \frac{1} {{\cal P}(q^2) \phi(q^2,q^2_0)} 
         \sum_{k=0}^{k_{max}} a_k(q^2_0) [z(q^2,q^2_0)]^k .
\end{eqnarray}
Here $(\Delta B/\Delta q^2)^{\rm data}$ is the measured spectrum, $f_+^{\rm lat}(q_l^2)$ are the
form-factor predictions from LQCD, and  
$(V_{ij}^{\rm data})^{-1}$ and $(V_{ij}^{\rm lat})^{-1}$ are the corresponding 
inverse covariance matrices for $(\Delta B/\Delta q^2)^{\rm data}$
and $G^2_F/(24 \pi^3) p_\pi^3(q^2_l) |f_+^{\rm lat}(q^2_l)|^2$, respectively.
The set of free parameters $\alpha$ of the fit function $g(q^2; \alpha)$ contains the
coefficients $a_k$ of the BGL parameterization and the normalization parameter $a_{\rm norm}$.

\begin{figure*}

\begin{tabular}{cc}
  \begin{minipage}{0.45\linewidth}
    \epsfig{file=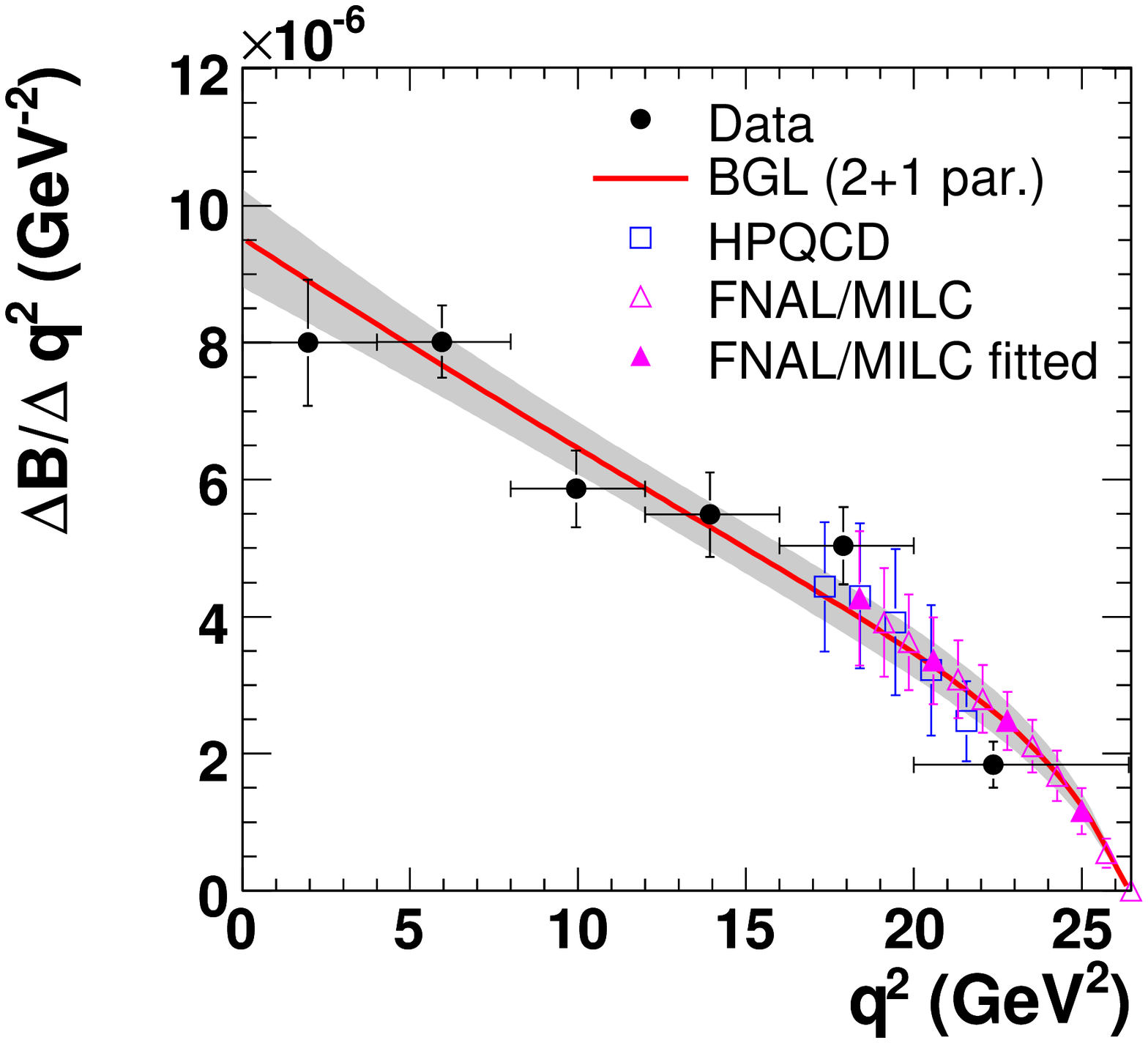, width=8cm} 
  \end{minipage}

  \begin{minipage}{0.45\linewidth}
    \epsfig{file=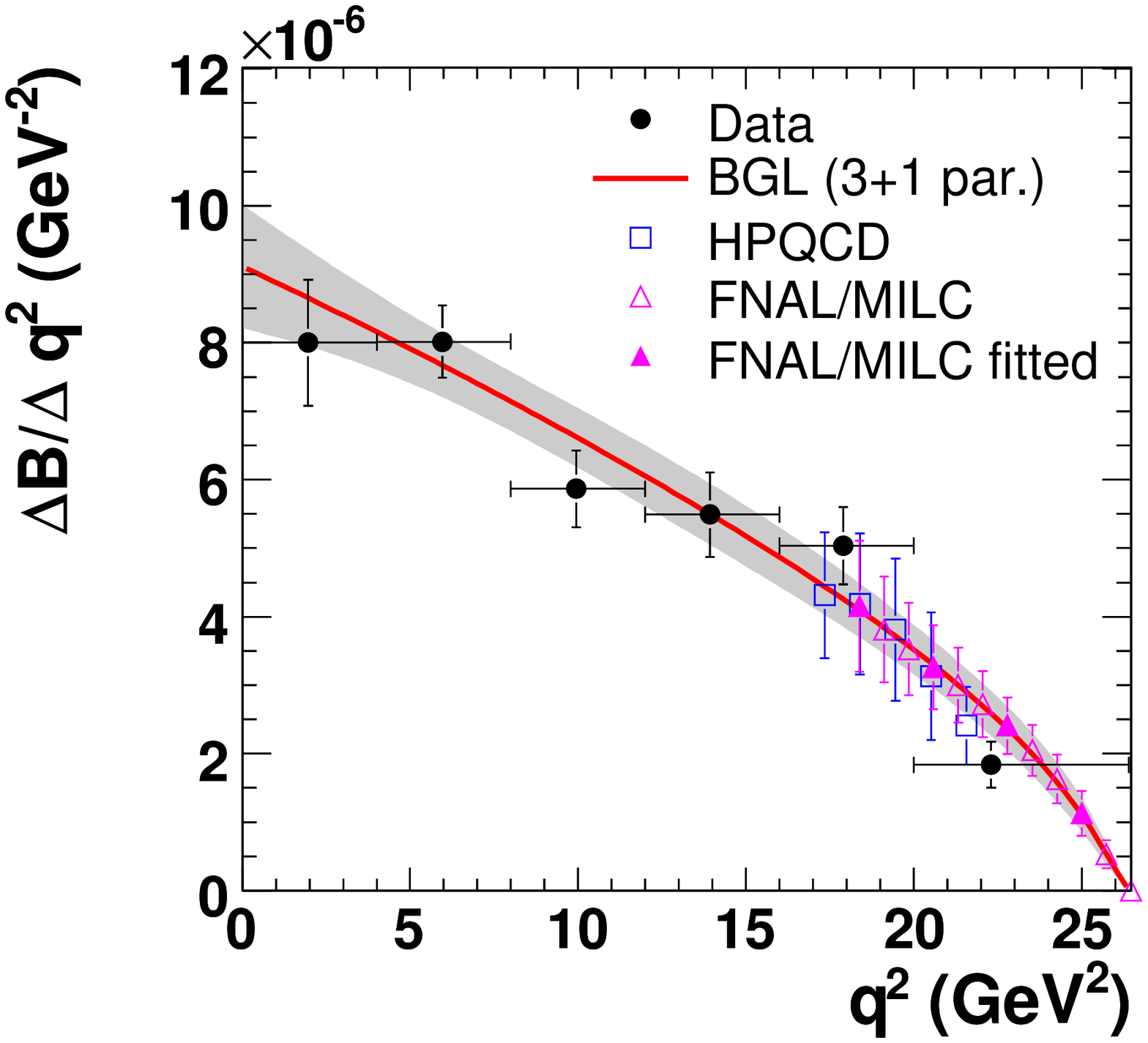, width=8cm} 
  \end{minipage}
\end{tabular}

  \caption{(color online)
    Simultaneous fits of the BGL parameterization to data (solid points with 
    vertical error bars representing the total experimental uncertainties) 
    and to four of the twelve points of the FNAL/MILC lattice prediction 
    (magenta, closed triangles). 
     Left: linear (2+1-parameter) BGL fit,
     right: quadratic (3+1-parameter) BGL fit. 
     The LQCD results are rescaled to the data according to the \Vub value obtained in the fit.
     The shaded band illustrates the uncertainty of the fitted function.
     For comparison, the HPQCD (blue, open squares) lattice results are also shown.
     They are used in an alternate fit. }
  \label{fig:simulFFfit}
\end{figure*}

From the FNAL/MILC~\cite{fnal09} lattice calculations, we use only subsets with six, four or three 
of the twelve predictions at different values of $q^2$, since neighboring points are very strongly correlated. All chosen subsets of LQCD points contain the point at lowest $q^2$. 
It has been checked that alternative choices of subsets give compatible results.  
From the HPQCD~\cite{hpqcd04} lattice calculations, we use only the point at lowest $q^2$ since the correlation matrix 
for the four predicted points is not available. 
For comparison, we also perform the corresponding fit using only the point at lowest $q^2$ from FNAL/MILC.
The data, the lattice predictions, and the fitted functions are shown in Figure~\ref{fig:simulFFfit}.
Table~\ref{tab:BGLfits} shows the numerical results of the fit.

\begin{table*}[hbt]
\renewcommand{\arraystretch}{1.2}
\caption{Results of simultaneous fits to data and LQCD
         calculations, based on the linear or quadratic BGL parameterizations.
         $\Vub$ is determined from the relative normalization of data and theory prediction.}
\begin{tabular}{lcccl} \hline\hline
Parameterization & Input       & $\chi^2$/ndf     & Prob($\chi^2$/ndf) & Fit parameters  \\ \hline                          
BGL (2+1 par.)  & Data        &$18.2/9$& 0.033   & $a_0     = (2.07 \pm 0.21)\times 10^{-2}$ \\  
                & + FNAL/MILC &        &         & $a_1/a_0 = -0.78 \pm 0.22 $     \\
                & (6 points)  &        &         & $\Vub = (3.04 \pm 0.38)\times 10^{-3}$ \\ \hline
BGL (2+1 par.)  & Data        & $7.1/7$& 0.415   & $a_0     = (2.16 \pm 0.19)\times 10^{-2}$ \\  
                & + FNAL/MILC &        &         & $a_1/a_0 = -0.93 \pm 0.20 $     \\
                & (4 points)  &        &         & $\Vub = (2.99 \pm 0.32)\times 10^{-3}$ \\ \hline
BGL (2+1 par.)  & Data        & $6.8/6$& 0.341   & $a_0     = (2.17\pm 0.19)\times 10^{-2}$ \\  
                & + FNAL/MILC &        &         & $a_1/a_0 = -0.93 \pm 0.20 $     \\
                & (3 points)  &        &         & $\Vub = (2.97 \pm 0.32)\times 10^{-3}$ \\ \hline
BGL (2+1 par.)  & Data        & $6.6/4$& 0.156   & $a_0     = (2.23 \pm 0.26)\times 10^{-2}$ \\
                & + FNAL/MILC &        &         & $a_1/a_0 = -0.94 \pm 0.20 $     \\
                & (1 point)   &        &         & $\Vub = (2.90 \pm 0.36)\times 10^{-3}$ \\ \hline
BGL (2+1 par.)  & Data        & $6.6/4$& 0.156   & $a_0     = (2.19 \pm 0.23)\times 10^{-2}$ \\
                &  + HPQCD    &        &         & $a_1/a_0 = -0.94 \pm 0.20 $     \\
                & (1 point)   &        &         & $\Vub = (2.94 \pm 0.34)\times 10^{-3}$ \\               \hline\hline

BGL (3+1 par.)  & Data        & $9.8/8$& 0.276   & $a_0     = (2.31 \pm 0.20)\times 10^{-2}$ \\  
                & + FNAL/MILC &        &         & $a_1/a_0 = -0.71 \pm 0.19 $     \\
                & (6 points)  &        &         & $a_2/a_0 = -2.33  \pm 0.84  $     \\                  
                &             &        &         & $\Vub = (2.87 \pm 0.28)\times 10^{-3}$ \\ \hline
BGL (3+1 par.)  & Data        & $6.6/6$& 0.355   & $a_0     = (2.22 \pm 0.21)\times 10^{-2}$ \\  
                & + FNAL/MILC &        &         & $a_1/a_0 = -0.86 \pm 0.23 $     \\
                & (4 points)  &        &         & $a_2/a_0 = -0.97 \pm 1.36 $     \\                  
                &             &        &         & $\Vub = (2.95 \pm 0.31)\times 10^{-3}$ \\ \hline
BGL (3+1 par.)  & Data        & $6.3/5$& 0.279   & $a_0     = (2.24 \pm 0.22)\times 10^{-2}$ \\  
                & + FNAL/MILC &        &         & $a_1/a_0 = -0.84 \pm 0.23 $     \\
                & (3 points)  &        &         & $a_2/a_0 = -1.01 \pm 1.40 $     \\                  
                &             &        &         & $\Vub = (2.93 \pm 0.31)\times 10^{-3}$ \\ \hline
BGL (3+1 par.)  & Data        & $6.3/3$& 0.100   & $a_0     = (2.24 \pm 0.26)\times 10^{-2}$ \\
                & + FNAL/MILC &        &         & $a_1/a_0 = -0.82 \pm 0.29 $     \\
                & (1 point)   &        &         & $a_2/a_0 = -1.14  \pm 1.81  $     \\
                &             &        &         & $\Vub = (2.92 \pm 0.37)\times 10^{-3}$ \\ \hline
BGL (3+1 par.)  & Data        & $6.3/3$& 0.100   & $a_0     = (2.19\pm 0.23)\times 10^{-2}$ \\
                &  + HPQCD    &        &         & $a_1/a_0 = -0.82 \pm 0.29 $     \\
                & (1 point)   &        &         & $a_2/a_0 = -1.14  \pm 1.81  $     \\
                &             &        &         & $\Vub = (2.99 \pm 0.35)\times 10^{-3}$ \\               \hline\hline
\end{tabular}
\label{tab:BGLfits}
\end{table*}

For the nominal fit we use the subset with four FNAL/MILC points
and assume a quadratic BGL parameterization. 
We refer to this fit as 3+1-parameter BGL fit
(three coefficients $a_k$ and the normalization parameter $a_{\rm norm}$).
As can be seen in Table~\ref{tab:FFfits} for the fit to data alone, the data are well 
described by a linear function with the normalization $a_0$ 
and a slope $a_1/a_0$.
This indicates that most of the variation of the form factor is due to well-understood
QCD effects that are parameterized by the functions ${\cal P}(q^2)$  and $\phi(q^2,q_0^2)$
in the BGL parameterization.    
If we include a curvature term in the fit, the slope $a_1/a_0=-0.82 \pm 0.29$ 
is fully consistent with the linear fit; the curvature $a_2/a_0$ is 
negative and consistent with zero. 
Since the $z$ distribution is almost linear, we also perform a linear fit (2+1-parameter BGL fit) 
for comparison. The results of the linear fits are also shown in Table~\ref{tab:BGLfits}.

The simultaneous fits provide very similar results, both for the BGL expansion
coefficients, which determine the shape of the spectrum, and for \Vub. 
The fitted values for the form-factor parameters are very similar to those obtained from the fits to data alone. This is not surprising, since the data dominate the fit results. Unfortunately the decay rate is lowest and the experimental errors are largest at large $q^2$,  
where the lattice calculation can make predictions.
We obtain from these simultaneous fits
\begin{eqnarray}
\Vub &=& (2.87 \pm 0.28)\times 10^{-3}  \quad {\rm FNAL/MILC~(6~points)} \ , \nonumber \\ 
\Vub &=& (2.95 \pm 0.31)\times 10^{-3}  \quad {\rm FNAL/MILC~(4~points)} \ , \nonumber  \\ 
\Vub &=& (2.93 \pm 0.31)\times 10^{-3}  \quad {\rm FNAL/MILC~(3~points)} \ , \nonumber  \\ 
\Vub &=& (2.92 \pm 0.37)\times 10^{-3}  \quad {\rm FNAL/MILC~(1~point)} \ , \nonumber  \\ 
\Vub &=& (2.99 \pm 0.35)\times 10^{-3}  \quad {\rm HPQCD~(1~point)} \ , \nonumber         
\end{eqnarray}
where the stated error is the combined experimental and theoretical 
error obtained from the fit. 
The coefficients $a_k$ are significantly smaller than 1, as predicted.  The sum of the squares of the first two coefficients, $\sum_{k=0}^{1} a_k^2=  (0.85 \pm 0.20) \times 10^{-3}$, is consistent with the tighter bounds set by Becher and Hill~\cite{BHill}. 

Since the total error of 10\% on \Vub\ results from the simultaneous fit to data and LQCD predictions, 
it is non-trivial to separate the error into contributions from experiment and theory.
We have estimated that the error contains contributions of $3\%$ from the branching-fraction measurement,
$5\%$ from the shape of the $q^2$ spectrum determined from data, and $8.5\%$ 
from the form-factor normalization obtained from theory.

We study the effect of variations of the isospin relations imposed in
the combined four-mode fit as stated in Eqs.~\ref{eq:isospin}.  
These relations are not expected to be exact, though the comparison of the single-mode fit results 
provides no indication for isospin breaking. 
The isospin-breaking effects are primarily due to $\piz-\eta$ and $\rho^0-\omega$ mixing in  \Bppizlnu\ and \Bprhozlnu\ decays, respectively.  They are expected to increase the branching fractions of the $B^+$ relative to the $B^0$ meson.  Given the masses and widths of the mesons involved, the impact of $\piz-\eta$ mixing is expected to be smaller than that of $\rho^0-\omega$ mixing. 

Detailed calculations have been performed to correct form-factor measurements and to extract $V_{us}$ from semileptonic decays of charged and neutral kaons~\cite{ckmbook}. These calculations account for isospin breaking due to $\piz-\eta$ mixing and should also be applicable to \Bppizlnu\ decays.  
For \Bppizlnu\ decays the effect is expected to be smaller by a factor of three, 
{\it i.e.}, the predicted increase is $(1.5\pm 0.2)$\%~\cite{vincenzo}. 
For \Bprhozlnu\ decays, 
calculations have not been carried out to the same precision. Based on 
the change in the $\pi^+\pi^-$ rate at the peak of the $\rho$ mass distribution, 
the branching fraction is predicted to increase by as much as 34\%~\cite{lopez}. 
However, an integration over the resonances weighted by the proper Breit-Wigner function and taking into account the masses and finite $\rho$ and $\omega$ widths results in a much smaller effect, an increase in the $\pi^+\pi^-$ branching fraction of 6\%~\cite{lange}.

We have assessed the impact of changes in the ratios of the branching fractions for charged and neutral $B$ mesons on the extraction of the differential decay rates due to adjustments of the MC default branching fractions of the $B^+$ decays in the combined 
four-mode fit. 
For a 1.5\% increase in the \Bppizlnu\ branching fraction, the fitted \Bpilnu\ partial branching fraction decreases by 0.5\%, while the \Brholnu\ rate increases by less than 0.1\%.
A 6\% increase in the \Bprhozlnu\ branching fraction results in a decrease of the  
\Brholnu\ rate by 3.1\% and a 0.14\% increase for the fitted \Bpilnu\ rate.
We observe a partial compensation to the change in the simulated \Bppizlnu\ rate due to
changes in the \Bprhozlnu\ background contribution, and vice versa. The observed changes
in the fitted yields depend linearly on the imposed branching-fraction changes and are independent of $q^2$. 

For a $1.5\%$  variation of the \Bppizlnu\ branching fraction, the value 
for \Vub\ extracted from the measured \Bpilnu\ spectrum decreases by $0.2\%$. 
A $+6\%$ variation of the \Bprhozlnu\ branching fraction 
increases the value of \Vub\ extracted from the same measured spectrum by $0.3\%$. 
\section{Conclusions}
\label{sec:conclusions}

In summary, we have measured the exclusive branching fractions
\BRBzpilnu\ and \BRBzrholnu\ as a function of $q^2$ and have
determined \Vub using recent form-factor calculations. 
We measure the total branching fractions, based on samples of charged and neutral 
$B$ mesons and isospin constraints, to be
\begin{eqnarray}
  \BRBzpilnu  &=& (1.41 \pm 0.05 \pm 0.07) \times 10^{-4} \ , \nonumber \\
  \BRBzrholnu &=& (1.75 \pm 0.15 \pm 0.27) \times 10^{-4} \ , \nonumber
\end{eqnarray}
where the first error is the statistical uncertainty of the fit employed 
to determine the signal and background yields and the second is the 
systematic uncertainty. 
The separate measurements of the branching fractions for charged and neutral 
$B$ mesons  are consistent within errors with the assumed isospin relations, 
\begin{eqnarray}
\frac{\BRBzpilnu}{\BRBppizlnu \times 2 \frac{\tau_0}{\tau_+}}
  &=& 1.03 \pm 0.09 \pm 0.06 \ , \nonumber \\
\frac{\BRBzrholnu}{\BRBprhozlnu \times 2 \frac{\tau_0}{\tau_+}}   
  &=& 1.06 \pm 0.16 \pm 0.08 \ .  \nonumber
\end{eqnarray}
We have assessed the sensitivity of the combined branching-fraction 
measurements to isospin violations due to $\piz - \eta$  
and $\rho^0 - \omega$ mixing in $B^+$ decays.  Based on the best estimates currently available,  the impact on the branching fractions is small compared to the total 
systematic errors.  We refrain from applying corrections, given the uncertainties 
in the size of the effects.

The measured branching fraction for \Bpilnu\ is more precise than any previous
measurement and agrees well with the current world average 
$\BRBzpilnu = (1.36 \pm 0.05 \pm 0.05) \times 10^{-4}$~\cite{HFAG2009}. 
The branching fraction for \Brholnu\ is also the most precise single
measurement to date based on a large signal event sample, although the 
Belle Collaboration~\cite{pilnu_belle_tag} has reported a smaller 
systematic error (by a factor of two),  based 
on a small signal sample of hadronically-tagged events~\cite{pilnu_belle_tag}.
The \Brholnu\ branching fraction presented here is significantly lower 
(by about $2.5~\sigma$) compared to the current world average 
$\BRBzrholnu = (2.77 \pm 0.18 \pm 0.16) \times 10^{-4}$~\cite{HFAG2009}.
The dominant uncertainty of this \Brholnu\ measurement is due to 
the limited knowledge of the normalization and shape of the
irreducible background from other \BXulnu\ decays.  
 
Within the sizable errors, the measured $q^2$ spectrum for \Brholnu\ agrees well with the predictions from light-cone sum rules~\cite{ball_rho} and the ISGW2~\cite{isgw2} quark model. Neither of these calculations includes an estimate of their uncertainties. 
In the future, it will require much cleaner data samples and considerably better understanding of other \bulnu\ decays to achieve significant improvements in the measurements of the form factors in $B$ decays to vector mesons. 

For \Bpilnu\ decays, the measured $q^2$ spectrum agrees best with the one predicted
by the HPQCD lattice calculations~\cite{hpqcd04}.
The measurement of the differential decay rates is consistent with earlier \babar\ measurements~\cite{pilnu_jochen,pilnu_cote} within the stated errors, 
though the yield at low $q^2$ is somewhat higher than previously measured.
This results in a smaller value of $\alpha_{BK}$,
the parameter introduced by Becirevic and Kaidalov~\cite{BK}, namely $\alpha_{BK}=0.31 \pm 0.09$.  Using the BGL ansatz, we determine a value 
$f_+(0)\Vub = (1.08 \pm 0.06)\times 10^{-3}$, which is larger than the value, $f_+(0)\Vub =(0.91 \pm 0.06\pm 0.3)\times 10^{-3}$~\cite{ball07}, based on the earlier 
\babar\ decay rate measurement~\cite{pilnu_cote} and an average branching fraction of $(1.37 \pm 0.06 \pm 0.07)\times 10^{-4}$~\cite{HFAG2009}.

We determine the CKM matrix element \Vub using two different approaches.
First, we use the traditional method to derive \Vub\ by combining the measured partial branching fractions with
the form-factor predictions based on different QCD calculations.  
The results, presented in Table~\ref{tab:vub}, agree within the sizable uncertainties of 
the form-factor predictions. 
For this approach we quote as a result the value of
\begin{equation}
  \Vub = (3.78 \pm 0.13 ^{+0.55}_{-0.40})\times 10^{-3}  \nonumber ,
\end{equation}
based on the most recent LCSR calculation for $q^2 < 12 \gev^2$.
Second, we extract \Vub from simultaneous fits to data and lattice predictions using the quadratic BGL parameterization 
for the whole $q^2$ range. These fits to data and the two most recent lattice calculations by the
FNAL/MILC~\cite{fnal09} and HPQCD~\cite{hpqcd04} Collaborations agree very well. We quote as a result the fitted value of
\begin{equation}
  \Vub = (2.95 \pm 0.31)\times 10^{-3}  \nonumber,
\end{equation}
based on the normalization predicted by the FNAL/MILC Collaboration.
The total error of 10\% is dominated by the theory error of 8.5\%.
This value of \Vub\ is smaller by one standard deviation compared to the results of a combined fit to earlier \babar\ measurements 
and the same recent FNAL/MILC lattice calculations~\cite{fnal09}.

The values of \Vub\ presented here appear to be sensitive to the $q^2$ range for which theory predictions and the 
measured spectrum can be compared. LCSR calculations are restricted to low values of $q^2$ and result 
in values of \Vub\ in the range of $(3.63 - 3.78) \times 10^{-3}$ with theoretical uncertainties 
of $^{+16}_{-11}\%$ and experimental errors of $3-4\%$.
LQCD predictions are available for $q^2 > 16 \gev^2$ and result in \Vub\ in the range
of $(2.95 - 3.21) \times 10^{-3}$ and experimental errors of $5-6\%$ 
for both the traditional method 
and the simultaneous fit to LQCD predictions and the measured spectrum.  
This fit combines the measured shape of the spectrum over the full $q^2$ range 
with the lattice QCD form-factor predictions at high $q^2$ and results in a 
reduced theoretical uncertainty of $8.5\%$, 
as compared to $^{+17}_{-11}\%$ for the traditional method. 

Both \Vub\ values quoted as results are also lower than most determinations of \Vub\ based on inclusive \bulnu\ decays, 
which are typically in the range $(4.0 - 4.5)\times 10^{-3}$. 
These inclusive measurements are very sensitive to the mass of the $b$ quark, which is extracted from fits to moments of 
inclusive \bclnu\ and $B \to X_s \gamma$ decay distributions~\cite{moments} and depends on higher-order QCD corrections.
Estimated theoretical uncertainties are typically $6\%$.

Global fits constraining the parameters of the CKM unitarity triangle performed 
by the CKMfitter~\cite{CKMfitter} and UTfit~\cite{UTfit} Collaborations currently predict 
values for \Vub\ that fall between the two results presented here,
$\Vub = 3.51^{+0.14}_{-0.16} \times 10^{-3}$  and $\Vub = 3.41\pm 0.18 \times 10^{-3}$,
respectively.

To permit more stringent tests of the CKM framework and its consistency with the standard model of electroweak interactions, further reductions in the experimental and theoretical uncertainties will be necessary.  For \Bpilnu\ decays this will require a reduction in the statistical errors and improved detector hermeticity to more effectively reconstruct the neutrino, which will reduce backgrounds from all sources. Further improvements in the precision of lattice and other QCD calculations will also be beneficial.  

\section{Acknowledgments}

We would like to thank A. Khodjamirian, A. Kronfeld, P. Mackenzie, T. Mannel, J. Shigemitsu, and R. Van de Water for their help with theoretical form-factor calculations. 
We are grateful for the 
extraordinary contributions of our \pep2\ colleagues in
achieving the excellent luminosity and machine conditions
that have made this work possible.
The success of this project also relies critically on the 
expertise and dedication of the computing organizations that 
support \babar.
The collaborating institutions wish to thank 
SLAC for its support and the kind hospitality extended to them. 
This work is supported by the
US Department of Energy
and National Science Foundation, the
Natural Sciences and Engineering Research Council (Canada),
the Commissariat \`a l'Energie Atomique and
Institut National de Physique Nucl\'eaire et de Physique des Particules
(France), the
Bundesministerium f\"ur Bildung und Forschung and
Deutsche Forschungsgemeinschaft
(Germany), the
Istituto Nazionale di Fisica Nucleare (Italy),
the Foundation for Fundamental Research on Matter (The Netherlands),
the Research Council of Norway, the
Ministry of Education and Science of the Russian Federation, 
Ministerio de Ciencia e Innovaci\'on (Spain), and the
Science and Technology Facilities Council (United Kingdom).
Individuals have received support from 
the Marie-Curie IEF program (European Union), the A. P. Sloan Foundation (USA) 
and the Binational Science Foundation (USA-Israel).



\clearpage
\section{Appendix}

\subsection{Systematic Uncertainties for One-Mode Fits}
\label{sec:appendix_systTable}
\begin{table*}[hbt]
\centering
\caption{Systematic errors in \%  for \BRBzpilnu\ from the \Bzpilnu\ and \Bppizlnu\ one-mode fits.}

\begin{tabular}{lccccccc|ccccccc} \hline\hline

                        & \multicolumn{7}{c}{\boldmath \Bzpilnu}
                        & \multicolumn{7}{c}{\boldmath \Bppizlnu} \\ \hline
$q^2$ range ($\gev^2$)  & 0-4 & 4-8 & 8-12 &12-16&16-20&$>$20& 0-26.4
                        & 0-4 & 4-8 & 8-12 &12-16&16-20&$>$20& 0-26.4  \\ \hline
Track efficiency        & 2.0  &  1.7  &  2.9  &  1.3  &  0.1  &  2.3  & 1.7 
                        & 7.3  &  1.8  &  3.3  &  1.6  &  1.8  &  6.8  & 3.7   \\
Photon efficiency       & 0.7  &  0.9  &  1.9  &  3.5  &  0.7  &  1.0  & 1.6 
                        & 2.9  &  1.8  &  5.3  &  2.4  &  8.9  &  10.2  & 4.8    \\ \hline

Lepton identification   & 4.1  &  1.7  &  1.8  &  1.8  &  2.1  &  2.4  & 1.8 
                        & 3.3  &  1.3  &  2.2  &  1.4  &  2.7  &  2.8  & 1.4        \\ \hline

$K_L$ efficiency        & 1.0  &  0.1  &  0.3  &  3.5  &  1.2  &  1.7  & 1.2 
                        & 1.2  &  0.5  &  2.4  &  3.6  &  1.8  &  1.1  & 1.8   \\
$K_L$ shower energy     & 0.1  &  0.1  &  0.1  &  0.7  &  1.4  &  2.8  & 0.7 
                        & 2.6  &  0.5  &  0.4  &  0.6  &  1.4  &  4.1  & 1.4   \\
$K_L$ spectrum          & 1.1  &  2.0  &  2.5  &  2.6  &  5.1  &  1.1  & 2.4 
                        & 2.7  &  0.7  &  2.0  &  4.5  &  4.4  &  4.2  & 2.9    \\ \hline

$\Bpilnu FF$ $f_+$      & 0.5  &  0.3  &  0.4  &  0.5  &  0.6  &  0.8  & 0.5 
                        & 0.5  &  0.5  &  0.6  &  0.7  &  1.8  &  2.6  & 1.0   \\ 
$\Brholnu FF A_1$       & 1.6  &  1.8  &  2.0  &  0.9  &  2.0  &  1.5  & 1.6 
                        & 2.7  &  2.2  &  3.6  &  2.0  &  1.6  &  5.1  & 2.8   \\
$\Brholnu FF A_2$       & 1.5  &  1.7  &  1.7  &  0.5  &  2.1  &  0.6  & 1.4 
                        & 2.5  &  2.0  &  3.0  &  1.4  &  1.9  &  2.7  & 2.2   \\
$\Brholnu FF V$         & 0.3  &  0.4  &  0.5  &  0.3  &  0.4  &  0.8  & 0.4 
                        & 1.1  &  0.7  &  1.2  &  1.0  &  0.9  &  3.0  & 1.2     \\ \hline

$\BRBzrholnu $          & 0.3  &  0.2  &  0.5  &  0.8  &  0.7  &  0.4  & 0.5 
                        & 0.4  &  0.2  &  0.3  &  0.4  &  0.9  &  1.6  & 0.5   \\
$\BRBpomlnu$            & 0.1  &  0.1  &  0.1  &  0.1  &  0.3  &  1.4  & 0.2 
                        & 0.6  &  0.3  &  0.6  &  0.6  &  0.9  &  4.8  & 1.1   \\
$\BRBpetalnu$           & 0.1  &  0.1  &  0.1  &  0.1  &  0.1  &  0.1  & 0.1 
                        & 0.1  &  0.1  &  0.1  &  0.1  &  0.5  &  0.6  & 0.2   \\
$\BRBpetaplnu$          & 0.1  &  0.1  &  0.1  &  0.1  &  0.1  &  0.3  & 0.1 
                        & 0.1  &  0.1  &  0.1  &  0.1  &  0.4  &  0.7  & 0.2   \\
$\BRXulnu$              & 0.1  &  0.1  &  0.1  &  0.1  &  0.3  &  0.9  & 0.2 
                        & 0.7  &  0.2  &  0.3  &  0.4  &  1.2  &  1.4  & 0.6   \\
$\BXulnu$ SF param.     & 0.4  &  0.2  &  0.1  &  0.4  &  0.6  &  2.3  & 0.5 
                        & 0.6  &  0.3  &  0.4  &  1.1  &  1.3  &  6.2  & 1.3   \\ \hline

$\B \to D\ell\nu$ FF $\rho^{2}_{D}$     & 0.1  &  0.1  &  0.4  &  0.3  &  0.1  &  0.3  & 0.2
                                        & 0.4  &  0.1  &  0.6  &  0.3  &  0.2  &  0.8  & 0.4   \\ 
$\B \to D^*\ell\nu$ FF $R_1$            & 0.2  &  0.4  &  0.8  &  0.9  &  0.3  &  0.9  & 0.6
                                        & 0.1  &  0.5  &  1.3  &  0.5  &  0.7  &  1.8  & 0.7   \\
$\B \to D^*\ell\nu$ FF $R_2$            & 0.5  &  0.1  &  0.2  &  0.1  &  0.1  &  0.2  & 0.2
                                        & 0.5  &  0.3  &  0.6  &  0.5  &  0.9  &  1.4  & 0.6   \\
$\B \to D^*\ell\nu$ FF $\rho^{2}_{D^*}$ & 0.9  &  0.4  &  0.5  &  1.0  &  0.2  &  0.8  & 0.6
                                        & 0.7  &  0.4  &  0.6  &  0.6  &  0.6  &  2.4  & 0.8   \\

$\BRDlnu $           & 0.2  &  0.1  &  0.2  &  0.3  &  0.4  &  0.1  & 0.2  
                     & 1.1  &  0.3  &  0.8  &  0.3  &  0.3  &  0.9  & 0.6 \\
$\BRDslnu $          & 0.5  &  0.1  &  0.3  &  0.4  &  0.3  &  0.3  & 0.3  
                     & 0.9  &  0.4  &  0.6  &  0.7  &  1.7  &  1.8  & 0.9 \\
$\BRDssnlnu $        & 0.3  &  0.1  &  0.1  &  0.4  &  0.1  &  0.1  & 0.2  
                     & 0.2  &  0.2  &  0.3  &  0.3  &  0.2  &  0.2  & 0.2 \\
$\BRDssblnu $        & 0.2  &  0.1  &  0.1  &  0.5  &  0.1  &  0.2  & 0.2  
                     & 0.3  &  0.3  &  0.3  &  0.4  &  0.1  &  0.6  & 0.3 \\
Secondary leptons    & 0.8  &  0.1  &  0.2  &  0.1  &  0.1  &  0.2  & 0.2  
                     & 0.5  &  0.2  &  0.3  &  0.4  &  0.8  &  0.9  & 0.5 \\ \hline 
Continuum            & 5.0  &  1.6  &  1.6  &  1.3  &  1.8  &  4.1  & 1.7  
                     & 8.8  &  1.8  &  6.1  &  4.0  &  4.8  &  13.1 & 4.2    \\ \hline

Bremsstrahlung       & 0.2  &  0.1  &  0.1  &  0.1  &  0.1  &  0.2  & 0.1  
                     & 0.2  &  0  &  0  &  0.1  &  0.2  &  0.2  & 0.1   \\
Radiative corrections& 0.4  &  0.1  &  0.1  &  0.3  &  0.5  &  0.9  & 0.3  
                     & 0.5  &  0.1  &  0.1  &  0.3  &  0.5  &  0.9  & 0.4   \\ \hline

$N_{\BB}$            & 1.1  &  1.2  &  1.1  &  1.2  &  1.2  &  1.0  & 1.1  
                     & 1.2  &  1.1  &  1.2  &  1.2  &  1.1  &  1.2  & 1.2   \\
$f_\pm / f_{00}$     & 1.5  &  1.3  &  1.3  &  1.3  &  1.0  &  0.8  & 1.3  
                     & 1.3  &  1.1  &  1.0  &  1.0  &  0.9  &  0.2  & 1.0   \\  \hline
Total                & 7.8  &  4.8  &  6.0  &  6.8  &  7.1  &  7.6  &  4.4 
                     & 13.7  &  5.0  &  11.1  &  8.6  &  12.7  &  22.4  & 7.5 \\ \hline\hline
\end{tabular}
\label{tab:syst_pi_oneMode}
\end{table*}

\begin{table*}[hbt]
\centering
\caption{Systematic errors in \%  for \BRBzrholnu\ from the \Bzrholnu\ and \Bprhozlnu\ one-mode fits.}

\begin{tabular}{lcccc|cccc} \hline\hline

                        & \multicolumn{4}{c}{\boldmath \Bzrholnu}
                        & \multicolumn{4}{c}{\boldmath \Bprhozlnu} \\ \hline
$q^2$ range ($\gev^2$)  & 0-8 & 8-16 & $>16$ & 0-20.3 
                        & 0-8 & 8-16 & $>16$ & 0-20.3 \\ \hline
Track efficiency        & 2.4  &  0.8  &  3.0  & 1.8  
                        & 2.4  &  4.0  &  1.0  & 2.9   \\
Photon efficiency       & 1.4  &  1.5  &  3.9  & 2.0  
                        & 5.5  &  3.5  &  3.5  & 4.2    \\ \hline
Lepton Identification   & 2.6  &  2.9  &  4.7  & 3.0   
                        & 4.3  &  3.2  &  4.6  & 3.7   \\ \hline

$K_L$ efficiency	& 8.2  &  1.0  &  6.5  & 4.3  
                        & 10.1  &  1.8  &  5.8 & 5.2    \\
$K_L$ shower energy     & 4.4  &  0.2  &  0.4  & 1.6  
                        & 1.0  &  0.7  &  0.8  & 0.8   \\
$K_L$ spectrum          & 9.3  &  8.2  &  9.7  & 8.8  
                        & 1.0  &  5.2  &  7.3  & 4.3   \\ \hline

$\Bpilnu$  FF $f_+$     & 1.2  &  0.3  &  2.8  & 1.1  
                        & 0.8  &  0.2  &  2.4  & 0.8   \\
$\Brholnu$ FF $A_1$      & 14.8  &  8.3  &  4.8  & 9.6 
                        & 14.3  &  8.3  &  4.3  & 9.4    \\
$\Brholnu$ FF $A_2$      & 11.3  &  5.1  &  0.8  & 6.2 
                        & 11.0  &  5.0  &  0.6  & 6.0    \\
$\Brholnu$ FF $V$       & 3.7  &  3.5  &  3.9  & 3.6  
                        & 3.8  &  3.6  &  3.5  & 3.6   \\ \hline

$\BRBzpilnu$            & 0.8  &  0.6  &  1.8  & 0.9  
                        & 0.6  &  0.1  &  1.5  & 0.5    \\
$\BRBpomlnu$            & 0.1  &  0.4  &  3.3  & 0.8  
                        & 0.9  &  0.9  &  3.3  & 1.4   \\
$\BRBpetalnu$           & 0.1  &  0.1  &  0.8  & 0.2  
                        & 0.1  &  0.1  &  0.6  & 0.1   \\
$\BRBpetaplnu$          & 0.1  &  0.1  &  1.2  & 0.3  
                        & 1.2  &  0.8  &  1.1  & 1.0   \\
$\BRXulnu$              & 4.0  &  5.3  &  10.3  & 5.9 
                        & 7.9  &  8.0  &  11.0  & 8.6   \\
$\BXulnu$ SF param.     & 6.9  &  6.5  &  13.1  & 7.9 
                        & 10.9  &  7.8  &  12.7  & 9.7  \\ \hline

$B\to D \ell\nu$ FF $\rho^2_{D}$     & 0.1  &  0.2  &  0.1  & 0.1
                                     & 0.6  &  0.1  &  0.1  & 0.2  \\ 
$B\to D^* \ell\nu $FF $R_1$          & 0.1  &  0.1  &  0.2  & 0.1 
                                     & 0.1  &  0.1  &  0.1  & 0.1   \\
$B\to D^* \ell\nu $FF $R_2$          & 0.2  &  0.1  &  0.3  & 0.2 
                                     & 0.6  &  0.1  &  0.2  & 0.3  \\
$B\to D^* \ell\nu $FF $\rho^2_{D^*}$ & 0.4  &  0.2  &  0.1  & 0.2 
                                     & 1.4  &  0.1  &  0.2  & 0.5  \\

$\BRDlnu $              & 0.6  &  0.3  &  0.2  & 0.4 
                        & 0.4  &  0.2  &  0.2  & 0.3  \\
$\BRDslnu $             & 0.3  &  0.1  &  0.2  & 0.2 
                        & 0.2  &  0.1  &  0.2  & 0.2  \\
$\BRDssnlnu $           & 0.7  &  0.1  &  0.1  & 0.2 
                        & 0.6  &  0.1  &  0.1  & 0.2  \\
$\BRDssblnu $           & 0.9  &  0.1  &  0.1  & 0.3 
                        & 0.9  &  0.1  &  0.1  & 0.4  \\

Secondary leptons       & 0.5  &  0.1  &  0.2  & 0.2 
                        & 0.7  &  0.1  &  0.1  & 0.2  \\ \hline
Continuum               & 5.9  &  3.4  &  6.3  & 3.2 
                        & 8.2  &  3.8  &  6.4  & 3.4  \\ \hline

Bremsstrahlung          & 0.2  &  0.1  &  0.2  & 0.2 
                        & 0.2  &  0.1  &  0.2  & 0.2   \\
Radiative corrections   & 0.6  &  0.1  &  0.8  & 0.4 
                        & 0.6  &  0.1  &  0.8  & 0.4   \\ \hline

$N_{\BB}$               & 1.7  &  1.9  &  2.9  & 2.0 
                        & 2.2  &  2.1  &  2.9  & 2.3  \\
$f_\pm / f_{00}$        & 1.9  &  1.4  &  1.0  & 1.5 
                        & 1.6  &  1.4  &  0.7  & 1.3  \\ \hline
Total                   & 25.6  &  16.5  &  23.9  & 19.4  
                        & 27.6  &  18.0  &  22.5  & 17.1   \\ \hline\hline

\end{tabular}
\label{tab:syst_rho_oneMode}
\end{table*}

\clearpage

\subsection{Correlation and Covariance Matrices}
\label{sec:appendix_corrMat}

Table~\ref{tab:fullCorrMat} shows the full correlation matrix
for all signal and background fit parameters in the four-mode maximum-likelihood fit
used to determine the signal yields, described in Section~\ref{sec:fit}.
This appendix also contains all statistical, systematic and total
correlation and covariance matrices for the 
\Bpilnu and \Brholnu $\Delta B/\Delta q^2$ measurements. 
The total correlation matrix is shown before and after 
unfolding of the $q^2$ spectrum. 
All covariance matrices are shown after $q^2$ unfolding.
The total covariance matrix for \Bpilnu in Table~\ref{tab:CovMatTotal_Unfolded_pi}
is used in the form-factor fits described in 
Eq.~\ref{eq:chi2} or~\ref{eq:chi2_Vubfit}.

\def\pDs{\ensuremath{D^{*}\ell\nu}\xspace}
\def\pulnu{\ensuremath{u\ell\nu}\xspace}
\def\ppilnu{\ensuremath{\pi\ell\nu}\xspace}
\def\prholnu{\ensuremath{\rho\ell\nu}\xspace}
\def\pip{\ensuremath{\pi^\pm}\xspace}
\def\piz{\ensuremath{\pi^0}\xspace}
\def\rhop{\ensuremath{\rho^\pm}\xspace}
\def\rhoz{\ensuremath{\rho^0}\xspace}

\begin{sidewaystable*}
\caption{Full correlation matrix of the four-mode maximum-likelihood fit, for all fit parameters (signal and background). 
The first column indicates the fit parameter that
corresponds to a certain row/column of the matrix.}
\scriptsize
\renewcommand{\arraystretch}{1.8}
\begin{tabular}{r|rrrrrrrrrrrrrrrrrrrrrrrrr} \hline\hline

                         &    1  &    2  &    3  &    4  &    5  &    6  &    7  &    8  &    9  &   10  &   11  &   12  &   13  &   14   &  15   &  16   
                         &   17  &   18  &   19  &   20  &   21  &   22  &   23  &   24  &   25  \\ \hline  
$p^{\qq}_{\pip}$        1&  1.000&  	 &	 &	 &	 &	 &	 &	 &	 &	 &	 &	 &	 &	  &	  &       
                         &       &  	 &	 &	 &	 &	 &	 &	 &	 \\
$p^{\pDs}_{\pip}$       2&  0.147&  1.000& 	 &	 &	 &       &	 &	 &	 &	 & 	 &       &       &	  &	  &
                         &       &  	 &	 &	 &	 &	 &	 &	 &	 \\
$p^{other\BB}_{\pip}$   3& -0.654& -0.449&  1.000& 	 &	 &	 &	 &	 &	 &	 &	 &	 &	 &	  &	  &
                         &       &  	 &	 &	 &	 &	 &	 &	 &	 \\
$p^{\pulnu}_{\pip,1}$   4& -0.096& -0.449& -0.225&  1.000&  	 &	 &	 &	 &	 &	 &	 &	 &	 &	  &	  &
                         &       &  	 &	 &	 &	 &	 &	 &	 &	 \\
$p^{\pulnu}_{\pip,2}$   5& -0.175& -0.119&  0.033&  0.214&  1.000& 	 &	 &	 &	 &	 &	 &	 &	 &	  &	  &
                         &       &  	 &	 &	 &	 &	 &	 &	 &	 \\
$p^{\prholnu}_{1}$      6&  0.021&  0.197& -0.148& -0.158& -0.080&  1.000&  	 &	 &	 &	 &	 &	 &	 &	  &	  &
                         &       &  	 &	 &	 &	 &	 &	 &	 &	 \\
$p^{\prholnu}_{2}$      7&  0.055&  0.060& -0.066& -0.158& -0.111&  0.264&  1.000&  	 &	 &	 &	 &	 &	 &	  &	  &
                         &       &  	 &	 &	 &	 &	 &	 &	 &	 \\
$p^{\prholnu}_{3}$      8&  0.033&  0.002& -0.015& -0.079& -0.564&  0.137&  0.189&  1.000& 	 &	 &	 &	 &	 &	  &	  &
                         &       &  	 &	 &	 &	 &	 &	 &	 &	 \\
$p^{\ppilnu}_{1}$       9& -0.263&  0.134&  0.030& -0.013&  0.044& -0.144& -0.033& -0.019&  1.000&  	 &	 &	 &	 &	  &	  &
                         &       &  	 &	 &	 &	 &	 &	 &	 &	 \\
$p^{\ppilnu}_{2}$      10& -0.105&  0.126& -0.007& -0.120&  0.032& -0.429& -0.050& -0.033&  0.191&  1.000&  	 &	 &	 &	  &	  &
                         &       &  	 &	 &	 &	 &	 &	 &	 &	 \\
$p^{\ppilnu}_{3}$      11&  0.086&  0.150& -0.110& -0.271& -0.044&  0.085& -0.156&  0.024&  0.050&  0.089&  1.000&  	 &	 &	  &	  &
                         &       &  	 &	 &	 &	 &	 &	 &	 &	 \\
$p^{\ppilnu}_{4}$      12&  0.004& -0.073&  0.110& -0.232& -0.025& -0.001& -0.267&  0.026& -0.005&  0.058&  0.197&  1.000&  	 &	  &	  &
                         &       &  	 &	 &	 &	 &	 &	 &	 &	 \\
$p^{\ppilnu}_{5}$      13& -0.162&  0.116&  0.244& -0.432&  0.216&  0.061&  0.059& -0.469&  0.068&  0.085&  0.127&  0.135&  1.000&  	  &	  &
                         &       &  	 &	 &	 &	 &	 &	 &	 &	 \\
$p^{\ppilnu}_{6}$      14& -0.124&  0.036&  0.079& -0.033& -0.602&  0.007& -0.005&  0.035&  0.057&  0.011&  0.005& -0.008&  0.032&  1.000 &	  &
                         &       &  	 &	 &	 &	 &	 &	 &	 &	 \\
$p^{\qq}_{\piz}$       15&  0.055& -0.015& -0.049&  0.029&  0.103&  0.031&  0.037& -0.020& -0.119&  0.059& -0.050&  0.056& -0.079& -0.163 & 1.000 &
                         &       &  	 &	 &	 &	 &	 &	 &	 &	 \\
$p^{\pDs}_{\piz}$      16& -0.075&  0.104& -0.009& -0.082& -0.043&  0.128& -0.019&  0.024&  0.156&  0.109&  0.086& -0.021&  0.095&  0.073 &-0.193 & 1.000
                         &       &  	 &	 &	 &	 &	 &	 &	 &	 \\
$p^{other\BB}_{ \piz}$ 17&  0.003& -0.086&  0.092&  0.007& -0.082& -0.143& -0.039&  0.059& -0.064& -0.127& -0.078&  0.033&  0.031&  0.067 &-0.438 &-0.532
                         &  1.000&  	 &	 &	 &	 &	 &	 &	 &	 \\ 
$p^{\pulnu}_{\piz,1}$  18&  0.024& -0.052& -0.045&  0.200&  0.098& -0.118& -0.119& -0.198&  0.006& -0.094& -0.078& -0.177& -0.142&  0.028 &-0.445 &-0.235
                         &  0.018&  1.000&  	 &       &	 &	 &	 &	 &       \\	 
$p^{\pulnu}_{\piz,2}$  19&  0.022& -0.002& -0.033&  0.066&  0.557& -0.077& -0.104& -0.617&  0.034&  0.003&  0.020& -0.052&  0.259& -0.377 &-0.237 &-0.020
                         & -0.089&  0.431&  1.000&  	 &	 &	 &	 &	 &       \\	 
$p^{\qq}_{\rhop}$      20& -0.006& -0.022& -0.002&  0.084&  0.015& -0.066& -0.230& -0.019&  0.011& -0.013&  0.032& -0.007& -0.091&  0.001 &-0.017 &-0.001
                         &  0.009&  0.060&  0.017&  1.000& 	 &	 &	 &	 &       \\	 
$p^{\pDs}_{\rhop}$     21& -0.004&  0.113& -0.071& -0.088& -0.002&  0.528&  0.069& -0.000& -0.063& -0.201&  0.076&  0.023&  0.077&  0.005 & 0.011 & 0.080
                         & -0.085& -0.048&  0.005& -0.145&  1.000& 	 &	 &	 &       \\	 
$p^{other\BB}_{\rhop}$ 22&  0.002& -0.109&  0.072&  0.080&  0.064& -0.537& -0.069& -0.111&  0.061&  0.216& -0.085& -0.006& -0.006& -0.010 &-0.001 &-0.082
                         &  0.076&  0.055&  0.061& -0.514& -0.671&  1.000& 	 &	 &       \\	 
$p^{\qq}_{\rhoz}$      23& -0.004& -0.005& -0.025&  0.106&  0.040&  0.013& -0.353& -0.058& -0.012& -0.049&  0.043&  0.015& -0.103& -0.001 &-0.006 & 0.015
                         & -0.011&  0.068&  0.038&  0.086&  0.027& -0.028&  1.000& 	 &       \\	 
$p^{\pDs}_{\rhoz}$     24&  0.012&  0.116& -0.093& -0.079&  0.020&  0.567&  0.156& -0.036& -0.082& -0.232&  0.041& -0.007&  0.075& -0.002 & 0.026 & 0.071
                         & -0.097& -0.045&  0.027& -0.044&  0.304& -0.294& -0.064&  1.000& 	 \\
$p^{other\BB}_{\rhoz}$ 25& -0.013& -0.119&  0.103&  0.057&  0.022& -0.598& -0.067& -0.043&  0.088&  0.257& -0.058&  0.008& -0.005& -0.002 &-0.021 &-0.080
                         &  0.096&  0.044&  0.021&  0.021& -0.319&  0.319& -0.472& -0.781&  1.000\\ 

\hline \hline

\end{tabular}
\normalsize
\label{tab:fullCorrMat}
\end{sidewaystable*}
\clearpage

\begin{table}[hbt]
\centering
\caption{Statistical (fit) correlation matrix of the \Bpilnu\ $\Delta B/\Delta q^2$ measurement for the four-mode fit.}

\begin{tabular}{lrrrrrr} \hline \hline

$q^2$ range ($\gev^2$)  & 0-4 & 4-8 & 8-12 &12-16&16-20&$>$20 \\ \hline  

 0-4    &   1.000  &    0.191    &    0.050  & -0.005 &    0.068 &  0.057 \\
 4-8    &   	   &    1.000    &    0.089  &  0.058 &    0.085 &  0.011 \\
 8-12   &          &    	 &    1.000  &  0.197 &    0.127 &  0.005 \\
 12-16  &  	   &   		 &    	     &  1.000 &    0.135 & -0.008 \\
 16-20  &  	   &    	 &           &        &    1.000 &  0.032 \\
 $>$20  &   	   &    	 &           &        &   	 &  1.000 \\
\hline\hline
\end{tabular}

\label{tab:CorrMatStat_pi}
\end{table}

\begin{table}[hbt]
\centering
\caption{Systematic correlation matrix of the \Bpilnu\ $\Delta B/\Delta q^2$ measurement for the four-mode fit.}

\begin{tabular}{lrrrrrr} \hline \hline

$q^2$ range ($\gev^2$)  & 0-4 & 4-8 & 8-12 &12-16&16-20&$>$20 \\ \hline  

 0-4    & 1.000  &  0.521  &   0.705  &   0.394  &  -0.052  &   0.075 \\
 4-8    &        &  1.000  &   0.853  &   0.687  &   0.605  &   0.478 \\
 8-12   &        &         &   1.000  &   0.652  &   0.366  &   0.439 \\
 12-16  &        &         &          &   1.000  &   0.637  &   0.367 \\
 16-20  &        &         &          &          &   1.000  &   0.509 \\
 $>$20  &        &         &          &          &          &   1.000 \\

\hline \hline

\end{tabular}

\label{tab:CorrMatSyst_pi}
\end{table}
                          
\begin{table}[hbt]
\centering

\caption{Total correlation matrix of the \Bpilnu\ $\Delta B/\Delta q^2$ measurement for the four-mode fit.}

\begin{tabular}{lrrrrrr} \hline \hline

$q^2$ range ($\gev^2$)  & 0-4 & 4-8 & 8-12 &12-16&16-20&$>$20 \\ \hline  

 0-4    & 1.000 &   0.337  &  0.401 &   0.212  &  0.015 &   0.066 \\
 4-8    &       &   1.000  &  0.430 &   0.343  &  0.272 &   0.205 \\
 8-12   &       &          &  1.000 &   0.443  &  0.227 &   0.219 \\
 12-16  &       &          &        &   1.000  &  0.350 &   0.180 \\
 16-20  &       &          &        &          &  1.000 &   0.221 \\
 $>$20  &       &          &        &          &        &   1.000 \\

\hline \hline

\end{tabular}

\label{tab:CorrMatTotal_pi}
\end{table}

\begin{table}[hbt]
\centering

\caption{Total correlation matrix of the \Bpilnu\ $\Delta B/\Delta q^2$ measurement for the four-mode fit
         after unfolding of the $q^2$ spectrum.}

\begin{tabular}{lrrrrrr} \hline \hline

$q^2$ range ($\gev^2$)  & 0-4 & 4-8 & 8-12 &12-16&16-20&$>$20 \\ \hline  

 0-4    &   1.000  &    0.272  &    0.331  &    0.216 &   -0.037  &    0.045 \\
 4-8    &          &    1.000  &    0.390  &    0.273 &    0.252  &    0.172 \\
 8-12   &          &           &    1.000  &    0.475 &    0.194  &    0.170 \\
 12-16  &          &           &           &    1.000 &    0.462  &    0.042 \\
 16-20  &          &           &           &          &    1.000  &    0.195 \\
 $>$20  &          &           &           &          &           &    1.000 \\

\hline \hline

\end{tabular}

\label{tab:CorrMatTotal_Unfolded_pi}
\end{table}


\begin{table}[hbt]
\centering

\caption{Statistical (fit) correlation matrix of the \Brholnu\ $\Delta B/\Delta q^2$ measurement for the four-mode fit.}

\begin{tabular}{lrrr} \hline \hline

$q^2$ range ($\gev^2$)  & 0-8 & 8-16 & $>$16 \\ \hline  

 0-8    &  1.000  & 0.264  &  0.137   \\
 8-16   &  	  & 1.000  &  0.189   \\
 $>$16  & 	  & 	   &  1.000   \\
\hline \hline

\end{tabular}

\label{tab:CorrMatStat_rho}
\end{table}
\begin{table}[hbt]
\centering

\caption{Systematic correlation matrix of the \Brholnu\ $\Delta B/\Delta q^2$ measurement for the four-mode fit.}

\begin{tabular}{lrrr} \hline \hline

$q^2$ range ($\gev^2$)  & 0-8 & 8-16 & $>$16 \\ \hline  

 0-8    & 1.000  & 0.339  & 0.692   \\
 8-16   & 	 & 1.000  & 0.296   \\
 $>$16  & 	 & 	  & 1.000   \\

\hline \hline
\end{tabular}

\label{tab:CorrMatSyst_rho}
\end{table}
\begin{table}[hbt]
\centering

\caption{Total correlation matrix of the \Brholnu\ $\Delta B/\Delta q^2$ measurement for the four-mode fit.}

\begin{tabular}{lrrr} \hline \hline

$q^2$ range ($\gev^2$)  & 0-8 & 8-16 & $>$16 \\ \hline  

 0-8    & 1.000  & 0.307  & 0.532   \\
 8-16   &        & 1.000  & 0.281   \\
 $>$16  &        & 	  & 1.000   \\

\hline \hline

\end{tabular}

\label{tab:CorrMatTotal_rho}
\end{table}
\begin{table}[hbt]
\centering

\caption{Total correlation matrix of the \Brholnu\ $\Delta B/\Delta q^2$ measurement for the four-mode fit 
         after unfolding of the $q^2$ spectrum.}

\begin{tabular}{lrrr} \hline \hline

$q^2$ range ($\gev^2$)  & 0-8 & 8-16 & $>$16 \\ \hline  

 0-8    & 1.000  & 0.574  & 0.380   \\
 8-16   &        & 1.000  & 0.389   \\
 $>$16  &        & 	  & 1.000   \\

\hline \hline

\end{tabular}

\label{tab:CorrMatTotal_unfolded_rho}
\end{table}

\clearpage

\begin{table}[hbt]
\centering

\caption{Statistical (fit) covariance matrix of the \Bpilnu\ $\Delta B/\Delta q^2$ measurement for the four-mode fit
         after unfolding of the $q^2$ spectrum in units of $10^{-13}$.}

\begin{tabular}{lrrrrrr} \hline \hline

$q^2$ range ($\gev^2$)  & 0-4 & 4-8 & 8-12 &12-16&16-20&$>$20 \\ \hline  

 0-4    &   4.039 &  0.436 & -0.134 & -0.020 & -0.015 &  0.116  \\
 4-8    &         &  1.861 &  0.135 & -0.027 &  0.104 &  0.018  \\
 8-12   &         &        &  1.462 &  0.404 &  0.110 & -0.018  \\
 12-16  &         &        &        &  1.720 &  0.534 & -0.157  \\
 16-20  &         &        &        &        &  1.995 &  0.014  \\
 $>$20  &         &        &        &        &        &  0.650  \\

\hline \hline

\end{tabular}

\label{tab:CovMatStat_Unfolded_pi}
\end{table}

\begin{table}[hbt]
\centering

\caption{Systematic covariance matrix of the \Bpilnu\ $\Delta B/\Delta q^2$ measurement for the four-mode fit
         after unfolding of the $q^2$ spectrum in units of $10^{-13}$.}

\begin{tabular}{lrrrrrr} \hline \hline

$q^2$ range ($\gev^2$)  & 0-4 & 4-8 & 8-12 &12-16&16-20&$>$20 \\ \hline  

 0-4    & 4.425 &  0.888 &  1.836 &  1.254 & -0.176 &  0.022 \\
 4-8    &       &  0.931 &  1.018 &  0.920 &  0.653 &  0.289 \\
 8-12   &       &        &  1.666 &  1.244 &  0.504 &  0.338 \\
 12-16  &       &        &        &  2.123 &  1.091 &  0.245 \\
 16-20  &       &        &        &        &  1.228 &  0.359 \\
 $>$20  &       &        &        &        &        &  0.488 \\

\hline \hline

\end{tabular}

\label{tab:CovMatSyst_Unfolded_pi}
\end{table}

\begin{table}[hbt]
\centering

\caption{Total covariance matrix of the \Bpilnu\ $\Delta B/\Delta q^2$ measurement for the four-mode fit
         after unfolding of the $q^2$ spectrum in units of $10^{-13}$.}

\begin{tabular}{lrrrrrr} \hline \hline

$q^2$ range ($\gev^2$)  & 0-4 & 4-8 & 8-12 &12-16&16-20&$>$20 \\ \hline  

 0-4    &  8.463 &  1.324 &  1.702 &  1.234 & -0.191 &  0.139 \\
 4-8    &        &  2.792 &  1.152 &  0.894 &  0.757 &  0.307 \\
 8-12   &        &        &  3.129 &  1.648 &  0.615 &  0.320 \\
 12-16  &        &        &        &  3.843 &  1.625 &  0.089 \\
 16-20  &        &        &        &        &  3.223 &  0.373 \\
 $>$20  &        &        &        &        &        &  1.138 \\

\hline \hline

\end{tabular}

\label{tab:CovMatTotal_Unfolded_pi}
\end{table}

\newpage
\begin{table}[hbt]
\centering

\caption{Statistical (fit) covariance matrix of the \Brholnu\ $\Delta B/\Delta q^2$ measurement for the four-mode fit 
         after unfolding of the $q^2$ spectrum in units of $10^{-12}$.}

\begin{tabular}{lrrr} \hline \hline

$q^2$ range ($\gev^2$)  & 0-8 & 8-16 & $>$16 \\ \hline  

 0-8    &   1.798 &  0.536 & -0.093  \\
 8-16   &         &  0.543 &  0.066  \\
 $>$16  &         &        &  0.269  \\

\hline \hline

\end{tabular}

\label{tab:CovMatStat_unfolded_rho}
\end{table}

\begin{table}[hbt]
\centering

\caption{Systematic covariance matrix of the \Brholnu\ $\Delta B/\Delta q^2$ measurement for the four-mode fit 
         after unfolding of the $q^2$ spectrum in units of $10^{-12}$.}

\begin{tabular}{lrrr} \hline \hline

$q^2$ range ($\gev^2$)  & 0-8 & 8-16 & $>$16 \\ \hline  

 0-8    & 2.500 &  1.653  &  1.230   \\
 8-16   &       &  2.837  &  0.968   \\
 $>$16  &       &         &  1.816   \\

\hline \hline

\end{tabular}

\label{tab:CovMatSyst_unfolded_rho}
\end{table}

\begin{table}[hbt]
\centering

\caption{Total covariance matrix of the \Brholnu\ $\Delta B/\Delta q^2$ measurement for the four-mode fit 
         after unfolding of the $q^2$ spectrum in units of $10^{-12}$.}

\begin{tabular}{lrrr} \hline \hline

$q^2$ range ($\gev^2$)  & 0-8 & 8-16 & $>$16 \\ \hline  

 0-8    &   4.298 &  2.188 &  1.137    \\
 8-16   &         &  3.381 &  1.034    \\
 $>$16  &         &        &  2.086    \\

\hline \hline

\end{tabular}

\label{tab:CovMatTotal_unfolded_rho}
\end{table}

\newpage


\begin{thebibliography}{10}
\bibitem{charge_conjugate}  
Charge conjugate decay modes are implicitly included.
\bibitem{q0sq}
For the BGL parameterization, we choose the value 
$q_0^2 =-0.65 m_-^2$ 
as proposed in Ref.~\cite{fnal09}.
For the BCL parameterization, we choose 
$q_0^2 = (M_B + m_{\pi})(\sqrt{M_B}-\sqrt{m_{\pi}})^2$,
as proposed in Ref.~\cite{BCL}.
For the $B$-meson and pion masses, 
we use $M_B = 5.279 \gev$ and $m_{\pi} = 0.1396 \gev$.
\bibitem{phi}
The function $\phi(q^2, q_0^2)$ in the BGL parameterization 
is given by the expression
\begin{eqnarray*}
\phi(q^2, q_0^2) &=& \sqrt{\frac{1}{32\pi\chi_J^{(0)}}} 
                    (\sqrt{m_+^2 - q^2} + \sqrt{m_+^2 - q_0^2}) \\
          &\times&  (\sqrt{m_+^2 - q^2} + \sqrt{m_+^2 - m_-^2})^{3/2} \\
          &\times&  (\sqrt{m_+^2 - q^2} + \sqrt{m_+^2})^{-5} \\
          &\times& \frac{(m_+^2 - q^2)}{(m_+^2 - q_0^2)^{1/4}}
\end{eqnarray*}
with numerical factor $\chi_J^{(0)}=6.889 \times 10^{-4}$.
\end{thebibliography}
\end{document}